\documentclass[12pt]{cernart}

\usepackage{graphicx}
\usepackage{amsmath}
\usepackage{amssymb}
\usepackage{times}
\usepackage{cite}
\usepackage{multirow}

\begin{document}
\setcounter{topnumber}{3}
\renewcommand{\topfraction}{0.999}
\renewcommand{\bottomfraction}{0.99}
\renewcommand{\textfraction}{0.0}
\setcounter{totalnumber}{6}

\begin{titlepage}
\docnum{CERN-PH-EP/2009-029}
\date{November 22, 2009}
%
\title{\large{Inclusive production of charged kaons in p+p collisions \\
              at 158 GeV/c beam momentum and a new evaluation \\
              of the energy dependence of kaon production \\
              up to collider energies}}
\begin{Authlist}
\vspace{2mm}
\noindent
T.~Anticic$^{12}$,
B.~Baatar$^{5}$, J.~Bartke$^{4}$, L.~Betev$^{6}$, H.~Bia{\l}kowska$^{11}$, B.~Boimska$^{11}$, J.~Bracinik$^{1,a}$,
V.~Cerny$^{1}$,  O.~Chvala$^{8,b}$,
J.~Dolejsi$^{8}$,
V.~Eckardt$^{7}$,
H.G.~Fischer$^{6}$, Z.~Fodor$^{3}$,
E.~G{\l}adysz$^{4}$,
K.~Kadija$^{12}$, A.~Karev$^{6}$, V.~Kolesnikov$^{5}$, M.~Kowalski$^{4}$, M.~Kreps$^{1,c}$,
M.~Makariev$^{10}$, A.~Malakhov$^{5}$, M.~Mateev$^{9}$, G.~Melkumov$^{5}$,
A.~Rybicki$^{4}$,
N.~Schmitz$^{7}$, P.~Seyboth$^{7}$, T.~Susa$^{12}$, P.~Szymanski$^{11}$,
V.~Trubnikov$^{11}$,
D.~Varga$^{2}$, G.~Vesztergombi$^{3}$,
S.~Wenig$^{6,}$\footnote{Corresponding author: Siegfried.Wenig@cern.ch}
\vspace*{2mm} 

\noindent
{\it (The NA49 Collaboration)}  \\
\vspace*{2mm}

\noindent
$^{1}$Comenius University, Bratislava, Slovakia\\
$^{2}$ E\"otv\"os Lor\'and University, Budapest, Hungary \\
$^{3}$KFKI Research Institute for Particle and Nuclear Physics, Budapest, Hungary\\
$^{4}$H. Niewodnicza\'nski Institute of Nuclear Physics,
      Polish Academy of Sciences, Cracow, Poland \\
$^{5}$Joint Institute for Nuclear Research, Dubna, Russia\\
$^{6}$CERN, Geneva, Switzerland\\
$^{7}$Max-Planck-Institut f\"{u}r Physik, Munich, Germany\\
$^{8}$Charles University, Faculty of Mathematics and Physics, Institute of
      Particle and Nuclear Physics, Prague, Czech Republic \\
$^{9}$Atomic Physics Department, Sofia University St. Kliment Ohridski, Sofia, Bulgaria\\
$^{10}$Institute for Nuclear Research and Nuclear Energy, BAS, Sofia, Bulgaria\\
$^{11}$Institute for Nuclear Studies, Warsaw, Poland\\
$^{12}$Rudjer Boskovic Institute, Zagreb, Croatia\\ 
$^{a}$now at School of Physics and Astronomy, University of Birmingham, Birmingham, UK \\
$^{b}$now at UC Riverside, Riverside, CA, USA\\
$^{c}$now at Institut fur Experimentelle Kernphysik, Karlsruhe, DE
\end{Authlist}

\vspace{10mm}
\begin{center}
{\small{\it to be published in EPJC }}
\end{center}
\vspace*{2mm} 
\clearpage

\normalsize
\begin{abstract}
\vspace{-3mm}
New data on the production of charged kaons in p+p interactions
are presented. The data come from a sample of 4.8 million
inelastic events obtained with the NA49 detector at the CERN SPS
at 158~GeV/c beam momentum. The kaons are identified by energy 
loss in a large TPC tracking system. Inclusive invariant cross
sections are obtained in intervals from 0 to 1.7~GeV/c in 
transverse momentum and from 0 to 0.5 in Feynman x. Using these
data as a reference, a new evaluation of the energy dependence 
of kaon production, including neutral kaons, is conducted over 
a range from 3~GeV to p+$\overline{\textrm{p}}$ collider energies.
\end{abstract}
 
\end{titlepage}

%
%

\section{Introduction} 
\vspace{3mm}
\label{sec:intro}

Following the detailed investigation of inclusive pion \cite{pp_pion} and 
baryon \cite{pp_proton} production in p+p interactions, the present paper 
concentrates on the study of charged kaons. It thus completes
a series of publications aimed at the exploration of final
state hadrons in p+p collisions by using a new set of high
precision data from the NA49 detector at the CERN SPS \cite{nim}.
The data have been obtained at a beam momentum of 158~GeV/c
corresponding to a center-of-mass system (cms) energy of 17.2~GeV. This matches
the highest momentum per nucleon obtainable with lead
beams at the SPS, permitting the direct comparison of
elementary and nuclear reactions. In addition, the chosen
cms energy marks, concerning kaon production, the transition   
from threshold-dominated effects with strong $s$-dependences
to the more gentle approach to higher energies where scaling
concepts become worth investigating. On the other hand the
characteristic differences between K$^+$ and K$^-$ production which
are directly related to the underlying production mechanisms, 
as for instance associate kaon+hyperon versus K+$\overline{\textrm{K}}$ pair production, are still
well developed at SPS energy. They are manifest in the strong
evolution of the K$^+$/K$^-$ ratio as a function of the kinematic
variables. One of the aims of this paper is in addition the
attempt to put the available results from other experiments
into perspective with the present data in order to come to
a quantitative evaluation of the experimental situation.  

A critical assessment of the complete $s$-dependence of kaon
production seems the more indicated as its evolution in
heavy ion interactions, especially in relation to pions,
is promulgated since about two decades
as a signature of "new" physics by the creation
of a deconfined state of matter in these interactions. As
all claims of this nature have to rely completely on a 
comparison with elementary collisions, the detailed study 
of the behaviour of kaon production in p+p reactions from
threshold up to RHIC and collider energies should be regarded
as a necessity in particular as the last global evaluation
of this type dates back by more than 30 years \cite{rossi}. A complete
coverage of phase space, as far as a comparison of different
experiments is concerned, is made possible in this paper,
as compared to pions \cite{pp_pion} and baryons \cite{pp_proton}, 
by the fact that there is no concern about feed-down corrections 
from weak hyperon decays, with the exception of $\Omega$ decay which 
is negligible for all practical purposes. 

This paper is arranged in the same fashion as the preceding
publications \cite{pp_pion,pp_proton}. A summary of the phase space coverage of the
available data from other experiments in Sect.~\ref{sec:exp_sit} is followed 
by a short presentation of the NA49 experiment, its acceptance
coverage and the corresponding binning scheme in Sect.~\ref{sec:na49}.
Section~\ref{sec:pid} gives details on the particle identification via energy loss
measurement as they are specific to the problem of kaon yield extraction.
The evaluation of the inclusive cross sections and of the
necessary corrections is described in Sect.~\ref{sec:corr}, followed by
the data presentation including a detailed data interpolation scheme in 
Sect.~\ref{sec:data}. K$^+$/K$^-$, K/$\pi$ and K/baryon ratios
are presented in Sect.~\ref{sec:ratios}. A first step of data comparison 
with data in the SPS/Fermilab energy range is taken in Sect.~\ref{sec:comp}.
Section~\ref{sec:ptint} deals with the data integrated over transverse momentum
and the total measured kaon yields. The data comparison is extended,
in a second step, over the range from $\sqrt{s}\sim$~3 to ISR, RHIC and
p+$\overline{\textrm{p}}$ collider energies in Sect.~\ref{sec:sdep}. 
Section~\ref{sec:kzero} concentrates on an evaluation of K$^0_S$ yields
in relation to charged kaons and on a discussion of total kaon multiplicities
as a function of $\sqrt{s}$.
A comment on the influence of resonance decay on the observed patterns
of $p_T$ and $s$ dependence is given in Sect.~\ref{sec:reson}.
In Sect.~\ref{sec:over} a global overview of charged and neutral kaon
yields as they result from the study of $s$-dependence in this 
paper is presented, both for the $p_T$ integrated invariant
yields at $x_F$~=~0 and for the total kaon multiplicities.
A summary of results and conclusion is given in Sect.~\ref{sec:concl}. 
 
%
%
\section{The experimental situation}
\vspace{3mm}
\label{sec:exp_sit}

This paper considers the double differential inclusive cross
sections of identified charged kaons,

\begin{equation}
  \frac{d^2\sigma}{dx_Fdp_T^2}   ,
\end{equation}
as a function of the phase space variables defined as transverse
momentum $p_T$ and reduced longitudinal momentum 

\begin{equation}
\label{eq:xf}
x_F = \frac{p_L}{\sqrt{s}/2}
\end{equation}
where $p_L$ denotes the longitudinal momentum component in the cms.

\begin{figure}[h]
  \begin{center}
  	\includegraphics[width=15.cm]{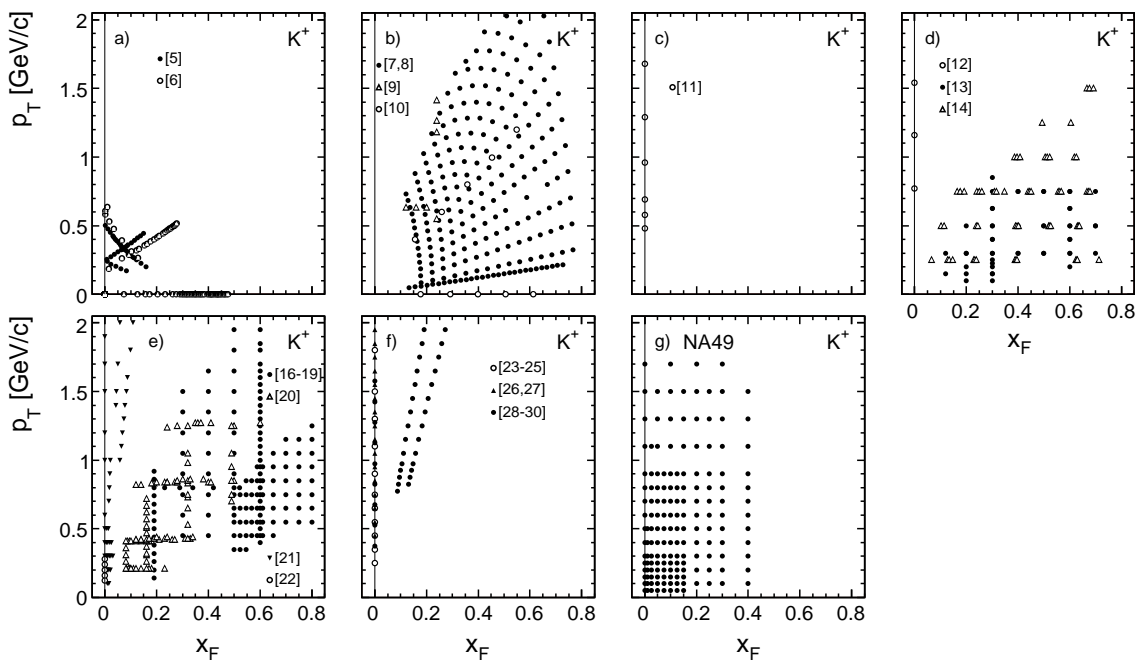}
  	\caption{Phase space coverage of the existing K$^+$ data: a) Cosmotron/PPA \cite{hogan,reed},
  	         b) PS/AGS \cite{allaby1,allaby2,akerlof,dekkers}, c) Serpukhov \cite{abramov},
  	         d) SPS/Fermilab \cite{cronin,brenner,johnson}, 
  	         e) ISR \cite{albrow2,albrow3,albrow4,albrow5,capi,alper,guettler},
  	         f) RHIC \cite{star1,star2,star3,phenix1,phenix2,brahms1,brahms2,brahms3},
  	         g) NA49}
  	\label{fig:phs_kap}
  \end{center}
\end{figure}

\begin{figure}[h]
  \begin{center}
  	\includegraphics[width=15.cm]{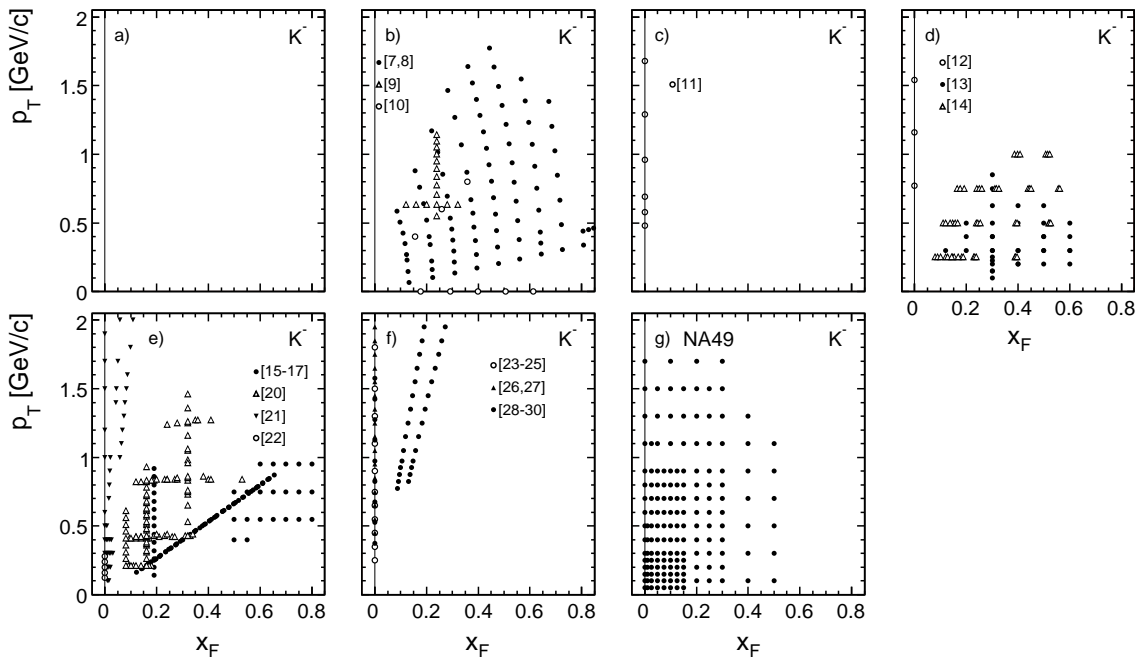}
  	\caption{Phase space coverage of the existing K$^-$ data: a) Cosmotron/PPA,
  	         b) PS/AGS \cite{allaby1,allaby2,akerlof,dekkers}, c) Serpukhov \cite{abramov},
  	         d) SPS/Fermilab \cite{cronin,brenner,johnson}, 
  	         e) ISR \cite{albrow1,albrow2,albrow3,capi,alper,guettler},
  	         f) RHIC \cite{star1,star2,star3,phenix1,phenix2,brahms1,brahms2,brahms3},
  	         g) NA49}
  	\label{fig:phs_kam}
  \end{center}
\end{figure}

If the phase space coverage of the existing data has been
shown to be incomplete and partially incompatible for pion and
baryon production in the preceding publications \cite{pp_pion,pp_proton}, 
the situation is even more unsatisfactory for charged kaons. A wide
range of data covering essentially the complete energy range 
from kaon threshold via the PS and AGS up to the ISR and RHIC energy has been considered
here. One advantage concerning the data comparison for kaons is
the absence of feed-down from weak decays with the exception
of $\Omega^-$ decay which can be safely neglected at least up to
ISR energies. An overview of the available data sets is given in
Fig.~\ref{fig:phs_kap} for K$^+$ and Fig.~\ref{fig:phs_kam} for K$^-$ in 
the $x_F$/$p_T$ plane.

The sub-panels a) through g) show successively the energy ranges 
of the Cosmotron/PPA \cite{hogan,reed}, PS/AGS \cite{allaby1,allaby2,akerlof,dekkers}, 
Serpukhov \cite{abramov}, SPS/Fermilab \cite{cronin,brenner,johnson}, 
ISR \cite{albrow1,albrow2,albrow3,albrow4,albrow5,capi,alper,guettler} and 
RHIC \cite{star1,star2,star3,phenix1,phenix2,brahms1,brahms2,brahms3} accelerators 
in comparison to the new data from NA49. The scarcity of data
in the important intermediate energy range around $\sqrt{s} \sim$~10~GeV
and the general lack of coverage in the low-$p_T$ and low-$x_F$ regions
are clearly visible. The coverage of the NA49 data, Figs.~\ref{fig:phs_kap}g 
and \ref{fig:phs_kam}g, is essentially only limited by counting statistics 
towards high $p_T$ and by limitations concerning particle identification towards 
high $x_F$, in particular for K$^+$, see Sect.~\ref{sec:pid} below.
 
The task of establishing data consistency over the wide range of 
energies considered here is a particularly ardent one for kaons,
as will be shown in the data comparison, see Sects.~\ref{sec:comp} and \ref{sec:sdep} below.
This concerns especially any attempt at establishing total
integrated yields where the existing efforts evidently suffer
from a gross under-estimation of systematic errors. Their relation
to the total yields of K$^0_S$ which are established with considerably
higher reliability up to SPS/Fermilab energies as well as their
eventual comparison with strangeness production in nuclear collisions 
should therefore be critically reconsidered.      

%
%
\section{The NA49 experiment, acceptance coverage and binning}
\vspace{3mm}
\label{sec:na49}

\begin{figure}[h]
  \begin{center}
  	\includegraphics[width=12.2cm]{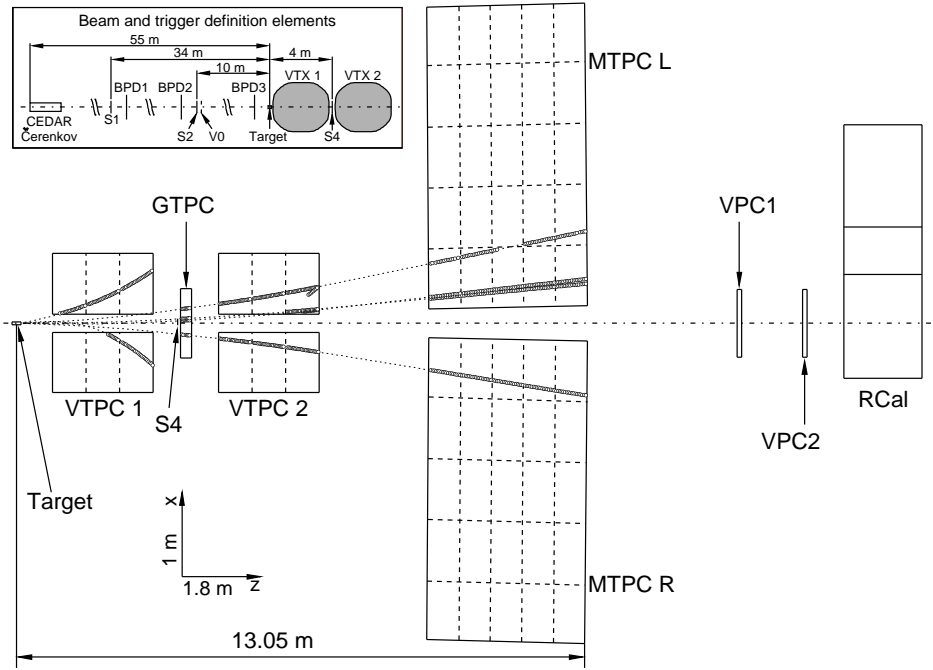}
  	\caption{NA49 detector layout and real tracks of a typical mean multiplicity
             p+p event. The open circles are the points registered in the TPC's, the dotted 
             lines are the interpolated trajectories between the track segments and the 
             extrapolations to the event vertex in the LH$_2$ target. The beam and trigger
             definition counters are presented in the inset}
  	\label{fig:exp}
  \end{center}
\end{figure}

The basic features of the NA49 detectors have been described in
detail in \cite{pp_pion,pp_proton, nim}. The top view shown in 
Fig.~\ref{fig:exp} recalls the main components.

The beam is a secondary hadron beam produced by 450~GeV/c primary
protons impinging on a 10~cm long Be target. It is defined by
a CEDAR Cerenkov counter, several scintillation counters
(S1, S2, V0) and a set of high precision proportional chambers
(BPD1-3). The hydrogen target is placed in front of two
superconducting Magnets (VTX1 and VTX2). Four large volume
Time Projection Chambers (VTPC1 and VTPC2 inside the magnetic
fields, MTPCL and MTPCR downstream of the magnets) provide
for charged particle tracking and identification. A smaller
Time Projection Chamber (GTPC) placed between the two magnets
together with two Multiwire Proportional Chambers (VPC1 and VPC2) 
in forward direction allows tracking in the high momentum region 
through the gaps between the principal track detectors. A Ring 
Calorimeter (RCal) closes the detector setup 18~m downstream 
of the target.

The phase space region accessible to kaon detection is essentially
only limited by the available number of 4.6~M inelastic events.
It spans a range of transverse momenta between 0.05 and 1.7~GeV/c
for K$^+$ and K$^-$ and Feynman $x_F$ between 0 and 0.5 for K$^-$. For K$^+$
a limitation to $x_F \leq$~0.4 is imposed by the constraints on
particle identification discussed in Sect.~\ref{sec:pid} below. 

\begin{figure}[h]
  \begin{center}
  	\includegraphics[width=15.cm]{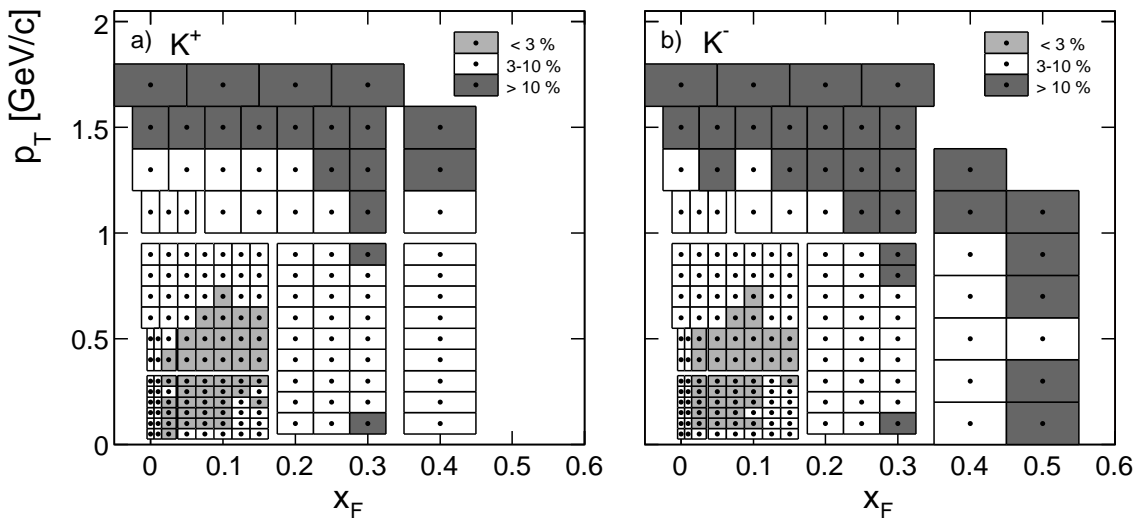}
 		\caption{Binning schemes in $x_F$ and $p_T$ for a) K$^{+}$ and b) K$^{-}$  
             together with information on the statistical errors}
  	\label{fig:accept}
  \end{center}
\end{figure}

These kinematical regions are subdivided into bins in the $x_F$/$p_T$
plane which vary according to the measured particle yields, effects
of finite bin widths being corrected for in the evaluation of the
inclusive cross sections (Sect.~\ref{sec:corr}). The resulting binning
schemes are shown in Fig.~\ref{fig:accept} also indicating different ranges of
the corresponding statistical errors.
 
%
%
\section{Particle identification}
\vspace{3mm}
\label{sec:pid}

The identification of kaons by their ionization energy loss
in the gas of the TPC detector system meets with specific problems
if compared to pion \cite{pp_pion} and baryon \cite{pp_proton} selection. This 
specificity has several reasons:

\begin{itemize}
\item Corresponding to the momentum range of the NA49 data the
      ionization energy loss has to be determined in the region
      of the relativistic rise of the energy deposit, with the kaon
      energy loss positioned in between the one for baryons and for pions.
\item The relative distance in $dE/dx$ between the different particle
      species is small and varies from only 4.5 to 7\% for kaons with 
      respect to protons and from 6.5 to 14\% with respect to pions,
      over the $x_F$ range of the present data, with an rms width of
      the energy loss distributions of typically 3\%. This creates
      an appreciable overlap problem over most of the phase space
      investigated. 
\item High precision in the determination of the absolute position
      of the mean truncated energy loss per particle species and
      of the corresponding widths is therefore mandatory.
\item The relative production yield of kaons is generally small as 
      compared to pions, with K/$\pi$ ratios on the level of  5--30\% for K$^+$ 
      and 5--20\% for K$^-$. In addition, for K$^+$ the 
      fast decrease of the K$^+$/p ratio from typically 1 at $x_F$~=~0 to 
      less than 5\% at $x_F$~=~0.4 finally imposes a limit on the 
      applicability of $dE/dx$ identification towards high $x_F$ values.
\end{itemize}

\begin{figure}[h]
  \begin{center}
  	\includegraphics[width=12cm]{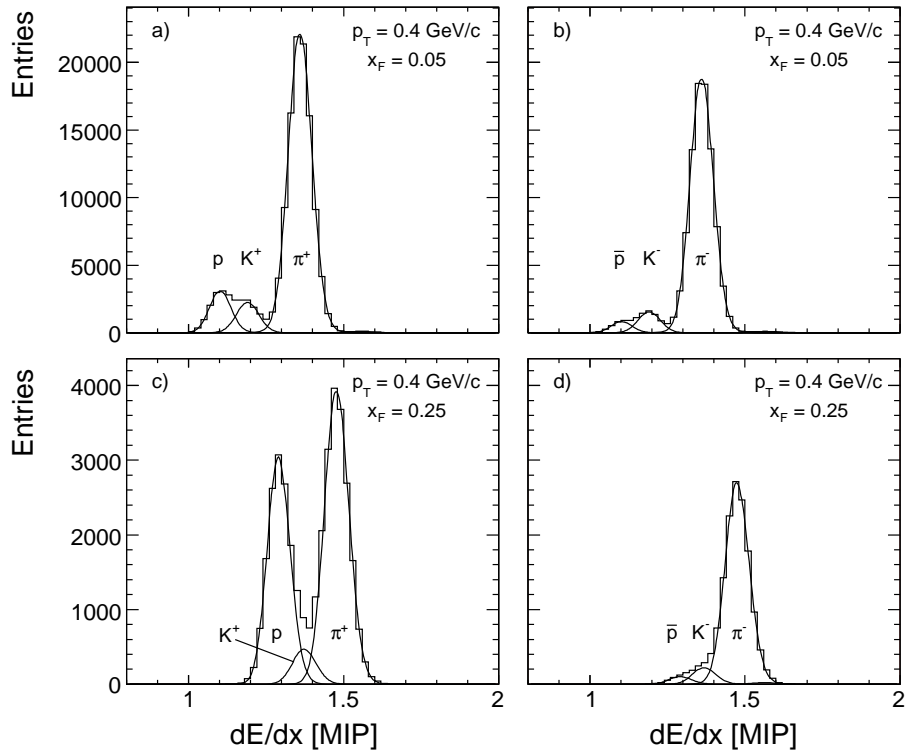}
 		\caption{$dE/dx$ distributions for K$^+$  and K$^-$ bins at $x_F$~=~0.05, $p_T$~=~0.4~GeV/c 
             and $x_F$~=~0.25, $p_T$~=~0.4~GeV/c superimposed with
             results of the fitted distributions}
  	\label{fig:dedxdist}
  \end{center}
\end{figure}

This general situation may be visualized by looking at a couple of 
typical $dE/dx$ distributions for different $x_F$ regions as 
shown in Fig.~\ref{fig:dedxdist}.

As already described in \cite{pp_proton} a considerable effort has been 
invested into the improved control of the analog response of the 
detector. Several aspects and results of this work, in particular 
as far as kaon identification is concerned, will be discussed in 
the following sub-sections.

%
%
\subsection{Non-Gaussian shape of the $\boldsymbol{dE/dx}$ distributions}
\vspace{3mm}
\label{sec:dedx_shape}  

Due to the small K/$\pi$ and K/p ratios mentioned above, the precise
description of the tails of the energy loss distributions of the 
dominant particle species becomes important. The extraction of kaon yields
becomes indeed sensitive to small deviations in the upper tail
of the proton and in the lower tail of the pion distributions
for the extreme yield ratios mentioned above, as is also apparent
from the examples shown in Fig.~\ref{fig:dedxdist}. Eventual asymmetries with respect
to the generally assumed Gaussian shape of the energy loss 
distributions have therefore to be carefully investigated as they
will influence both the fitted central position and the extracted
yields of the kaons. A detailed study of the shape of the $dE/dx$
distributions has therefore been performed both experimentally and 
by analytical calculation.

\begin{figure}[h]
  \begin{center}
  	\includegraphics[width=14cm]{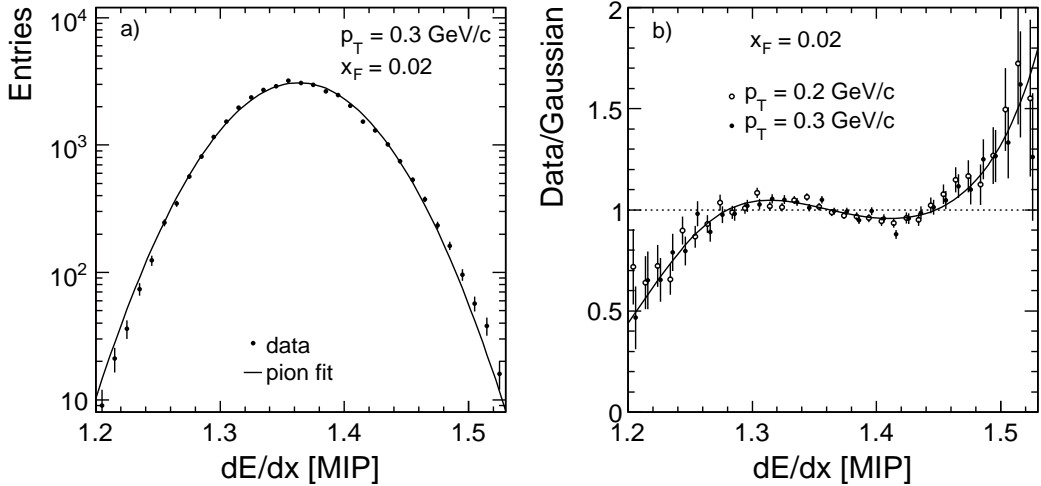}
 		\caption{a) Conventional Gaussian fit of the MTPC $dE/dx$
             distribution, for tracks with pion selection using the VTPC $dE/dx$;
             b) Ratio of data and fit function}
  	\label{fig:dedxshape}
  \end{center}
\end{figure}

By selecting long tracks in the NA49 TPC system which pass both 
through the VTPC and the MTPC detectors one may use the energy deposit
in one of the TPC's to sharply select a specific particle type of
high yield, for instance pions or protons. The $dE/dx$ deposit in the 
other TPC will then allow a precise shape determination. An example
is shown in Fig.~\ref{fig:dedxshape} for the selection of pions at $x_F$~=~0.02 and 
$p_T$~=~0.3~GeV/c in the VTPC. The corresponding distribution of the
truncated mean for 90 samples in the MTPC is presented in Fig.~\ref{fig:dedxshape}a 
together with a Gaussian fit.

The small but very evident skewness of the truncated energy loss 
distribution is expressed in Fig.~\ref{fig:dedxshape}b by the ratio of the experimental 
data to a Gaussian fit. This ratio may be described by a cubic polynomial 
form with one normalization parameter $Z$, shown as the full line in
Fig.~\ref{fig:dedxshape}b.

\begin{equation}
   (\textrm{Data})/(\textrm{Gaussian}) \approx 1+Z(g^3-3g)  ,
\end{equation}
where $g$ is the distance from the mean of the
$dE/dx$ distribution, normalized to the rms of the Gaussian fit,
 
\begin{equation}                 
   g = \frac{1}{\sigma} \left[ \left(\frac{dE}{dx}\right) - \left\langle \frac{dE}{dx} \right\rangle \right] .
\end{equation}

The parameter $Z$ is related to the number of measured
points, $N_p$, on each track, and the central $dE/dx$ value by the
relation

\begin{equation}                
   Z = z_0N_p^{-\beta}\left(\frac{dE}{dx}\right)^{\gamma} ,
\end{equation}
with $\beta$ and $\gamma$ experimentally determined to 0.5 \cite{pp_pion} and
0.4$\pm$0.2, respectively. Together with the relation:

\begin{equation}    
  \label{eq:resol}          
   \frac{\sigma}{(dE/dx)} = \sigma_0N_p^{-\beta}\left(\frac{dE}{dx}\right)^{\alpha},
\end{equation}
assuming $\alpha$~=~$\gamma$ which is a safe assumption regarding the sizeable
error in the determination of $\gamma$, $z_0$ is obtained as

\begin{equation}              
\begin{split}              
   z_0 &= 0.215 \pm 0.02 \qquad \textrm{for the VTPC}  \\
   z_0 &= 0.21 \,\,\,\pm 0.02  \qquad \textrm{for the MTPC.}
\end{split}              
\end{equation}

A Monte Carlo simulation based on the Photon Absorption Ionization
(PAI) model \cite{pai} confirmed these results, demonstrating that
the shape distortion is indeed a remnant of the basically asymmetric 
Landau distribution of ionization energy loss.

%
%
\subsection{Position and width of the energy loss distributions}
\vspace{3mm}
\label{sec:dedx_shift}

\begin{figure}[h]
  \begin{center}
  	\includegraphics[width=8cm]{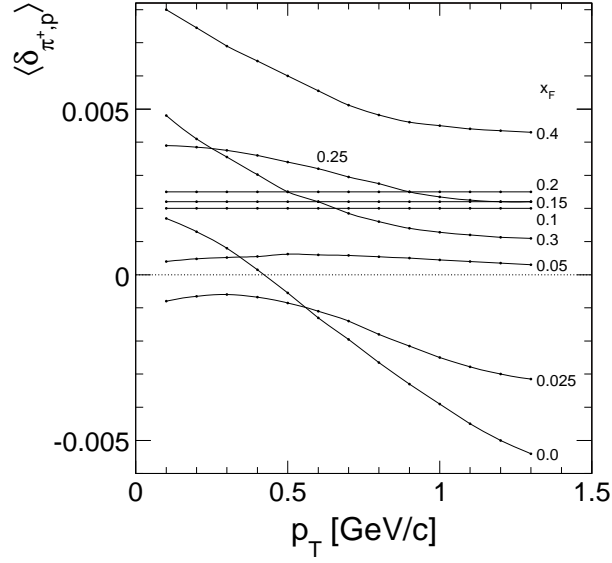}
 		\caption{Mean deviation $\langle \delta_{\pi^+,p}\rangle$, in units of minimum 
                   ionization, of $\pi^+$ and proton 
 		         $dE/dx$ with respect to the Bethe-Bloch parametrization as a function 
 		         of $p_T$ for different values of $x_F$}
  	\label{fig:dedxshift}
  \end{center}
\end{figure}

Particle identification proceeds, in each defined bin of phase
space, via a $\chi^2$ optimization procedure between the measured
energy loss distributions and four single particle $dE/dx$ 
distributions of known shape but a priori unknown positions and
widths for electrons, pions, kaons and protons, respectively. Due
to the generally small fraction of electrons and their position
in the density plateau of the energy loss function, and due to the
known dependence of the $dE/dx$ resolution on the $dE/dx$ value for
each particle species \cite{pp_pion}, (Eq.~\ref{eq:resol}), the problem reduces 
in practice to the determination of eight quantities: three absolute positions of
the energy loss of $\pi$, K, p, one width parameter and four yield values 
which correspond to the particle
cross sections to be determined. If the fit of the pre-dominant
particle species like pions and protons in general presents no
problems, the situation is more critical for the kaons. Here it
is in principle the central kaon position and the overall rms width 
of the $dE/dx$ distributions which are liable to create systematic
yield variations. In the ideal case, the detector response should
reproduce exact scaling in the $p$/$m$ variable as implied by the 
Bethe-Bloch function of ionization energy loss (BB), with $p$ the lab
momentum and $m$ the particle mass. As shown in \cite{pp_pion,pp_proton, nim} this 
scaling is fulfilled for pions and protons in the NA49 detector 
on the sub-percent level. The precision of the $dE/dx$ fitting procedure
allows for a quantification of the remnant deviations $\delta$ with respect
to the Bethe-Bloch parametrization as a function of $x_F$ and $p_T$

\begin{equation}
\delta(x_F,p_T) = \frac{dE}{dx}(x_F,p_T) - BB
\end{equation}
in units of minimum ionization (MIP), where $dE/dx$ is the mean truncated
energy loss \cite{pp_pion}.
This is presented in Fig.~\ref{fig:dedxshift} for the mean deviation 
of $\pi^+$ and protons.

\begin{figure}[h]
  \begin{center}
  	\includegraphics[width=13cm]{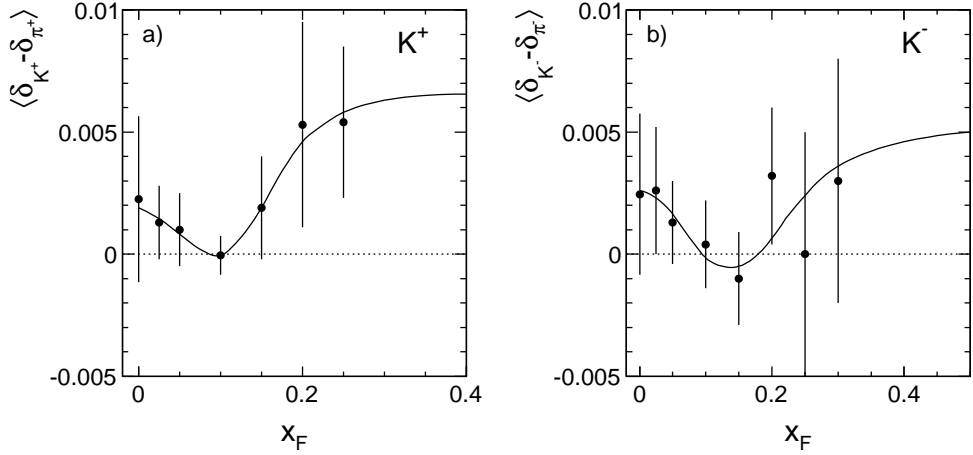}
 		\caption{Mean deviations in units of minimum ionization of a) K$^+$  and b) K$^-$  with respect to the pion
                 position $\langle \delta_{\textrm{K}^{\pm}}-\delta_{\pi^{\pm}}\rangle$ 
                 as a function of $x_F$, averaged over $p_T$}
  	\label{fig:dedxkashift}
  \end{center}
\end{figure}

The observed deviations are due to residual errors in the 
calibration of the detector response and in the transformation
between the Bethe-Bloch parametrizations of the different gases used
in the VTPC and MTPC detectors \cite{nim}. They stay in general below the 
level of $\pm$0.005. The fitted shifts of the kaon position, as 
characterized by their difference to the pion position $\delta_{\textrm{K}}-\delta_{\pi}$, 
are shown in Fig.~\ref{fig:dedxkashift} as a function of $x_F$ and averaged over $p_T$, 
the error bars representing the rms deviation of the averages. 

Evidently the measured positions fall well within the margin of
$\pm$0.005  in units of minimum ionization as obtained for pions and protons. The similarity, within
errors, between the results for K$^+$ and K$^-$ indicates systematic 
detector response effects as the principle source of the measured deviations.

\begin{figure}[h]
  \begin{center}
  	\includegraphics[width=6.cm]{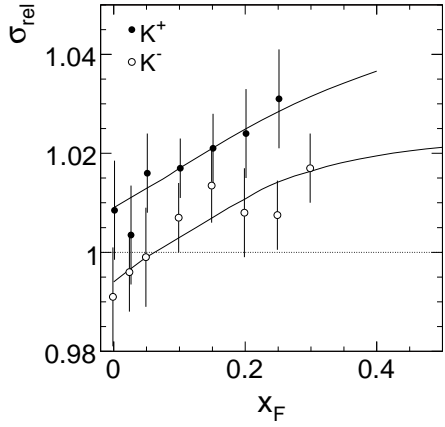}
 		\caption{Relative rms width $\sigma_{\textrm{rel}}$ as a function of $x_F$ for K$^+$ 
             and K$^-$, averaged over $p_T$}
  	\label{fig:relsig}
  \end{center}
\end{figure}

The fitted rms widths of the $dE/dx$ distributions, characterized by their 
relative deviation from the calculated expectation value (Eq.~\ref{eq:resol} above), 
are shown in Fig.~\ref{fig:relsig} as a function of $x_F$, after averaging over $p_T$.

The results show that the predicted widths are reproduced with an
accuracy within a few percent of the expected values, with a slight
systematic upwards trend as a function of $x_F$ closely similar for K$^+$ and K$^-$.
 
%
%
\subsection{Estimation of systematic errors}
\vspace{3mm}
\label{sec:dedx_sys}

\begin{figure}[h]
  \begin{center}
  	\includegraphics[width=6.cm]{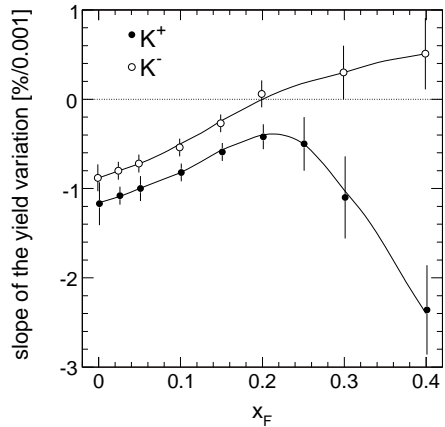}
 	\caption{Slope of the yield variation given in \% per assumed kaon shift of 0.001 for 
             K$^+$ and K$^-$ as a function of $x_F$, averaged over $p_T$}
  	\label{fig:shift_slope}
  \end{center}
\end{figure}

\begin{figure}[h]
  \begin{center}
  	\includegraphics[width=6.cm]{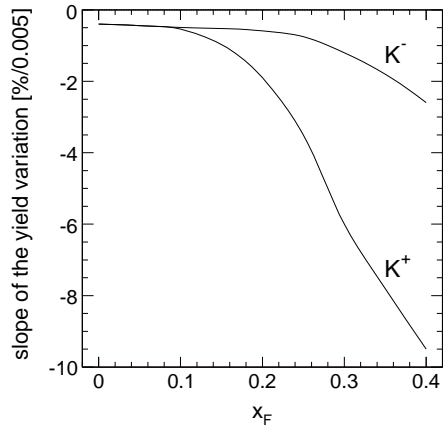}
    \caption{Slope of the yield variation given in \% per assumed change of $\sigma_{\textrm{rel}}$ 
 		     of 0.005 for K$^+$ and K$^-$ as a function of $x_F$, averaged over $p_T$}
  	\label{fig:relsig_slope}
  \end{center}
\end{figure}

The dependence of the fitted kaon yields on the four parameters
mentioned above, namely the positions of pions, kaons, protons,
and the relative rms width of the fits, has been studied in detail.
It appears that only two of these parameters are liable to produce 
noticeable systematic effects. These are the kaon position and the
rms width. By enforcing a range of fixed values of these parameters, 
their influence on the extraction of kaon yields may be obtained. This is
demonstrated in Fig.~\ref{fig:shift_slope} for the dependence on kaon position and in 
Fig.~\ref{fig:relsig_slope} for the dependence on the relative rms width, the error bars in
each plot indicating the rms size of the $p_T$ dependence.

Several aspects of this study are noteworthy:

\begin{itemize}
\item As far as the influence of the kaon position uncertainty is concerned, and
      taking into account the size of the measured deviations from
      pions and protons and their rms fluctuation (see Fig.~\ref{fig:dedxkashift}) the
      related errors stay on the level of less than 1\%
      up to $x_F$~=~0.2. Above this value the K$^+$ yield reacts very
      critically on the fitted position. This is related to the
      proton yield which becomes rapidly overwhelming towards high
      $x_F$.
\item Concerning the rms width the situation is somewhat more critical
      especially for K$^+$. Here, allowing for a systematic error of 
      about 0.5\% in the fitted relative rms, Fig.~\ref{fig:relsig}, the corresponding 
      yield error reaches values of about 2\% at $x_F$~=~0.2 and about 10\% 
      at $x_F$~=~0.4. This is again measuring the influence of the large
      proton fraction. For K$^-$ on the other hand, the systematic error
      stays below the 2\% level for the whole $x_F$ region investigated.
\end{itemize}

The systematic errors estimated here have been included in the
error estimation in Table~\ref{tab:sys}.  
 
%
%
\subsection{Fit stability and $\boldsymbol{x_F}$ limit for kaon yield extraction}
\vspace{3mm}
\label{sec:dedx_fit}

The fitting procedure described above results in stable values
for all eight parameters involved for $x_F$ values below about 
0.25 for K$^+$ and below 0.3 for K$^-$. This is to be understood in the
sense that the $\chi^2$ optimization procedure converges to a 
well-defined minimum in all variables with reasonable values for the 
ratio of $\chi^2$ over the degrees of freedom. For higher $x_F$ values 
the fits tend to become unstable in the sense that certain variables
tend to "run away" into unphysical configurations. In the present
case of extraction of kaon yields this concerns basically only the kaon
position in the $dE/dx$ variable and the rms width parameter of
the energy loss distributions, as the pion and proton positions
are always well constrained even in the critical regions of
phase space. The problem is of course connected to the high
sensitivity of the extracted kaon yield on these two parameters
in relation to the small K/$\pi$ and K/p ratios as discussed in the 
preceding section.

As the evolution of both the kaon position and the rms width with
the phase space variables $x_F$ and $p_T$ shows no indication of any
rapid variation up to the limits of fitting stability, and as
indeed the geometrical configuration of the tracks in the TPC
detectors shows a smooth and slow dependence on the track 
momenta in the regions concerned, it has been decided to extend
the $x_F$ range up to 0.4 for K$^+$ and to 0.5 for K$^-$ by imposing constraints
on the two critical parameters. This is realized by constraining
the kaon position to fixed values with respect to the pions,
as indicated by the extrapolated lines in Fig.~\ref{fig:dedxkashift}, and by also
fixing the rms widths to the values following from the smooth
extrapolation indicated in Fig.~\ref{fig:relsig}. The expected statistical error margins,
allowing for reasonable values for the uncertainties in the
quantities concerned, see Figs.~\ref{fig:shift_slope} and \ref{fig:relsig_slope}, 
have been added in quadrature to the statistical errors.  

%
%
\subsection{Estimation of statistical errors}
\vspace{3mm}
\label{sec:dedx_stat}

\begin{figure}[h]
  \begin{center}
  	\includegraphics[width=12cm]{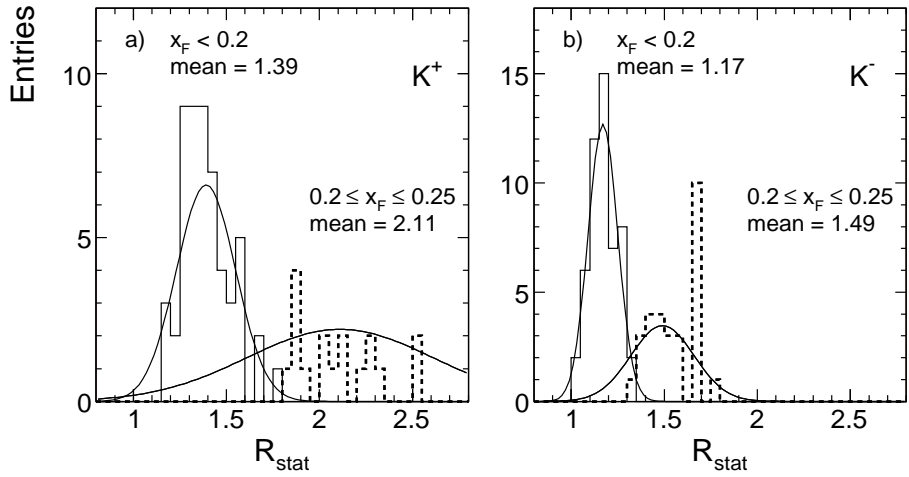}
 		\caption{$R_{\textrm{stat}}= \sigma_{\textrm{stat}}/(1/\sqrt{N})$ for the bins 
             $x_F <$~0.2 (solid line) and 0.2~$\leq x_F \geq$~0.25 (dashed line);
             a) K$^+$ and b) K$^-$ }
  	\label{fig:statfactor}
  \end{center}
\end{figure}

It has been shown in \cite{pp_proton} that the estimation of the statistical 
error of the extracted particle yields has to take into account the 
dependence of the fit result on all parameters fitted via the
covariance matrix. This means that the inverse square root of the
predicted numbers of each particle species is only a first
approximation to the relative statistical error. The fluctuations
of the fitted particle positions discussed above and their
contributions to the error of the yield parameters are intercorrelated
with the particle ratios and with the relative distances of the
energy deposits in the $dE/dx$ variable. The method outlined in \cite{pp_proton}
has been applied to all extracted kaon yields and results in the
statistical errors quoted in the data tables, Sect.~\ref{sec:data} below.
The ratio $R_{\textrm{stat}}$ between the full statistical error and the inverse
square root of the extracted yields is a sensitive indicator of
the fluctuations inherent in the fitting method itself. It can
vary drastically over phase space according to the correlation
with the particle ratios and the relative positions with respect
to the Bethe-Bloch function. This is visible in the distributions
of the ratio $R_{\textrm{stat}}$ defined above and shown in Fig.~\ref{fig:statfactor} 
for K$^+$ and K$^-$ in two different regions of $x_F$.

$R_{\textrm{stat}}$ is in general bigger for K$^+$ than for K$^-$ due to the large
p/$\overline{\textrm{p}}$ ratio. In both cases the forward bins in $x_F$ show
a strong increase in $R_{\textrm{stat}}$ which indicates the approach to
the limit of stability of the fit procedure in particular for
K$^+$. In the higher $x_F$ bins, $x_F$~=~0.3 and $x_F$~=~0.4 the constraints
imposed on some fit parameters, Sect.~\ref{sec:dedx_fit}, limit of course
also the range of the possible statistical fluctuations. Here,
the problem has to be tracked by the evaluation of the corresponding
systematic errors.    
   
%
%
\section{Evaluation of invariant cross sections and corrections}
\vspace{3mm}
\label{sec:corr}

The experimental evaluation of the invariant cross section

\begin{equation}
  f(x_F,p_T) = E(x_F,p_T) \cdot \frac{d^3\sigma}{dp^3} (x_F,p_T)
\end{equation}
follows the methods described in \cite{pp_pion}. This includes the absolute normalization
via the measured trigger cross section of 28.23~mb and the number of events
originating from the liquid hydrogen target. The
trigger is defined by a system of scintillation 
counters and proportional chambers on the incoming beam
plus a downstream scintillator vetoing non-interacting
beam particles.

%
%
\subsection{Empty target correction}
\vspace{3mm}
\label{sec:empty}

\begin{figure}[h]
  \begin{center}
  	\includegraphics[width=6cm]{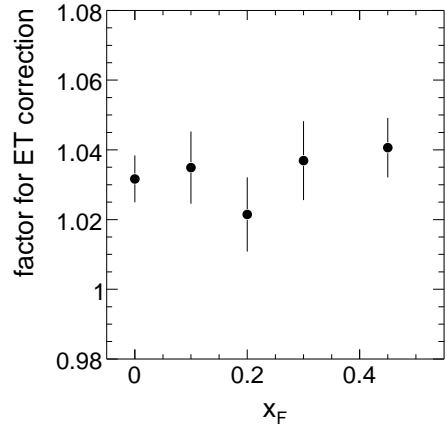}
  	\caption{Empty target correction for K$^+$ and K$^-$ as
             a function of $x_F$, averaged over $p_T$}
  	\label{fig:empty}
  \end{center}
\end{figure}

Due to the small empty/full target ratio of 9\% and the larger
fraction of zero prong events in the empty target sample, 
the empty target contribution may be treated as a small
correction as argued in \cite{pp_pion}. This correction is, within the statistical 
errors, equal for K$^+$ and K$^-$ and independent on $p_T$ and $x_F$. It 
is compatible with the one given for pions \cite{pp_pion} and protons \cite{pp_proton}
and is presented in Fig.~\ref{fig:empty} as a function of	 $x_F$. 
  
%
%
\subsection{Trigger bias correction}
\vspace{3mm}
\label{sec:s4}

\begin{figure}[h]
  \begin{center}
  	\includegraphics[width=10cm]{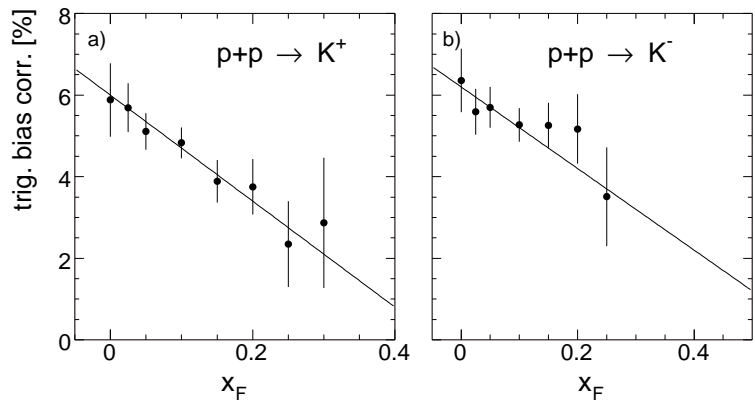}
  	\caption{Trigger bias correction as a function of $x_F$ for a) K$^+$ and b) K$^-$. 
             The lines correspond to the parametrization of the correction}
  	\label{fig:s4}
  \end{center}
\end{figure}

This correction is necessitated by the interaction trigger which
uses a small scintillator placed between the two magnets (S4 in
Fig.~\ref{fig:exp}) in anti-coincidence with the beam signal. This trigger
vetoes events with fast forward particles and thereby necessitates
a trigger bias correction which can in principle depend both on
particle type and on the kinematic variables. As described in
detail in \cite{pp_pion} the correction is quantified experimentally
by increasing the diameter of the S4 veto counter off-line and
extrapolating the observed change in cross sections to diameter
zero. For the case of kaons, the correction turns out to be
within errors independent on $p_T$ and similar for K$^+$ and K$^-$.
Its $x_F$ dependence is shown in Fig.~\ref{fig:s4}.

%
%
\subsection{Re-interaction in the target}
\vspace{3mm}
\label{sec:reint}

\begin{figure}[h]
  \begin{center}
  	\includegraphics[width=6cm]{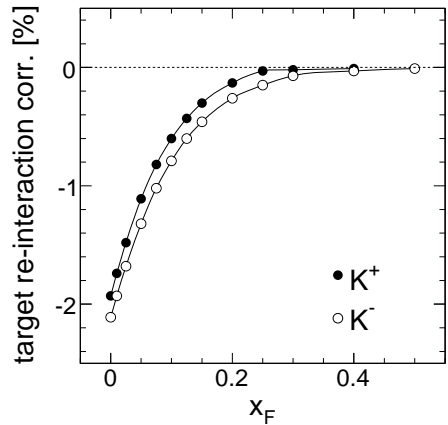}
  	\caption{Target re-interaction correction as a function of $x_F$ }
  	\label{fig:reint}
  \end{center}
\end{figure}

This correction has been evaluated \cite{pp_pion} using the PYTHIA event 
generator. It is $p_T$ independent within the available event
statistics. The $x_F$ dependence is shown in Fig.~\ref{fig:reint}.

%
%
\subsection{Absorption in the detector material}
\vspace{3mm}
\label{sec:absorp}

\begin{figure}[h]
  \begin{center}
  	\includegraphics[width=6cm]{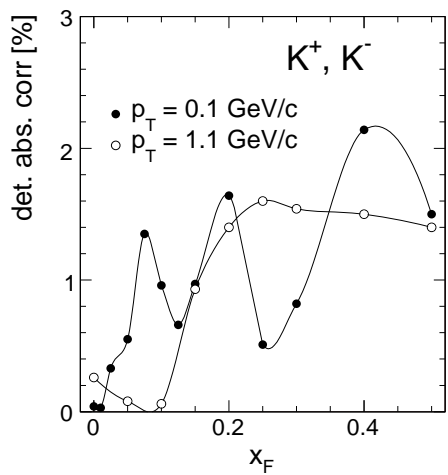}
  	\caption{Correction due to the absorption of produced kaons
             in the downstream detector material as a function of $x_F$ for two $p_T$ values. 
             The lines are shown to guide the eye}
  	\label{fig:absorp}
  \end{center}
\end{figure}

The correction for kaons interacting in the detector material
downstream of the target is determined using the GEANT simulation 
of the NA49 detector, taking account of the K$^+$ and K$^-$ inelastic
cross sections in the mostly light nuclei (Air, Plastic foils, Ceramic rods). 
Due to the non-homogeneous material distribution
the correction shows some structure both in $p_T$ and $x_F$ as presented
in Fig.~\ref{fig:absorp}.

%
%
\subsection{Kaon weak decays}
\vspace{3mm}
\label{sec:decay}

\begin{figure}[h]
  \begin{center}
  	\includegraphics[width=6.cm]{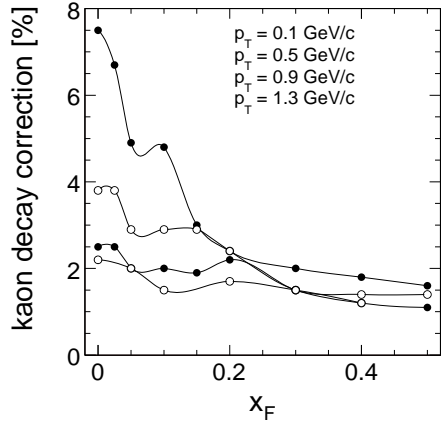}
  	\caption{Decay correction as a function of $x_F$ at different $p_T$ values. 
             The lines are shown to guide the eye}
  	\label{fig:kdecay}
  \end{center}
\end{figure}

Due to their decay length of about 30~m at the lowest lab momentum
studied here, the weak decay of kaons necessitates corrections of
up to 7\% for for kaons produced in the
lowest measured $x_F$ range. Due to the high $Q$ value of the
decay channels and unlike the weak decay of pions, the decay
products are not reconstructed to the primary vertex. This has
been verified by detailed eyescans using identified kaons with
visible decays inside the TPC system. The decay correction is
therefore determined for those kaons which decay before having
passed the necessary number of pad rows for reconstruction and 
identification. The resulting corrections are presented in Fig.~\ref{fig:kdecay}.
 
%
%
\subsection{Binning correction}
\vspace{3mm}
\label{sec:bin}

\begin{figure}[h]
  \begin{center}
  	\includegraphics[width=11.cm]{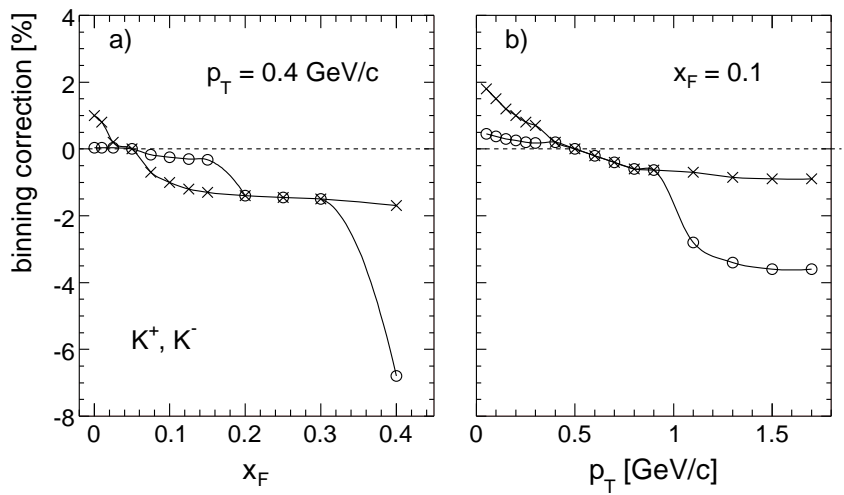}
  	\caption{Binning correction, a) as a function of $x_F$
             for $p_T$~=~0.4~GeV/c and b) as a function of $p_T$ for $x_F$~=~0.1.
             The crosses represent the corrections for fixed values
             of $\delta x_F$~=~0.05 and $\delta p_T$~=~0.1~GeV/c, respectively, and
             the open circles give the corrections for the used bin widths. The lines are 
             shown to guide the eye}
  	\label{fig:bincor}
  \end{center}
\end{figure}

The effect of finite bin sizes on the extracted inclusive cross
sections is determined using the second derivatives of the $x_F$
or $p_T$ distributions, as discussed in detail in \cite{pp_pion}. The
associated corrections are within the statistical errors equal
for K$^+$ and K$^-$. Examples are shown in Fig.~\ref{fig:bincor} as a function of $x_F$ at
fixed $p_T$ and as a function of $p_T$ at fixed $x_F$.

%
%
\subsection{Systematic errors}
\vspace{3mm}
\label{sec:sys}

\begin{figure}[h]
  \begin{center}
  	\includegraphics[width=13.cm]{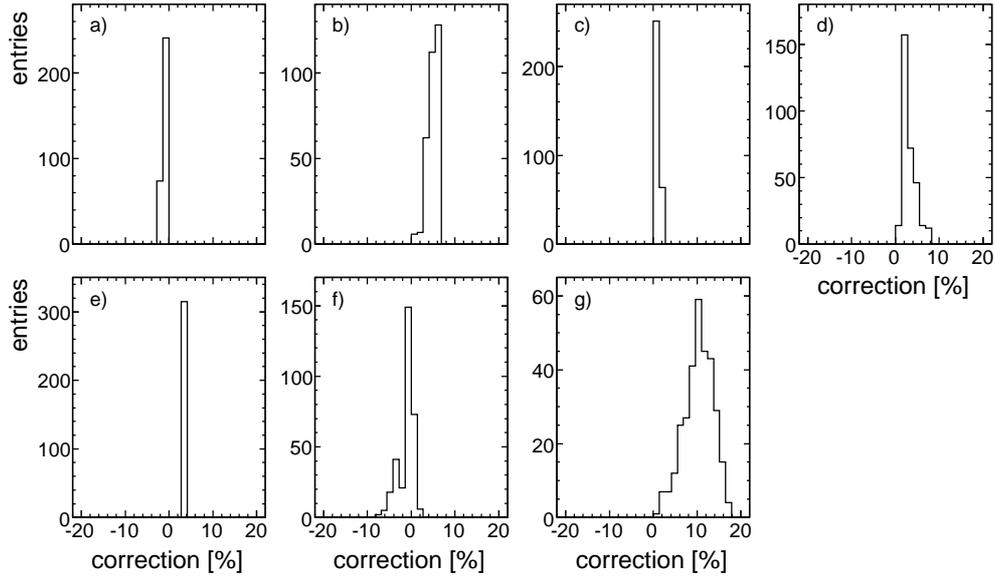}
  	\caption{Distribution of correction for a) target re-interaction, 
             b) trigger bias, c) absorption in detector material, 
	           d) kaon decay, e) empty target contribution,  
	           f) binning, g) total correction}
  	\label{fig:corr}
  \end{center}
\end{figure}

The systematic errors of the extracted cross sections are defined
by the uncertainties of the normalization and correction procedures
and by a contribution from particle identification as described in 
Sect.~\ref{sec:pid}. In particular the uncertainties due to the 
corrections may be well estimated from their distributions over all 
measured bins presented in Fig.~\ref{fig:corr}.
The corresponding error estimates are given in Table~\ref{tab:sys}.

\begin{table}[h]
  \begin{center}
    \begin{tabular}{|lc@{\hspace{8mm}}c|cc|}
      \hline
                        &                & $x_F \leq$~0.2 & \multicolumn{2}{|c|}{$x_F \geq$~0.25}\\  
      \hline
                              &                & K$^+$,K$^-$          & K$^+$        &  K$^-$ \\  \hline
      Normalization           &                &  1.5\%               &     1.5\%    &     1.5\% \\ 
      Tracking efficiency     &                &  0.5\%               &     0.5\%    &     0.5\% \\ 
      Particle identification &                &  0.0\%               &   4--12\%    &   0--6\% \\
      Trigger bias            &                &  1.0\%               &     1.0\%    &   1.0\%  \\ 
      Detector absorption     & \multirow{3}{0.mm}{\large $\biggr\}$}           
                              & \hspace{-10mm} \multirow{3}{0mm}{1.0\%}  
                              & \hspace{-10mm} \multirow{3}{0mm}{1.0\%}  
                              & \hspace{-10mm} \multirow{3}{0mm}{1.0\%} \\
      Kaon decay              &                 &                      &             &    \\
      Target re-interaction   &                 &                      &             &  \\ 
      Binning                 &                 &  0.5\%               & 0.5\%       & 0.5\%  \\ \hline 
      Total(upper limit)      &                 &  4.5\%               & 8.5--16.5\% &  4.5--10.5\%  \\ 
      Total(quadratic sum)    &                 &  2.2\%               & 4.6--12.2\% &  2.2--6.4\%  \\      
      \hline
    \end{tabular}
  \end{center}
  \vspace{-2mm}
  \caption{Summary of systematic errors}
  \label{tab:sys}
\end{table}

The linear sum of these estimations gives an upper limit of 4.5\%,
the quadratic sum an effective error of 2.2\% for $x_F \leq$~0.2. These values are
close to the estimations obtained for pions \cite{pp_pion} and
baryons \cite{pp_proton}. In the $x_F$ region above 0.25, however, the upper 
limit (quadratic sum) can reach 16.5\%(12.2\%) for K$^+$ and 10.5\%(6.4\%) 
for K$^-$. The spread of the corrections over all selected
bins of phase space may be visualized in Fig.~\ref{fig:corr} which also
gives the distribution of the sum of all corrections.

%
%
\section{Results on double differential cross sections}
\vspace{3mm}
\label{sec:data}

%
%
\subsection{Data tables}
\vspace{3mm}
\label{sec:tables}

The binning scheme described in Sect.~\ref{sec:na49} results in 158 data
points each for K$^+$ and K$^-$. The corresponding cross sections
are presented in Tables~\ref{tab:kaplus} and \ref{tab:kaminus}.

\begin{table}[h]
\renewcommand{\tabcolsep}{0.22pc} 
\renewcommand{\arraystretch}{1.3}
\footnotesize
\begin{center}
\begin{tabular}{|c|cr|cr|cr|cr|cr|cr|}
\hline
\multicolumn{13}{|c|}{$f(x_F,p_T), \Delta f$} \\ 
\hline
$p_T \backslash x_F$ & \multicolumn{2}{|c|}{0.0} & \multicolumn{2}{|c|}{0.01} & \multicolumn{2}{|c|}{0.025} & \multicolumn{2}{|c|}{0.05} & \multicolumn{2}{|c|}{0.075} & \multicolumn{2}{|c|}{0.1}\\ \hline
0.05 & 2.78&7.23 & 2.73&5.30 & 3.174&2.83 & 2.797&3.33 & 2.438&3.54 & 2.169&4.47\\
0.1  & 2.96&4.96 & 3.22&4.22 & 2.799&2.27 & 2.572&2.23 & 2.344&2.62 & 2.046&2.61\\
0.15 & 2.56&6.29 & 2.53&5.10 & 2.598&2.06 & 2.460&1.93 & 2.070&2.28 & 1.955&2.71\\
0.2  & 2.35&5.00 & 2.27&5.37 & 2.294&1.87 & 2.219&1.69 & 2.093&1.98 & 1.738&2.06\\
0.25 & 1.95&6.80 & 2.11&5.14 & 2.014&3.09 & 1.904&1.73 & 1.789&1.94 & 1.563&2.38\\
0.3  & 1.748&4.82 & 1.692&5.51 & 1.762&2.70 & 1.625&1.66 & 1.517&1.92 & 1.395&1.86\\
0.4  & 1.289&3.62 & 1.232&4.22 & 1.224&2.55 & 1.177&1.54 & 1.074&1.56 & 0.963&1.72\\
0.5  & 0.839&4.23 & 0.916&5.12 & 0.819&3.14 & 0.804&2.46 & 0.740&1.82 & 0.680&2.00\\
0.6  & 0.530&3.89 && & 0.539&3.67 & 0.539&3.15 & 0.487&2.10 & 0.4462&2.09\\
0.7  & 0.371&4.47 && & 0.333&4.22 & 0.320&4.16 & 0.323&3.32 & 0.2760&2.73\\
0.8  & 0.241&5.33 && & 0.216&5.32 & 0.215&5.09 & 0.1903&4.51 & 0.1812&3.98\\
0.9  & 0.1412&6.75 && & 0.1358&6.63 & 0.1286&6.72 & 0.1357&5.34 & 0.1217&4.96\\
1.1  & 0.0580&6.25 && & 0.0595&6.75 & 0.0498&6.76 && & 0.0485&4.98\\
1.3  & 0.0231&8.04 && && & 0.0194&9.08 && & 0.0184&8.03\\
1.5  & 0.0106&12.8 && && & 0.0092&12.7 && & 0.00856&11.6\\
1.7  & 0.00354&17.3 && && && && & 0.00291&13.8\\
\hline
$p_T \backslash x_F$ & \multicolumn{2}{|c|}{0.125} & \multicolumn{2}{|c|}{0.15} & \multicolumn{2}{|c|}{0.2} & \multicolumn{2}{|c|}{0.25} & \multicolumn{2}{|c|}{0.3} & \multicolumn{2}{|c|}{0.4}\\ \hline
0.05 & 1.87&5.51 & 1.60&6.47 && && && &&\\
0.1 & 1.758&4.20 & 1.469&4.09 & 1.220&4.21 & 0.941&7.94 & 0.796&7.00 & 0.475&5.95\\
0.15 & 1.813&3.21 & 1.396&4.02 && && && &&\\
0.2 & 1.603&3.11 & 1.390&2.96 & 0.995&3.34 & 0.843&6.62 & 0.720&5.00 & 0.408&4.56\\
0.25 & 1.379&2.85 & 1.313&3.44 && && && &&\\
0.3 & 1.243&2.75 & 1.057&2.76 & 0.842&3.24 & 0.729&5.29 & 0.546&5.00 & 0.300&4.35\\
0.4 & 0.912&2.19 & 0.809&2.46 & 0.666&3.06 & 0.540&5.19 & 0.450&5.00 & 0.253&4.11\\
0.5 & 0.617&2.40 & 0.574&2.59 & 0.493&3.04 & 0.394&5.00 & 0.392&5.00 & 0.2014&4.15\\
0.6 & 0.410&2.82 & 0.385&2.96 & 0.339&3.64 & 0.269&4.22 & 0.208&6.00 & 0.1632&4.23\\
0.7 & 0.2593&3.52 & 0.2453&3.56 & 0.2046&4.35 & 0.1756&4.84 & 0.143&7.00 & 0.1042&4.95\\
0.8 & 0.1709&4.97 & 0.1528&4.34 & 0.1340&5.01 & 0.1130&7.62 & 0.1040&8.00 & 0.0583&7.14\\
0.9 & 0.1102&6.16 & 0.1019&5.55 & 0.0754&6.94 & 0.0753&8.23 & 0.0488&10.0 & 0.0372&8.47\\
1.1 && & 0.0400&5.15 & 0.0339&6.85 & 0.0267&8.90 & 0.0200&11.0 & 0.0148&8.65\\
1.3 && & 0.0184&8.29 & 0.0143&9.77 & 0.0118&11.7 & 0.0091&14.0 & 0.00640&12.3\\
1.5 && & 0.00458&18.5 & 0.0055&20.6 & 0.00437&16.7 & 0.00384&19.0 & 0.00165&23.0\\
1.7 && && & 0.00257&16.0 & 0.00232&23.2 & 0.00141&30.3 &&\\
\hline
\end{tabular}
\end{center}
\caption{Invariant cross section, $f(x_F,p_T)$, in mb/(GeV$^2$/c$^3$) for  K$^+$ in p+p collisions at 
         158~GeV/c beam momentum. The relative statistical errors, $\Delta f$, are given in \%, the 
         transverse momentum $p_T$ in GeV/c}
\label{tab:kaplus}
\end{table}

\begin{table}[b]
\renewcommand{\tabcolsep}{0.22pc} 
\renewcommand{\arraystretch}{1.4}
\footnotesize
\begin{center}
\begin{tabular}{|c|cr|cr|cr|cr|cr|cr|cr|}
\hline
\multicolumn{15}{|c|}{$f(x_F,p_T), \Delta f$} \\ 
\hline
$p_T \backslash x_F$ & \multicolumn{2}{|c|}{0.0} & \multicolumn{2}{|c|}{0.01} & \multicolumn{2}{|c|}{0.025} & \multicolumn{2}{|c|}{0.05} & \multicolumn{2}{|c|}{0.075} & \multicolumn{2}{|c|}{0.1} & \multicolumn{2}{|c|}{0.125}\\ \hline
0.05 & 1.90&6.06 & 2.04&5.04 & 2.270&3.05 & 1.911&3.47 & 1.530&4.29 & 1.294&5.22 & 1.089&6.03\\
0.1 & 2.153&4.25 & 2.151&4.54 & 2.041&2.07 & 1.847&2.17 & 1.540&2.92 & 1.234&2.84 & 1.081&4.32\\
0.15 & 1.93&5.43 & 1.885&4.93 & 1.907&1.94 & 1.735&2.09 & 1.422&2.52 & 1.219&3.08 & 0.934&3.84\\
0.2 & 1.600&4.65 & 1.683&5.19 & 1.717&1.69 & 1.501&1.74 & 1.333&2.26 & 1.067&2.22 & 0.842&3.60\\
0.25 & 1.599&5.57 & 1.661&5.04 & 1.522&2.82 & 1.296&1.91 & 1.147&2.22 & 0.949&2.69 & 0.811&3.25\\
0.3 & 1.301&4.39 & 1.188&5.58 & 1.291&2.49 & 1.141&1.89 & 0.997&2.13 & 0.864&2.00 & 0.694&3.30\\
0.4 & 0.909&3.51 & 0.907&4.41 & 0.905&2.35 & 0.796&1.54 & 0.736&1.67 & 0.615&1.76 & 0.515&2.54\\
0.5 & 0.588&4.10 & 0.551&5.88 & 0.572&2.97 & 0.538&2.54 & 0.4751&2.02 & 0.4115&2.02 & 0.3734&2.62\\
0.6 & 0.380&4.39 && & 0.378&3.45 & 0.357&3.14 & 0.3073&2.40 & 0.2709&2.31 & 0.2263&3.21\\
0.7 & 0.235&4.62 && & 0.2374&4.15 & 0.2038&4.32 & 0.1779&4.06 & 0.1626&2.93 & 0.1546&3.54\\
0.8 & 0.1502&5.35 && & 0.1391&5.54 & 0.1356&5.34 & 0.1258&4.67 & 0.1055&4.20 & 0.0958&5.36\\
0.9 & 0.0910&7.10 && & 0.0833&7.06 & 0.0770&6.93 & 0.0614&7.47 & 0.0687&5.29 & 0.0554&7.24\\
1.1 & 0.0340&6.66 && & 0.0303&7.23 & 0.0303&6.79 && & 0.0245&5.65 &&\\
1.3 & 0.0113&9.68 && && & 0.0112&11.0 && & 0.00758&8.00 &&\\
1.5 & 0.00435&15.4 && && & 0.00487&16.8 && & 0.00341&15.9 &&\\
1.7 & 0.00211&18.8 && && && && & 0.00172&15.7 &&\\
\hline
$p_T \backslash x_F$ & \multicolumn{2}{|c|}{0.15} & \multicolumn{2}{|c|}{0.2} & \multicolumn{2}{|c|}{0.25} & \multicolumn{2}{|c|}{0.3} & \multicolumn{2}{|c|}{0.4} & \multicolumn{2}{|c|}{0.5}\\ \cline{0-12}
0.05 & 0.774&7.78 && && && && &&\\
0.1 & 0.759&4.52 & 0.514&4.54 & 0.290&7.50 & 0.208&11.4 & 0.1318&7.14 & 0.0323&30.7\\
0.15 & 0.742&4.75 && && && && &&\\
0.2 & 0.709&3.37 & 0.478&3.24 & 0.299&6.11 & 0.217&7.72 && &&\\
0.25 & 0.647&3.96 && && && && &&\\
0.3 & 0.597&3.00 & 0.381&3.15 & 0.243&5.46 & 0.183&7.46 & 0.0966&4.77 & 0.0226&11.0\\
0.4 & 0.424&2.67 & 0.2623&3.63 & 0.212&5.06 & 0.1268&7.87 && &&\\
0.5 & 0.3022&2.90 & 0.2044&3.64 & 0.1415&5.88 & 0.0936&8.14 & 0.0389&5.84 & 0.0172&9.78\\
0.6 & 0.1900&3.36 & 0.1524&3.91 & 0.0957&6.08 & 0.0706&8.41 && &&\\
0.7 & 0.1306&3.81 & 0.0921&4.86 & 0.0553&7.57 & 0.0445&9.76 & 0.0165&7.64 & 0.0110&10.4\\
0.8 & 0.0780&4.80 & 0.0600&5.63 & 0.0381&8.89 & 0.0252&12.4 && &&\\
0.9 & 0.0480&5.98 & 0.0402&6.72 & 0.0257&9.77 & 0.0187&13.3 & 0.00878&9.26 & 0.00363&16.0\\
1.1 & 0.0191&6.32 & 0.01204&8.14 & 0.00875&10.8 & 0.00736&13.3 & 0.00246&16.2 & 0.00106&27.7\\
1.3 & 0.00805&10.2 & 0.00578&10.5 & 0.00302&16.9 & 0.00163&27.8 & 0.00080&26.5 &&\\
1.5 & 0.00271&17.0 & 0.00198&18.5 & 0.00189&19.9 & 0.00110&31.4 && &&\\
1.7 && & 0.00102&18.3 && & 0.00054&30.3 && &&\\
\cline{0-12}
\end{tabular}
\end{center}
\caption{Invariant cross section, $f(x_F,p_T)$, in mb/(GeV$^2$/c$^3$) for  K$^-$ in p+p collisions at 
         158~GeV/c beam momentum. The relative statistical errors, $\Delta f$, are given in \%, the 
         transverse momentum $p_T$ in GeV/c}
\label{tab:kaminus}
\end{table}
\clearpage

%
%
\subsection{Interpolation scheme}
\vspace{3mm}
\label{sec:interp}

\begin{figure}[h]
  \begin{center}
  	\includegraphics[width=13cm]{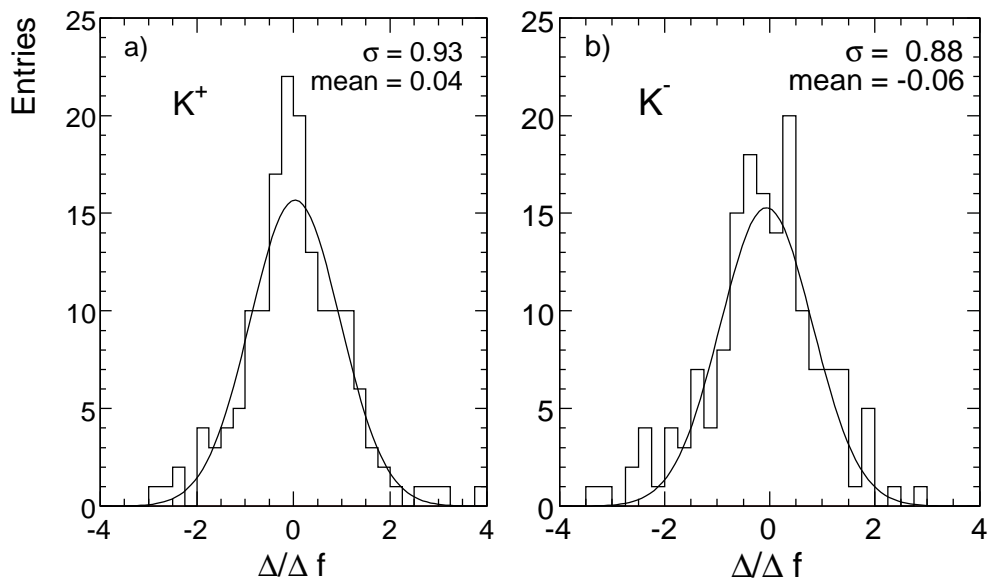}
  	\caption{Difference $\Delta$ between the measured invariant cross sections and the
             corresponding interpolated values divided by the experimental uncertainty 
             $\Delta f$ for a) K$^+$ and b) K$^-$}
  	\label{fig:diff}
  \end{center}
\end{figure}

As in the preceding publications concerning p+p and p+C interactions
\cite{pp_pion,pp_proton,pc_pion} a two-dimensional interpolation based on a multi-step 
recursive method using eyeball fits has been applied. The distribution
of the differences of the measured points with respect to 
this interpolation, divided by the given statistical error should
be Gaussian with mean zero and variance unity if the interpolation
is bias-free and if the estimation of the statistical errors, see
Sect.~\ref{sec:dedx_stat} above, is correct. The corresponding distributions
shown in Fig.~\ref{fig:diff} comply with this expectation.

As to first order the 8 first neighbours and to the 
second the 24 second neighbours to each data point contribute to the establishment
of the interpolation, a reduction of the local statistical fluctuations
of a factor of 3 to 4 may be expected. The authors therefore advise to
use the data interpolation which is available under \cite{site} for data
comparison and analysis purposes. On the point-by-point level, the statistical error of the 
interpolated cross sections has been estimated as the mean value of 
the statistical errors of each measured point plus the 8 surrounding 
points in the $p_T$/$x_F$ plane, divided by the (conservative) factor of 2.
The systematic uncertainties are of course not touched by this procedure,
in addition they are of mostly non-local origin.

%
%
\subsection{Dependence of invariant cross sections on $\mathbf {x_F}$ and $\mathbf {p_T}$ }
\vspace{3mm}
\label{sec:dist}

\begin{figure}[t!]
  \begin{center}
  	\includegraphics[width=15.4cm]{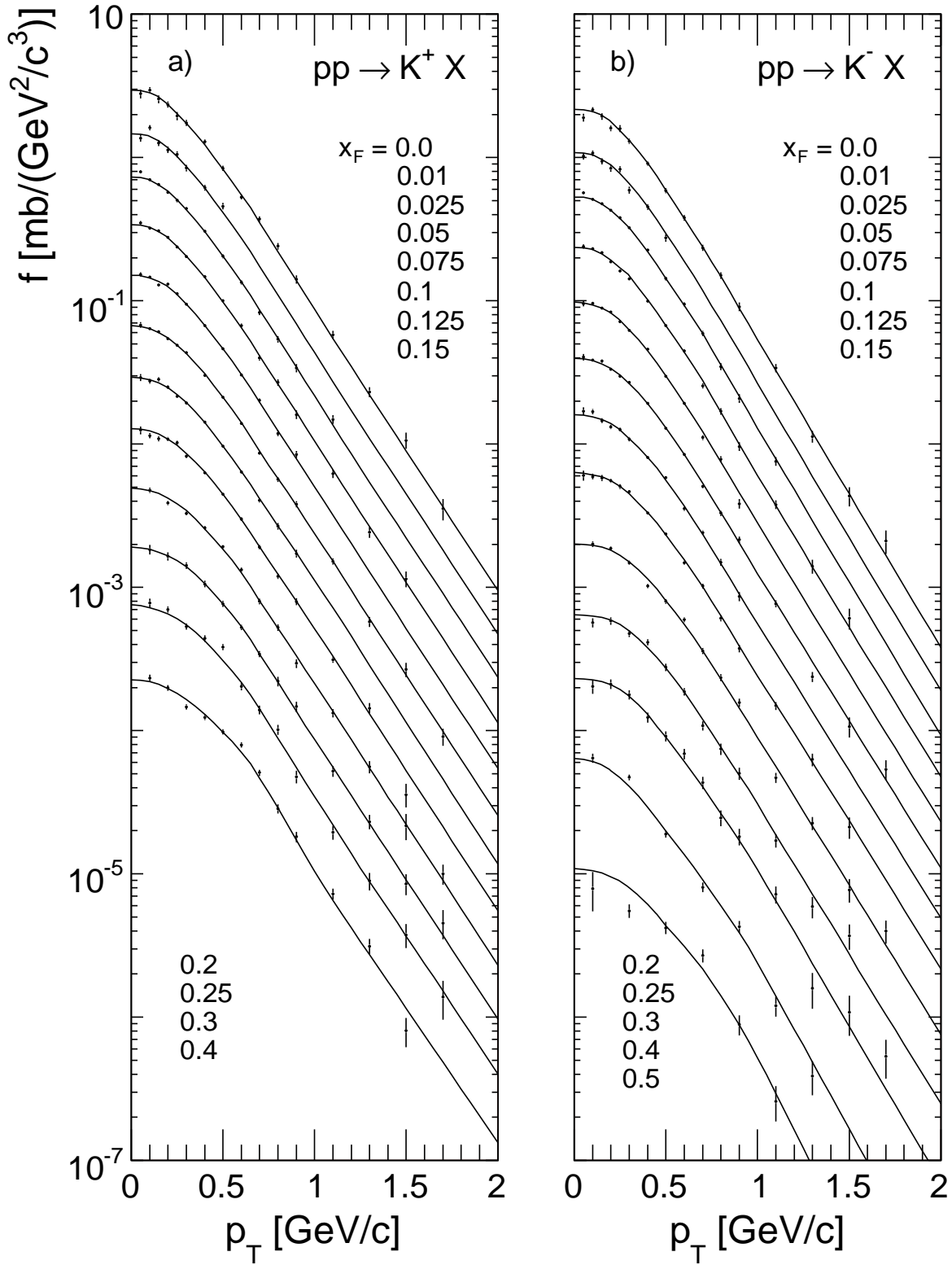}
  	\caption{Double differential invariant cross section $f(x_F,p_T)$ [mb/(GeV$^2$/c$^3$)] as a 
             function of $p_T$ at fixed $x_F$ for a) K$^+$ and b) K$^-$  produced in p+p 
             collisions at 158~GeV/c beam momentum. The distributions for different $x_F$ 
             values are successively scaled down by 0.5 for better separation. The lines show the 
             result of the data interpolation, Sect.~\ref{sec:interp}}
  	\label{fig:ptdist}
  \end{center}
\end{figure}

\begin{figure}[t!]
  \begin{center}
  	\includegraphics[width=15.8cm]{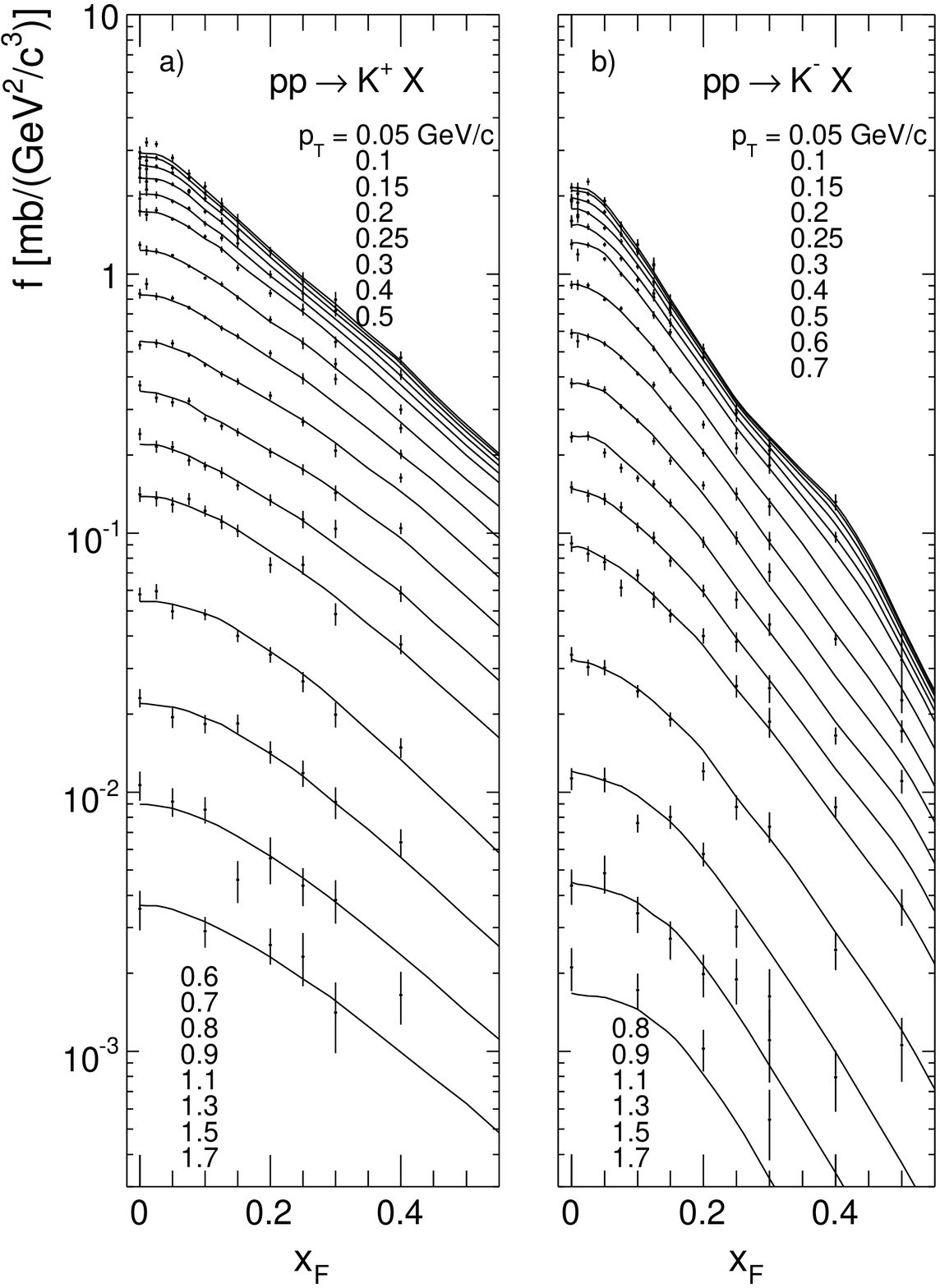}
  	\caption{Double differential invariant cross section $f(x_F,p_T)$ [mb/(GeV$^2$/c$^3$)] as a 
             function of $x_F$ at fixed $p_T$ for a) K$^+$ 
             and b) K$^-$  produced in p+p collisions at 158~GeV/c beam momentum. The lines show the 
             result of the data interpolation, Sect.~\ref{sec:interp}}
  	\label{fig:xfdist}
  \end{center}
\end{figure}

Shapes of the invariant cross sections as functions of $p_T$ and
$x_F$ are shown in Figs.~\ref{fig:ptdist} and \ref{fig:xfdist} including the data interpolation
presented above. In order to clearly demonstrate the shape
evolution and to avoid overlap of plots and error bars, subsequent
$p_T$ distributions have been multiplied by factors of 0.5 (Fig.~\ref{fig:ptdist}).

%
%
\subsection{Rapidity and transverse mass distributions}
\vspace{3mm}
\label{sec:rap}
 
\begin{figure}[t!]
  \begin{center}
  	\includegraphics[width=15.8cm]{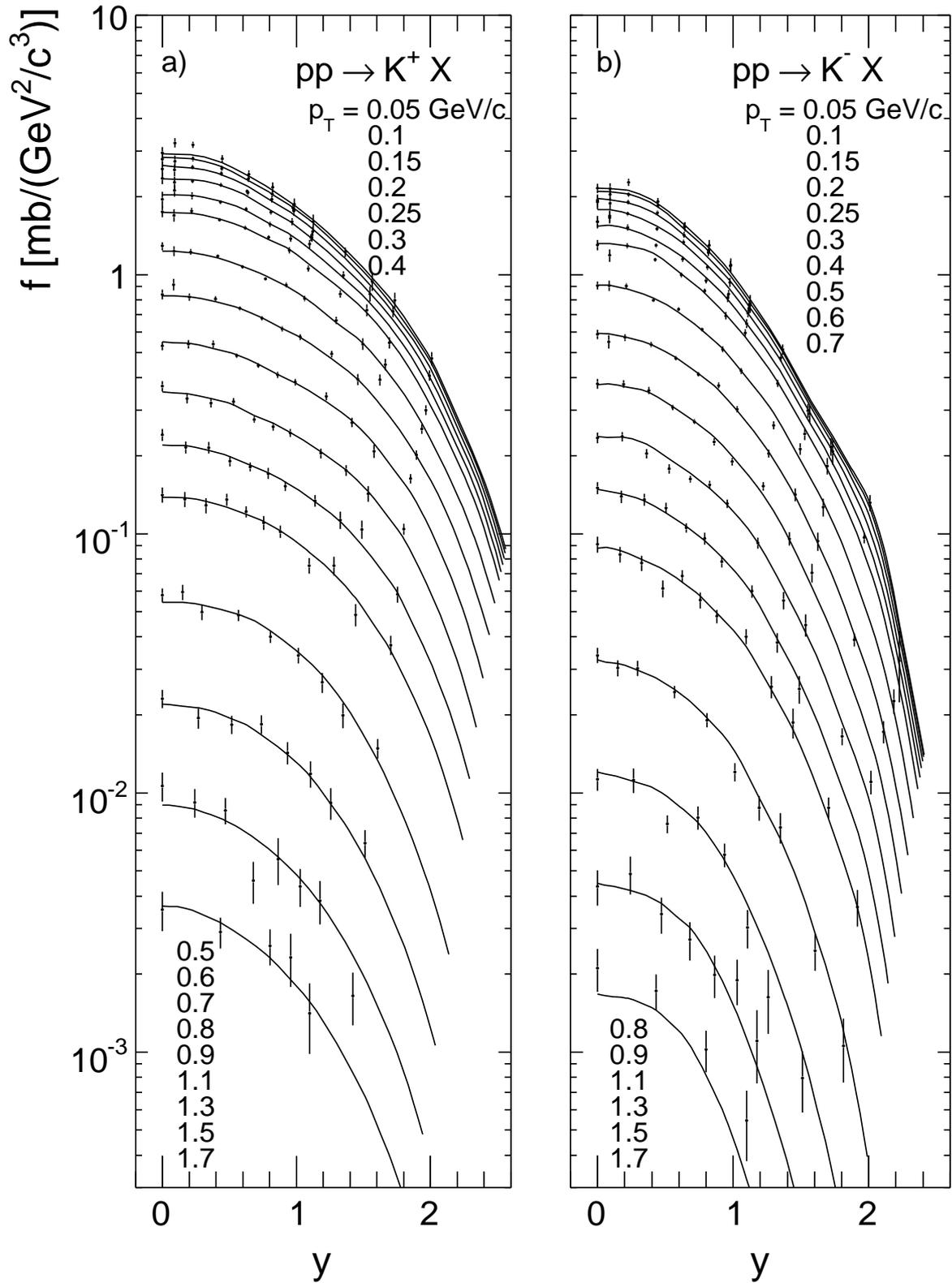}
  	\caption{Double differential invariant cross section $f(x_F,p_T)$ [mb/(GeV$^2$/c$^3$] as a 
             function of $y$ at fixed $p_T$ for a) K$^+$ 
             and b) K$^-$  produced in p+p collisions at 158~GeV/c beam momentum. The lines show the 
             result of the data interpolation, Sect.~\ref{sec:interp}}
  	\label{fig:ydist}
  \end{center}
\end{figure}

As in the preceding publications \cite{pp_pion,pp_proton,pc_pion} data and interpolation are also
shown as functions of rapidity at fixed $p_T$ in Fig.~\ref{fig:ydist}.

\begin{figure}[t]
  \begin{center}
  	\includegraphics[width=10.1cm]{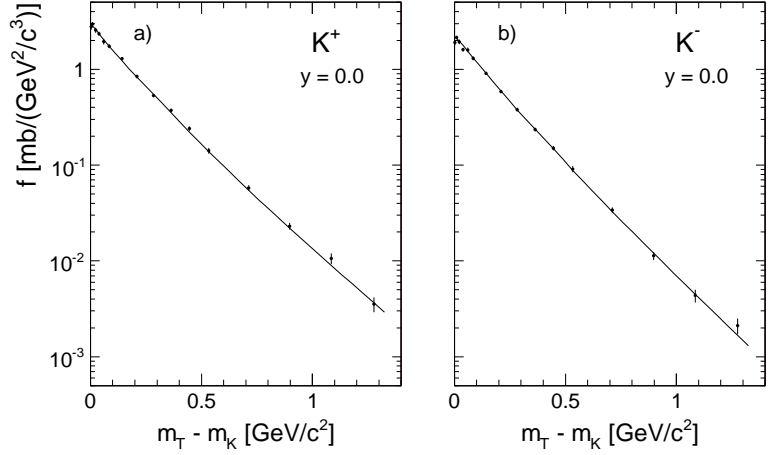}
  	\caption{Invariant cross section as a function of $m_T - m_K$ for a) K$^+$  and
             b) K$^-$  produced at $y$~=~0.0. The lines show the result
             of the data interpolation, Sect.~\ref{sec:interp}}
  	\label{fig:mtdist}
  \end{center}
\end{figure}

\begin{figure}[h]
  \begin{center}
  	\includegraphics[width=10.2cm]{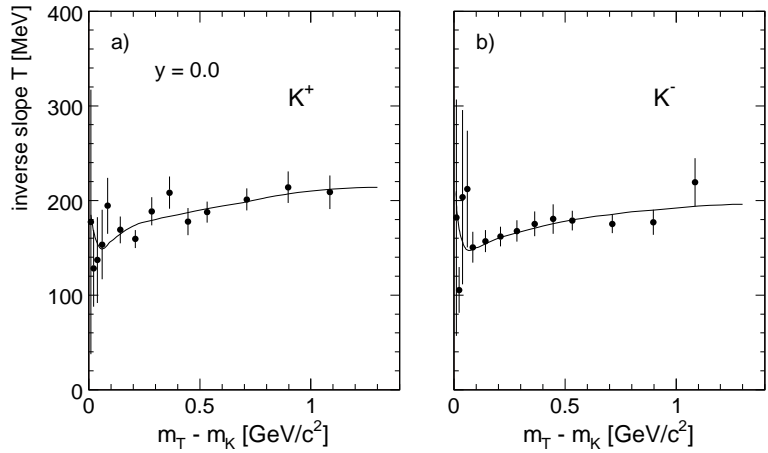}
  	\caption{Local inverse slope of the $m_T$ distribution as a function of $m_T - m_K$
             for a) K$^+$  and b) K$^-$. The lines correspond to the data
             interpolation, Sect.~\ref{sec:interp}}
  	\label{fig:mtslope}
  \end{center}
\end{figure}

Transverse mass distributions at $x_F$~=~$y$~=~0, with
$m_T = \sqrt{m_{\textrm{K}}^2+p_T^2}$, are presented in Fig.~\ref{fig:mtdist}
again including the data interpolation. The corresponding
dependence of the inverse slopes of these distributions on $m_T - m_K$
is shown in Fig.~\ref{fig:mtslope} together with the results from the data 
interpolation. The local slope values are defined by three successive data points.

K$^+$ and K$^-$ show a similar behaviour of the inverse slope parameters
which fall from values around 180~MeV at low $m_T - m_K$ to a
minimum of 150~MeV at $m_T - m_K \sim$~0.05~GeV/c$^2$ ($p_T \sim$~0.220~GeV/c),
see also results from ISR (Fig.~\ref{fig:slope_isrxf0}). They
then rise steadily towards higher $m_T$ and reach 180~MeV at $p_T \sim$~0.6~(0.9)~GeV/c 
and 200~MeV at $p_T \sim$~1.0~(1.8)~GeV/c for K$^+$ and K$^-$,
respectively. These trends resemble the ones observed for pions \cite{pp_pion}.

%
%
\section{Particle ratios}
\vspace{3mm}
\label{sec:ratios}

The present data on charged kaon production offer, together
with the already available pion \cite{pp_pion} and baryon \cite{pp_proton} 
cross sections with similar phase space coverage and precision,
a unique possibility to study particle ratios, in particular
their evolution with transverse momentum and $x_F$. This section
will therefore not only deal with K$^+$/K$^-$  but will also
address K/$\pi$ and K/baryon ratios.

%
%
\subsection{K$^+$/K$^-$ ratios}
\vspace{3mm}
\label{sec:p2m_rat}

\begin{figure}[h]
  \begin{center}
  	\includegraphics[width=12.cm]{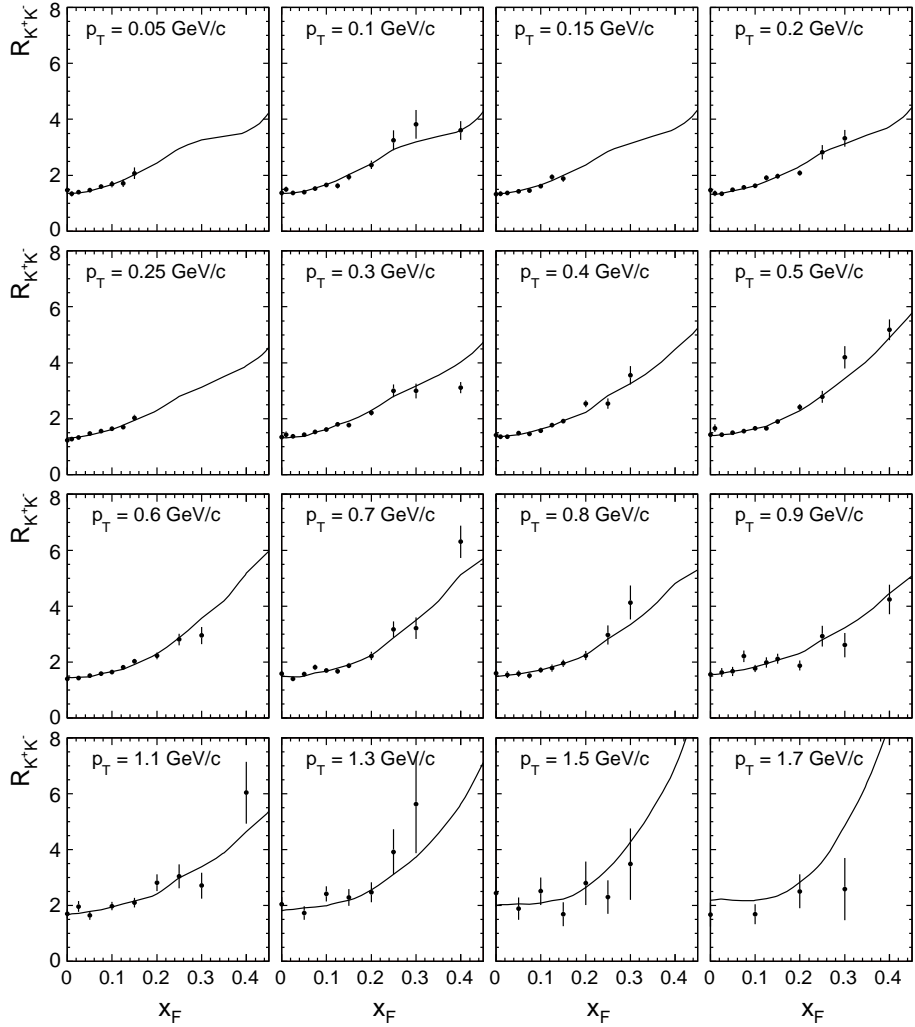}
  	\caption{$R_{\textrm{K}^+\textrm{K}^-}$  as a function of $x_F$ at fixed $p_T$. The lines 
  	         show the result of the data interpolation, Sect.~\ref{sec:interp}}
  	\label{fig:p2m_xf}
  \end{center}
\end{figure}

\begin{figure}[h]
  \begin{center}
  	\includegraphics[width=12.cm]{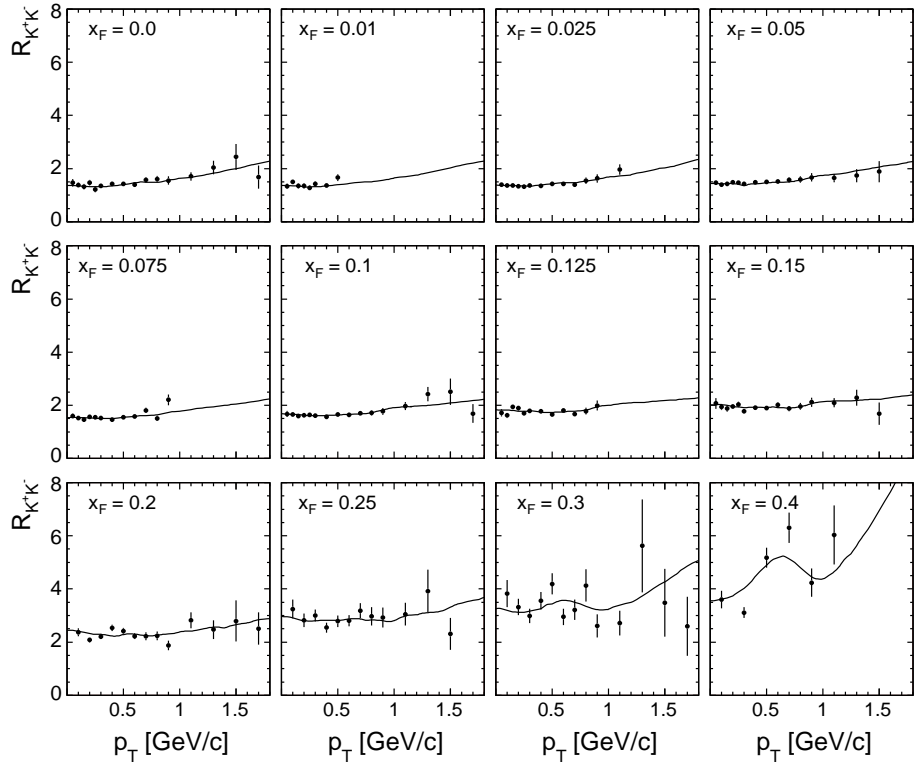}
  	\caption{$R_{\textrm{K}^+\textrm{K}^-}$ as a function of $p_T$ at fixed $x_F$. The lines 
  	         show the result of the data interpolation, Sect.~\ref{sec:interp}}
  	\label{fig:p2m_pt}
  \end{center}
\end{figure}

The ratio of the inclusive K$^+$ and K$^-$ cross sections,

\begin{equation}
   R_{\textrm{K}^+\textrm{K}^-} = f(\textrm{K}^+)/f(\textrm{K}^-) 
\end{equation}
is shown in Fig.~\ref{fig:p2m_xf} as a function of $x_F$ at fixed $p_T$ and in 
Fig.~\ref{fig:p2m_pt} as a function of $p_T$ at fixed values of $x_F$.


\begin{figure}[h]
  \begin{center}
  	\includegraphics[width=12cm]{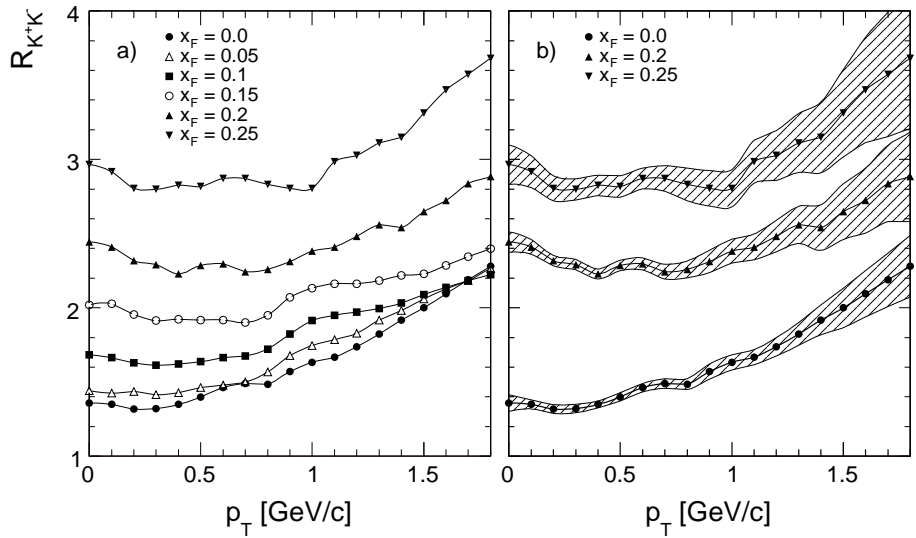}
  	\caption{a) $R_{\textrm{K}^+\textrm{K}^-}$ for the data interpolation as a function 
             of $p_T$ for different $x_F$; b) Error bands expected for data interpolation}
  	\label{fig:p2m_sum}
  \end{center}
\end{figure}

In each panel the corresponding ratio of the data interpolations,
Sect.~\ref{sec:interp}, is superimposed to the data points as a solid line.
The basic features of the data may be described as a steady increase
of $R_{\textrm{K}^+\textrm{K}^-}$ over the available  $x_F$ range by about 
a factor of three (Fig.~\ref{fig:p2m_xf}) with some structure visible at certain $x_F$ and $p_T$ values. 
The $p_T$ dependence (Fig.~\ref{fig:p2m_pt}) reveals a more detailed evolution. The increase
of $R_{\textrm{K}^+\textrm{K}^-}$ in the interval 0~$< p_T <$~1.7~GeV/c which 
amounts to  about 60\% at 
low $x_F$ flattens out in the $x_F$ range 0.1 -- 0.2 to only 20\% before it
increases again towards higher $x_F$. This may be visualized in Fig.~\ref{fig:p2m_sum}
where the ratios of the interpolated cross sections are shown as
a function of $p_T$ for several $x_F$ values on an enlarged vertical scale.  
Fig.~\ref{fig:p2m_sum}b gives an estimate of the statistical uncertainty of 
$R_{\textrm{K}^+\textrm{K}^-}$ to be expected for data interpolation, 
characterized by the hatched area around the nominal values.

%
%
\subsection{K/$\boldsymbol{\pi}$ ratios}
\vspace{3mm}
\label{sec:k2pi_rat}

\begin{figure}[h]
  \begin{center}
  	\includegraphics[width=9.5cm]{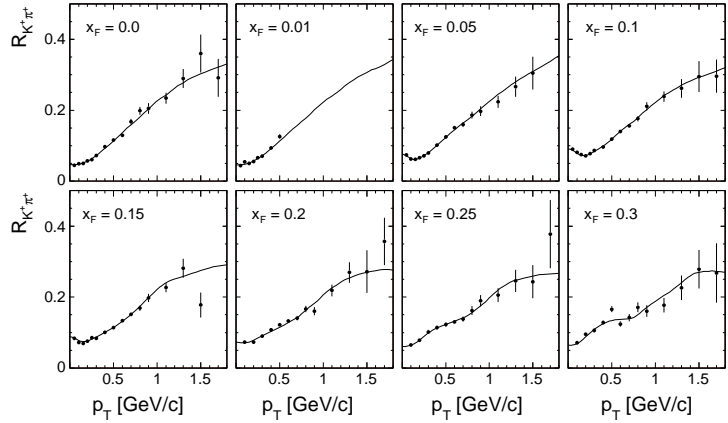}
  	\caption{$R_{\textrm{K}^+\pi^+}$ as a function of $p_T$ at fixed $x_F$. The lines 
  	         show the result of the data interpolation, Sect.~\ref{sec:interp}}
  	\label{fig:k2pi_pos_pt}
  \end{center}
\end{figure}

\begin{figure}[h]
  \begin{center}
  	\includegraphics[width=9.5cm]{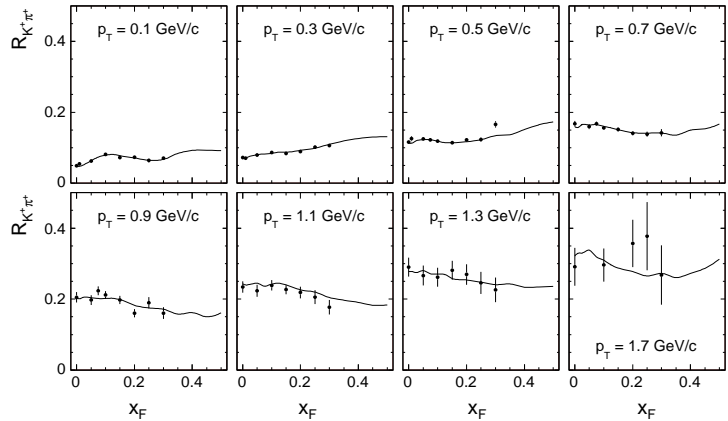}
  	\caption{$R_{\textrm{K}^+\pi^+}$ as a function of $x_F$ at fixed $p_T$. The lines 
  	         show the result of the data interpolation, Sect.~\ref{sec:interp}}
  	\label{fig:k2pi_pos_xf}
  \end{center}
\end{figure}

\begin{figure}[h]
  \begin{center}
  	\includegraphics[width=5.cm]{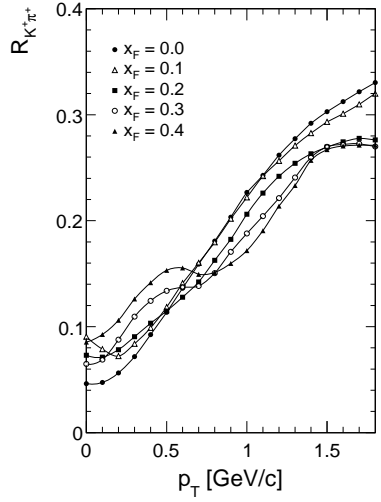}
  	\caption{$R_{\textrm{K}^+\pi^+}$ as a function of $p_T$ for different $x_F$}
  	\label{fig:k2pi_pos_sum}
  \end{center}
\end{figure}

\begin{figure}[h]
  \begin{center}
  	\includegraphics[width=9.5cm]{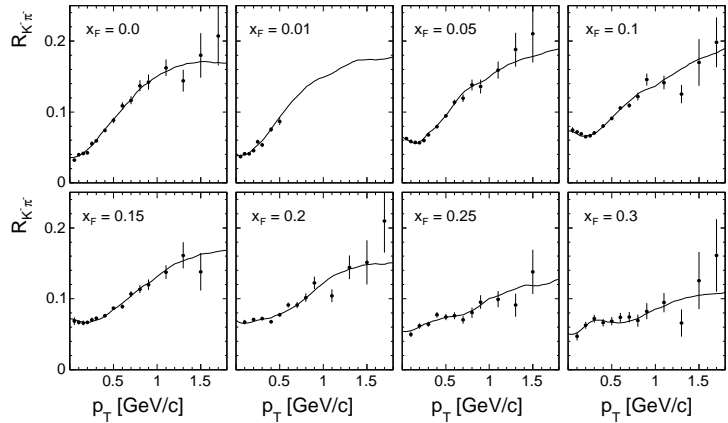}
  	\caption{$R_{\textrm{K}^-\pi^-}$ as a function of $p_T$ at fixed $x_F$. The lines 
  	         show the result of the data interpolation, Sect.~\ref{sec:interp}}
  	\label{fig:k2pi_neg_pt}
  \end{center}
\end{figure}

\begin{figure}[h]
  \begin{center}
  	\includegraphics[width=9.5cm]{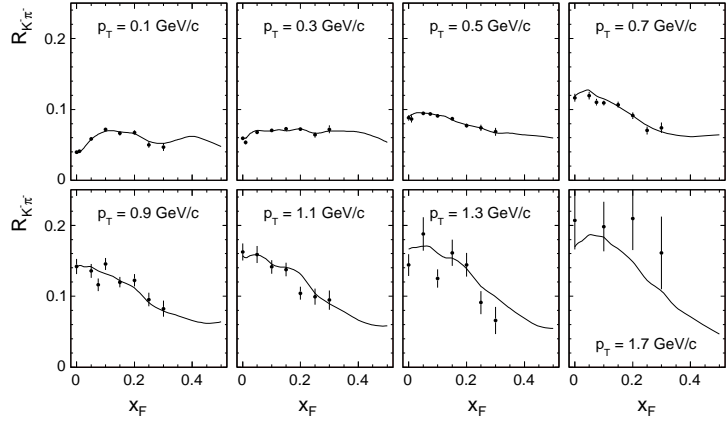}
  	\caption{$R_{\textrm{K}^-\pi^-}$ as a function of $x_F$ at fixed $p_T$. The lines 
  	         show the result of the data interpolation, Sect.~\ref{sec:interp}}
  	\label{fig:k2pi_neg_xf}
  \end{center}
\end{figure}

\begin{figure}[h]
  \begin{center}
  	\includegraphics[width=5.cm]{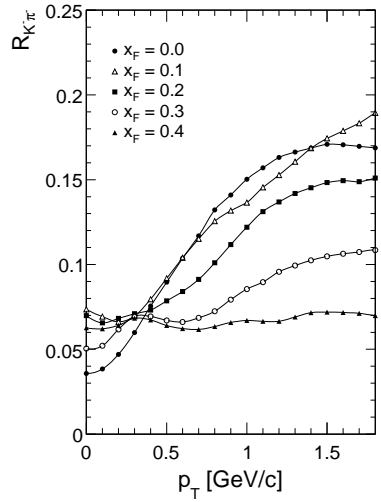}
  	\caption{$R_{\textrm{K}^-\pi^-}$ as a function of $p_T$ for different $x_F$}
  	\label{fig:k2pi_neg_sum}
  \end{center}
\end{figure}

The K/$\pi$ ratios shown here make use of the pion data and the
corresponding interpolation published in \cite{pp_pion}.
The ratios of the invariant inclusive cross sections

\begin{align}
   R_{\textrm{K}^+\pi^+} &= f(\textrm{K}^+)/f(\pi^+)  \\
   R_{\textrm{K}^-\pi^-} &= f(\textrm{K}^-)/f(\pi^-) 
\end{align}
are presented in Figs.~\ref{fig:k2pi_pos_pt} to \ref{fig:k2pi_neg_sum}.

$R_{\textrm{K}^+\pi^+}$ is shown in Fig.~\ref{fig:k2pi_pos_pt} as a function of $p_T$ for fixed $x_F$ and 
in Fig.~\ref{fig:k2pi_pos_xf} as a function of $x_F$ for fixed $p_T$.

Here the salient features are the strong increase with $p_T$ which
is rather independent on $x_F$ and reaches values of about 6 relative to low $p_T$
at 1.7~GeV/c, and the rather small $x_F$ dependence with a slight 
increase at low $p_T$ and a comparable small decrease in the high $p_T$ region.
These features are again shown in the composite Fig.~\ref{fig:k2pi_pos_sum} where the
$p_T$ dependence of $R_{\textrm{K}^+\pi^+}$ from the interpolated data is plotted for
the full range of $x_F$ values.

In Fig.~\ref{fig:k2pi_pos_sum} the "cross-over" point at $p_T \sim$~0.5--0.7~GeV/c where the full 
relative variation of $R_{\textrm{K}^+\pi^+}$ with $x_F$ is on the level of only 20\% of the
ratio, and the
practically parallel evolution of $R_{\textrm{K}^+\pi^+}$ with $p_T$ for different $x_F$
over a wide range of transverse momentum should be pointed out.

$R_{\textrm{K}^-\pi^-}$ is shown in Fig.~\ref{fig:k2pi_neg_pt} as a function of $p_T$ at fixed 
$x_F$ and in Fig.~\ref{fig:k2pi_neg_xf} as a function of $x_F$ for fixed $p_T$.

Also for $R_{\textrm{K}^-\pi^-}$ a strong increase with $p_T$ and the independence
on $x_F$ for low $p_T$ followed by a decrease with $x_F$ at high 
$p_T$ are evident. This is visualized in the composite Fig.~\ref{fig:k2pi_neg_sum} where
the $p_T$ dependence for several $x_F$ values is plotted for the
interpolated data values.

Again a "cross-over" point in $p_T$ with a practically complete $x_F$
independence, for $R_{\textrm{K}^-\pi^-}$ at $p_T \sim$~0.3~GeV/c should be mentioned,
together with the more pronounced decrease at higher $p_T$. A 
general remark concerns the low $p_T$ regions of Figs.~\ref{fig:k2pi_pos_pt}, \ref{fig:k2pi_pos_sum}, 
\ref{fig:k2pi_neg_pt} and ~\ref{fig:k2pi_neg_sum}. The rapid variation of the K/$\pi$ ratios below 
$p_T \sim$~0.2~GeV/c with some minima at $p_T \sim$~0.15~GeV/c are due to the structure of
the $\pi^+$ and $\pi^-$ cross sections observed in this region \cite{pp_pion}. This
structure is more pronounced for $\pi^+$ than for $\pi^-$ and has been
explained by resonance decay \cite{pp_pion,site:S8}.

%
%
\subsection{K/baryon ratios}
\vspace{3mm}
\label{sec:k2p_rat}

\begin{figure}[h]
  \begin{center}
  	\includegraphics[width=9.5cm]{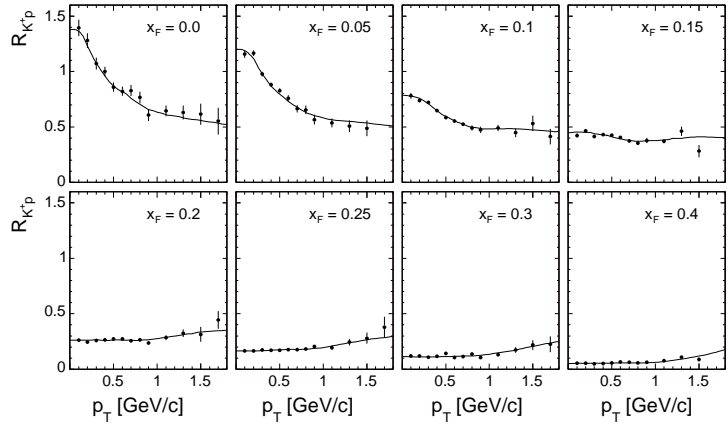}
  	\caption{$R_{\textrm{K}^+\textrm{p}}$ as a function of $p_T$ at fixed $x_F$. The lines 
  	         show the result of the data interpolation, Sect.~\ref{sec:interp}}
  	\label{fig:k2p_pos_pt}
  \end{center}
\end{figure}

\begin{figure}[h]
  \begin{center}
  	\includegraphics[width=9.5cm]{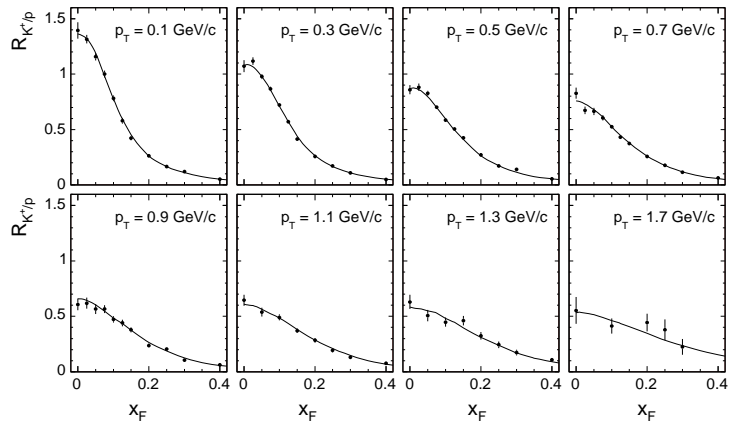}
  	\caption{$R_{\textrm{K}^+\textrm{p}}$ as a function of $x_F$ at fixed $p_T$. The lines 
  	         show the result of the data interpolation, Sect.~\ref{sec:interp}}
  	\label{fig:k2p_pos_xf}
  \end{center}
\end{figure}

\begin{figure}[h]
  \begin{center}
  	\includegraphics[width=10.5cm]{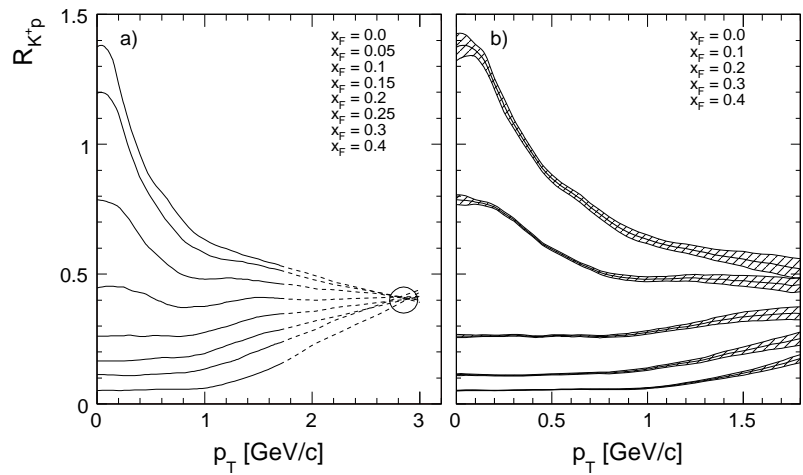}
  	\caption{a) $R_{\textrm{K}^+\textrm{p}}$ for the data interpolation as a function 
             of $p_T$ for different $x_F$; b) Error bands expected for data interpolations}
  	\label{fig:k2p_pos_sum}
  \end{center}
\end{figure}

\begin{figure}[h]
  \begin{center}
  	\includegraphics[width=5cm]{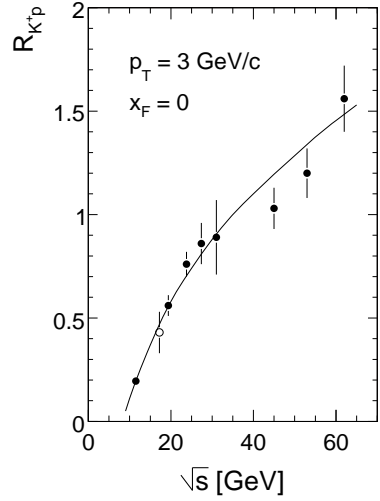}
  	\caption{$s$-dependence of $R_{\textrm{K}^+\textrm{p}}$ at $p_T$~=~3~GeV/c and $x_F$~=~0. 
  	         The open circle corresponds to the NA49 extrapolation, Fig.~\ref{fig:k2p_pos_sum}}
  	\label{fig:k2p_sdep}
  \end{center}
\end{figure}

\begin{figure}[h]
  \begin{center}
  	\includegraphics[width=9.5cm]{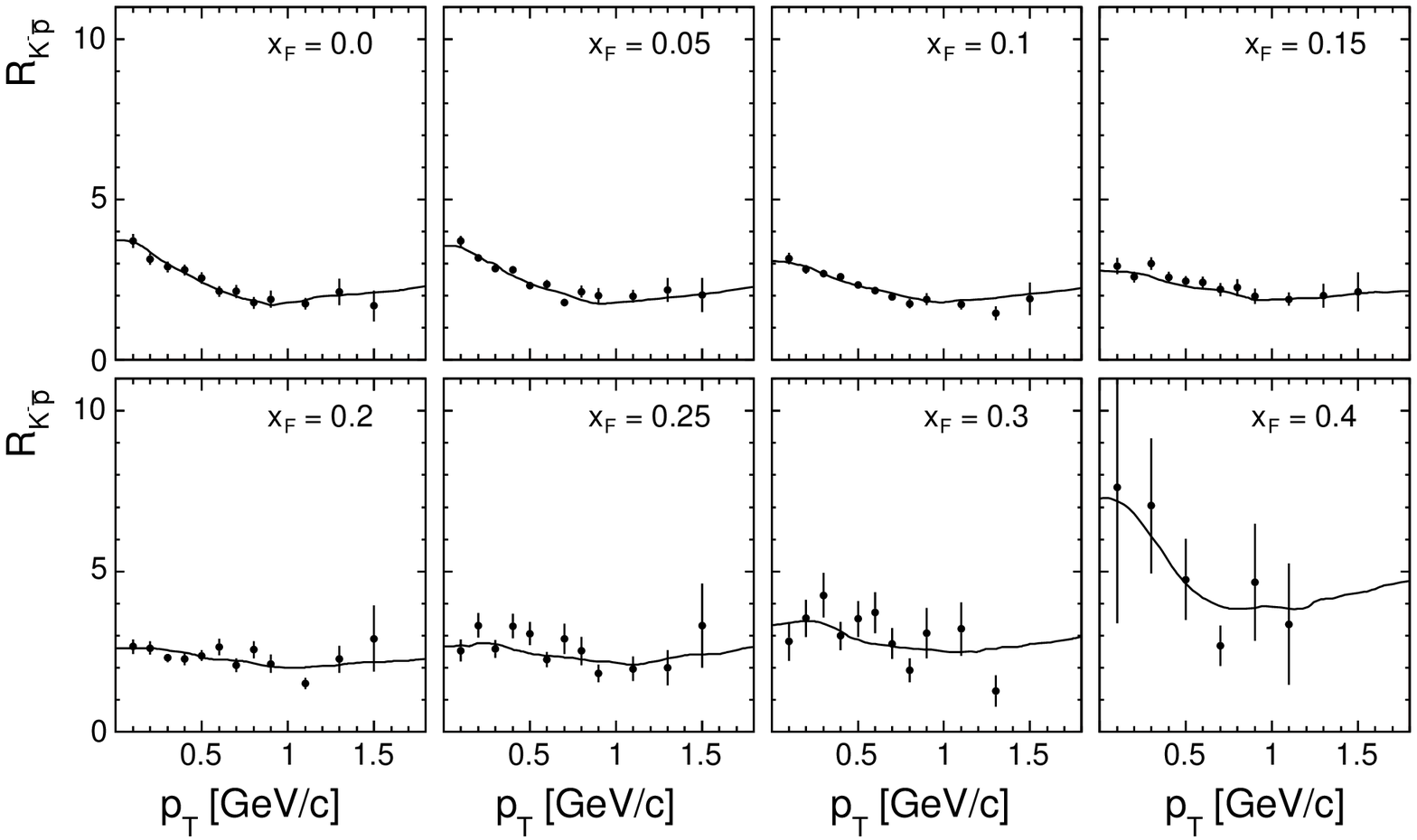}
  	\caption{$R_{\textrm{K}^-\overline{\textrm{p}}}$ as a function of $p_T$ at fixed $x_F$. The lines 
  	         show the result of the data interpolation, Sect.~\ref{sec:interp}}
  	\label{fig:k2p_neg_pt}
  \end{center}
\end{figure}

\begin{figure}[h]
  \begin{center}
  	\includegraphics[width=9.5cm]{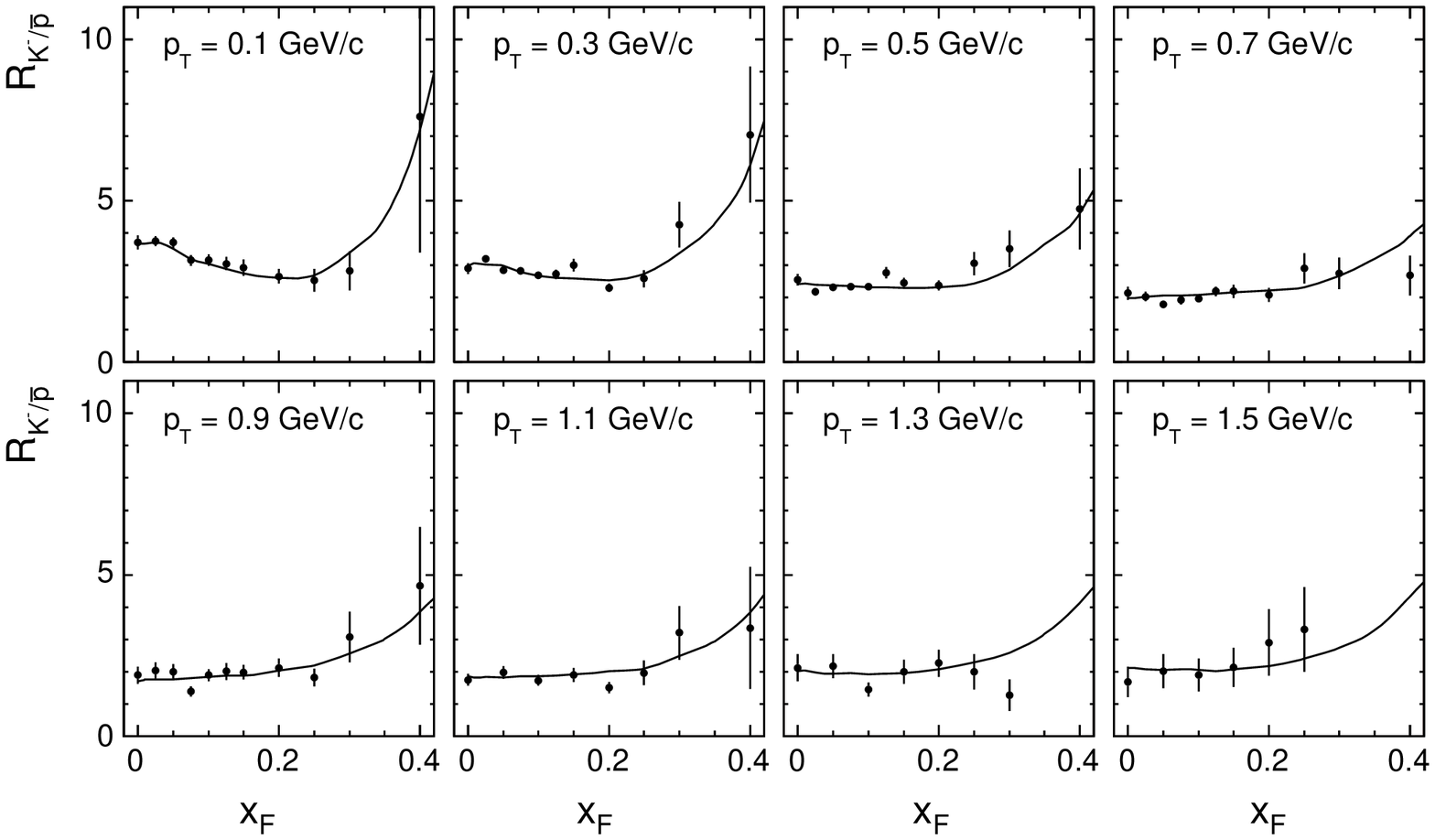}
  	\caption{$R_{\textrm{K}^-\overline{\textrm{p}}}$ as a function of $x_F$ at fixed $p_T$. The lines 
  	         show the result of the data interpolation, Sect.~\ref{sec:interp}}
  	\label{fig:k2p_neg_xf}
  \end{center}
\end{figure}

\begin{figure}[h]
  \begin{center}
  	\includegraphics[width=13.cm]{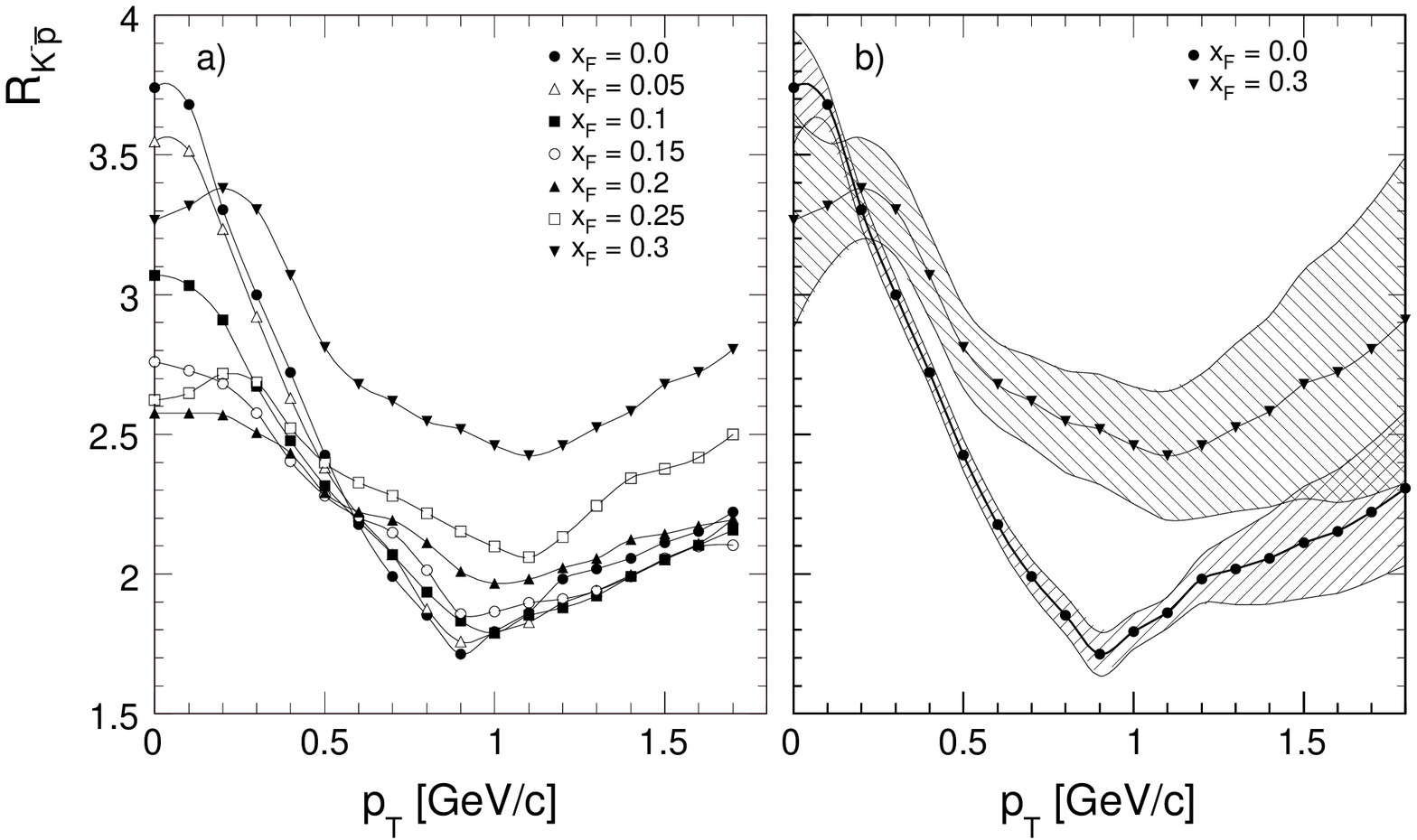}
  	\caption{a) $R_{\textrm{K}^-\overline{\textrm{p}}}$ for the data interpolation as a function 
             of $p_T$ for different $x_F$; b) Error bands expected for data interpolations}
  	\label{fig:k2p_neg_sum}
  \end{center}
\end{figure}

The K/baryon ratios shown below use the new data on proton and
anti-proton production published in \cite{pp_proton}.
The ratios of the invariant inclusive cross sections

\begin{align}
   R_{\textrm{K}^+\textrm{p}} &= f(\textrm{K}^+)/f(\textrm{p}) \\
   \label{eq:kamin}
   R_{\textrm{K}^-\overline{\textrm{p}}} &= f(\textrm{K}^-)/f(\overline{\textrm{p}}) 
\end{align}
are presented in Figs.~\ref{fig:k2p_pos_pt} to \ref{fig:k2p_neg_sum}.

$R_{\textrm{K}^+\textrm{p}}$ is shown in Fig.~\ref{fig:k2p_pos_pt} as a function of $p_T$ 
for fixed $x_F$ and in Fig.~\ref{fig:k2p_pos_xf} as a function of $x_F$ for fixed $p_T$.

Fig.~\ref{fig:k2p_pos_pt} indicates a strong, rapidly decreasing K$^+$ component at low $p_T$ 
and $x_F \lesssim$~0.15, superimposed on an almost $p_T$ independent contribution
which shows a marked decrease with increasing $x_F$ but also a slight
increase with $p_T$ at $x_F >$~0.2. This corresponds to the strong $x_F$
dependence at low $p_T$ in Fig.~\ref{fig:k2p_pos_xf} which flattens out rapidly with
increasing $p_T$. The composite Fig.~\ref{fig:k2p_pos_sum} joins these trends using the
ratio of the data interpolations as a function of $p_T$ for different 
$x_F$ values.

$R_{\textrm{K}^+\textrm{p}}$ seems to converge towards high $p_T$ to an $x_F$ independent value
of about 0.4 -- 0.5 as indicated in Fig.~\ref{fig:k2p_pos_sum} by the dashed extrapolated
lines for the different $x_F$ values. This is reminiscent of a similar behaviour
for the p/$\langle \pi \rangle$ ratio pointed out in \cite{pp_proton}. As the point of convergence
seems to lie close to $p_T \sim$~3~GeV/c it is tempting to use the available
data at this transverse momentum from different $\sqrt{s}$, although the
detailed study of the $s$-dependence of $R_{\textrm{K}^+\textrm{p}}$ is outside the scope of this work.
The analysis of the existing data at $p_T$~=~3~GeV/c and $x_F$~=~0 from Serpukhov
energy \cite{abramov} via Fermilab \cite{cronin} to 
ISR  \cite{albrow1,albrow2,albrow3,capi,alper,guettler}, Fig.~\ref{fig:k2p_sdep}, shows indeed
consistency within errors with the value from the extrapolation shown
above, indicating at the same time the very strong $s$-dependence of
this particle ratio at high $p_T$.

It should be remarked here that the Fermilab data have been
corrected for a systematic effect of 20\% concerning the proton yields
discussed in \cite{pp_proton} and all ISR ratios by 10\% to account for the
expected amount of proton feed-down from strange baryons.    

$R_{\textrm{K}^-\overline{\textrm{p}}}$ is shown in Fig.~\ref{fig:k2p_neg_pt} as a function 
of $p_T$ for fixed $x_F$ and in Fig.~\ref{fig:k2p_neg_xf} as a function of $x_F$ for fixed $p_T$.

As for $R_{\textrm{K}^+\textrm{p}}$ the sizeable $p_T$ dependence at low $x_F$ flattens out
at medium $x_F$, 0.15~$< x_F <$~0.25, and re-appears towards $x_F$~=~0.4.
The $x_F$ dependence, Fig.~\ref{fig:k2p_neg_xf}, is very different from the one of $R_{\textrm{K}^+\textrm{p}}$.
There is no strong enhancement at low $x_F$, $R_{\textrm{K}^-\overline{\textrm{p}}}$ being rather 
independent on $x_F$ up to $x_F \sim$~0.3. Beyond this value the ratio
increases rapidly towards values between 5 and 6 at the maximum
accessible $x_F$. Fig.~\ref{fig:k2p_neg_sum} shows these trends using the ratio of the
interpolated data as a function of $p_T$ for different $x_F$.

The small structures described above are clearly visible, together
with the strong increase at $x_F >$~0.25 and a minimum at
$p_T$ values between 0.9 and 1.1~GeV/c.

%
%
\section{Comparison to Fermilab data}
\vspace{3mm}
\label{sec:comp}

In a first step of data comparison, the NA49 data will be
compared to the existing, double differential cross sections
at neighbouring energies in order to control data consistency
with only small necessary corrections for $s$ dependence.
A wider range of comparisons ranging from kaon threshold to RHIC 
and collider energies will be performed in Sect.~\ref{sec:sdep} below.
For the case of kaons all comparisons are facilitated by the
absence of feed-down corrections from weak decays of strange
particles.

%
%
\subsection{The Brenner et al. data \cite{brenner}}
\vspace{3mm}
\label{sec:comp_bren}

\begin{figure}[h]
  \begin{center}
  	\includegraphics[width=15cm]{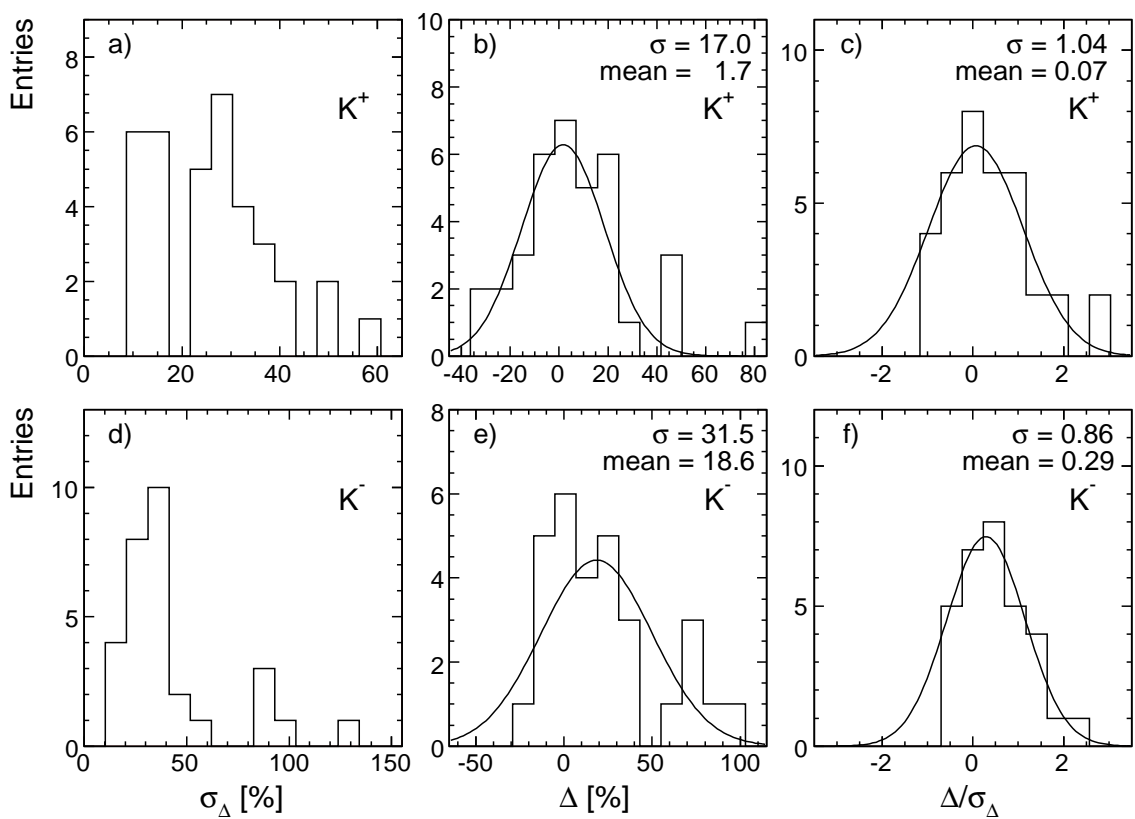}
  	\caption{Statistical analysis of the difference between the measurements of  \cite{brenner} 
  	         and NA49 for K$^+$ (upper three panels) and K$^-$ (lower three panels): 
  	         a) and d) error of the difference of the measurements; b) and e) difference of 
  	         the measurements; c) and f) difference divided by the error}
  	\label{fig:bren_stat}
  \end{center}
\end{figure}

This experiment which has shown a good agreement on the level
of the double differential cross sections for pions \cite{pp_pion} and
baryons \cite{pp_proton}, offers 37 data points for K$^+$ and 32 points for K$^-$ 
at the two beam momenta of 100 and 175~GeV/c. The average
statistical errors of these data are unfortunately rather
large for the kaon samples, with about 25\% for K$^+$ and 40\%
for K$^-$. This is shown in the error distributions of Fig.~\ref{fig:bren_stat},
panels a) and d). Although the $\sqrt{s}$ values at the two
beam energies are, with 13.5 and 18.1~GeV, close to the NA49 
energy, an upwards correction of 8\% (12\%) at the lower energy
and a downwards correction of -2\% (-5\%) at the higher energy
has been applied for K$^+$ and K$^-$, respectively, see Sect.~\ref{sec:sdep}
for a more detailed discussion of $s$ dependence.

The statistical analysis of the differences between the Brenner 
et al. data and the interpolated NA49 results is presented in 
Fig.~\ref{fig:bren_stat}. Although the relative differences, dominated by the
statistical errors of \cite{brenner}, are very sizeable, see panels 
b) and e), the differences normalized to the given statistical 
errors, panels c) and f) show reasonable agreement between the
two data sets, in particular for K$^+$ where the normalized differences 
are centered at $\Delta/\sigma$~=~0 with the expected variance of unity.
The K$^-$ show a positive offset of 0.3 standard deviations which
corresponds to an average difference of 19\%.

\begin{figure}[h]
  \begin{center}
  	\includegraphics[width=15cm]{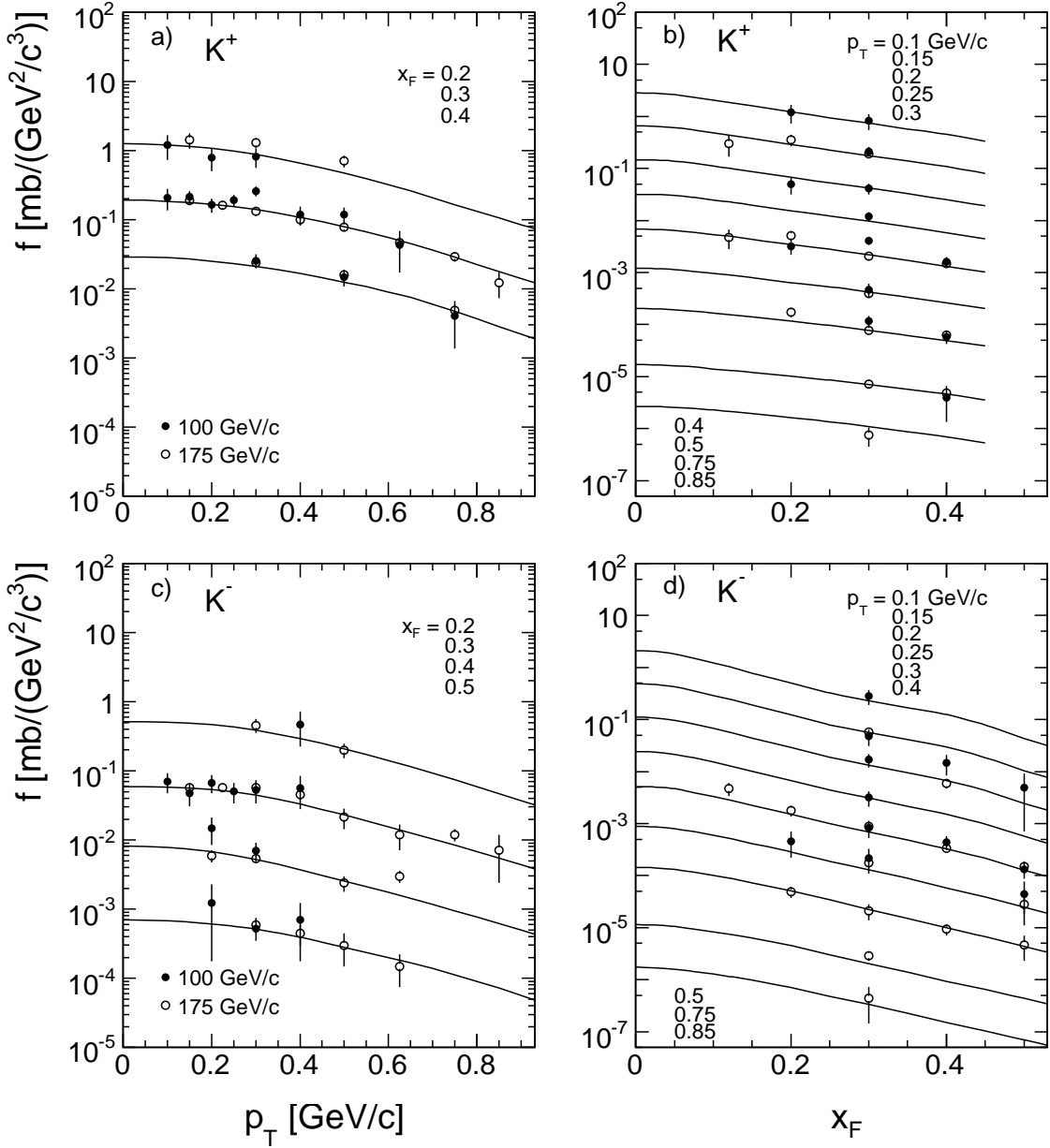}
  	\caption{Comparison of invariant cross section between NA49 (lines) and measurements from 	
  	         \cite{brenner} at 100 (full circles) and 175~GeV/c (open circles) for K$^+$ as a function 
  	         of a) $p_T$ at fixed $x_F$ and b) $x_F$ at fixed $p_T$, and for K$^-$ as a function of 
  	         c) $p_T$ at fixed $x_F$ and d) $x_F$ at fixed $p_T$. The data were successively divided by
  	         4 for better separation}
  	\label{fig:bren_comp}
  \end{center}
\end{figure}

A visualization of the Brenner data with respect to the interpolated
NA49 results and their distribution in the $x_F$ and $p_T$ variables is
given in Fig.~\ref{fig:bren_comp}.

Taking into account the comparison of all measured particle 
species for the two experiments \cite{pp_pion,pp_proton} it may be stated that 
a rather satisfactory overall agreement, within the limits of
the respective systematic and statistical errors, has been 
demonstrated.

%
%
\subsection{The Johnson et al. data \cite{johnson}}
\vspace{3mm}
\label{sec:comp_john}

\begin{figure}[h]
  \begin{center}
  	\includegraphics[width=15cm]{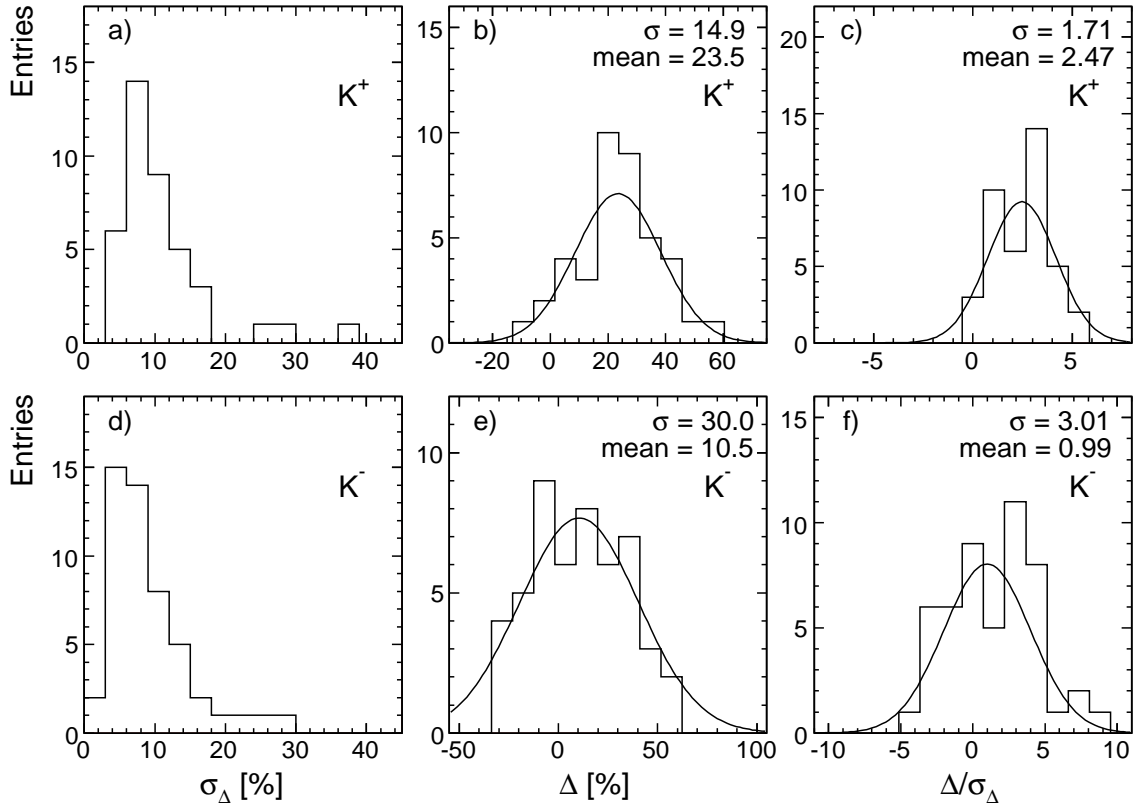}
  	\caption{Statistical analysis of the difference between the measurements of  \cite{johnson} 
  	         and NA49 for K$^+$ (upper three panels) and K$^-$ (lower three panels): 
  	         a) and d) error of the difference of the measurements; b) and e) difference of 
  	         the measurements; c) and f) difference divided by the error}
  	\label{fig:john_stat}
  \end{center}
\end{figure}

This experiment gives 40 data points for K$^+$ and 50 points for K$^-$
within the range of the NA49 data obtained at 100, 200 and 400~GeV/c
beam momentum. For comparison purposes the data have been corrected
to 158~GeV/c beam momentum using the $s$-dependence established in
Sect.~\ref{sec:sdep} below. The distribution of the relative statistical 
errors is shown in Fig.~\ref{fig:john_stat} panel a) for K$^+$ and panel d) for K$^-$, with 
mean values of 12\% and 9\%, respectively. This is substantially below
the errors of \cite{brenner}.

The statistical analysis of the differences with respect to the
interpolated cross sections of NA49 is also given in Fig.~\ref{fig:john_stat} in terms
of the distribution of the relative difference $\Delta$, panels b),e)
and of the difference normalized to the statistical error $\Delta/\sigma$,
panels c) and f). Two main features are apparent from this comparison:
an upwards shift of about 23\% (10\%) corresponding to 2.5 (1.0)
standard deviations and large fluctuations corresponding to 1.7 (3.0)
standard deviations for K$^+$ and K$^-$, respectively. As similar observations
have been made for pions \cite{pp_pion} and baryons \cite{pp_proton} one may state that
a general offset of 10 -- 20\% seems to be present which is compatible
with the normalization uncertainty given in \cite{johnson}. The fact 
that the proton data show a smaller offset might be connected with
their $x_F$ coverage which is mostly at large negative $x_F$ (low lab momenta). 
On the other hand, the underestimation of the point by point 
fluctuations by a factor of 2 to 4 with respect to the claimed
statistical errors, for all particle species, has to remain unresolved. 

The phase space distribution of the data of \cite{johnson} is shown
in Fig.45 as a function of $x_F$ at fixed values of $p_T$ in comparison
with the interpolated NA49 cross sections.

\begin{figure}[h]
  \begin{center}
  	\includegraphics[width=15.cm]{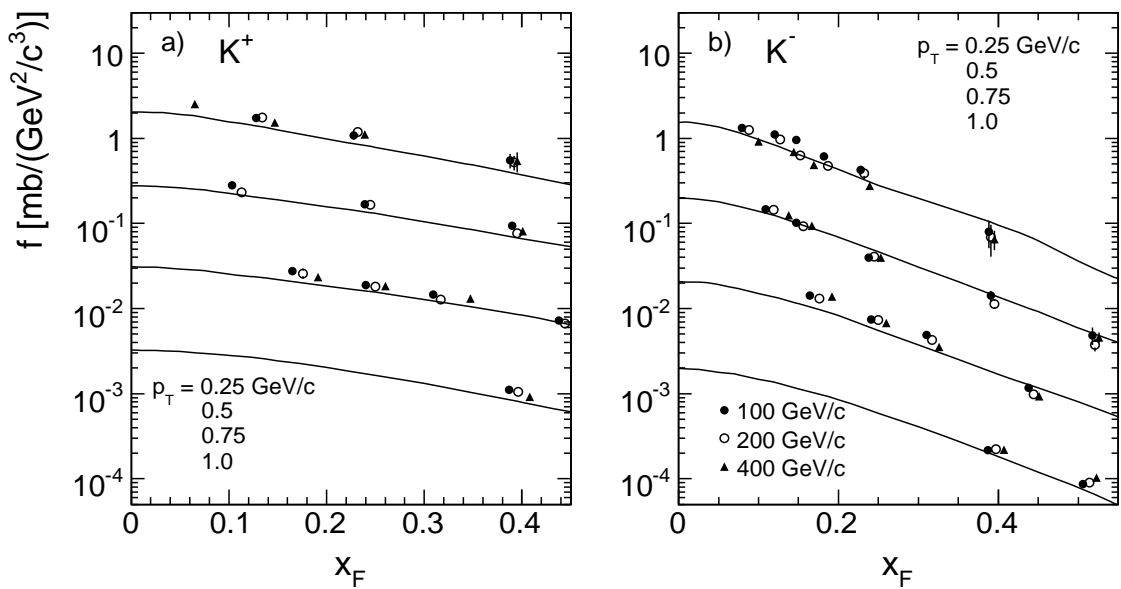}
  	\caption{Comparison of invariant cross section between NA49 (lines) and measurements from 	
  	         \cite{johnson} at 100 (full circles), 200 (open circles) and 400~GeV/c (full triangles) 
  	         as a function of $x_F$ at fixed $p_T$ for a) K$^+$ and b) K$^-$. The data 
  	         were successively divided by 3 for better separation}
  	\label{fig:john_comp}
  \end{center}
\end{figure}
%
%
\subsection{The Antreasyan et al. data \cite{cronin}}
\vspace{3mm}
\label{sec:comp_cron}

It is only the low-$p_T$ part of this experiment which can be compared
to the NA49 data, at $x_F$ close to 0. Due to the fact that the
spectrometer of \cite{cronin} was set to a constant lab angle for all beam
energies and particle species, the given cross sections have to
be compared at their proper $x_F$ values as given in Table~\ref{tab:cron_comp}, see
also the corresponding arguments in \cite{pp_proton}.

\begin{table}[h]
\footnotesize
\begin{center}
\begin{tabular}{ccr@{}lr@{}lr@{}l}
\hline
 \multirow{2}{16mm}{$p_T$~[GeV/c]} 
  &$p_{\textrm{beam}}$ [GeV/c]&\multicolumn{2}{c}{200}&\multicolumn{2}{c}{300}&\multicolumn{2}{c}{400}\\        
  &    $\sqrt{s}$ [GeV]    & \multicolumn{2}{c}{19.3}& \multicolumn{2}{c}{23.7}& \multicolumn{2}{c}{27.3} \\
  \hline 
 \multirow{3}{16mm}{0.77}                                            
  &    $x_F$                &  -0.0&054            &   -0.0&11         &  -0.0&20    \\ 
  &   $R_{\textrm{K}^+}$   &  0.826&$\pm$0.12     & 1.026&$\pm$0.16   &  1.110&$\pm$0.20  \\  
  &   $R_{\textrm{K}^-}$   &  0.966&$\pm$0.12     & 1.217&$\pm$0.18   &  1.164&$\pm$0.18  \\ 
     \hline 
 \multirow{3}{16mm}{1.54}                                            
  &    $x_F$               &     0.0&302          &    -0.0&31        &   -0.0&20    \\
  &    $R_{\textrm{K}^+}$  & 0.796&$\pm$0.05      & 1.080&$\pm$0.08   & 1.260&$\pm$0.12  \\  
  &    $R_{\textrm{K}^-}$  & 0.791&$\pm$0.06      & 1.240&$\pm$0.06   & 1.616&$\pm$0.14  \\ \hline
 \end{tabular}
\end{center}
\caption{Offset in $x_F$ at different $\sqrt{s}$ and $p_T$. The cross section 
         ratio $R_{\textrm{K}^{\pm}}$ between the data from \cite{cronin} and NA49.}
\label{tab:cron_comp}
\end{table}

\begin{figure}[h]
  \begin{center}
  	\includegraphics[width=8.8cm]{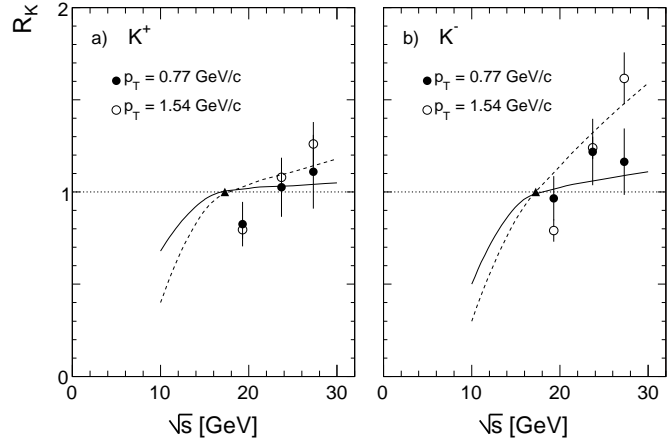}
  	\caption{The cross section ratios between the data from \cite{cronin} and NA49  
  	         as a function of $\sqrt{s}$ for two values of $p_T$ for a) K$^+$ and b) K$^-$. 
  	         In both of the panels the NA49 point is indicated with full triangle. The full and 
             dashed lines represent the result of the $s$-dependence at $x_F$~=~0 established 
             in Sect.~\ref{sec:sdep} below at $p_T$~=~0.77 and 1.54~GeV/c, respectively }
  	\label{fig:cron_comp}
  \end{center}
\end{figure}

The cross section ratios $R_{\textrm{K}^+}$ and $R_{\textrm{K}^-}$ are shown in Fig.~\ref{fig:cron_comp} as
a function of $\sqrt{s}$ at fixed $p_T$, together with the $s$-dependence
extracted in Sect.~\ref{sec:sdep} below from data at $x_F$~=~0 at Serpukhov
energy \cite{abramov} and ISR energy \cite{alper,guettler}. 

Evidently the data \cite{cronin} comply, within their sizeable statistical
errors, with the $s$-dependence as established by the other experiments.
However, three of the four points at 200~GeV/c beam momentum are low
by about two standard deviations. This would, by using the data \cite{cronin}
alone to establish the $s$-dependence, lead to a large underestimation
of the kaon yields at lower $s$. See also the discussion in \cite{pp_proton}
for baryons.

%
%
\subsection{Comparison of particle ratios}
\vspace{3mm}
\label{sec:comp_rat}

\begin{figure}[b]
  \begin{center}
  	\includegraphics[width=15cm]{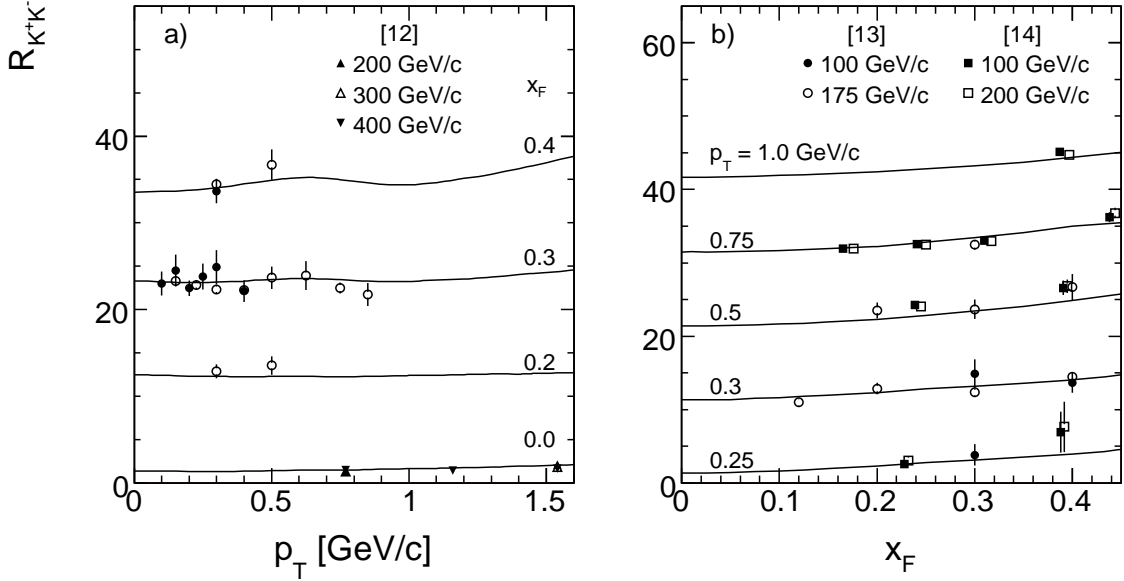}
  	\caption{Comparison of $R_{\textrm{K}^+\textrm{K}^-}$ between \cite{cronin} (triangles), 
             \cite{brenner} (circles), \cite{johnson} (squares) and NA49 (lines) as a function 
             of a) $x_F$ and b) $p_T$. The data 
  	         were successively shifted by 10 for better separation}
  	\label{fig:p2m_comp}
  \end{center}
\end{figure}

\begin{figure}[t]
  \begin{center}
  	\includegraphics[width=15.cm]{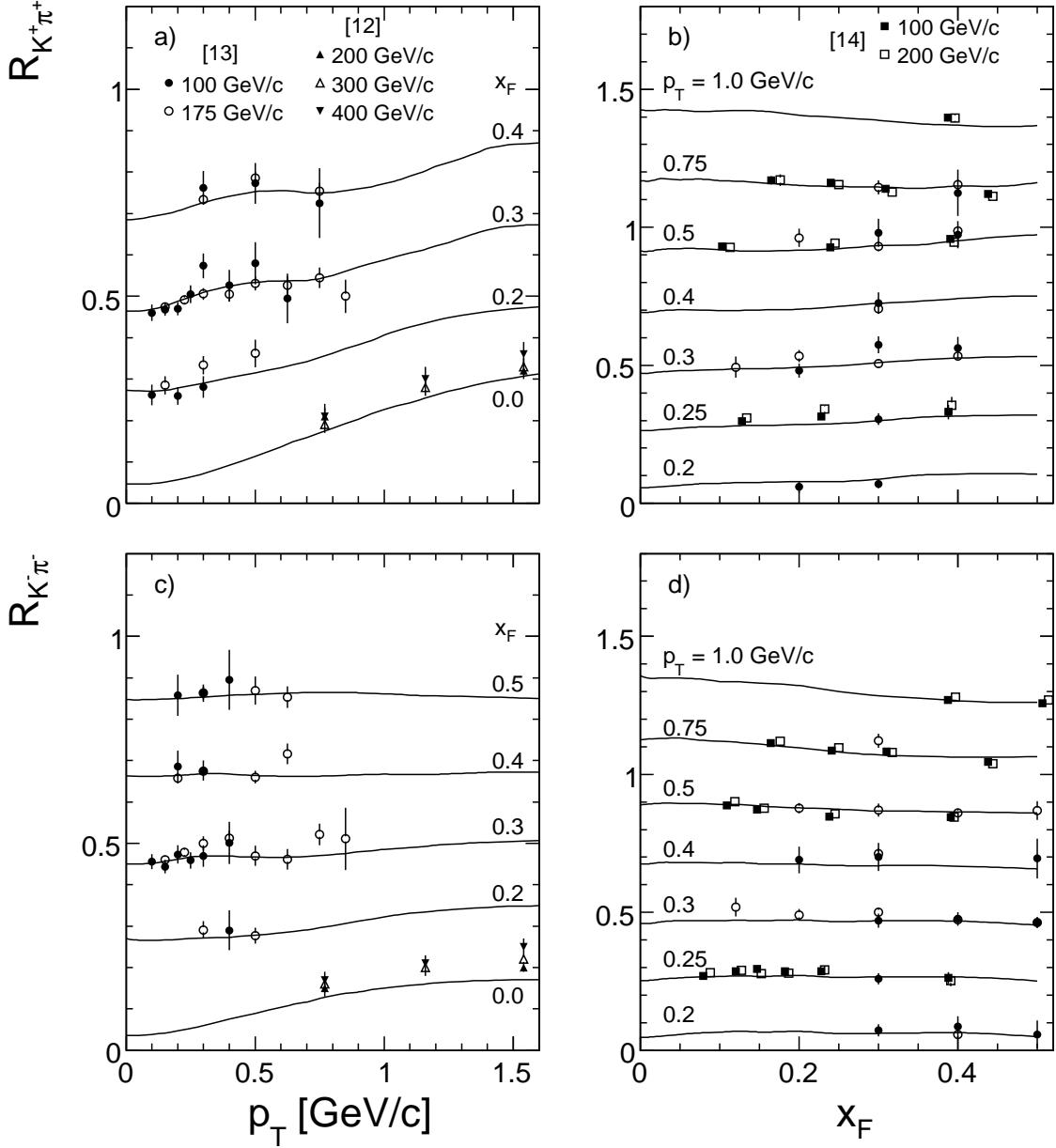}
  	\caption{Comparison between \cite{cronin} (triangles), \cite{brenner} (circles), 
             \cite{johnson} (squares) and NA49 (lines) of $R_{\textrm{K}^+\pi^+}$ as a 
             function of a) $p_T$ and b) $x_F$ and $R_{\textrm{K}^-\pi^-}$ as a function 
             of c) $p_T$ and d) $x_F$. The data were successively shifted by 0.2 for better separation}
  	\label{fig:k2pi_comp}
  \end{center}
\end{figure}

\begin{figure}[h]
  \begin{center}
  	\includegraphics[width=15.cm]{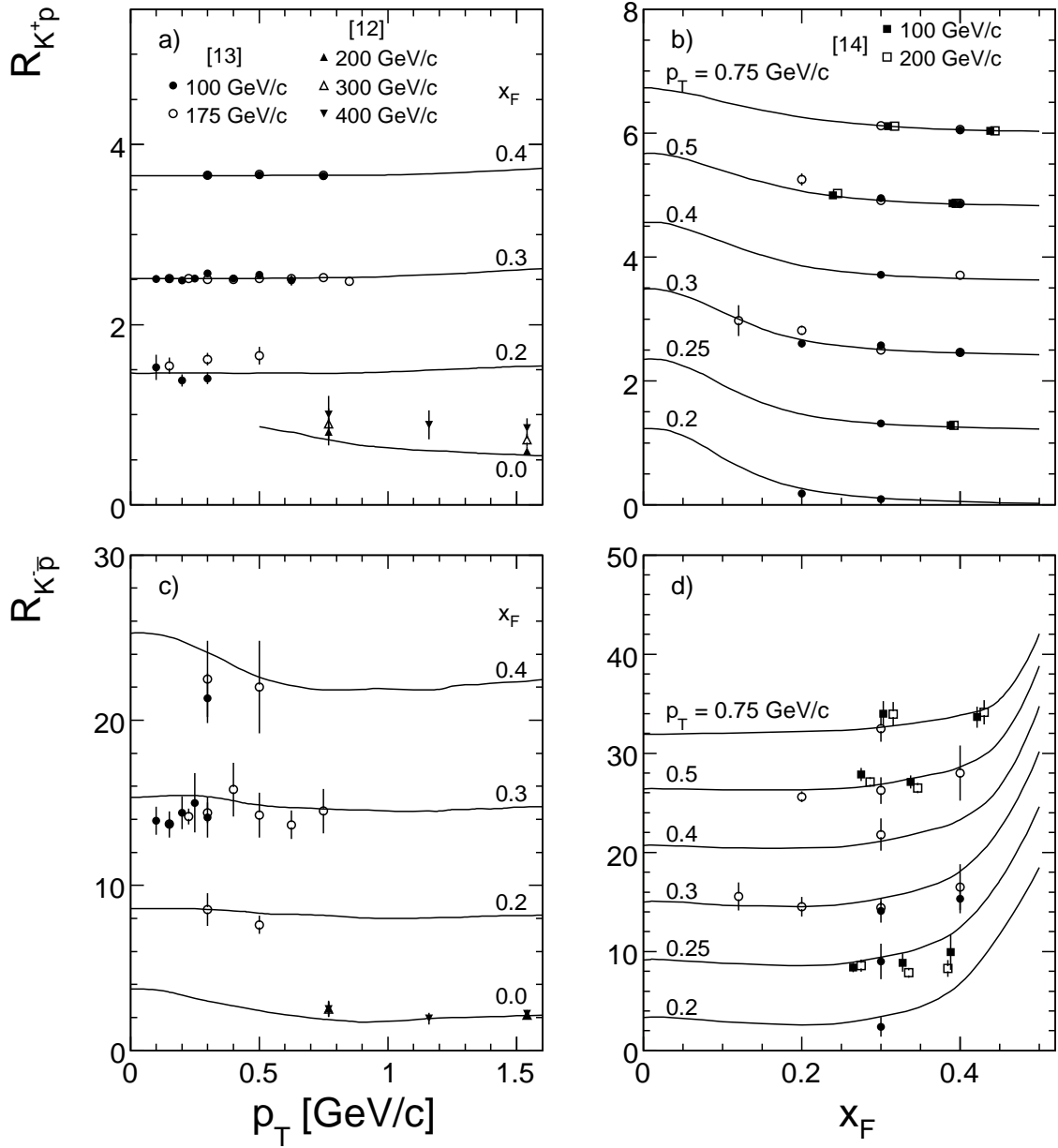}
  	\caption{Comparison between \cite{cronin} (triangles), \cite{brenner} (circles), 
             \cite{johnson} (squares) and NA49 (lines) of $R_{\textrm{K}^+\textrm{p}}$ as a function 
             of a) $p_T$ and b) $x_F$ and $R_{\textrm{K}^-\overline{\textrm{p}}}$ as a function 
             of c) $p_T$ and d) $x_F$. The data 
  	         were successively shifted by 1.2 for $R_{\textrm{K}^+\textrm{p}}$ and
             by 6 for $R_{\textrm{K}^-\overline{\textrm{p}}}$ for better separation}
  	\label{fig:k2p_comp}
  \end{center}
\end{figure}

As systematic effects tend in general to be reduced
in particle ratios, it is interesting to also look at the consistency
of the corresponding ratios from \cite{cronin,brenner,johnson} with the NA49 data
shown in Sect.~\ref{sec:ratios} of this paper. This is shown in Fig.~\ref{fig:p2m_comp}
for $R_{\textrm{K}^+\textrm{K}^-}$
ratios, in Fig.~\ref{fig:k2pi_comp} for $R_{\textrm{K}^\pm\pi^\pm}$  
and in Fig.~\ref{fig:k2p_comp} for K/baryon ratios.

%
%
\subsection{Conclusion from data comparison at Fermilab energies}
\vspace{3mm}
\label{sec:comp_concl}

In conclusion of the detailed comparisons in the Fermilab/SPS
energy range shown above it may be stated that a mutually consistent
picture for kaon production from several independent experiments 
has been established, with the exception of some offsets in the absolute
cross section especially for \cite{cronin} and \cite{johnson}. These offsets tend to
cancel in the particle ratios $R_{\textrm{K}^+\textrm{K}^-}$ for both
\cite{cronin} and \cite{johnson}. The ratios $R_{\textrm{K}\pi}$ and
$R_{\textrm{Kp}}$ are consistent for \cite{brenner} and \cite{johnson} within
their statistical uncertainties, whereas for \cite{cronin} the systematic
effects discussed in \cite{pp_proton} for baryons and in Sect.~\ref{sec:comp_cron}
for kaons persist for $R_{\textrm{K}\pi}$ and $R_{\textrm{Kp}}$. What is also
important to note is the apparent absence of systematic deviations as a function
of kinematic variables $x_F$ and $p_T$. This lends, as none of the existing experiments
has on its own sufficient phase space coverage, some confidence to
the establishment of $p_T$ integrated and total yields from the NA49
measurements alone, as discussed below.

%
%
\section{Integrated data}
\vspace{3mm}
\label{sec:ptint}

In a first step the data interpolation, Sect.~\ref{sec:interp}, will be used 
to perform an integration over transverse momentum. In a second
step the total charged kaon yields will be determined. These can
be used, in conjunction with the total pion and baryon yields
published before \cite{pp_pion,pp_proton} to control the total charged multiplicity
with respect to the precision data from bubble chamber experiments.

%
%
\subsection{$\boldsymbol{p_T}$ integrated distributions}
\vspace{3mm}
\label{sec:ptint_dist}

The $p_T$ integrated non-invariant and invariant kaon yields are
defined by:

\begin{align}
\label{eq:int}
  \frac{dn}{dx_F} &= \frac{\pi}{\sigma_{\textrm{inel}}} \frac{\sqrt{s}}{2} \int{\frac{f}{E}\; dp_T^2} \nonumber \\
  F &= \int{f \; dp_T^2}  \\
  \frac{dn}{dy} &= \frac{\pi}{\sigma_{\textrm{inel}}} \int{f \; dp_T^2} \nonumber  
\end{align} 
with $f=E \cdot d^3\sigma/dp^3$, the invariant double differential cross section.
The integrations are performed numerically using the two-dimensional
data interpolation (Sect.~\ref{sec:interp}) which is available in steps of 
0.05~GeV/c in transverse momentum.

\vspace{3mm}
\begin{table}[h]
\scriptsize
\renewcommand{\tabcolsep}{0.11pc} 
\renewcommand{\arraystretch}{1.1} 
\begin{center}
\begin{tabular}{|l|cc|cc|cc|cc||cc|cc|cc|cc||c|c|c|}
\hline
 & \multicolumn{8}{|c||}{K$^+$ } &\multicolumn{8}{|c||}{K$^-$ } & & K$^+$  & K$^-$  \\
\hline
 $x_F$& $F$& $\Delta$& $dn/dx_F$& $\Delta$& $\langle p_T \rangle$& $\Delta$
      & $\langle p_T^2 \rangle$ & $\Delta$
      & $F$& $\Delta$& $dn/dx_F$& $\Delta$& $\langle p_T \rangle$& $\Delta$
      & $\langle p_T^2 \rangle $& $\Delta$
      & $y$& $dn/dy$ & $dn/dy$ \\ \hline
 0.0   & 0.6715  & 1.17 & 0.8531   & 1.30 & 0.4157 & 0.65 & 0.2427 & 1.15 &
         0.4762  & 1.04 & 0.6166   & 1.50 & 0.4002 & 0.68 & 0.2227 & 1.29 &
 0.0   & 0.06635 & 0.04729 \\ 
 0.01  & 0.6688  & 1.54 & 0.8417   & 1.49 & 0.4165 & 0.78 & 0.2435 & 1.39 &
         0.4760  & 1.90 & 0.6096   & 1.78 & 0.4007 & 0.87 & 0.2228 & 1.60 &
 0.2   & 0.06597 & 0.04693 \\ 
 0.025 & 0.6648  & 0.87 & 0.7985   & 0.78 & 0.4208 & 0.53 & 0.2480 & 1.04 &
         0.4644  & 0.82 & 0.5666   & 0.76 & 0.4053 & 0.45 & 0.2277 & 0.91 &
  0.4  & 0.06496 & 0.04536 \\
 0.05  & 0.6344  & 0.73 & 0.6633   & 0.67 & 0.4343 & 0.40 & 0.2629 & 0.86 &
         0.4286  & 0.66 & 0.4547   & 0.63 & 0.4198 & 0.45 & 0.2427 & 1.03 &
  0.6  & 0.06258 & 0.04216 \\
 0.075 & 0.5906  & 0.63 & 0.5260   & 0.59 & 0.4509 & 0.39 & 0.2814 & 0.76 &
         0.3745  & 0.68 & 0.3364   & 0.72 & 0.4378 & 0.35 & 0.2627 & 0.70 &
  0.8  & 0.05910 & 0.03819 \\
  0.1  & 0.5374  & 0.64 & 0.4077   & 0.61 & 0.4657 & 0.42 & 0.2990 & 0.91 &
         0.3210  & 0.70 & 0.2449   & 0.68 & 0.4542 & 0.46 & 0.2815 & 0.99 &
  1.0  & 0.05458 & 0.03339 \\
 0.125 & 0.4923  & 0.81 & 0.3219   & 0.80 & 0.4776 & 0.47 & 0.3136 & 0.89 &
         0.2730  & 0.90 & 0.1792   & 0.90 & 0.4690 & 0.46 & 0.2989 & 0.88 &
  1.2  & 0.04904 & 0.02803 \\
  0.15 & 0.4449  & 0.87 & 0.2542   & 0.86 & 0.4881 & 0.49 & 0.3261 & 1.02 &
         0.2267  & 1.05 & 0.1298   & 1.04 & 0.4826 & 0.63 & 0.3156 & 1.30 &
  1.4  & 0.04328 & 0.02227 \\
  0.2  & 0.3614  & 1.07 & 0.1635   & 1.06 & 0.5037 & 0.69 & 0.3449 & 1.58 &
         0.1568  & 1.31 & 0.07101  & 1.22 & 0.5006 & 0.71 & 0.3369 & 1.44 &
  1.6  & 0.03680 & 0.01677 \\
  0.25 & 0.2965  & 1.79 & 0.1104   & 1.70 & 0.5127 & 0.94 & 0.3551 & 1.77 &
         0.1041  & 1.90 & 0.03880  & 2.10 & 0.5080 & 1.08 & 0.3460 & 2.19 &
  1.8  & 0.03030 & 0.01182 \\
  0.3  & 0.2373  & 1.90 & 0.07491  & 1.90 & 0.5184 & 0.97 & 0.3620 & 1.98 &
         0.07112 & 2.53 & 0.02248  & 2.53 & 0.5059 & 1.44 & 0.3445 & 2.83 &
  2.0  & 0.02389 & 0.00774 \\
  0.4  & 0.1481  & 1.44 & 0.03569  & 1.43 & 0.5259 & 0.96 & 0.3705 & 1.92 &
         0.03296 & 2.51 & 0.007959 & 2.51 & 0.4933 & 1.49 & 0.3305 & 2.88 &
  2.2  & 0.01788 & 0.00485 \\      
  0.5  &  &&  &&  &&  &&
         0.01370 & 5.36 & 0.002667 & 5.37 & 0.5117 & 2.87 & 0.3440 & 4.37 &  
  2.4  & 0.01253 & 0.00281 \\
       & &&  &&  &&  &&  
         &&  &&  &&  &&  
  2.6  & 0.00754 & 0.00139 \\ 
       & &&  &&  &&  &&  
         &&  &&  &&  &&  
  2.8  & 0.00353 & 0.00045 \\ 
       & &&  &&  &&  &&  
         &&  &&  &&  &&  
  3.0  & 0.00113 & 0.00007 \\  \hline
\end{tabular}
\end{center}
\caption{$p_T$ integrated invariant cross section $F$ [mb$\cdot$c],
         density distribution $dn/dx_F$, mean transverse momentum $\langle p_T \rangle $
         [GeV/c], mean transverse momentum squared $\langle p_T^2 \rangle $ 
         [(GeV/c)$^2$] as a function of $x_F$, as well as density distribution 
         $dn/dy$ as a function of $y$ for K$^+$ and K$^-$. The statistical 
         uncertainty $\Delta$ for each quantity is given in \% as
         an upper limit considering the full statistical
         error of each measured $p_T$/$x_F$ bin}
\label{tab:integr}
\end{table}

The statistical uncertainties of the integrated quantities
given in Table~\ref{tab:integr} are upper limits obtained by using the full 
statistical fluctuations over the measured bins. As such 
they are equivalent, for the kaon yields, to the statistical 
error of the total number of kaons contained in each $x_F$ bin.
    
\begin{figure}[h]
  \begin{center}
  	\includegraphics[width=15cm]{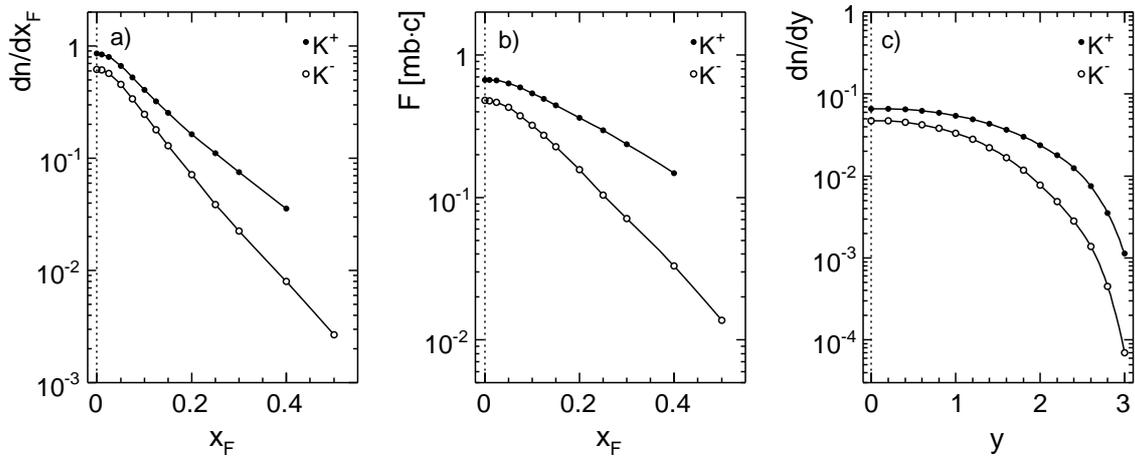}
  	\caption{Integrated distributions of K$^+$ and K$^-$ produced in p+p 
             interactions at 158~GeV/c:
             a) density distribution $dn/dx_F$ as a function of $x_F$;
             b) invariant cross section $F$ as a function of $x_F$;
             c) density distribution $dn/dy$ as a function of $y$}
  	\label{fig:ptint}
  \end{center}
\end{figure}

\begin{figure}[h]
  \begin{center}
  	\includegraphics[width=16cm]{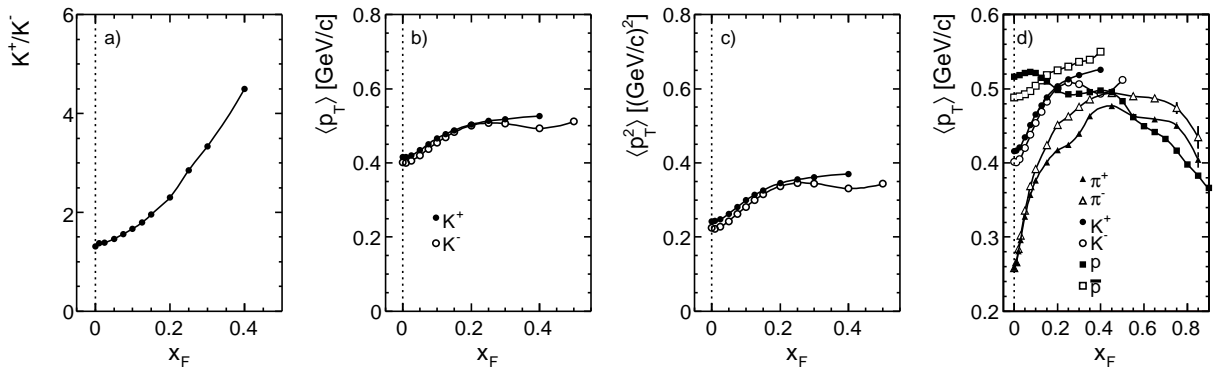}
  	\caption{a) K$^+$/K$^-$ ratio, b) mean $p_T$, and c) mean $p_T^2$ as a function 
             of $x_F$ for K$^+$ and  K$^-$ produced in p+p interactions at 158~GeV/c; 
             d) mean $p_T$ for $\pi^+$, $\pi^-$, K$^+$, K$^-$, p, $\overline{\textrm{p}}$
             on an enlarged vertical scale}
  	\label{fig:ratint}
  \end{center}
\end{figure}

The resulting distributions are shown in Fig.~\ref{fig:ptint} for K$^+$ and
K$^-$ as a function of $x_F$ and $y$. The relative statistical errors of all quantities 
are generally below the percent level. They increase towards the high end
of the available $x_F$ region essentially defined by the available
event number and, especially for K$^+$, by limits concerning particle
identification (Sect.~\ref{sec:pid}). The K$^+$/K$^-$ ratio, $\langle p_T \rangle$ and 
$\langle p_T^2 \rangle$ for kaons as a function of $x_F$ are presented in Fig.~\ref{fig:ratint}a--c.
Fig.~\ref{fig:ratint}d shows the mean transverse momentum 
of all measured particle species in a single panel in order to allow
a general overview of the interesting evolution of this quantity
with $x_F$ which demonstrates that $\langle p_T \rangle$ is equal to within 0.05~GeV/c for
all particles at $x_F \sim$~0.3 -- 0.4.

%
%
\subsection{Comparison to other data}
\vspace{3mm}
\label{sec:ptint_comp}

\begin{figure}[t]
  \begin{center}
  	\includegraphics[width=12.cm]{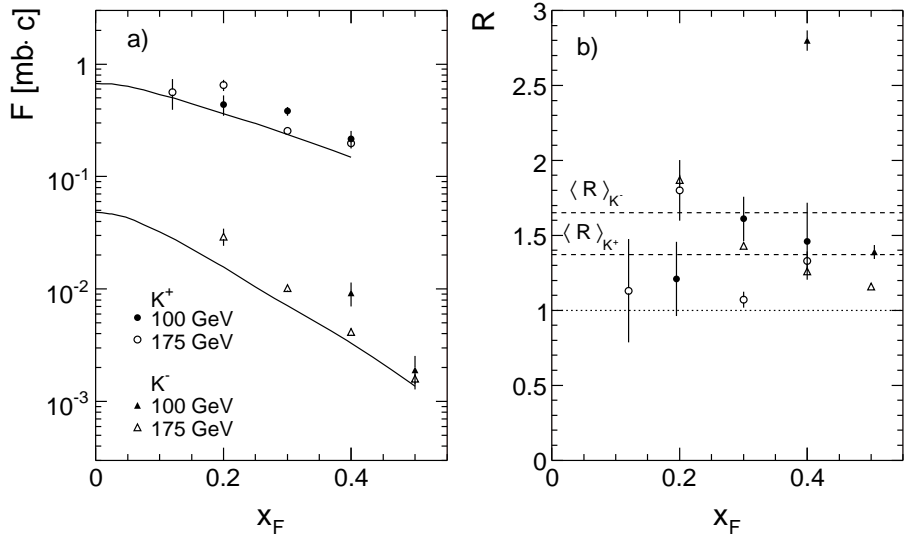}
  	\caption{a) Comparison of $p_T$ integrated invariant cross section $F$ as a function of 
  	         $x_F$ for K$^+$ and K$^-$ measured by \cite{brenner} to 
  	         NA49 results (represented as lines). The data for K$^-$ are multiplied by 0.1; 
             b) Ratio $R$ between measurements of \cite{brenner} and NA49 results. The
             mean ratios for K$^+$ and K$^-$ are presented with dashed lines}
  	\label{fig:bren_int_comp}
  \end{center}
\end{figure}

As in Sect.~\ref{sec:comp}, a first stage of the comparison is limited to the 
SPS/Fermilab energy range where only two experiments provide integrated
cross sections. The data of Brenner et al. \cite{brenner} are obtained from
a limited set of double differential cross sections, using basically
exponential fits to the measured points. The resulting invariant
cross sections $F(x_F)$ are shown in Fig.~\ref{fig:bren_int_comp} in comparison to the NA49
data.

As already remarked for the case of pions and protons, very
sizeable deviations are visible in the distributions of Fig.~\ref{fig:bren_int_comp}a,
which are quantified by the ratio of the two measurements shown
in Fig.~\ref{fig:bren_int_comp}b. If the relative differences in $F$ were limited to about
$\pm$40\% for pions and protons \cite{pp_pion,pp_proton}, the factors are even bigger
for kaons, with a mean deviation of about 50\%. 
This again demonstrates the danger of using oversimplified algebraic parametrizations of
double differential data which comply with the NA49 measurements on the point-by-point level
within their statistical errors (Sect.~\ref{sec:comp_bren}).

\begin{figure}[h]
  \begin{center}
  	\includegraphics[width=11.cm]{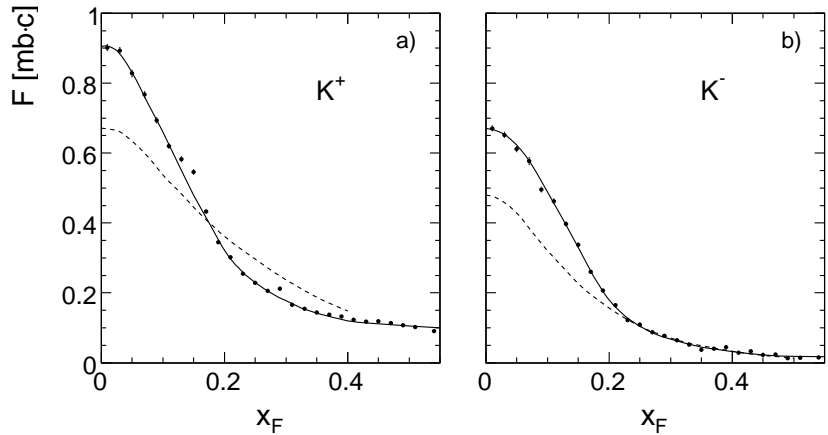}
  	\caption{Comparison of $p_T$ integrated invariant cross section $F$ as a function of 
             $x_F$ for a) K$^+$ and b) K$^-$  measured by \cite{ehs} to NA49 results 
             (represented as dashed lines) }
  	\label{fig:ehs_comp}
  \end{center}
\end{figure}

The EHS experiment \cite{ehs} at the CERN SPS, using a 400~GeV/c proton
beam, offers $p_T$ integrated data which are directly comparable in
all quantities defined in Eq.~\ref{eq:int}. In view of the $s$-dependence which
is enhanced at low $x_F$ in the quantity $dn/dx_F$ \cite{pp_pion,pp_proton} and of the
important shape change to be expected in the rapidity distributions,
only the invariant integrated cross section $F$ is plotted in Fig.~\ref{fig:ehs_comp}
in comparison to the NA49 data.

\begin{figure}[h]
  \begin{center}
  	\includegraphics[width=11cm]{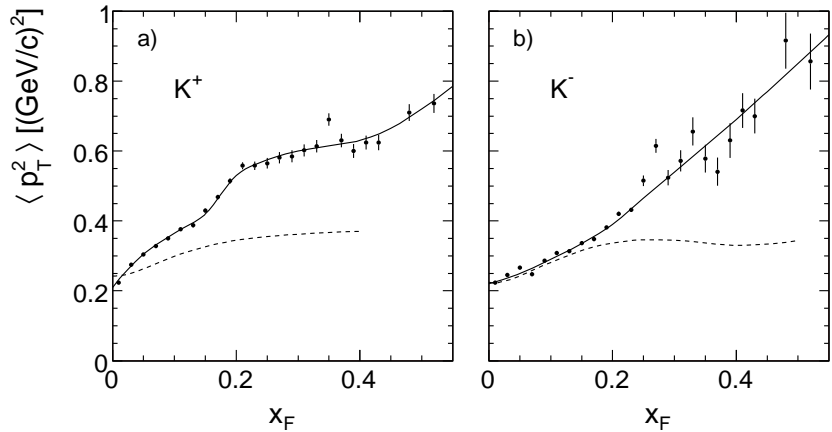}
  	\caption{Comparison $\langle p_T^2 \rangle$ as a function of $x_F$ for a) K$^+$ and 
             b) K$^-$  measured by \cite{ehs} to NA49 results 
             (represented as dashed lines) }
  	\label{fig:ehs_mpt2}
  \end{center}
\end{figure}

Some remarks are in place here. The EHS K$^+$ data show an enhancement
at low $x_F$ of about 35\% which is substantially above the expected 
$s$-dependence, see also the discussion of the kaon data in 
Sect.~\ref{sec:over}. After a local deviation from a smooth $x_F$ dependence  
at $x_F \sim$~0.15 the distribution cuts, however, below the NA49 data 
in the region 0.175~$< x_F <$~0.45. This decrease cannot be explained 
by any known $s$-dependence. For K$^-$ the situation is qualitatively
similar. Here at $x_F$~=~0 an enhancement of 41\% is observed, with
an $x_F$ dependence which smoothly approaches the NA49 data to become
equal to these cross sections within errors at $x_F >$~0.22. Again
such behaviour contradicts the expected $s$-dependence. A possible
explanation might be contained in the mean $p_T^2$ data shown in
Fig.~\ref{fig:ehs_mpt2}. If the results on $\langle p_T^2 \rangle$ agree at $x_F$~=~0 
within the respective errors, the EHS 
data deviate rapidly upwards from the NA49 measurements with
increasing $x_F$. For K$^+$ the instability in the cross sections at
$x_F$~=~0.15 is seen as a break in the $x_F$ dependence of $\langle p_T^2 \rangle$
at the same $x_F$ value. The $x_F$ dependence then flattens in the
region 0.2~$< x_F <$~0.45 which corresponds to the depletion of the 
cross section, rising again steeply to very large values at $x_F$
beyond the range accessible to NA49. A similar behaviour is
observed for K$^-$ where $\langle p_T^2 \rangle$ shows reasonable consistency
up to $x_F \sim$~0.2 with a slight increase over the NA49 data which
is however inconsistent with the $s$-dependence in Sect.~\ref{sec:coll_mpt}.
Above $x_F \sim$~0.2,
however, there is again a strong almost linear increase of $\langle p_T^2 \rangle$
with $x_F$ with values in excess of 0.8~(GeV/c)$^2$ in the high $x_F$
region. One may speculate that both the behaviour of the invariant
cross sections and the one of $\langle p_T^2 \rangle$ are of the same origin if
one assumes that there are detection losses for kaons with 
increasing $x_F$ and at transverse momenta below the mean value. 
This would reduce the observed cross sections and enhance 
the mean $p_T^2$.

In conclusion of the comparisons with the EHS experiment which
have been carried out with some precision for pions \cite{pp_pion}, 
baryons \cite{pp_proton} and here for kaons, a somewhat unsatisfactory and
partially inconsistent picture emerges. In general it may be
stated that sizeable relative differences, even after taking
into account possible $s$-dependences, emerge at a level of
typically $\pm$10 -- 30\% which cannot be explained by a common factor
like normalization uncertainties. In addition there seems to be
a general tendency of unphysical behaviour in the EHS data for
$x_F$ values above about 0.2 both in the cross sections and, more
extremely, for the behaviour of mean $p_T^2$.
 
%
%
\subsection{Total kaon yields and mean charged multiplicity}
\vspace{3mm}
\label{sec:total}

For the $x_F$ integration of the $dn/dx_F$ distributions presented
in Table~\ref{tab:integr} an exponential extrapolation into the unmeasured
region at high $x_F$ has been used. This is well justified by the
shape of the distributions within the measured region and by
the fact that only 4\% (0.3\%) of the total yields are beyond
the experimental limits for K$^+$ and K$^-$, respectively. The
resulting total kaon yields are:

\begin{equation}
	\begin{split}
     \langle n_{\textrm{K}^+} \rangle  &= 0.2267 \\
     \langle n_{\textrm{K}^-} \rangle   &= 0.1303 \\
     \langle n_{\textrm{K}^+} \rangle/\langle n_{\textrm{K}^-} \rangle  &= 1.740 \\
     \frac{\langle n_{\textrm{K}^+} \rangle + \langle n_{\textrm{K}^-} \rangle}{2}  &= 0.1785
	\end{split}
\end{equation}

The statistical errors of these yields may be estimated
by the total number of kaons extracted from the 4.8M
events of this experiment. These are 260k for K$^+$ and
170k for K$^-$. From these numbers follows, including the
additional statistical errors from particle identification,
Sect.~\ref{sec:dedx_stat} Fig.~\ref{fig:statfactor}, an error of 0.27\% for K$^+$ 
and 0.28\% for K$^-$ which is about one order of magnitude below the smallest
estimated systematic error (Table~\ref{tab:sys}).
    
These numbers, together with the results for pions \cite{pp_pion} and
baryons \cite{pp_proton} can be used to establish the mean charged multiplicity
as it results from this experiment. The respective numbers are
given in Table~\ref{tab:mean_ch} below.

\begin{table}[h]
\renewcommand{\arraystretch}{1.2} 
\normalsize
\begin{center}
\begin{tabular}{cccc}
\hline
                                         & positives  &  negatives &  total    \\ \hline
     $\langle n_{\pi} \rangle$           &    3.018   &    2.360   &  5.378    \\
     $\langle n_{\textrm{K}} \rangle$    &    0.227   &    0.130   &  0.357    \\
     $\langle n_{\textrm{p}} \rangle$    &    1.162   &    0.039   &  1.201    \\ \hline
     $\langle n \rangle$                 &    4.407   &    2.529   &  6.936    \\ \hline
\end{tabular}
\end{center}
\caption{Mean multiplicities of charged particles}
\label{tab:mean_ch}
\end{table}

In order to establish the total charged multiplicity and to
be able to compare to the results from bubble chamber work
where the charged hyperons are included as on-vertex tracks,
an estimation of $\Sigma^+$ and $\Sigma^-$ yields has to be performed.
Several measurements of $\Sigma^+$, $\Sigma^-$ and $\Sigma^0$ are available
in the energy range 3~$< \sqrt{s} <$~27~GeV, all with rather big
relative statistical errors of typically 15 to 50\%. For the
present purpose where the charged hyperons constitute a
correction of about 1\%, this is nevertheless acceptable since
all results stem from bubble chamber experiments with small
systematic uncertainties. Three quantities are interesting
and necessary for the present comparison:

\begin{enumerate}
  \item the $\Sigma^0$/$\Lambda$ ratio 
  \item the $\Sigma^+$/$\Sigma^-$ ratio
  \item the ratio ($\Sigma^-$ + $\Sigma^0$ + $\Sigma^+$)/$\Lambda$
\end{enumerate}

The $\Sigma^0$/$\Lambda$ ratio has been obtained by 5 experiments
\cite{ansorge,alex,bierman,jaeger1,ammosov} with values between
0.1 and 0.74 with an average of 0.4. This value may be
used to obtain the ratio ($\Sigma^-$ + $\Sigma^0$ + $\Sigma^+$)/$\Lambda$
\cite{fesef,alpgard1,kichimi} which varies between 0.83 and 1.09
with an average of 0.99. The ratio $\Sigma^+$/$\Sigma^-$ 
\cite{jaeger1,fesef,alpgard1,kichimi,oh}shows a variation from 2 to 5.2 
with an average of 3.3.

\begin{figure}[h]
  \begin{center}
  	\includegraphics[width=6.5cm]{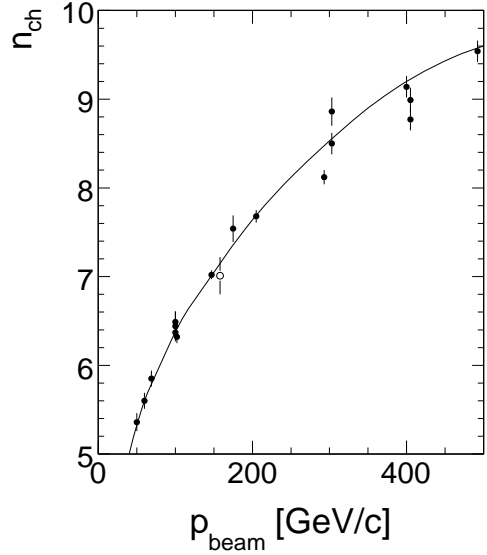}
  	\caption{$\langle n_{\textrm{ch}} \rangle$ as a function of beam momentum $p_{\textrm{beam}}$.
             The NA49 measurement is indicated with an open circle}
  	\label{fig:tot_ch}
  \end{center}
\end{figure}

Adopting the average values for the ratios (1) and (3) the
combined yield $\Sigma^+$ + $\Sigma^-$ may be obtained at $\sqrt{s}$~=~17.2~GeV 
by interpolating the well-established total yield of $\Lambda$ 
\cite{eisner,jaeger1,blobel,fesef,boggild,alpgard1,bartke,bogo,ammosov,
      alston,chapman,brick,allday,jaeger2,sheng,lopinto,dao,bailly,kass,kichimi}
to $\langle n_{\Lambda} \rangle$~=~0.12 per inelastic
event at this energy. This results in a contribution of 0.07
per inelastic event from charged hyperons and gives a total
charged multiplicity

\begin{equation}
  \langle n_{\textrm{ch}} \rangle = 7.01
\end{equation}
from this experiment. This multiplicity may be compared to the existing
measurements essentially from Bubble Chamber experiments taken
from \cite{wroblewski} and presented in Fig.~\ref{fig:tot_ch}.

The full line in Fig.~\ref{fig:tot_ch} represents a hand interpolation of the
measurements in the range from 50 to 300~GeV/c beam momentum.
It coincides incidentally, at $\sqrt{s}$~=~17.2~GeV, with the 
parametrization

\begin{equation}
   \langle n_{\textrm{ch}} \rangle = -4.8 + 10/\sqrt{s} + 2.0\ln{s}
\end{equation}
given by \cite{whitmore} which predicts

\begin{equation}
   \langle n_{\textrm{ch}} \rangle = 7.15
\end{equation}

The relative deviation of the summed integrated yields given 
above from this value corresponds to -2\%. It is certainly
governed by the systematic uncertainties of the dominant
pion and proton yields for which the systematic error estimation
\cite{pp_pion,pp_proton} gave 4.8\% (5\%) for the linear sum and 2\% (2.5\%) for the
more optimistic quadratic sum of the contributions,respectively.
Allowing for a typical error of about 1\% of the bubble chamber
data, it may be stated that the observed deviation is within the
error estimate for the NA49 data.

At this point it is indicated to also check the charge balance
of the NA49 results where the difference between positive and
negative particle yields should give two units from charge
conservation. Using the total charged hyperon yield estimated
above and the average $\Sigma^+$/$\Sigma^-$ ratio of 3.3 the following
yields are obtained:

\begin{equation}
  \begin{split}
     \langle n_{\Sigma^+} \rangle &= 0.054 \\
     \langle n_{\Sigma^-} \rangle &= 0.016  \\
     \langle n_{\textrm{pos}} \rangle    &= 4.461  \\
     \langle n_{\textrm{neg}} \rangle    &= 2.545  \\
     \langle n_{\textrm{pos}} \rangle - \langle n_{\textrm{neg}} \rangle  &= 1.916
  \end{split}
\end{equation}

This means that the charge balance is off by 0.08 units or about
4\% of its nominal value. In order to put this number into
perspective it should be realized that a systematic downwards
deviation of the $\pi^+$ yield by 1.5\% accompanied by an upward shift
of the $\pi^-$ yield by the same relative amount is sufficient to 
explain this imbalance. Therefore it may be stated that also 
the charge conservation of the NA49 results is established within 
the stated systematic errors. 

%
%
\section{A new evaluation of $\boldsymbol{s}$ dependence}
\vspace{3mm}
\label{sec:sdep}

The new set of kaon data presented and discussed above has
been used, in connection with existing data at other cms
energies, to re-assess the experimental situation as far 
as the $s$-dependence, in particular also for integrated yields,
is concerned. It is indeed rather surprising that the very
first attempt in this direction by Rossi et al. \cite{rossi}
which dates from 1975, is still being used as a reference
for rather far-reaching conclusions with respect to kaon 
production in heavy ion interactions \cite{na61}. This is 
especially concerning the admitted systematic uncertainties
which are given in \cite{rossi} as only 15\% for their estimated
total yields. In view of the rather sparse phase space coverage
of most of the preceding data sets, see Sect.~\ref{sec:exp_sit} and Fig.~\ref{fig:phs_kap}
above, it is in fact for most cms energies quite difficult  
to establish integrated yields with 
defendable reliability. In this context it is interesting to also 
look at the available data on K$^0_S$ production which, coming for the
$s$-range up to medium ISR energies exclusively from bubble chamber
experiments, have well defined systematic errors in particular for
integrated yields, notwithstanding their in general rather 
limited statistical significance. Here, the relation between
charged and neutral kaon production deserves special attention
as it is directly sensitive to the respective production mechanisms.
In the following section, five energy ranges from Cosmotron
up to RHIC and collider energies will be inspected in an attempt
at establishing some coherence with respect to $s$-dependence.

%
%
\subsection{The K$^+$ data of Hogan et al. \cite{hogan} and 
            Reed et al. \cite{reed} at $\boldsymbol{\sqrt{s}}$~=~2.9~GeV}
\vspace{3mm}
\label{sec:hogan-reed}

\begin{figure}[h]
  \begin{center}
  	\includegraphics[width=12cm]{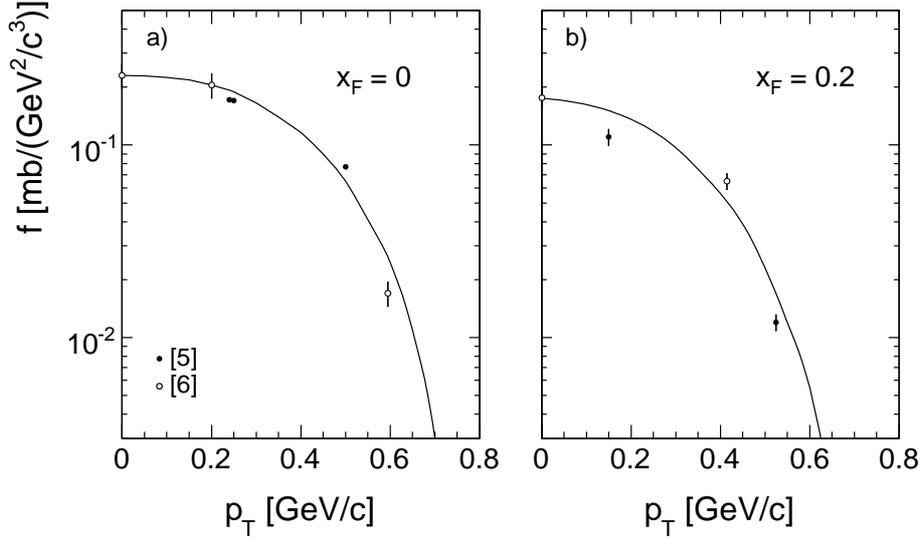}
  	\caption{$f$ as a function of $p_T$ for a) $x_F$~=~0 and b) $x_F$~=~0.2.
             The data points from Hogan et al. \cite{hogan} and Reed et al. \cite{reed}
             are given together with the data interpolation (full lines)}
  	\label{fig:hr}
  \end{center}
\end{figure}

\begin{figure}[h]
  \begin{center}
  	\includegraphics[width=5.5cm]{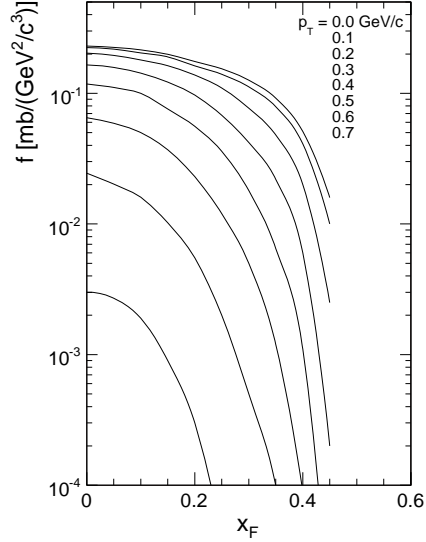}
  	\caption{Interpolated invariant cross sections as a 
             function of $x_F$ for fixed values of $p_T$}
  	\label{fig:hr_xf}
  \end{center}
\end{figure}

These early experiments at the Princeton-Penn (PPA) and
BNL Cosmotron accelerators use a range of beam momenta 
from 3.2 to 3.9~GeV/c with a common point at
about 3.7~GeV/c. The data at this energy have been used in
order to establish a maximum of combined phase space coverage 
in the ranges 0~$< x_F <$~0.4 and 0~$< p_T <$~0.6~GeV/c, see Fig.~\ref{fig:phs_kap}a. It should be 
mentioned here that the definition of $x_F$ (Eq.~\ref{eq:xf}) has been used
throughout although it progressively limits the available
$x_F$ range at low interaction energies due to energy-momentum
conservation, see \cite{allaby2} for a detailed discussion. At
$\sqrt{s}$~=~3~GeV this means rather sharp cut-offs in production
cross section towards $x_F \sim$~0.5 and $p_T \sim$~0.7~GeV/c. Within these
limits, reasonable inter- or extra-extrapolation may be performed
in order to establish approximate $p_T$ and $x_F$ dependences.
It should be stressed that throughout this paper no arithmetic
parametrizations of $x_F$ or $p_T$ distributions have been used as
those would introduce large systematic biases which are difficult to control. 
Instead, two-dimensional interpolation by multi-step eyeball
fits, as discussed in Sect.~\ref{sec:interp} above, have been applied.
Two examples of this procedure are shown in Fig.~\ref{fig:hr} for 
selected $x_F$ and $p_T$ values, where the available, interpolated
or slightly extrapolated data points are indicated. The 
resulting interpolation of cross sections over the complete $x_F$ 
and $p_T$ ranges is presented in Fig.~\ref{fig:hr_xf}.

\begin{figure}[h]
  \begin{center}
  	\includegraphics[width=13cm]{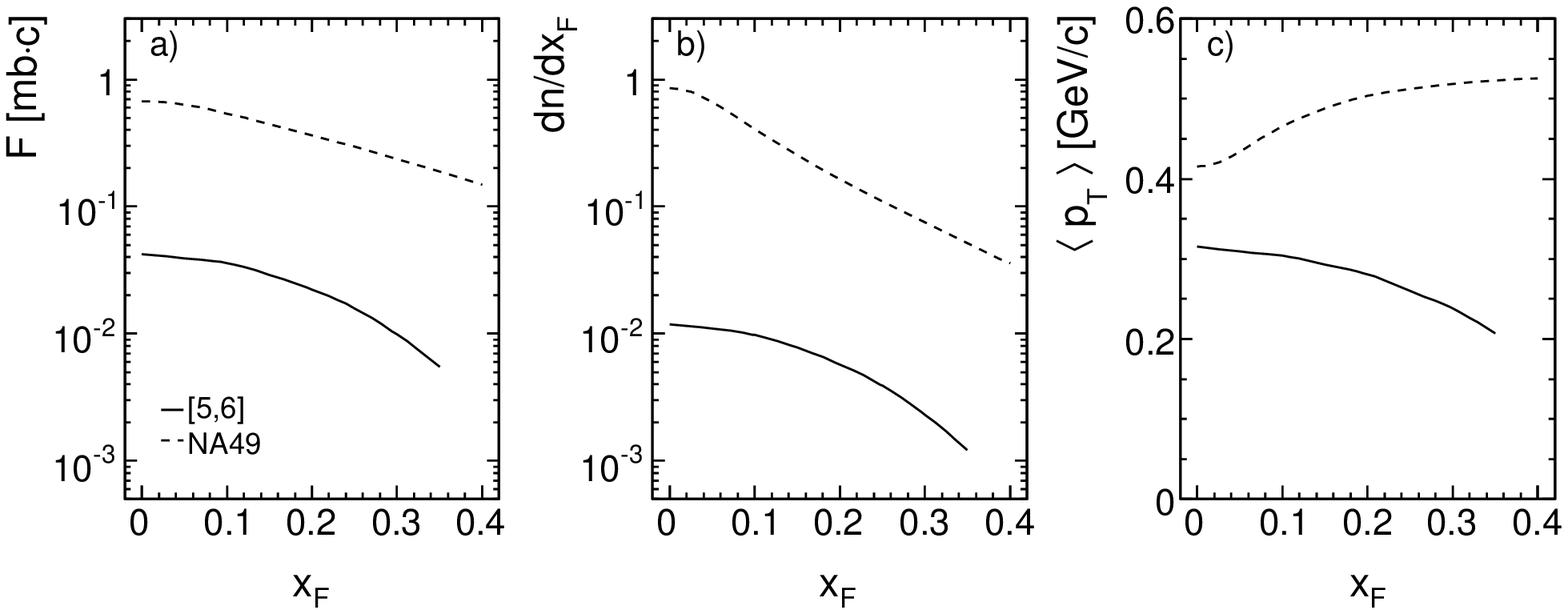}
  	\caption{$p_T$ integrated a) $F$, b) $dn/dx_F$ and c) $\langle p_T \rangle$ distributions as
             a function of $x_F$. The results obtained at $\sqrt{s}$~=~17.2~GeV (dashed lines)
             are also shown for comparison}
  	\label{fig:hr_ptint}
  \end{center}
\end{figure}

The interpolation shown in Fig.~\ref{fig:hr_xf} may be $p_T$ integrated in order
to obtain the $F$, $dn/dx_F$ and $\langle p_T \rangle$ dependences shown in Fig.~\ref{fig:hr_ptint}.

In a second step the integration over $x_F$ may be performed resulting
in an average K$^+$ multiplicity of $\langle n_{\textrm{K}^+} \rangle$~=~0.00481. 
This value is 8.3\% (6.0\%) higher than the multiplicities 
$\langle n_{\textrm{K}^+} \rangle$~=~0.00441$\pm$17\% and 
$\langle n_{\textrm{K}^+} \rangle$~=~0.00452$\pm$23\% 
obtained by \cite{hogan} and \cite{reed}, respectively. 
These groups imposed isotropy (S-wave decay) in the cms system in order 
to be able to carry out the data integration. There is also a 
bubble chamber experiment from the BNL Cosmotron at the same beam 
momentum \cite{louttit} which gives 
$\langle n_{\textrm{K}^+} \rangle$~=~0.00462$\pm$19\% for the K$^+$ 
multiplicity which is only 4\% lower than the result obtained above. 
In conclusion a statistically consistent K$^+$ yield from 3 independent 
experiments may be claimed at $\sqrt{s}$~=~3~GeV/c which is about 55\% above 
the one elaborated in \cite{rossi}.  

\begin{figure}[h]
  \begin{center}
  	\includegraphics[width=6.5cm]{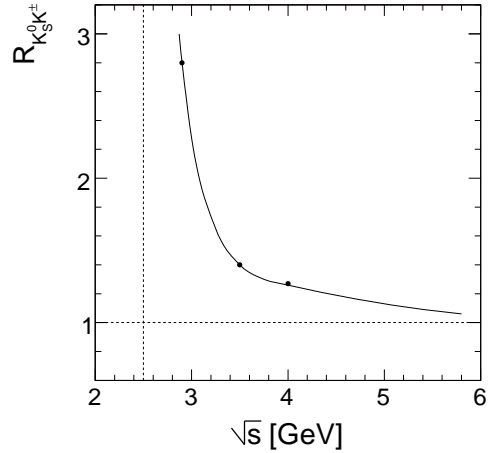}
  	\caption{Ratio $R_{\textrm{K}^0_S\textrm{K}^\pm}$ between the average charged
             kaon and $\textrm{K}^0_S$ yields as a function of $\sqrt{s}$. The
             threshold of kaon production is indicated at about
             $\sqrt{s} \sim$~2.5~GeV}
  	\label{fig:kr}
  \end{center}
\end{figure}

The bubble chamber experiment \cite{louttit} also gives the K$^0$ multiplicity
as $\langle n_{\textrm{K}^0} \rangle$~=~0.00165. With the usual assumption 
$\langle n_{\textrm{K}^0_S} \rangle$~=~0.5$\langle n_{\textrm{K}^0} \rangle$
this corresponds to $\langle n_{\textrm{K}^0_S} \rangle$~=~0.000824. The ratio

\begin{equation}
   \label{eq:rk}
   R_{\textrm{K}^0_S\textrm{K}^\pm} = \frac{0.5\,(\langle n_{\textrm{K}^+} \rangle + 
                                                 \langle n_{\textrm{K}^-} \rangle )} 
                                          {\langle n_{\textrm{K}^0_S} \rangle} 
\end{equation}
is therefore, with $\langle n_{\textrm{K}^+} \rangle$ from \cite{louttit} as given above, 2.8
which is substantially above the value $R_{\textrm{K}^0_S\textrm{K}^\pm}$~=~1
expected from isospin invariance. Inspecting the $\textrm{K}^0_S$
and $\textrm{K}^\pm$ data of \cite{alex} and \cite{fire}, $R_{\textrm{K}^0_S\textrm{K}^\pm}$ 
is determined to 1.4 at $\sqrt{s}$~=~3.5~GeV and 1.27 at $\sqrt{s}$~=~4~GeV.
This indicates a steep deviation from isospin invariance
in kaon production as the threshold is approached from
above, as shown in Fig.~\ref{fig:kr}.

Evidently $R_{\textrm{K}^0_S\textrm{K}^\pm}$ approaches unity rather quickly with 
increasing energy so that $R_{\textrm{K}^0_S\textrm{K}^\pm}$~=~1 may be assumed within
a few percent error margin at $\sqrt{s} >$~5~GeV, see Sect.~\ref{sec:kzero}
below for a more detailed discussion.

\begin{figure}[h]
  \begin{center}
  	\includegraphics[width=11cm]{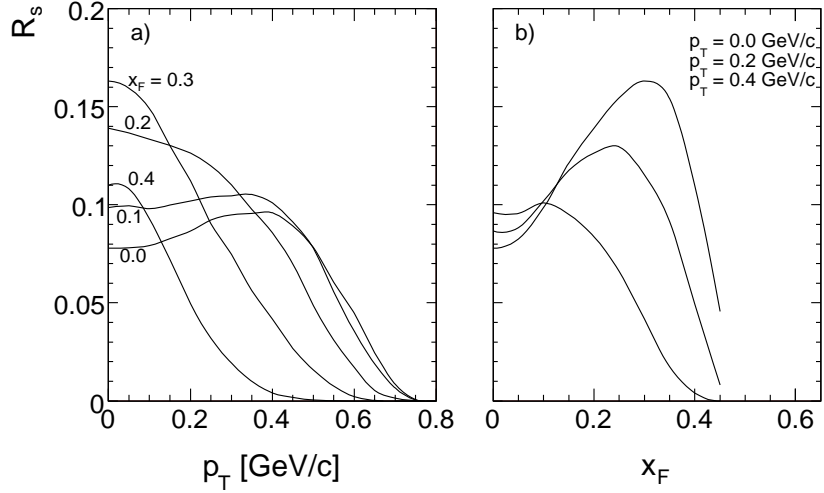}
  	\caption{Ratio $R_s$ as a function of a) $p_T$ at fixed $x_F$ and b) $x_F$ at fixed $p_T$ }
  	\label{fig:hr_rat}
  \end{center}
\end{figure}

It is also interesting to compare the differential data of \cite{hogan} and \cite{reed} 
directly to the NA49 data. The ratio of the invariant inclusive cross sections,

\begin{equation}
   R_s = \frac{f( x_F, p_T, \sqrt{s} = \textrm{3~GeV} )}{f( x_F, p_T, \sqrt{s} = \textrm{17.2~GeV} )}
\end{equation}
is shown in Fig.~\ref{fig:hr_rat} as a function of $p_T$ at constant $x_F$ and as a 
function of $x_F$ at constant values of $p_T$. 

Evidently the total yield ratio of 0.021 does not translate into
a common suppression factor for the differential distributions
but the local cross section ratios show a strong and complex
dependence on the kinematical variables. If the complete suppression 
of K$^+$ production for $p_T \gtrsim$~0.7~GeV/c and $x_F \gtrsim$~0.5 is a trivial
consequence of energy-momentum conservation, the local structures
as for instance the maximum at $x_F \sim$~0.3 and low $p_T$ are a consequence
of the evolution of different production mechanisms with increasing
interaction energy.  

%
%
\subsection{Data in the PS/AGS energy range}
\vspace{3mm}
\label{sec:ps-ags}

\begin{figure}[h]
  \begin{center}
  	\includegraphics[width=12cm]{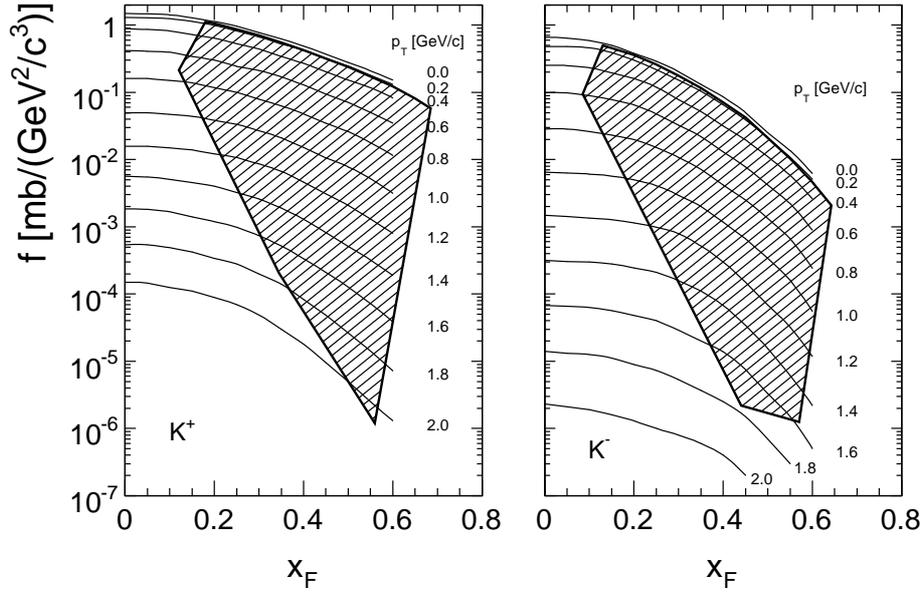}
  	\caption{Invariant, inter/extrapolated cross sections 
             as a function of $x_F$ for fixed values of $p_T$ for a) K$^+$ and
             b) K$^-$. The $x_F$, $p_T$ regions covered by data are indicated
             as the hatched areas}
  	\label{fig:allaby_area}
  \end{center}
\end{figure}

In this subsection data in a range from 12.5 to 24~GeV/c beam 
momentum are grouped together, again in an effort to consolidate
the available information and to quantify the consistency of the
different data sets. This concerns the double differential data
by Akerlof et al. \cite{akerlof} at 12.5~GeV/c beam momentum, of
Dekkers et al. \cite{dekkers} at 18.8 and 23.1~GeV/c, and the extensive data 
sets of the CERN/Rome group, Allaby et al. \cite{allaby1,allaby2} at 14.2, 19.2 and
24~GeV/c beam momentum. The data sets from all these groups have
been tabulated conveniently by Diddens and Schl\"upmann in Landoldt-
B\"ornstein \cite{landolt}. As the overview of Fig.~\ref{fig:phs_kap}b shows, there is
a fair coverage of phase space and some mutual overlap, unfortunately
again \cite{pp_pion,pp_proton} with the exception of the low $x_F$ region, $x_F <$~0.1--0.15,
at all $p_T$. In a first step, the Allaby et al. data \cite{allaby1,allaby2} are transformed
to the standard $x_F$ values following Eq.~\ref{eq:xf} and interpolated
using the two-dimensional, multistep eyeball method described in
sect.~\ref{sec:interp}. An extrapolation into the non-measured phase space
areas is then attempted in order to allow the establishment of
integrated yields. The situation may be judged from Fig.~\ref{fig:allaby_area} where
the interpolated/extrapolated cross sections are shown as a
function of $x_F$ at fixed values of $p_T$ for K$^+$, Fig.~\ref{fig:allaby_area}a, and K$^-$,
Fig.~\ref{fig:allaby_area}b. Here the regions with available measurements from \cite{allaby1,allaby2}
are marked by the hatched areas.

\begin{figure}[h]
  \begin{center}
  	\includegraphics[width=8.8cm]{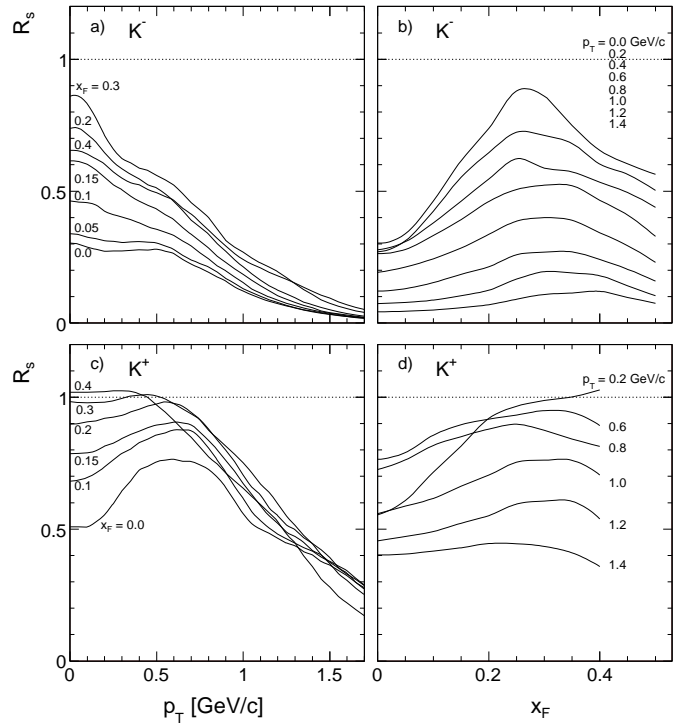}
  	\caption{$R_s$ a) as a function of $p_T$ at fixed $x_F$
             and b) as a function of $x_F$ for fixed $p_T$ for K$^-$ and
             c) as a function of $p_T$ at fixed $x_F$
             and d) as a function of $x_F$ for fixed $p_T$ for K$^+$}
  	\label{fig:allaby_comp}
  \end{center}
\end{figure}

Clearly, the above remark concerning the problems with data extrapolation
is well in place here, especially for the higher $p_T$ regions. 
However, at least towards low $p_T$ there is not much freedom 
of choice, as well as for the $p_T$ region 0~$< p_T <$~1~GeV/c towards
$x_F$~=~0. Up to $p_T \sim$~1~GeV/c it is hard to imagine an extrapolation 
which would be off by more than, say, 10-20\% from the lines
shown at $x_F$~=~0. It is also clear that the increasing error margin
towards higher $p_T$ will not contribute too much to the integrated
cross sections. The ratio

\begin{equation}
   R_s = \frac{f( x_F, p_T, \sqrt{s} = \textrm{6.8~GeV} )}{f( x_F, p_T, \sqrt{s} = \textrm{17.2~GeV} )}
\end{equation}
is shown in Fig.~\ref{fig:allaby_comp} for K$^-$ and K$^+$ as a function of $p_T$ and $x_F$.

\begin{figure}[h]
  \begin{center}
  	\includegraphics[width=12cm]{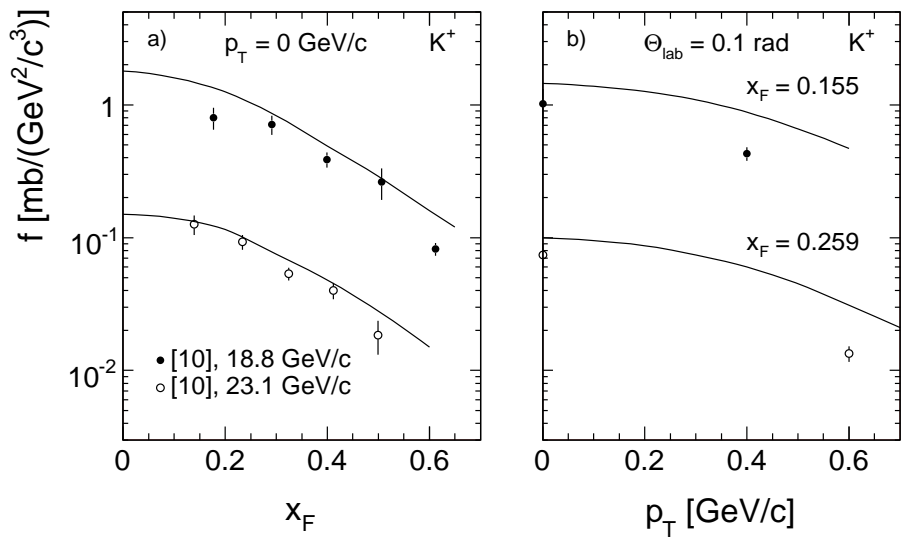}
  	\caption{K$^+$ comparison Dekkers \cite{dekkers} and Allaby
             a) K$^+$, $p_T$~=~0~GeV/c as a function of $x_F$. Full lines \cite{allaby1,allaby2} 
             extrapolation at 19.2 and 24~GeV/c beam momentum, data points 
             from \cite{dekkers} at 18.8 and 23.1~GeV/c
             b) Dekkers data and Allaby interpolation at 18.8 (19.2)~GeV/c
             and 100~mrad lab angle as a function of $x_F$. Full lines Allaby
             interpolation, data points from Dekkers. The data at 23.1~GeV/c and lines at 24~GeV/c
             are multiplied by 0.1 for better separation}
  	\label{fig:dek_all_kaplus}
  \end{center}
\end{figure}

\begin{figure}[h]
  \begin{center}
  	\includegraphics[width=12cm]{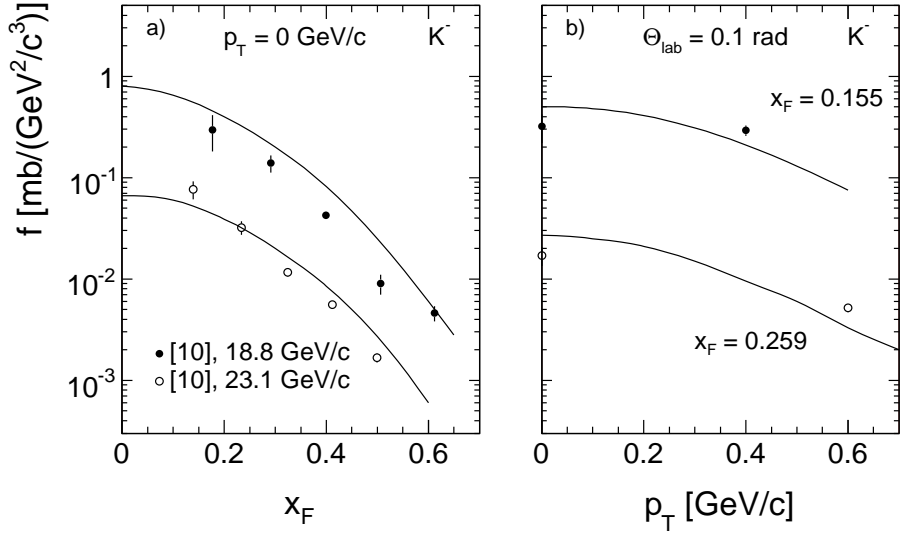}
  	\caption{K$^-$ comparison Dekkers \cite{dekkers} and Allaby
             a) K$^-$, $p_T$~=~0~GeV/c as a function of $x_F$. Full lines \cite{allaby1,allaby2} 
             extrapolation at 19.2 and 24~GeV/c beam momentum, data points 
             from \cite{dekkers} at 18.8 and 23.1~GeV/c
             b) Dekkers data and Allaby interpolation at 18.8 (19.2)~GeV/c
             and 100~mrad lab angle as a function of $x_F$. Full lines Allaby
             interpolation, data points from Dekkers. The data at 23.1~GeV/c and lines at 24~GeV/c
             are multiplied by 0.1 for better separation}
  	\label{fig:dek_all_kaminus}
  \end{center}
\end{figure}

As already visible for the low energy data, Fig.~\ref{fig:hr_rat} above, very
clear structures are apparent with local maxima at $p_T <$~0.5~GeV/c
and $x_F$ around 0.2 -- 0.4. Evidently the $s$-dependence varies over 
phase space by more than an order of magnitude. The apparent 
$s$-independence, within 5\%, of the K$^+$ cross sections in the 
region of $x_F$ around 0.3 for $p_T$ from 0 up to 0.6~GeV/c, should 
however be seen as an indicator of systematic problems. The 
fact that the data of \cite{allaby1,allaby2} are apparently over-estimating 
the K$^+$ yields in this region can be shown in comparison with
data from other experiments \cite{akerlof,dekkers}. Although these data do not
offer enough coverage to permit a complete interpolation, they
may be used to bring out local mutual inconsistencies between
the experimental results. As the data of Dekkers et al. \cite{dekkers}
have been obtained at 18.8 and 23.1~GeV/c beam momenta, at only 
two fixed lab angles of 0 and 100~mrad, also the Allaby et al.
data at 19.2 GeV/c beam momentum, which offer much inferior
phase space coverage, have been interpolated in order to permit
direct comparison for a maximum of data points. The results
are presented in Figs.~\ref{fig:dek_all_kaplus} and ~\ref{fig:dek_all_kaminus}.

If the 0 degree data (Fig.~\ref{fig:dek_all_kaplus}a) show all the Dekkers points
below the Allaby extrapolation, with a mean relative difference of
20\%, the values at higher $p_T$, Fig.~\ref{fig:dek_all_kaplus}b, are all far below the 
interpolation by a factor of about 2.
The same comparison for K$^-$ is shown in Fig.~\ref{fig:dek_all_kaminus}.

Also for K$^-$ the Dekkers data at $p_T$~=~0~GeV/c are below the Allaby data,
here by 30\%. This might indicate a general offset between the two
data sets of 20-30\% which does not seem to be excluded by the
systematic uncertainties given for the respective experiments.
In contrast to the situation for K$^+$ the data interpolation
at 19.2~GeV/c at higher $p_T$  is bracketing the Dekkers data for 
K$^-$ such that the mean deviation over the given $p_T$ scale tends 
to be small, Fig.~\ref{fig:dek_all_kaminus}b.

\begin{figure}[h]
  \begin{center}
  	\includegraphics[width=12cm]{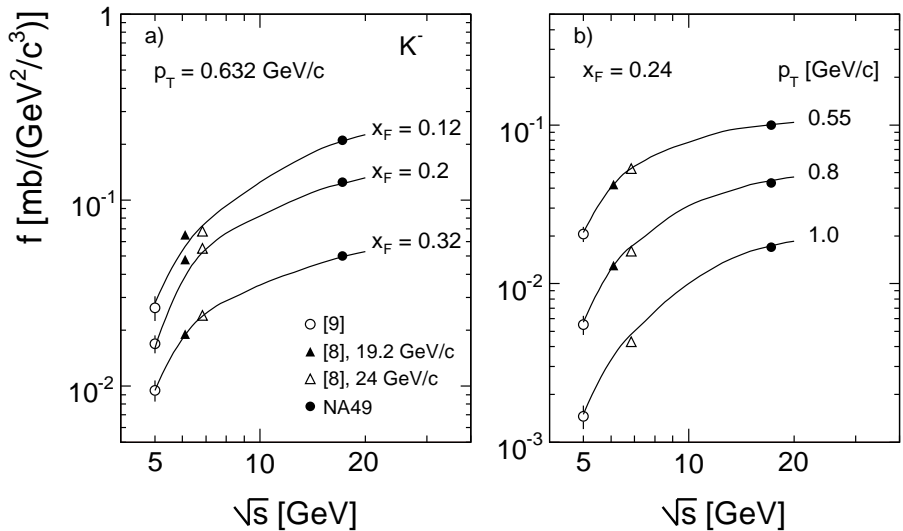}
  	\caption{K$^-$ comparison as a function of $\sqrt{s}$,
             a) $p_T$~=~0.632~GeV/c for $x_F$~=~0.12, 0.2 and 0.32
             b) $x_F$~=~0.24 for $p_T$~=~0.55, 0.8 and 1.0~GeV/c}
  	\label{fig:allaby_kaminus_sdep}
  \end{center}
\end{figure}

A further possibility of controlling  $s$-dependence is given by 
the data of Akerlof et al. \cite{akerlof}  which were obtained with a 12.5~GeV/c 
proton beam at the Argonne ZGS. Although only 7 points for K$^+$ and 17
points for K$^-$ have been measured, the $s$-dependence between these
data is revealing if compared to the 19.2 and 24~GeV/c data of
Allaby et al. Starting with K$^-$, the Akerlof data allow comparison
at fixed $p_T$ of 0.632~GeV/c in the $x_F$ range from 0.12 to 0.32,
and at fixed $x_F$~=~0.24 for $p_T$ between 0.55 and 1.14~GeV/c. Fig.~\ref{fig:allaby_kaminus_sdep}
shows the $s$-dependence for fixed $x_F$ (panel a) and fixed $p_T$ (panel b)
including the data from Allaby and NA49. 

\begin{figure}[h]
  \begin{center}
  	\includegraphics[width=12cm]{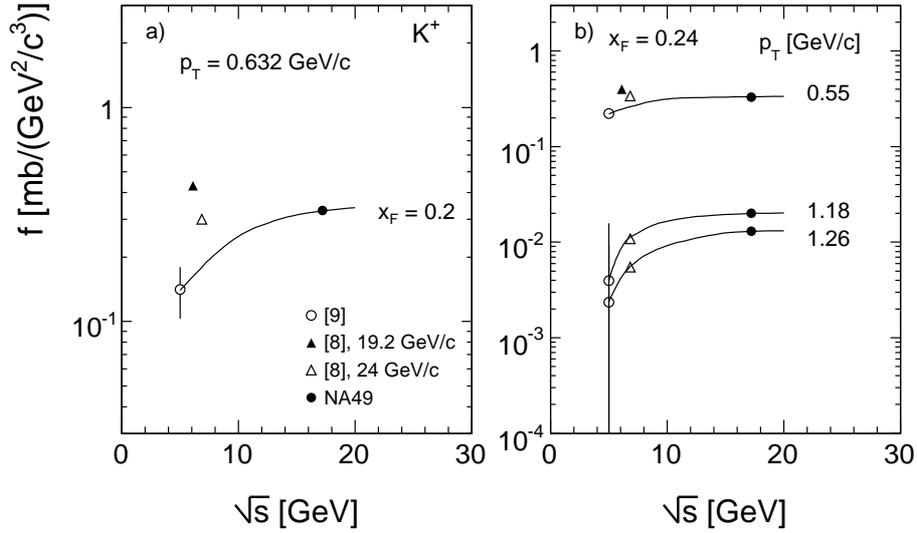}
  	\caption{K$^+$ comparison as a function of $\sqrt{s}$
             a) $p_T$~=~0.632~GeV/c for $x_F$~=~0.2
             b) $x_F$~=~0.24 for $p_T$~=~0.55, 1.18 and 1.26~GeV/c}
  	\label{fig:allaby_kaplus_sdep}
  \end{center}
\end{figure}

For all $x_F$/$p_T$ combinations, a smooth $s$-dependence between
the four data sets is observed. A different picture emerges for
K$^+$, Fig.~\ref{fig:allaby_kaplus_sdep}, where only 3 points in $\sqrt{s}$ 
at $p_T$~=~0.632~GeV/c and $x_F$~=~0.24 are available.

\begin{figure}[h]
  \begin{center}
  	\includegraphics[width=7cm]{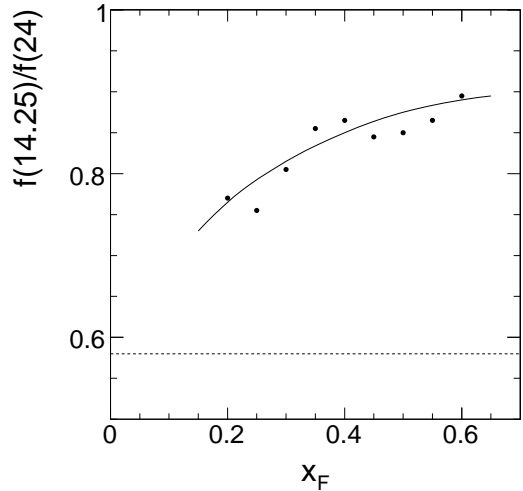}
  	\caption{$f(\textrm{14.25~GeV/c})$/$f(\textrm{24~GeV/c})$ as a function of $x_F$}
  	\label{fig:allaby_rat}
  \end{center}
\end{figure}

As already apparent from Fig.~\ref{fig:allaby_comp}, the Allaby et al data at $p_T$~=~0.632~GeV/c, 
$x_F$~=~0.2 (Fig.~\ref{fig:allaby_kaplus_sdep}a) and $x_F$~=~0.24, $p_T$~=~0.55~GeV/c
(Fig.~\ref{fig:allaby_kaplus_sdep}b) are on the same 
level as the NA49 data for 24~GeV/c beam momentum, and are even 
higher for 19.2~GeV/c. Compared to this the Akerlof data show the 
expected decrease at $\sqrt{s}$~=~5~GeV. The resulting $s$-dependence
looks definitely unphysical indicating an excess of the order of
60\% in the K$^+$ yields of Allaby et al. A similar problem is present
in the Allaby et al. data at 14.25~GeV/c beam momentum which only
exist for K$^+$ at a lab angle of 12~mrad, thus covering the low $p_T$
region from 0.04 to 0.11~GeV/c for 0.25~$< x_F <$~0.6. Those data may be compared
to the interpolation at 24~GeV/c beam momentum. The cross section
ratio $f(\textrm{14.25~GeV/c})$/$f(\textrm{24~GeV/c})$ is shown in Fig.~\ref{fig:allaby_rat} 
as a function of $x_F$.

Evidently there is a very small $s$-dependence also in this
low-$p_T$ region, with an average relative factor of only 0.85$\pm$0.05
where factors of 0.5 -- 0.6 should be expected, see also the discussion
in Sect.~\ref{sec:kzero} below. 

\begin{figure}[h]
  \begin{center}
  	\includegraphics[width=11cm]{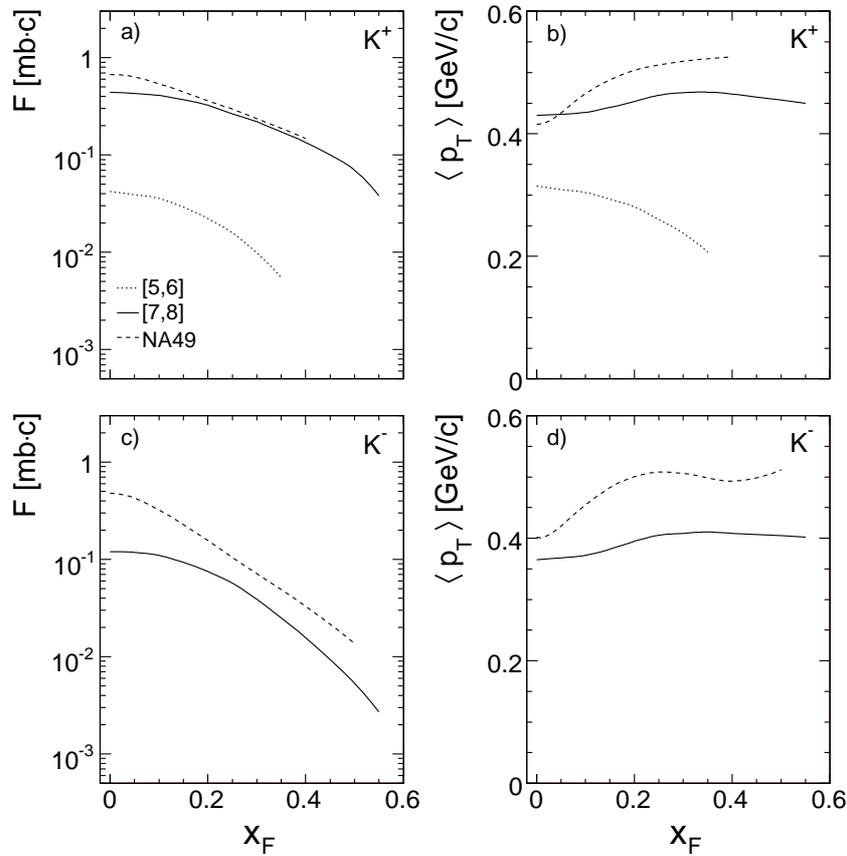}
  	\caption{$F(\textrm{24~GeV/c})$ and $\langle p_T \rangle$ as a function of $x_F$ compared to the
             NA49 (dashed lines) and $\sqrt{s}$~=~3~GeV (dotted lines) results; 
             a) and b) for K$^+$, and c) and d) for K$^-$}
  	\label{fig:allaby_ptint}
  \end{center}
\end{figure}

Notwithstanding the apparent problems with the K$^+$ measurements,
the interpolated data at 24~GeV/c beam momentum may be integrated
in order to obtain $p_T$ integrated invariant cross sections,
mean transverse momenta and total kaon multiplicities. The resulting
$p_T$ integrated invariant $x_F$ distributions and mean transverse momenta
are presented in Fig.~\ref{fig:allaby_ptint} in comparison to the NA49 results.

For the $p_T$ integrated distributions, Fig.~\ref{fig:allaby_ptint}a, the approach
of the lower energy data to the NA49 results for K$^+$, to within 10\%
at $x_F$~=~0.3, confirms the statements made above concerning $s$-dependence.
Also the behaviour of $\langle p_T \rangle$ for K$^+$ and K$^-$, 
Fig.~\ref{fig:allaby_ptint}b and d, raises questions, in particular
if compared to the results at $\sqrt{s}$~=~3~GeV also shown in Fig.~\ref{fig:allaby_ptint}b.

The total integrated kaon yields at $\sqrt{s}$~=~6.84~GeV, as they
result from the data interpolation, are

\begin{equation}
	\begin{split}
     \langle n_{\textrm{K}^+} \rangle &= 0.107 \\
     \langle n_{\textrm{K}^-} \rangle  &= 0.0262 .
	\end{split}
\end{equation}

This is 10\% above and 60\% below the values fitted by Rossi et al. \cite{rossi}
for K$^+$ and K$^-$, respectively. A more detailed discussion
of total yields is given in Sects.~\ref{sec:kzero} and \ref{sec:over} below.

%
%
\subsection{Data at Serpukhov energy}
\vspace{3mm}
\label{sec:serpukhov}

In the range of 30 to 70~GeV/c beam momentum accessible at the 
Serpukhov accelerator only a single double differential measurement
of kaons is available at 70~GeV/c \cite{abramov}. This measurement has been 
performed at a constant lab angle of 160~mrad which corresponds 
approximately to $x_F$~=~0 and in a transverse momentum range of 
$p_T >$~0.46~GeV/c. In consequence there is no possibility to establish 
a reliable $p_T$ integrated yield at $x_F$~=~0, not to speak of the total 
production cross section. A comparison with the NA49 data has 
been performed taking account of the dependence of $x_F$ on $p_T$ shown 
in Table~\ref{tab:abr} together with the invariant cross sections

\begin{table}[h]
\renewcommand{\arraystretch}{1.2} 
\normalsize
\begin{center}
\begin{tabular}{cc|ccc|ccc}
\hline
   $p_T$ [GeV/c]& $|x_F|$ & $f(\textrm{\cite{abramov}})$ & $f(NA49)$ & $R$  & $f(\textrm{\cite{abramov}})$ & $f(NA49)$ & $R$ \\
\hline
         &         &    \multicolumn{3}{|c|}{K$^+$}           & \multicolumn{3}{c}{K$^-$}  \\ \hline
   0.48  & 0.0405  & 0.739   & 0.879   & 0.841  & 0.398   & 0.586   & 0.679  \\
   0.58  & 0.0329  & 0.483   & 0.582   & 0.830  & 0.255   & 0.392   & 0.651  \\
   0.69  & 0.0270  & 0.313   & 0.361   & 0.867  & 0.186   & 0.244   & 0.762  \\
   0.96  & 0.0178  & 0.0805  & 0.1050  & 0.767  & 0.0374  & 0.0641  & 0.583  \\
   1.29  & 0.0111  & 0.0157  & 0.0232  & 0.678  & 0.00610 & 0.0126  & 0.484  \\
   1.55  & 0.0076  & 0.00377 & 0.00716 & 0.526  & 0.00121 & 0.00350 & 0.346  \\
   1.68  & 0.0061  & 0.00182 & 0.00400 & 0.455  & 0.00060 & 0.00194 & 0.310  \\
   1.75  & 0.0053  & 0.00134 & 0.00294 & 0.456  & 0.00043 & 0.00133 & 0.324  \\
   1.99  & 0.0031  & 0.00039 & 0.00097 & 0.400  & 0.00012 & 0.00035 & 0.343  \\
\hline                 
\end{tabular}
\end{center}
\caption{Relation between $p_T$ and $x_F$ for \cite{abramov}}
\label{tab:abr}
\end{table}

\begin{figure}[h]
  \begin{center}
  	\includegraphics[width=6.5cm]{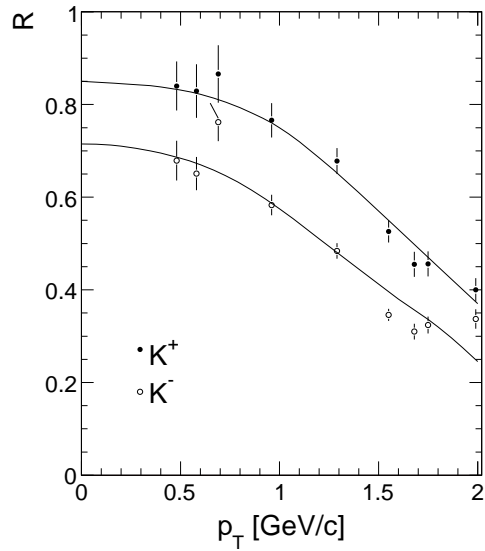}
  	\caption{Cross section ratio $R$ of \cite{abramov} to NA49
             for K$^+$ and K$^-$}
  	\label{fig:serp_ptdist}
  \end{center}
\end{figure}

The data of NA49 have been interpolated to the respective $p_T$/$x_F$
combinations, see Table~\ref{tab:abr}, in order to obtain the ratios of 
invariant cross sections presented in Table~\ref{tab:abr} and Fig.~\ref{fig:serp_ptdist}.

If one interpolates the cross section ratios as shown in
Fig.~\ref{fig:serp_ptdist} and if one takes the courage to extrapolate these
curves down to $p_T$~=~0~GeV/c as also shown in this Figure one may obtain
the invariant cross sections at $\sqrt{s}$~=~11.5~GeV, $x_F$~=~0, from 
the ones of the interpolated NA49 data and integrate over
$p_T$. This yields the invariant cross sections (Eq.~\ref{eq:int})

\begin{equation}
	\begin{split}
         F( \textrm{K}^+, x_F = 0) &= 0.549 \\ 
         F( \textrm{K}^-, x_F = 0) &= 0.322.
	\end{split}
\end{equation}

\begin{figure}[h]
  \begin{center}
  	\includegraphics[width=6.5cm]{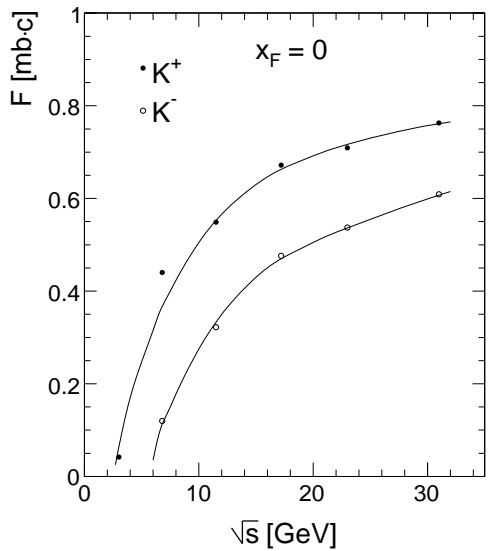}
  	\caption{$F(x_F = 0)$ for K$^+$ and K$^-$ as a function
             of $\sqrt{s}$ for 6 different energies}
  	\label{fig:serp_sdep}
  \end{center}
\end{figure}

These values are plotted in Fig.~\ref{fig:serp_sdep} together with the cross
sections determined in Sects.~\ref{sec:hogan-reed} and \ref{sec:ps-ags} above,
with the NA49 data and the lower range of ISR energies 
(see Sect.~\ref{sec:isr} below) in order to get a first view on $s$-dependence.
  
It is interesting to compare this dependence to the $p_T$ integrated
cross sections $F(x_F)$ and $d\sigma/dy(y)$ available from Zabrodin et al.
\cite{zabrodin} for K$^-$ only, at the lower Serpukhov energy of
32~GeV/c beam momentum or $\sqrt{s}$~=~7.85~GeV. They give $F(x_F = 0) \sim$~0.6
and $dn/dy(y=0)$~=~0.066. These values are about 30\% higher than the 
ones obtained by NA49 where on the other hand a decrease by 2.8
would be expected from the $s$-dependence, Fig.~\ref{fig:serp_sdep}. It is therefore
concluded that (contrary to the pion and baryon cross section
provided by \cite{zabrodin}) their extracted K$^-$ yields are flawed,
especially as the integration over their rapidity distribution
gives a total K$^-$ yield of 0.21, about 60\% higher than the value
from NA49 at $\sqrt{s}$~=~17.2~GeV/c.
            
%
%
\subsection{Data at ISR energy}
\vspace{3mm}
\label{sec:isr}

The ISR data on kaon production may be separated into three regions
of $x_F$. A first region at $x_F$~=~0 is covered by \cite{alper,guettler},
the region from $x_F$~=~0.08 to 0.49 by \cite{capi}, and finally the
data of \cite{albrow1,albrow2,albrow3,albrow4,albrow5} reach from 
$x_F \sim$~0.2 to 0.7. For the purpose of the present work these data 
are exploited in a phase space region (see Fig.~\ref{fig:phs_kap}) 
in $p_T$ up to 1.9~GeV/c and in $x_F$ up to 0.6 in order to
allow for a comparison to the NA49 data with reasonably small
extrapolations. This makes available a substantial set of 383 points
for K$^+$ and 335 points for K$^-$.

%
%
\subsubsection{The central region, \cite{alper,guettler}}
\vspace{3mm}
\label{sec:central}

\begin{figure}[h]
  \begin{center}
  	\includegraphics[width=16cm]{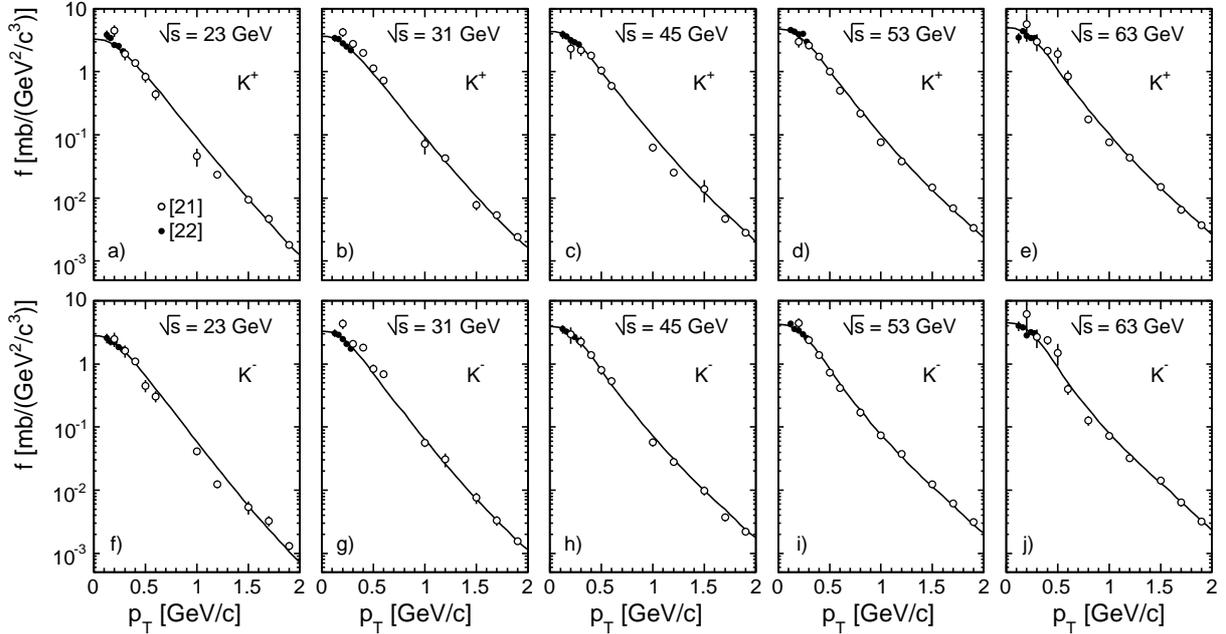}
  	\caption{Invariant cross sections at $x_F$~=~0 from \cite{alper,guettler}
             as functions of $p_T$ at five ISR energies, a) to e) for K$^+$, 
             f) to j) for K$^-$. The data interpolations are superimposed on the data points}
  	\label{fig:ptdist_isrxf0}
  \end{center}
\end{figure}

The data of Alper et al. \cite{alper} and Guettler et al. \cite{guettler} follow
each other with several years difference. The later data \cite{guettler}
are extending (and superseding) the earlier work \cite{alper} at low $p_T$,
in a range from 0.123 to 0.280~GeV/c. They feature statistical errors
in the 5 to 10\% range, exceptionally small for ISR standards, and
are probably the best controlled data as far as normalization and
internal consistency are concerned. The combined data sets are
presented in Fig.~\ref{fig:ptdist_isrxf0} at the five standard ISR energies from $\sqrt{s}$~=~23 
to $\sqrt{s}$~=~63~GeV. 

\begin{figure}[h]
  \begin{center}
  	\includegraphics[width=10.5cm]{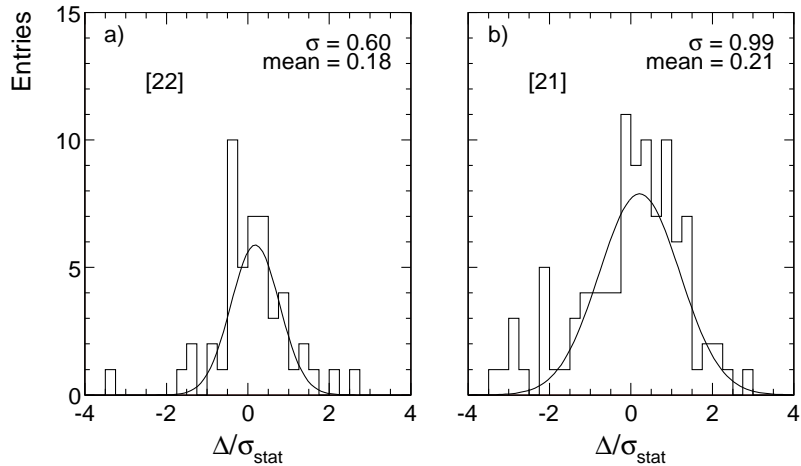}
  	\caption{Normalized differences between data and interpolation
             for a) \cite{guettler} b) \cite{alper}}
  	\label{fig:dist_isrxf0}
  \end{center}
\end{figure}

In order to eliminate some of the larger fluctuations in the Alper
et al. data \cite{alper}, a multistep eyeball interpolation imposing smoothness
both in $p_T$ and in the $s$-dependence has been performed,
again (see Sect.~\ref{sec:interp}) avoiding any kind of arithmetic fitting. The
resulting $p_T$ dependences are superimposed on the data in Fig.~\ref{fig:ptdist_isrxf0}. 
The distributions of the differences between data points and 
interpolation, normalized to the statistical errors, are shown in
Fig.~\ref{fig:dist_isrxf0} separately for \cite{alper} and \cite{guettler}.

The Gaussian fit to the differences shows an upwards shift of 0.18
or about 1\% and an rms of 0.6 for the data of Guettler et al. \cite{guettler},
indicating a certain overestimation of their statistical uncertainties.
For Alper et al. \cite{alper} the upwards shift is equivalent to 3--5\%, with an rms
compatible with unity. The accumulation of entries at $\Delta/\sigma$
in the region +0.5 and +1.5 corresponds to the data points in the low
$p_T$ region of the Alper et al. data \cite{alper} visible in 
Fig.~\ref{fig:ptdist_isrxf0}b), e), g) and j).
These points are in clear disagreement with the later precision data.
Other points of \cite{alper}, deviating far below the interpolation (region
of $\Delta/\sigma <$~-2) and partially even falling below the NA49 data, are 
visible notably in Fig.~\ref{fig:ptdist_isrxf0}a), c), and f). This demonstrates again a
certain instability in the absolute normalization of the earlier ISR
data also visible in Sects.~\ref{sec:capi} and \ref{sec:albrow} and discussed 
for protons in \cite{pp_proton}.

\begin{figure}[h]
  \begin{center}
  	\includegraphics[width=13cm]{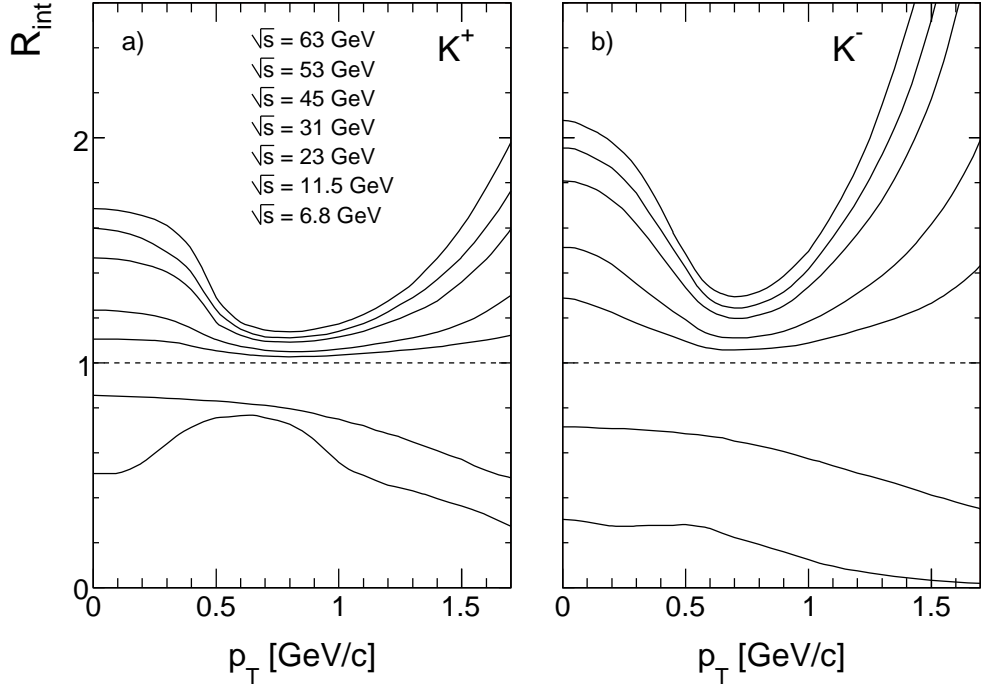}
  	\caption{$R_{\textrm{int}}( p_T, \sqrt{s})$ as a function of $p_T$ 
             including data from \cite{allaby2,abramov}}
  	\label{fig:isr_factor}
  \end{center}
\end{figure}

In dividing the interpolation of \cite{alper,guettler} by the one for NA49
(Sect.~\ref{sec:interp}) one may define the ratios 

\begin{equation}
  \label{eq:rint}
  R_{\textrm{int}}(x_F = 0, p_T, \sqrt{s}) = \frac{f^{\textrm{ISR}}(x_F = 0, p_T, \sqrt{s})}
                                                  {f^{\textrm{NA49}}(x_F = 0, p_T, \textrm{17.2~GeV} )}
\end{equation}
shown in Fig.~\ref{fig:isr_factor} as a function of $p_T$ for the five ISR energies. For 
comparison, also the corresponding ratios for the inter/extrapolation 
of the Serpukhov \cite{abramov} and PS \cite{allaby2} data are included. 

\begin{figure}[h]
  \begin{center}
  	\includegraphics[width=16cm]{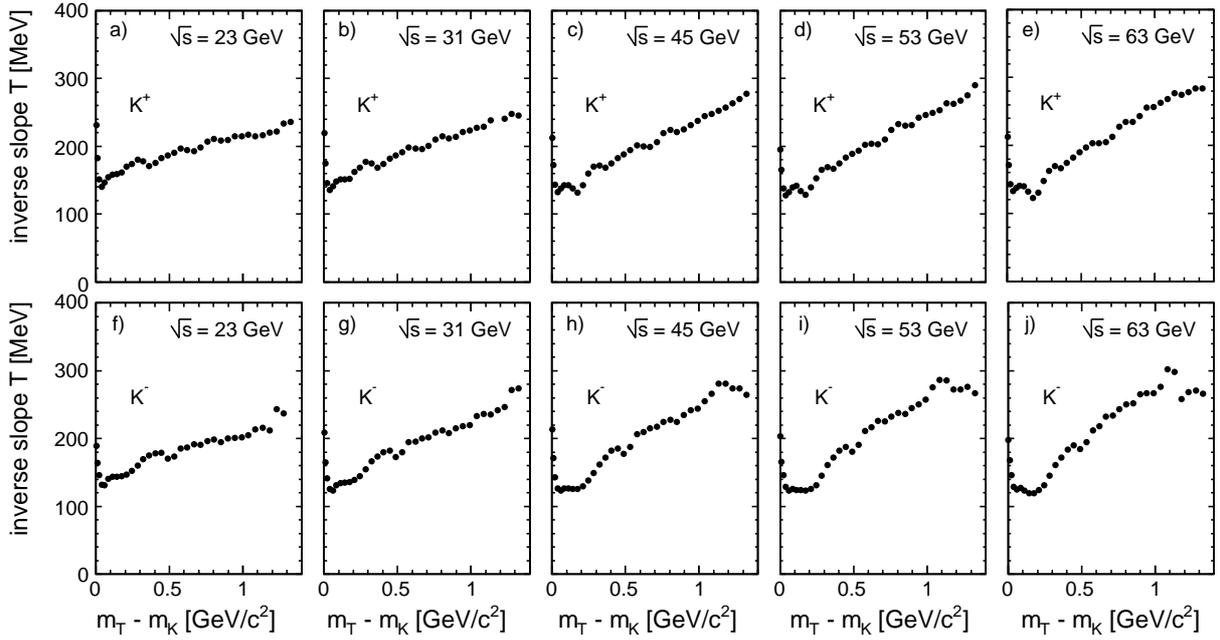}
  	\caption{Inverse slopes of the interpolated $m_T$ distributions as function 
             of $m_T-m_K$ for the 5 ISR energies and for K$^+$ ( panels a) to e) ) 
             and for K$^-$ ( panels f) to j) )}
  	\label{fig:slope_isrxf0}
  \end{center}
\end{figure}

A remarkable picture emerges. Evidently there is a very strong change 
of the $s$-dependence with $p_T$, with three clearly distinguished regions
of $p_T$. A first, low $p_T$ region extends up to $p_T \sim$~0.6~GeV/c. The strong
increase with $s$ already stressed in \cite{guettler} as "rising plateau" is
completely concentrated in this limited area. A second region at
0.6~$< p_T <$~1~GeV/c shows in contrast a rather small $s$-dependence, limited
here to a relative increase of only 10\% (20\%) for K$^+$ and K$^-$,
respectively, over the complete range from $\sqrt{s}$~=~17 to 63~GeV.
A third region at $p_T \gtrsim$~1.2~GeV/c shows again a strong $s$-dependence with
increasing $p_T$ up to factors of 2 (3.6) for K$^+$ and K$^-$, respectively,
over the before-mentioned range of $\sqrt{s}$. These features reflect
in the inverse slope parameters of the $m_T$ distributions (see Sect.~\ref{sec:rap})
presented in Fig.~\ref{fig:slope_isrxf0}.

\begin{figure}[b]
  \begin{center}
  	\includegraphics[width=14cm]{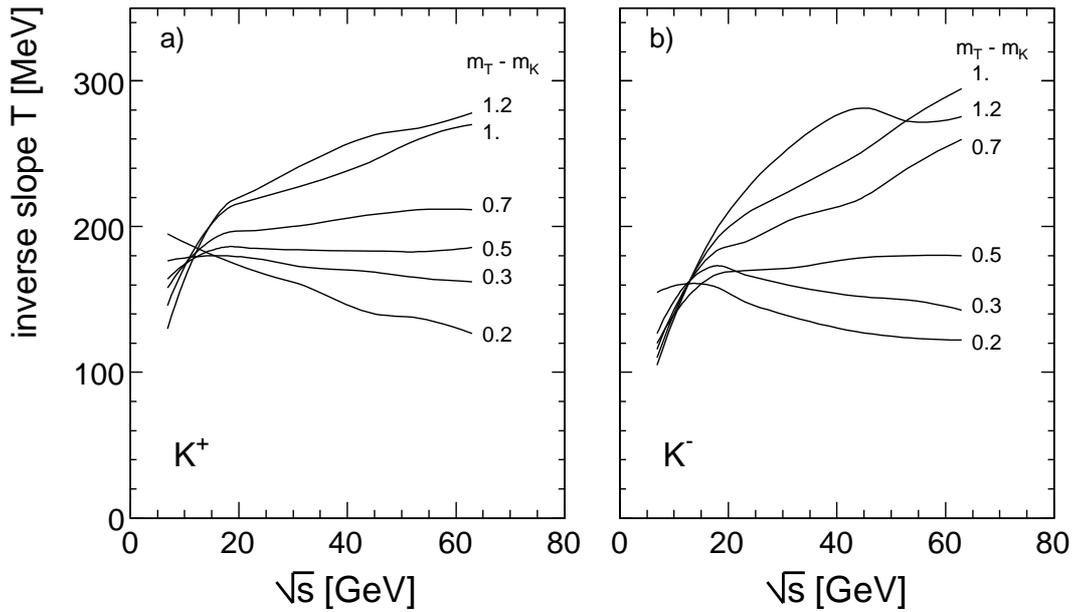}
  	\caption{Inverse slopes at fixed $p_T$ as a function of $\sqrt{s}$ for a) K$^+$ and b) K$^-$}
  	\label{fig:slope_sdep}
  \end{center}
\end{figure}

In plotting the extracted inverse slopes at fixed $p_T$ as a function
of $\sqrt{s}$, Fig.~\ref{fig:slope_sdep}, and extending the $s$-range to Serpukhov and
PS energies, the strong evolution of this "hadronic temperature"
both with $p_T$ and with $\sqrt{s}$ and thereby the sense (or, rather,
non-sense) of thinking in terms of a fixed "temperature" in
soft hadronic production, becomes evident, see also Sect.~\ref{sec:reson} below. Only (by accident)
in the region of Serpukhov energies there is a concentration
of inverse slopes in a small interval around 180~MeV for K$^+$ and 160~MeV for K$^-$. 

%
%
\subsubsection{The intermediate $x_F$ region, \cite{capi}}
\vspace{3mm}
\label{sec:capi}

In the following Sects.~\ref{sec:capi} and \ref{sec:albrow}, the NA49 data
are compared to ISR results at $x_F \neq$~0. In this comparison, in addition
to the ratio

\begin{equation}
  R(x_F, p_T, \sqrt{s}) = \frac{f^{\textrm{ISR}}(x_F, p_T, \sqrt{s})}
                               {f^{\textrm{NA49}}(x_F, p_T, \textrm{17.2~GeV} )} 
\end{equation} 
the ratio $R_{\textrm{int}}(x_F = 0, p_T, \sqrt{s})$ (Eq.~\ref{eq:rint}) 
which describes the $s$-dependence at $x_F$~=~0 is used in order to make 
a prediction at all $x_F \neq$~0. As shown in these two sections, the 
ISR data in forward direction are well described by the NA49 results 
multiplied by $R_{\textrm{int}}(x_F = 0, p_T, \sqrt{s})$ at all $x_F$.
This non-trivial result shows that the $s$-dependence has no major
change with $x_F$.  

\begin{figure}[h]
  \begin{center}
  	\includegraphics[width=14cm]{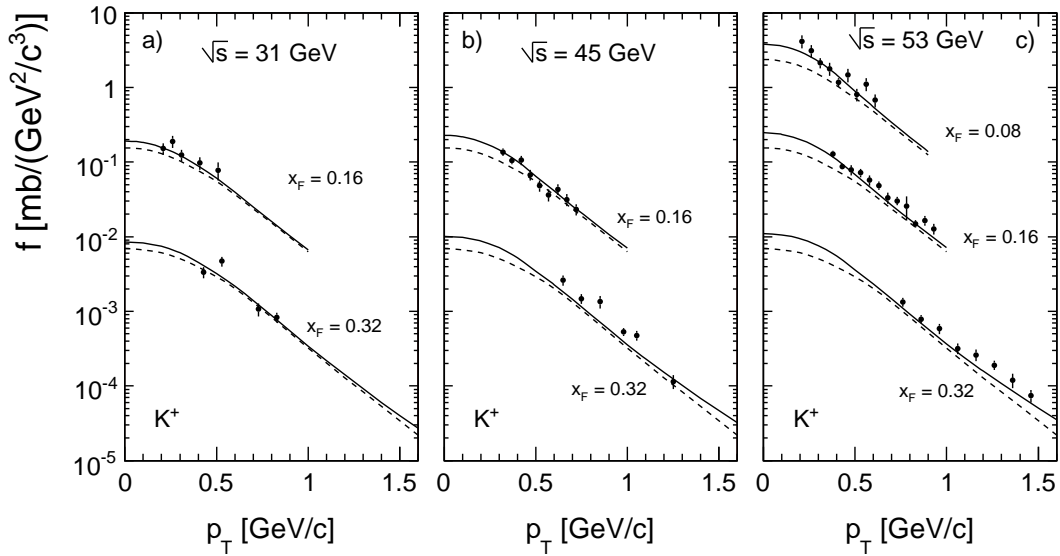}
  	\caption{Invariant K$^+$ cross sections \cite{capi} as a function of $p_T$ at fixed $x_F$ 
             for a) $\sqrt{s}$~=~31~GeV, b) $\sqrt{s}$~=~45~GeV and c) $\sqrt{s}$~=~53~GeV
             in comparison to the NA49 data (dashed lines) and to the evolution
             with $p_T$ and $\sqrt{s}$ as measured at $x_F$~=~0 (solid lines). The results at
             $x_F$~=~0.16 and 0.32 are multiplied by 0.1 and 0.01, respectively, for better 
             separation}
  	\label{fig:capi_kaplus}
  \end{center}
\end{figure}

\begin{figure}[h]
  \begin{center}
  	\includegraphics[width=14cm]{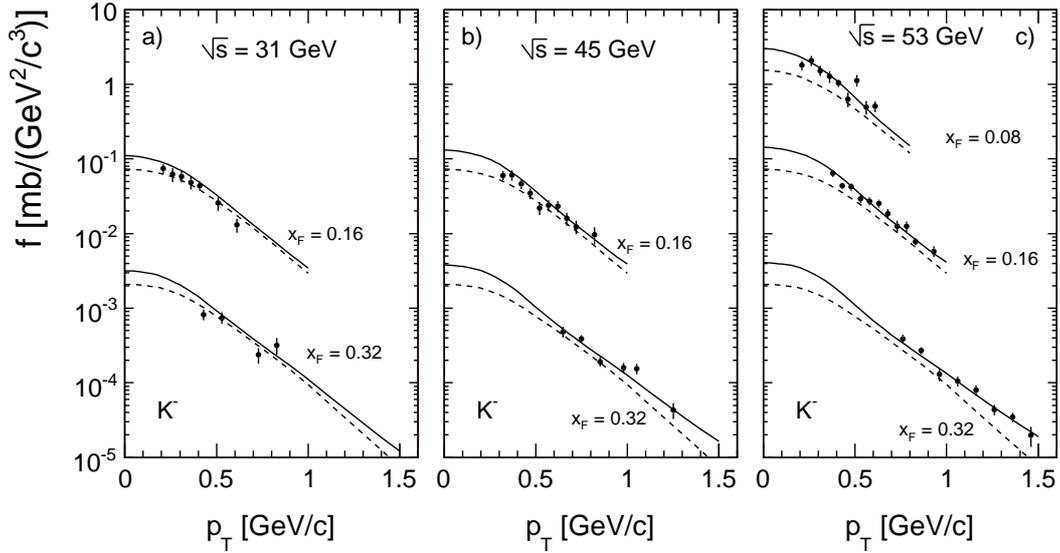}
  	\caption{Invariant K$^-$ cross sections \cite{capi} as a function of $p_T$ at fixed $x_F$ 
             for a) $\sqrt{s}$~=~31~GeV, b) $\sqrt{s}$~=~45~GeV and c) $\sqrt{s}$~=~53~GeV
             in comparison to the NA49 data (dashed lines) and to the evolution
             with $p_T$ and $\sqrt{s}$ as measured at $x_F$~=~0 (solid lines). The results at
             $x_F$~=~0.16 and 0.32 are multiplied by 0.1 and 0.01, respectively, for better 
             separation}
  	\label{fig:capi_kaminus}
  \end{center}
\end{figure}

The data of Capiluppi et al. \cite{capi} cover the ranges of 0.08~$< x_F <$~0.5
and 0.2~$< p_T <$~1.5~GeV/c, both as a function of $p_T$ for fixed $x_F$ and
as a function of $x_F$ for fixed $p_T$. An overview over the $p_T$ dependence
is presented in Fig.~\ref{fig:capi_kaplus} for K$^+$ and in 
Fig.~\ref{fig:capi_kaminus} for K$^-$. In both Figures
the invariant cross sections for the three $x_F$ values 0.08, 0.16 and
0.32 are plotted separately for the three $\sqrt{s}$ values of 31, 45 
and 63~GeV. Also shown are the NA49 cross sections at these $x_F$ 
values and, in addition, the evolution of the cross sections at $x_F$~=~0
with respect to NA49,
see Fig.~\ref{fig:isr_factor}, for these $\sqrt{s}$ values, as a function of $p_T$.

Evidently the $s$-dependence at $x_F$~=~0 is also describing the evolution
in the $x_F$ region up to 0.3 within the sizeable statistical errors
of typically 15--30\%, with some exceptions notably for K$^+$ at $x_F$~=~0.32.
The additional systematic uncertainties which can reach the same
size as the statistical fluctuations have to be taken into account here.

\begin{figure}[h]
  \begin{center}
  	\includegraphics[width=10cm]{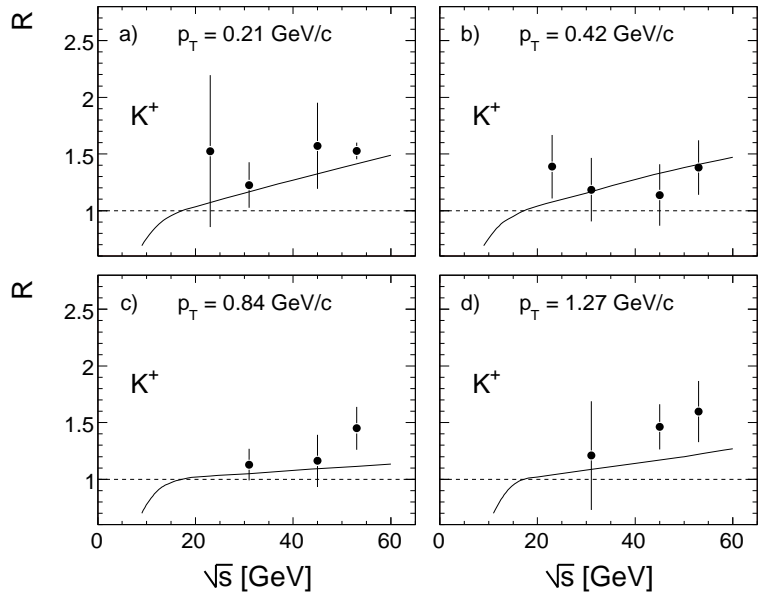}
  	\caption{Cross section ratio $R$ with respect to the NA49 data
             at a) $p_T$~=~0.21~GeV/c, b) $p_T$~=~0.42~GeV/c, c) $p_T$~=~0.84~GeV/c, 
             d) $p_T$~=~1.27~GeV/c, averaged over the $x_F$ ranges of \cite{capi}, as
             a function of $\sqrt{s}$, for K$^+$. Superimposed is the $s$-dependence
             measured at $x_F$~=~0}
  	\label{fig:capi_kaplus_sdep}
  \end{center}
\end{figure}

\begin{figure}[h]
  \begin{center}
  	\includegraphics[width=10cm]{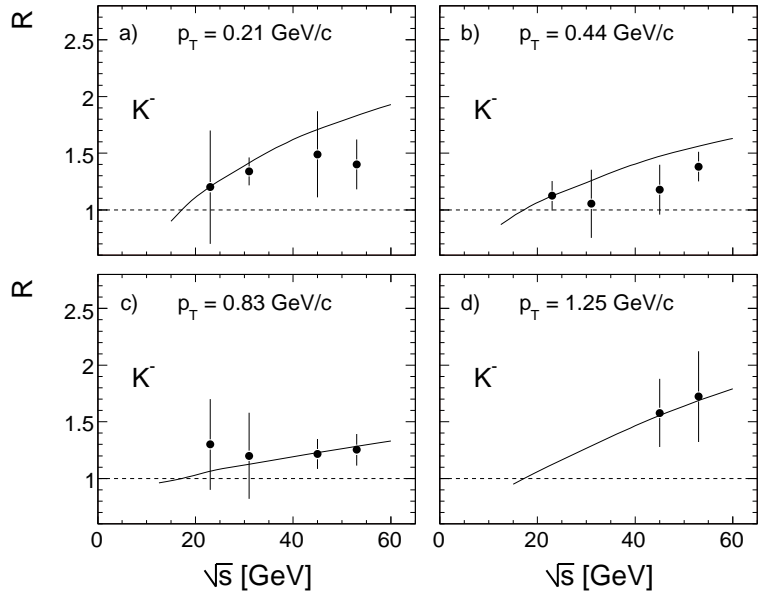}
  	\caption{Cross section ratio $R$ with respect to the NA49 data
             at a) $p_T$~=~0.21~GeV/c, b) $p_T$~=~0.44~GeV/c, c) $p_T$~=~0.83~GeV/c, 
             d) $p_T$~=~1.25~GeV/c, averaged over the $x_F$ ranges of \cite{capi}, as
             a function of $\sqrt{s}$, for K$^-$. Superimposed is the $s$-dependence
             measured at $x_F$~=~0}
  	\label{fig:capi_kaminus_sdep}
  \end{center}
\end{figure}

A similar picture emerges for the data sets obtained at fixed $p_T$
as a function of $x_F$. Here the ratio to the NA49 data, averaged
over the $x_F$ ranges of \cite{capi} from 0.1 to 0.4, is presented in 
Figs.~\ref{fig:capi_kaplus_sdep} and \ref{fig:capi_kaminus_sdep} as 
a function of $\sqrt{s}$, for the four $p_T$ values 0.21, 0.42, 0.82
and 1.27~GeV/c.

In both Figures the $s$-dependence extracted at these $p_T$ values for
$x_F$~=~0, Fig.~\ref{fig:isr_factor}, is shown as the full line. Again the data follow
this $s$-dependence within their statistical uncertainty.

In conclusion it may be stated that the data of \cite{capi} in the
intermediate $x_F$ range from 0.08 to about 0.4 are reasonably well
described by the NA49 data supplemented with the $s$-dependence
extracted at $x_F$~=~0.  

%
%
\subsubsection{The forward data of Albrow et al. \cite{albrow1,albrow2,albrow3,albrow4,albrow5}}
\vspace{3mm}
\label{sec:albrow}

\begin{figure}[h]
  \begin{center}
  	\includegraphics[width=9cm]{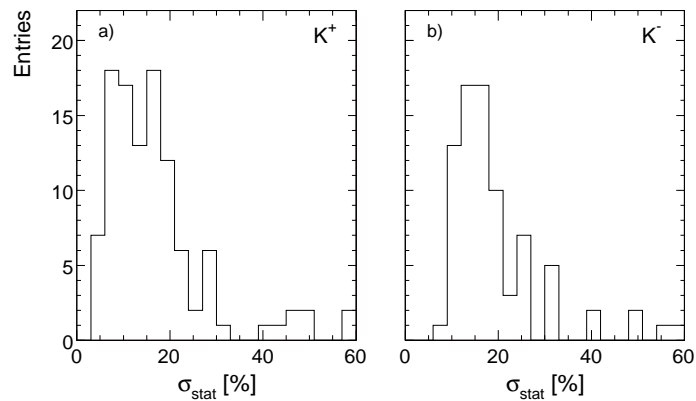}
  	\caption{Distributions of the statistical errors for the data of 
             \cite{albrow1,albrow2,albrow3,albrow4,albrow5} for a) K$^+$ and b) K$^-$}
  	\label{fig:alb_stat}
  \end{center}
\end{figure}

The CHLM collaboration has produced rich data sets for pions \cite{pp_pion}
and protons \cite{pp_proton}, in the latter case with far more than thousand
cross section values. For kaons, however, the situation is less
favourable. In fact only less than 100 data points for each charge
fall into the $x_F$ region below 0.5 usable for comparison purposes.
On the other hand there is good overlap with the data \cite{capi} thus
allowing for meaningful cross checks although the distributions
of the statistical errors, Fig.~\ref{fig:alb_stat}, show wide spreads around
mean values of about 15\%.

\begin{figure}[h]
  \begin{center}
  	\includegraphics[width=12cm]{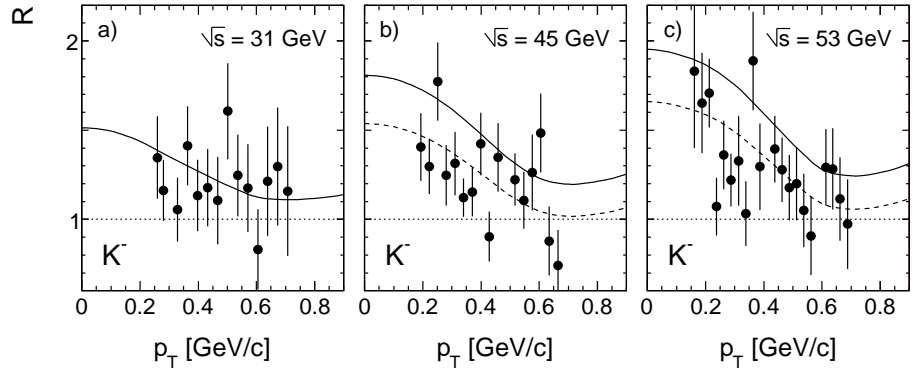}
  	\caption{Ratio $R$ between data from \cite{albrow1} and NA49 as a function of $p_T$
             for a) $\sqrt{s}$~=~31~GeV, b) $\sqrt{s}$~=~45~GeV and c) $\sqrt{s}$~=~53~GeV.
             The $p_T$ dependence at $x_F$~=~0, Fig.~\ref{fig:isr_factor}, are included as 
             full lines. The broken lines in panels b) and c) indicate the result of a 
             15\% downwards normalization error}
  	\label{fig:alb_ptdist}
  \end{center}
\end{figure}

For K$^-$ three data sets are available. A first set \cite{albrow1} covers, at fixed
cms angle, a $p_T$ range from 0.16 to 0.7~GeV/c in an $x_F$ window from
0.12 to 0.5, the upper cut-off being imposed here by the range of the
NA49 data. The relation between $x_F$ and $p_T$ is given by $p_T$~=~1.33~$x_F$.  
The ratio to the NA49 data is shown in Fig.~\ref{fig:alb_ptdist} as a
function of $p_T$ separately for the three available cms energies averaged over $x_F$.

\begin{figure}[h]
  \begin{center}
  	\includegraphics[width=5cm]{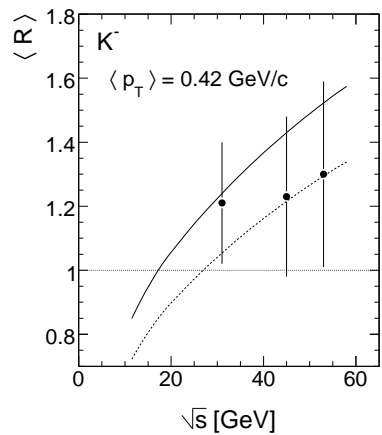}
  	\caption{Ratio $\langle R \rangle$ between \cite{albrow1} and NA49 averaged over the
             intervals 0.2~$< p_T <$~0.7~GeV/c and 0.15~$< x_F <$~0.5 as a function of $\sqrt{s}$.
             Full line: behaviour at $x_F$~=~0, broken line: 15\% normalization error}
  	\label{fig:alb_aver}
  \end{center}
\end{figure}

The corresponding $p_T$ dependences at $x_F$~=~0 are given as full lines in 
Fig.~\ref{fig:alb_ptdist}. As in the case of the data from \cite{capi}, see Sect.~\ref{sec:capi},
the general downward trend of the ratio with increasing $p_T$ is well
described by the ratios at $x_F$~=~0, although for $\sqrt{s}$~=~45 and 53~GeV
the data fall below (in contrast to \cite{capi}) by about 15\%. This
order of magnitude is definitely within the normalization errors
typical of ISR data, as discussed in some detail in \cite{pp_proton}.
The averaged ratios over 0.2~$< p_T <$~0.7~GeV/c and 0.15~$< x_F <$~0.5 are given
in Fig.~\ref{fig:alb_aver} as a function of $\sqrt{s}$, indicating again the trend
at $x_F$~=~0 as the full line and the reduced ratio corresponding to
a 15\% normalization error as the broken line.

\begin{figure}[h]
  \begin{center}
  	\includegraphics[width=5.cm]{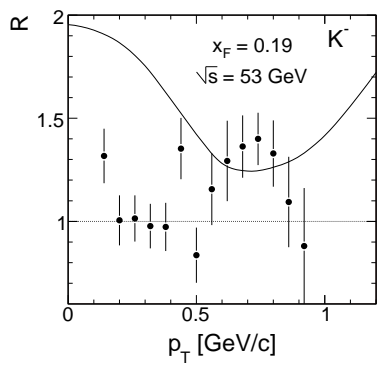}
  	\caption{Ratio $R$ between \cite{albrow2} and NA49 for K$^-$ at $x_F$~=~0.19 as a function of
             $p_{T}$. The full line gives the behaviour at $x_F$~=~0}
  	\label{fig:alb019_kaminus}
  \end{center}
\end{figure}

The second data set for K$^-$ \cite{albrow2} is obtained at the fixed $x_F$ value
of 0.19 in the $p_T$ range from 0.14 to 0.92~GeV/c, at $\sqrt{s}$~=~53~GeV. 
The ratio to the NA49 data, Fig.~\ref{fig:alb019_kaminus}, shows a structure which is 
very probably of systematic origin and equal for K$^-$ and K$^+$. At
$p_T$ below about 0.5~GeV/c the ratios are compatible with no
s-dependence from 17.2 to 53~GeV, whereas for $p_T$ above 0.6~GeV/c
the values are compatible with the $p_T$ dependence observed at
$x_F$~=~0, full line in Fig.~\ref{fig:alb019_kaminus}.

\begin{figure}[h]
  \begin{center}
  	\includegraphics[width=5.5cm]{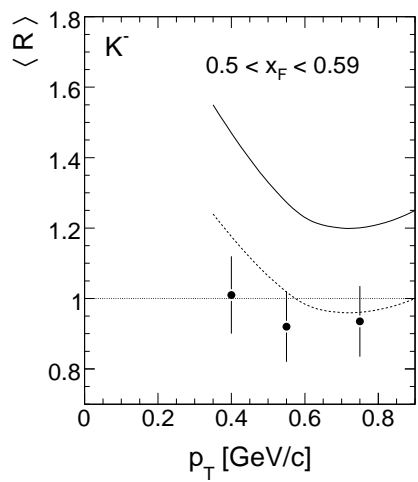}
  	\caption{Ratio $\langle R \rangle$ between \cite{albrow3} and NA49 K$^-$ as a function of
             $p_T$ averaged over the $x_F$ region from 0.5 to 0.59. The full line 
             gives the behaviour at $x_F$~=~0, the broken
             line shows the consequence of a 20\% normalization error}
  	\label{fig:alb3_kaminus}
  \end{center}
\end{figure}

The third and last data set available for K$^-$ \cite{albrow3} at 
$\sqrt{s}$~=~45~GeV covers the high $x_F$ range above 0.5 for the three $p_T$ values
0.4, 0.55 and 0.75~GeV/c. Here the comparison is extended up to
$x_F$~=~0.59 using a slight extrapolation of the NA49 K$^-$ data. The
ratio between \cite{albrow3} and NA49 has been averaged over this $x_F$ 
window and is shown in Fig.~\ref{fig:alb3_kaminus} as a function of $p_T$. 

Remarkably, the ratio turns out to be at or slightly below unity.
This would again indicate no $s$-dependence between $\sqrt{s}$~=~17.2
and 45~GeV, as compared to the behaviour at $x_F$~=~0 characterized
by the full line in Fig.~\ref{fig:alb3_kaminus}. This highly improbable case could
be explained by a 20\% normalization uncertainty, see the broken
line in Fig.~\ref{fig:alb3_kaminus}. After all it should be kept in mind that the
$s$-dependence in the intermediate $p_T$ range where most of the
comparison data are found is rather small and 10\% effects may make
all the difference in interpretation. 

\begin{figure}[h]
  \begin{center}
  	\includegraphics[width=5.5cm]{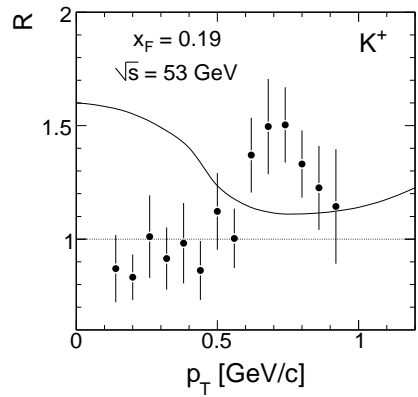}
  	\caption{Ratio $R$ between \cite{albrow2} and NA49 for K$^+$ at $x_F$~=~0.19  as a function of
             $p_{T}$. The full line gives the behaviour at $x_F$~=~0}
  	\label{fig:alb019_kaplus}
  \end{center}
\end{figure}

For K$^+$ four sets of data \cite{albrow2,albrow3,albrow4,albrow5}, 
with only partial overlap with the K$^-$ 
results, are available. A first set \cite{albrow2} is obtained
at fixed $x_F$~=~0.19 in the $p_T$ interval 0.14~$< p_T <$~0.92~GeV/c at $\sqrt{s}$~=~53~GeV.
As for K$^-$, the ratio between \cite{albrow2} and NA49, Fig.~\ref{fig:alb019_kaplus}, shows a structure 
indicating systematic problems, with no $s$-dependence below $p_T$~=~0.6~GeV/c 
followed by an increase of about 50\% at $p_T \sim$~0.75~GeV/c bracketing the
behaviour at $x_F$~=~0 shown as a full line in Fig.~\ref{fig:alb019_kaplus}.

\begin{figure}[h]
  \begin{center}
  	\includegraphics[width=5.5cm]{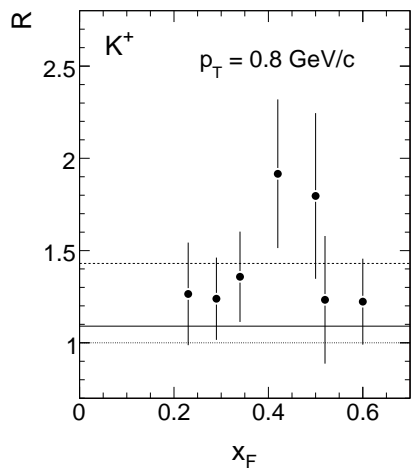}
  	\caption{Ratio $R$ between \cite{albrow4} and NA49 for K$^+$ as a function
             of $x_F$ at $p_T$~=~0.8~GeV/c and $\sqrt{s}$~=~45~GeV. The dashed
             line shows the mean ratio of the measured points and the full line
             shows the ratio at $x_F$~=~0}
  	\label{fig:alb4_kaplus}
  \end{center}
\end{figure}

A second data set \cite{albrow4} has been obtained at constant $p_T$~=~0.8~GeV/c
in a range of $x_F$ from 0.23 to 0.8 at $\sqrt{s}$~=~45~GeV. The $x_F$ range
has been cut at the upper limit of 0.6 for comparison purposes
with the (partially extrapolated) NA49 data. The ratio between
\cite{albrow4} and NA49 is shown in Fig.~\ref{fig:alb4_kaplus} as a function of $x_F$.

The ratio averaged over the seven $x_F$ values shown is 1.43
as compared to 1.09 as measured at $x_F$~=~0. This large value
is in internal disagreement with the other CHLM measurements
discussed here.

\begin{figure}[h]
  \begin{center}
  	\includegraphics[width=11cm]{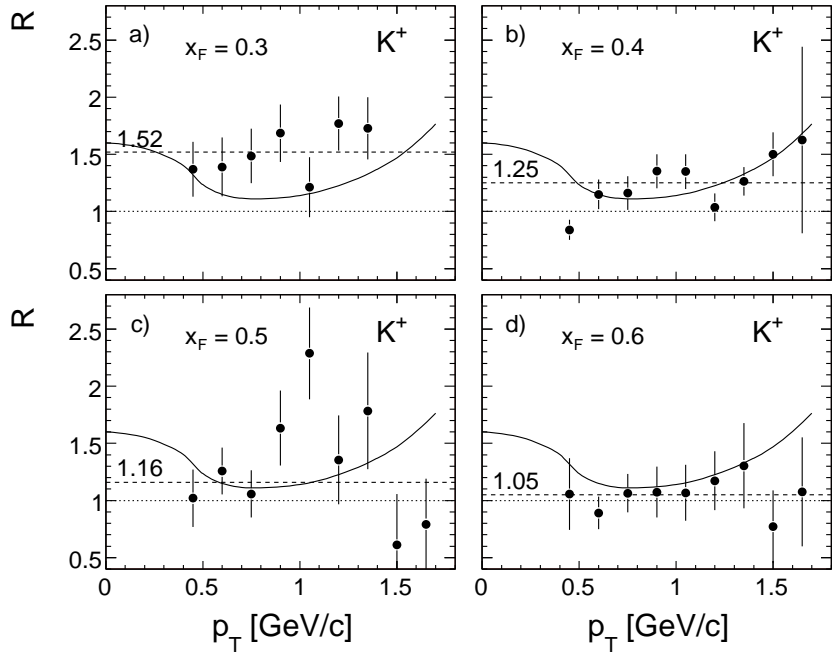}
  	\caption{Ratio $R$ between \cite{albrow5} and NA49 as a function
             of $p_T$ at a) $x_F$~=~0.3, b) $x_F$~=~0.4, c) $x_F$~=~0.5, d) $x_F$~=~0.6. 
             The $p_T$ dependence at $x_F$~=~0 is shown as the full line in each panel}
  	\label{fig:alb5_kaplus}
  \end{center}
\end{figure}

The third data set \cite{albrow5} contains data at fixed $x_F$ in $p_T$ ranges
from 0.4 to about 1.7~GeV/c, at $\sqrt{s}$~=~31, 45 and 53~GeV. At
$\sqrt{s}$~=~53~GeV the four $x_F$ values of 0.3, 0.4, 0.5 and 0.6
are available, whereas $\sqrt{s}$~=~31 and 45~GeV are limited to $x_F$~=~0.6
only, a bit uncomfortable with respect to the range of the NA49
data. Therefore only the data at $\sqrt{s}$~=~53~GeV are compared here,
as shown in Fig.~\ref{fig:alb5_kaplus}.

Within the large error margins of \cite{albrow5} the ratios are
compatible with flat $p_T$ dependences and mean values of 1.52,
1.25, 1.16 and 1.05~GeV/c for $x_F$~=~0.3, 0.4, 0.5 and 0.6, respectively.
Except for $x_F$~=~0.3, they are also compatible with the dependence
at $x_F$~=~0, shown as solid lines in Fig.~\ref{fig:alb5_kaplus}. The large mean ratio
at $x_F$~=~0.3 is in contradiction with \cite{capi,alper,guettler}.

\begin{figure}[h]
  \begin{center}
  	\includegraphics[width=5.5cm]{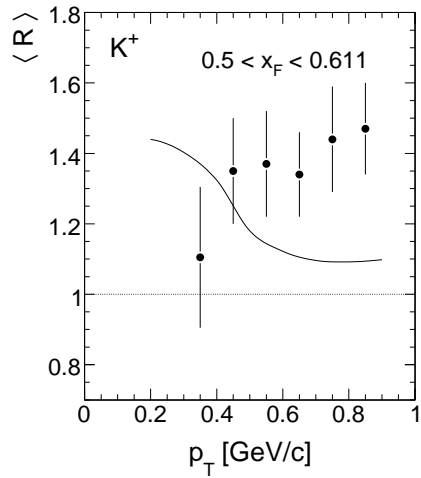}
  	\caption{Ratio $\langle R \rangle$ between \cite{albrow3} and NA49, averaged over $x_F$,
             as a function of $p_T$. The $p_T$ dependence at $x_F$~=~0 is given as
             the full line}
  	\label{fig:alb3_kaplus}
  \end{center}
\end{figure}

Finally, the K$^+$ data of \cite{albrow3} at $\sqrt{s}$~=~45~GeV cover the
range in $x_F$ from 0.5 to 0.611 for $p_T$ from 0.35 to 0.93~GeV/c.
The ratio between \cite{albrow3} and NA49 is shown in Fig.~\ref{fig:alb3_kaplus}, averaged
over the relatively small $x_F$ window, as a function of $p_T$.

Again the dependence is compatible with a constant ratio at 1.35,
but incompatible for $p_T >$~0.5~GeV/c with the
behaviour at $x_F$~=~0 \cite{albrow1,albrow2,albrow3,albrow4,albrow5}.

In conclusion of this sub-chapter concerning the CHLM data, a
certain frustration over the apparent sizeable systematic effects
contained in these data should be admitted. This precludes a
definite statement about the $x_F$ and $p_T$ dependences in the medium
to forward region of longitudinal momentum. This is the more
regrettable as no experiments are in view to produce new, more
precise data, especially not at the high energy colliders including
of course the LHC. Nevertheless it may be stated that the observed 
patterns are compatible, taking all data of Sects.~\ref{sec:capi} and \ref{sec:albrow}
together, with the behaviour observed at $x_F$~=~0 within the given
statistical errors, allowing also for the known systematic 
uncertainties (see also \cite{pp_proton}).

%
%
\subsubsection{Extrapolation of SPS and ISR data to $\sqrt{s}$~=~200~GeV}
\vspace{3mm}
\label{sec:isr_extr}

\begin{figure}[h]
  \begin{center}
  	\includegraphics[width=10cm]{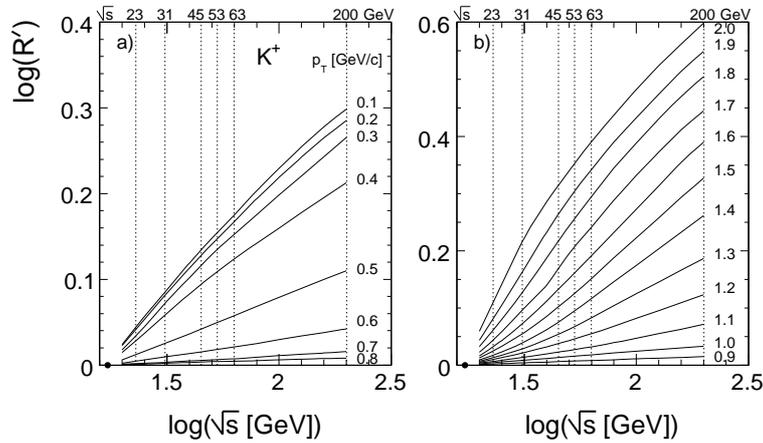}
  	\caption{$R'$ as a function of $\sqrt{s}$ for K$^+$ at a) $p_T$~=~0.1~$\div$~0.8~GeV/c and
  	         b) $p_T$~=~0.9~$\div$~2.0~GeV/c. The values of $\sqrt{s}$ are indicated with dotted
  	         lines. The NA49 point is marked with circle}
  	\label{fig:kaplus_inter}
  \end{center}
\end{figure}

\begin{figure}[h]
  \begin{center}
  	\includegraphics[width=10cm]{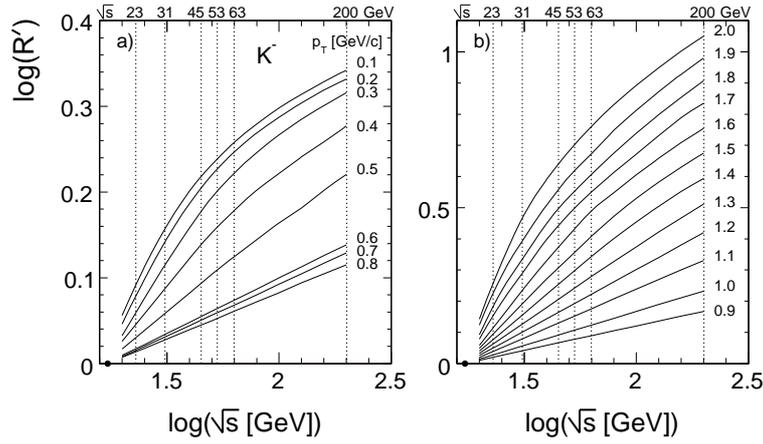}
  	\caption{$R'$ as a function of $\sqrt{s}$ for K$^-$ at a) $p_T$~=~0.1~$\div$~0.8~GeV/c and
  	         b) $p_T$~=~0.9~$\div$~2.0~GeV/c. The values of $\sqrt{s}$ are indicated with dotted
  	         lines. The NA49 point is marked with circle}
  	\label{fig:kaminus_inter}
  \end{center}
\end{figure}

In view of the scrutiny of kaon production at cms energies
above the ISR at RHIC and the p+$\overline{\textrm{p}}$ colliders in the
following sub-sections, and in view of the evident problems 
encountered with these higher energy data, it seems indicated
to perform an extrapolation of the combined SPS and ISR data
at least to RHIC energy, that is, to $\sqrt{s}$~=~200~GeV. This
attempt looks feasible given the dense coverage of the
$\sqrt{s}$ scale between 17 and 63~GeV and the smooth behaviour
of the cross sections as a function of $\sqrt{s}$. This is
evident from Figs.~\ref{fig:kaplus_inter} and \ref{fig:kaminus_inter} where 
the ratios of kaon densities per inelastic event at $x_F$~=~0

\begin{equation}
  \label{eq:rat_inel}
  R'=\frac{(f(x_F,p_T)/\sigma_{\textrm{inel}})^{\textrm{ISR}}}                 
          {(f(x_F,p_T)/\sigma_{\textrm{inel}})^{\textrm{NA49}}}
    = R\frac{\sigma_{\textrm{inel}}^{\textrm{NA49}}}{\sigma_{\textrm{inel}}^{\textrm{ISR}}},
\end{equation}
are shown as a function of $\sqrt{s}$ for fixed values of $p_T$ including
an extrapolation to $\sqrt{s}$~=~200~GeV for K$^+$ and K$^-$, respectively.
This extrapolation is extending the eyeball fits to the lower
energy data presented in Sect.~\ref{sec:central} without using arithmetic
formulations.    

The extrapolation is facilitated by the fact that over most of the
$p_T$ range covered the dependence on $\sqrt{s}$ is approximately
linear in the double-logarithmic plots of Figs.~\ref{fig:kaplus_inter} 
and \ref{fig:kaminus_inter}, which means a power-law behaviour of $R'$ 
as a function of $\sqrt{s}$. Noteable exceptions from this simple behaviour 
are visible both at low $p_T <$~0.5~GeV/c and at high $p_T >$~1.5~GeV/c. In both
regions the energy dependence flattens out with increasing $\sqrt{s}$. 

\begin{figure}[h]
  \begin{center}
  	\includegraphics[width=10cm]{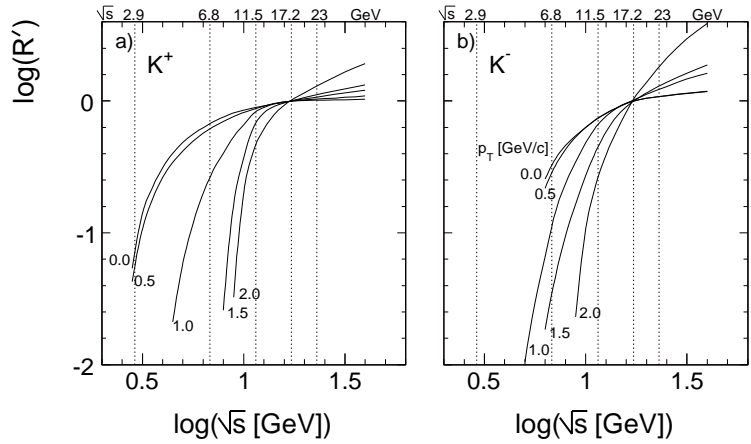}
  	\caption{$R'$ as a function of $\sqrt{s}$ for a) K$^+$ and b) K$^-$ at $p_T$~=~0, 0.5, 1, 1.5, 2~GeV/c.
  	         The values of $\sqrt{s}$ are indicated with dotted lines}
  	\label{fig:kainter_low}
  \end{center}
\end{figure}

In this context it is also interesting to look at the $\sqrt{s}$
dependence of $R'$ towards lower energies which is shown in Fig.~\ref{fig:kainter_low}
for a few $p_T$ values down to $\sqrt{s}$~=~3~GeV for K$^+$ (Sect.~\ref{sec:hogan-reed})
and 6.8~GeV for K$^-$ (Sect.~\ref{sec:ps-ags}).

As $R'$ must vanish at production threshold, the wide spread
of the $\sqrt{s}$ dependences for constant $p_T$ below SPS energies indicates
a corresponding and very characteristic spread of kaon production 
thresholds with transverse momentum. This spread is charge
dependent and reaches from $\sqrt{s}$ about 2.5 to 10~GeV for K$^+$
and from about 5 to 10~GeV for K$^-$, for $p_T$ from 0 to 2~GeV/c.
The lower effective threshold for K$^+$ is following from the
prevailing associate kaon-hyperon decays of non-strange 
baryonic resonances at low $\sqrt{s}$ whereas K$^-$ can only stem 
from heavy strange hyperons or heavy meson decay corresponding to a 
higher overall mass scale of the resonances involved.

\begin{figure}[h]
  \begin{center}
  	\includegraphics[width=10cm]{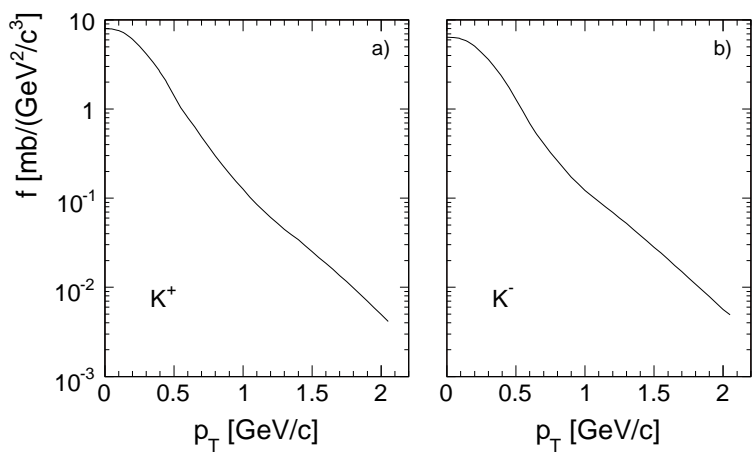}
  	\caption{$f(x_F = 0,p_T)$ as a function of $p_T$ extrapolated to $\sqrt{s}$~=~200~GeV for a) K$^+$, b) K$^-$}
  	\label{fig:kainter200}
  \end{center}
\end{figure}

The distributions of the invariant cross sections at $x_F$~=~0 as
a function of $p_T$ at $\sqrt{s}$~=~200~GeV, as they are resulting from
the extrapolation shown in Figs.~\ref{fig:kaplus_inter} and \ref{fig:kaminus_inter}, 
are presented in Fig.~\ref{fig:kainter200} for K$^+$ and K$^-$, respectively.

%
%
\subsubsection{$p_{T}$ integrated kaon yields in the ISR and RHIC energy range}
\vspace{3mm}
\label{sec:isr_ptint}

In view of the statements concerning the ISR data made above,
it might seem rather daring to attempt the integration of the
available double-differential cross sections into $p_T$ integrated
and even total kaon yields. Several facts encourage, nevertheless,
a new attempt based on purely experimental considerations:

\begin{itemize}
 \item The NA49 data offer a relatively precise starting point
       in the neighbourhood of the lowest ISR energy.
 \item Following the discussion in the preceding sections, the
       evolution of the double differential cross sections from
       SPS to ISR energies may be considered as experimentally
       established within error limits of about 10-30\%, depending on
       the $x_F$ range under study
 \item It is interesting to compare the integrated yields of charged
       kaons to the ones of K$^0_S$, the latter ones being rather 
       precisely determined well into the ISR energy range by bubble
       chamber experiments, see Sect.~\ref{sec:kzero} below.
 \item The extrapolation to RHIC energy described above will permit
       a comparison of the $p_T$ integrated results and an estimation 
       of the total kaon yields at $\sqrt{s}$~=~200~GeV.
\end{itemize}

The following approach has been followed. The detailed dependence
of the invariant cross sections on $p_T$ and $\sqrt{s}$ established at
$x_F$~=~0, Sect.~\ref{sec:central}, as characterised by the two-dimensional set
of factors $R_{\textrm{int}}$ relative to the NA49 data, Fig.~\ref{fig:isr_factor}, 
has been extended to the full range of $x_F$. This is motivated by the comparisons with 
all available data discussed in Sects.~\ref{sec:capi} and \ref{sec:albrow} above.
This means that the invariant cross sections at each energy are
obtained from the double differential NA49 data as follows:

\begin{equation}
   f(x_F,p_T,\sqrt{s}) = f(x_F,p_T,\sqrt{s} = \textrm{17.2~GeV})\,R(x_F=0,p_T,\sqrt{s})
\end{equation}

A set of cross sections at each ISR energy covering the major part
of the available phase space is thus established allowing the
extraction of $p_T$ integrated yields and total kaon multiplicities.
It should be stressed here that this approach avoids the use of
arithmetic formulations \cite{rossi} which would introduce systematic 
uncertainties beyond the statistical and systematic fluctuations 
of the data.
 
The resulting $p_T$ integrated quantities $F$, $dn/dx_F$ and $\langle p_T \rangle$, see
Sect.~\ref{sec:ptint_dist} for the definitions and the results from NA49, are presented in
Table~\ref{tab:kaplus_inter} for K$^+$ and in Table~\ref{tab:kaminus_inter} for K$^-$, as 
functions of $x_F$ for the five ISR energies and the extrapolation to $\sqrt{s}$~=~200~GeV.

\begin{table}[h]
\scriptsize
\renewcommand{\arraystretch}{1.04} 
\begin{center}
\begin{tabular}{|l|ccc|ccc|ccc|}
\hline
 $x_F$& $F$ & $dn/dx_F$ & $\langle p_T \rangle$& $F$ & $dn/dx_F$ & $\langle p_T \rangle$& 
        $F$ & $dn/dx_F$ & $\langle p_T \rangle$ \\ \hline & \multicolumn{3}{|c|}{$\sqrt{s}$~=~23~GeV} &\multicolumn{3}{|c|}{$\sqrt{s}$~=~31~GeV} &
   \multicolumn{3}{|c|}{$\sqrt{s}$~=~45~GeV}  \\
\hline

 0.0    & 0.718046 & 1.194105  &  0.4103 & 0.766959 & 1.697890  &  0.4040 & 0.860732 & 2.688205  &  0.3937 \\
 0.01   & 0.715161 & 1.169576  &  0.4117 & 0.763902 & 1.640579  &  0.4067 & 0.857354 & 2.514922  &  0.3990 \\
 0.025  & 0.710924 & 1.074290  &  0.4187 & 0.759372 & 1.429303  &  0.4177 & 0.852268 & 1.980574  &  0.4165 \\
 0.05   & 0.678220 & 0.835444  &  0.4365 & 0.724279 & 1.018700  &  0.4396 & 0.812600 & 1.255044  &  0.4409 \\
 0.075  & 0.631185 & 0.626612  &  0.4548 & 0.673675 & 0.724205  &  0.4581 & 0.755126 & 0.844196  &  0.4575 \\
 0.1    & 0.574058 & 0.468567  &  0.4696 & 0.612394 & 0.524937  &  0.4719 & 0.685878 & 0.594947  &  0.4692 \\
 0.125  & 0.525629 & 0.361256  &  0.4808 & 0.560492 & 0.397062  &  0.4818 & 0.627328 & 0.442880  &  0.4774 \\
 0.15   & 0.474791 & 0.280543  &  0.4904 & 0.505994 & 0.304496  &  0.4903 & 0.565758 & 0.336136  &  0.4846 \\
 0.2    & 0.385352 & 0.176750  &  0.5042 & 0.410305 & 0.189023  &  0.5024 & 0.458033 & 0.206202  &  0.4952 \\
 0.25   & 0.315988 & 0.118011  &  0.5119 & 0.336238 & 0.125195  &  0.5091 & 0.374951 & 0.135715  &  0.5010 \\
 0.3    & 0.252790 & 0.079474  &  0.5166 & 0.268889 & 0.083915  &  0.5132 & 0.299626 & 0.090630  &  0.5045 \\
 0.35   & 0.199830 & 0.054189  &  0.5193 & 0.212508 & 0.057048  &  0.5154 & 0.236701 & 0.061472  &  0.5064 \\
 0.4    & 0.157647 & 0.037561  &  0.5231 & 0.167566 & 0.039451  &  0.5189 & 0.186460 & 0.042417  &  0.5097 \\
 \hline 
 & \multicolumn{3}{|c|}{$\sqrt{s}$~=~53~GeV} &\multicolumn{3}{|c|}{$\sqrt{s}$~=~63~GeV} &
   \multicolumn{3}{|c|}{$\sqrt{s}$~=~200~GeV}  \\
\hline
 0.0    & 0.907310 & 3.283262  &  0.3896 & 0.958774 & 4.053594  &  0.3872 & 1.361199 & 15.780248  &  0.3701 \\
 0.01   & 0.903767 & 3.003092  &  0.3967 & 0.955065 & 3.593640  &  0.3968 & 1.356041 & 8.327144  &  0.4108 \\
 0.025  & 0.898397 & 2.235482  &  0.4172 & 0.949393 & 2.506432  &  0.4201 & 1.347942 & 3.865488  &  0.4343 \\
 0.05   & 0.856458 & 1.349593  &  0.4416 & 0.905015 & 1.445081  &  0.4436 & 1.284164 & 1.896418  &  0.4443 \\
 0.075  & 0.795580 & 0.891062  &  0.4571 & 0.840513 & 0.939197  &  0.4576 & 1.190655 & 1.179139  &  0.4518 \\
 0.1    & 0.722386 & 0.622589  &  0.4678 & 0.763056 & 0.651694  &  0.4675 & 1.079359 & 0.803331  &  0.4587 \\
 0.125  & 0.660523 & 0.461285  &  0.4754 & 0.697609 & 0.481094  &  0.4744 & 0.985550 & 0.587377  &  0.4641 \\
 0.15   & 0.595472 & 0.349070  &  0.4823 & 0.628735 & 0.363226  &  0.4809 & 0.886712 & 0.440633  &  0.4697 \\
 0.2    & 0.481787 & 0.213413  &  0.4924 & 0.508482 & 0.221500  &  0.4906 & 0.715118 & 0.266658  &  0.4787 \\
 0.25   & 0.394216 & 0.140205  &  0.4980 & 0.415966 & 0.145333  &  0.4959 & 0.583725 & 0.174178  &  0.4838 \\
 0.3    & 0.314933 & 0.093530  &  0.5014 & 0.332236 & 0.096874  &  0.4992 & 0.465636 & 0.115805  &  0.4870 \\
 0.35   & 0.248755 & 0.063399  &  0.5033 & 0.262389 & 0.065634  &  0.5010 & 0.367471 & 0.078339  &  0.4887 \\
 0.4    & 0.195886 & 0.043719  &  0.5065 & 0.206553 & 0.045234  &  0.5042 & 0.288794 & 0.053873  &  0.4919 \\
 \hline
\end{tabular}
\end{center}
\caption{$p_T$ integrated quantities $F$, $dn/dx_F$ and $\langle p_T \rangle$ as functions
         of $x_F$ for K$^+$ at the five ISR energies and the extrapolation to $\sqrt{s}$~=~200~GeV}
\label{tab:kaplus_inter}
\end{table}

\vspace{10mm}
\begin{table}[h]
\scriptsize
\renewcommand{\arraystretch}{1.04} 
\begin{center}
\begin{tabular}{|l|ccc|ccc|ccc|}
\hline
 $x_F$& $F$ & $dn/dx_F$ & $\langle p_T \rangle$& $F$ & $dn/dx_F$ & $\langle p_T \rangle$& 
        $F$ & $dn/dx_F$ & $\langle p_T \rangle$ \\ \hline
 & \multicolumn{3}{|c|}{$\sqrt{s}$~=~23~GeV} &\multicolumn{3}{|c|}{$\sqrt{s}$~=~31~GeV} &
   \multicolumn{3}{|c|}{$\sqrt{s}$~=~45~GeV}  \\
\hline
 0.0    & 0.546957 & 0.929748  &  0.3890 & 0.614254 & 1.389848  &  0.3823 & 0.714212 & 2.266652  &  0.3775 \\
 0.01   & 0.545661 & 0.912020  &  0.3901 & 0.612784 & 1.344296  &  0.3846 & 0.712491 & 2.121444  &  0.3823 \\
 0.025  & 0.532248 & 0.819621  &  0.3972 & 0.597648 & 1.144077  &  0.3955 & 0.694807 & 1.630612  &  0.3996 \\
 0.05   & 0.490776 & 0.612451  &  0.4159 & 0.550799 & 0.782306  &  0.4184 & 0.640128 & 0.992766  &  0.4257 \\
 0.075  & 0.428353 & 0.428707  &  0.4359 & 0.480435 & 0.519142  &  0.4392 & 0.558142 & 0.624896  &  0.4458 \\
 0.1    & 0.366644 & 0.300807  &  0.4527 & 0.410926 & 0.353239  &  0.4557 & 0.477167 & 0.414099  &  0.4612 \\
 0.125  & 0.311384 & 0.214662  &  0.4674 & 0.348679 & 0.247363  &  0.4697 & 0.404603 & 0.285632  &  0.4742 \\
 0.15   & 0.258263 & 0.152846  &  0.4807 & 0.288983 & 0.173995  &  0.4825 & 0.335162 & 0.199064  &  0.4863 \\
 0.2    & 0.178228 & 0.081804  &  0.4974 & 0.199154 & 0.091757  &  0.4980 & 0.230714 & 0.103833  &  0.5004 \\
 0.25   & 0.118186 & 0.044161  &  0.5038 & 0.131992 & 0.049151  &  0.5033 & 0.152835 & 0.055310  &  0.5048 \\
 0.3    & 0.080848 & 0.025436  &  0.5001 & 0.090330 & 0.028199  &  0.4982 & 0.104614 & 0.031643  &  0.4983 \\
 0.35   & 0.055669 & 0.015110  &  0.4931 & 0.062224 & 0.016712  &  0.4901 & 0.072072 & 0.018720  &  0.4890 \\
 0.4    & 0.037560 & 0.008959  &  0.4849 & 0.042014 & 0.009898  &  0.4810 & 0.048682 & 0.011078  &  0.4789 \\
 \hline 
 & \multicolumn{3}{|c|}{$\sqrt{s}$~=~53~GeV} &\multicolumn{3}{|c|}{$\sqrt{s}$~=~63~GeV} &
   \multicolumn{3}{|c|}{$\sqrt{s}$~=~200~GeV}  \\
\hline
 0.0    & 0.765590 & 2.807327  &  0.3762 & 0.811005 & 3.464805  &  0.3762 & 1.191690 &13.608795  &  0.3839 \\
 0.01   & 0.763732 & 2.568326  &  0.3828 & 0.809016 & 3.071368  &  0.3852 & 1.188510 & 7.248433  &  0.4278 \\
 0.025  & 0.744740 & 1.866148  &  0.4033 & 0.788890 & 2.091635  &  0.4089 & 1.158889 & 3.315935  &  0.4553 \\
 0.05   & 0.686089 & 1.083758  &  0.4298 & 0.726775 & 1.161676  &  0.4352 & 1.068634 & 1.576865  &  0.4704 \\
 0.075  & 0.598198 & 0.670376  &  0.4492 & 0.633740 & 0.708087  &  0.4539 & 0.933621 & 0.924127  &  0.4845 \\
 0.1    & 0.511375 & 0.440723  &  0.4641 & 0.541804 & 0.462577  &  0.4682 & 0.799559 & 0.594897  &  0.4974 \\
 0.125  & 0.433532 & 0.302680  &  0.4766 & 0.459325 & 0.316632  &  0.4803 & 0.678664 & 0.404384  &  0.5089 \\
 0.15   & 0.359087 & 0.210408  &  0.4886 & 0.380485 & 0.219703  &  0.4922 & 0.563037 & 0.279733  &  0.5211 \\
 0.2    & 0.247095 & 0.109417  &  0.5022 & 0.261793 & 0.114001  &  0.5055 & 0.387693 & 0.144548  &  0.5337 \\
 0.25   & 0.163661 & 0.058196  &  0.5063 & 0.173381 & 0.060565  &  0.5093 & 0.256741 & 0.076602  &  0.5366 \\
 0.3    & 0.112020 & 0.033267  &  0.4992 & 0.118662 & 0.034597  &  0.5017 & 0.175373 & 0.043612  &  0.5266 \\
 0.35   & 0.077167 & 0.019669  &  0.4895 & 0.081727 & 0.020444  &  0.4916 & 0.120507 & 0.025690  &  0.5139 \\
 0.40   & 0.052124 & 0.011636  &  0.4790 & 0.055198 & 0.012090  &  0.4808 & 0.081217 & 0.015151  &  0.5011 \\
 \hline
\end{tabular}
\end{center}
\caption{$p_T$ integrated quantities $F$, $dn/dx_F$ and $\langle p_T \rangle$ as functions
         of $x_F$ for K$^-$ at the five ISR energies and the extrapolation to $\sqrt{s}$~=~200~GeV}
\label{tab:kaminus_inter}
\end{table}

\begin{figure}[t]
  \begin{center}
  	\includegraphics[width=10cm]{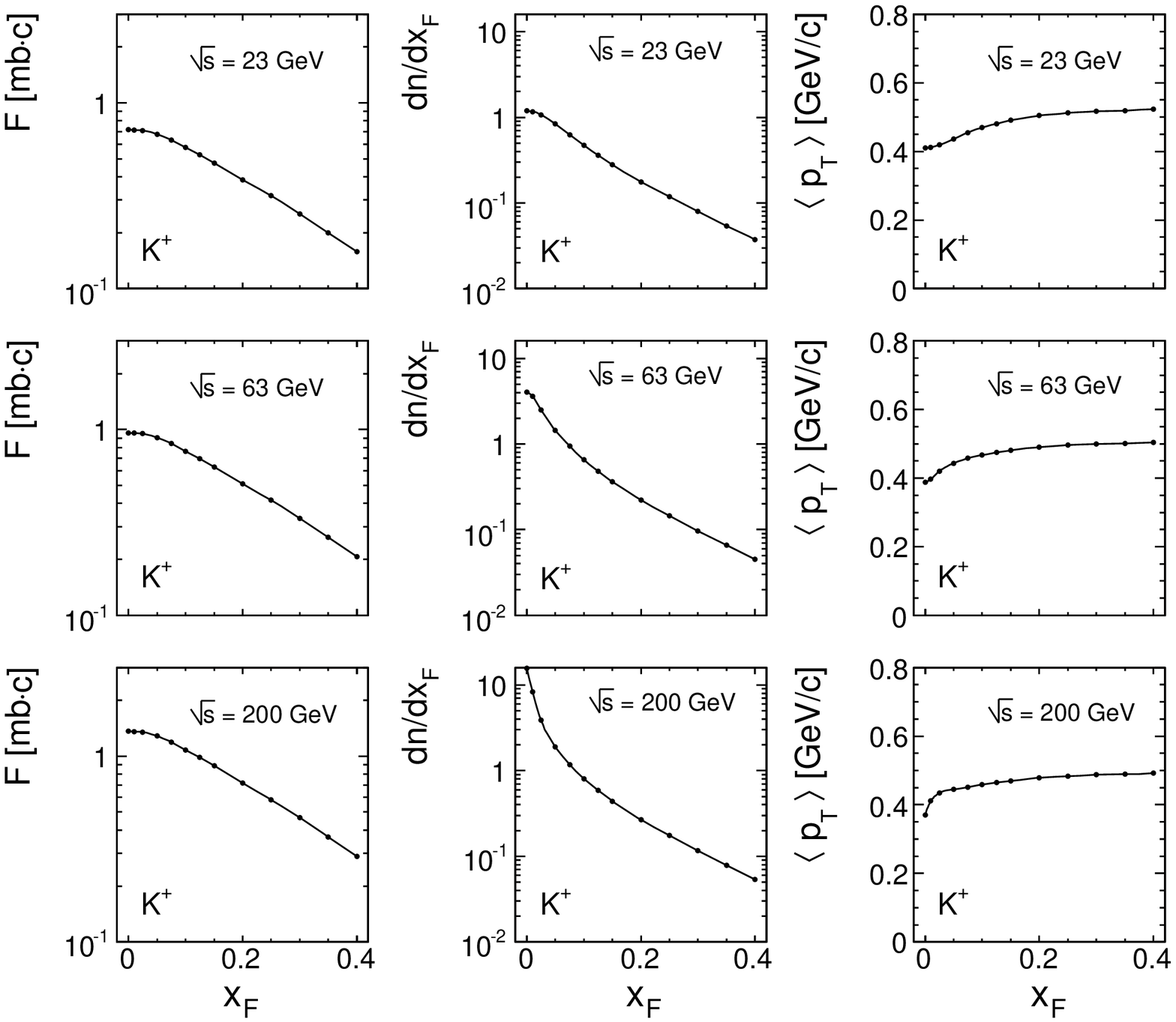}
  	\caption{$F$, $dn/dx_F$ and $\langle p_T \rangle$ as functions of $x_F$ for K$^+$ for 
             $\sqrt{s}$~=~23, 63 and 200~GeV}
  	\label{fig:isr_ptint_kaplus}
  \end{center}
\end{figure}

\begin{figure}[b]
  \begin{center}
  	\includegraphics[width=10cm]{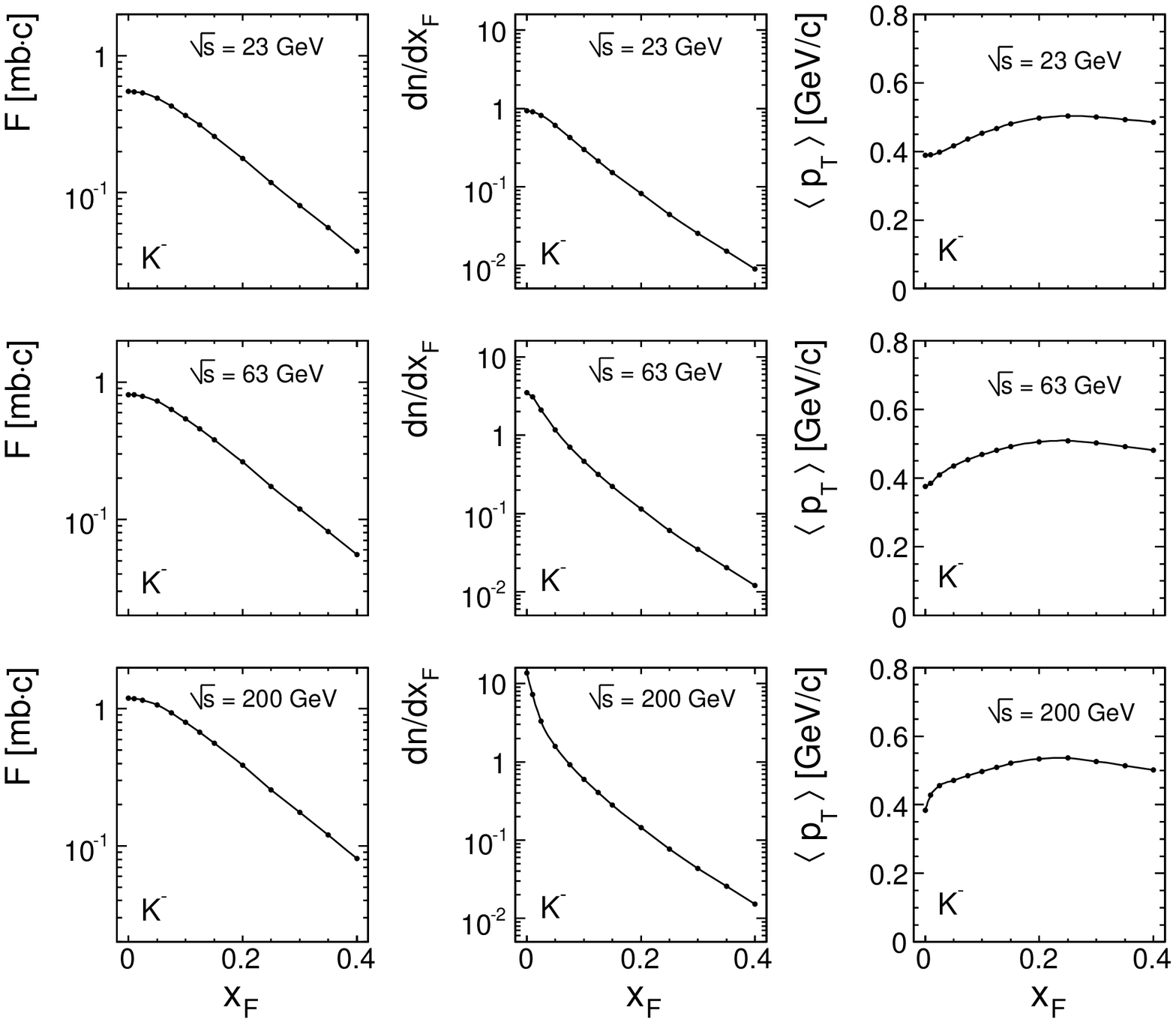}
  	\caption{$F$, $dn/dx_F$ and $\langle p_T \rangle$ as functions of $x_F$ for K$^-$ for 
             $\sqrt{s}$~=~23, 63 and 200~GeV}
  	\label{fig:isr_ptint_kaminus}
  \end{center}
\end{figure}

Corresponding plots of these quantities are shown in Figs.~\ref{fig:isr_ptint_kaplus}
and \ref{fig:isr_ptint_kaminus} for the three energies 23, 63 and 200~GeV and for K$^+$
and K$^-$, respectively. Salient features of these results are
the relatively slow and smooth increase of $F$ with energy
in comparison to the fast increase of $dn/dx_F$ at low $x_F$ which
is practically proportional to $\sqrt{s}$, see Eq.~\ref{eq:int} above,
and the quasi-invariance of mean $p_T$ with energy. The latter
feature is explained by the increase of the invariant cross
sections both at low $p_T$ and at high $p_T$, compensating each other
for the mean value, and the relatively small $s$-dependence in
the intermediate $p_T$ region.

%
%
\subsubsection{Total kaon yields}
\vspace{3mm}
\label{sec:tot_yield}

Integration over $x_F$ of the $dn/dx_F$ distributions results in the 
total multiplicities given in Table~\ref{tab:isr} together with the
mean kaon yields and the total K$^+$/K$^-$ ratios.

\begin{table}[h]
 \small
 \begin{center}
  \begin{tabular}{ccccc}
   \hline
   $\sqrt{s}$ [GeV] &  $\langle n_{\textrm{K}^+} \rangle$ & $\langle n_{\textrm{K}^-} \rangle$ &
    $(\langle n_{\textrm{K}^+} \rangle$ + $\langle n_{\textrm{K}^-} \rangle)/2$& 
    $\langle n_{\textrm{K}^+} \rangle$/$\langle n_{\textrm{K}^-} \rangle$ \\
   \hline
       23           &   0.2734   &  0.1709  &  0.2222  &  1.600  \\
       31           &   0.3269   &  0.2204  &  0.2737  &  1.483  \\
       45           &   0.4087   &  0.2901  &  0.3494  &  1.409  \\
       53           &   0.4482   &  0.3277  &  0.3880  &  1.367  \\
       63           &   0.4928   &  0.3625  &  0.4277  &  1.359  \\
      200           &   0.8189   &  0.6511  &  0.7350  &  1.258  \\
   \hline                 
  \end{tabular}
 \end{center}
 \caption{Total kaon multiplicities, mean charged yields and K$^+$/K$^-$ ratio at ISR energies 
          and extrapolation to $\sqrt{s}$~=~200 GeV}
 \label{tab:isr}
\end{table}

\begin{figure}[h]
  \begin{center}
  	\includegraphics[width=14cm]{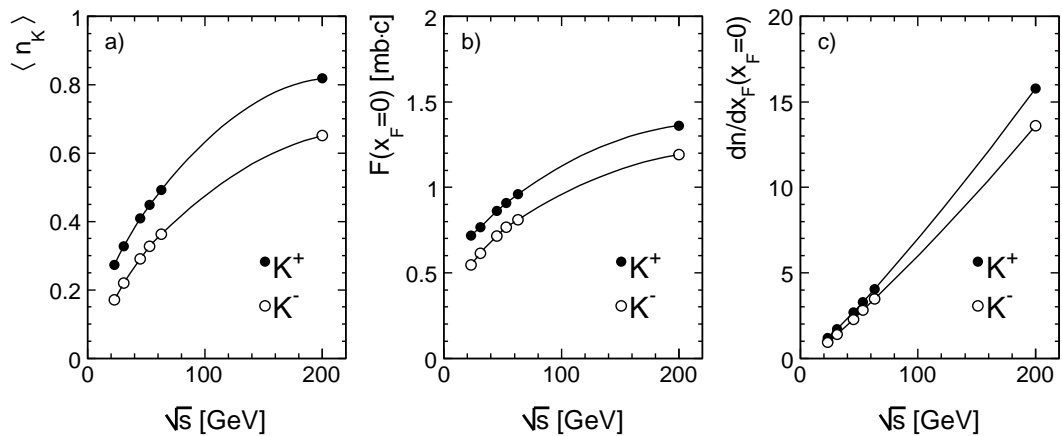}
  	\caption{a) $\langle n_K \rangle$, b) $F(x_F=0)$ and c) $dn/dx_F(x_F=0)$ as functions 
             of $\sqrt{s}$ for K$^+$ and K$^-$}
  	\label{fig:kainters}
  \end{center}
\end{figure}

These multiplicities are plotted as functions of $\sqrt{s}$ in Fig.~\ref{fig:kainters}a.
They are supplemented in Figs.~\ref{fig:kainters}b and \ref{fig:kainters}c
by the quantities $F(x_F~=~0,\sqrt{s})$ and $dn/dx_F(x_F~=~0,\sqrt{s})$, 
respectively.

It is difficult to define an error estimation for these quantities.
In fact all values with the exception of the bubble chamber 
experiment \cite{louttit} and the NA49 data for which the systematic errors
are within a bracket of 2--12\%, see Table~\ref{tab:sys} above, have been obtained
using rather important inter- and extrapolations.
It would therefore be advisable when performing comparisons or
predictions, in particular in connection with heavy ion interactions,
to allow for error margins of at least 20\% both at energies
below and above the SPS range.

%
%
\subsection{Data at RHIC}
\vspace{3mm}
\label{sec:rhic}

Only rather limited experimental information is available to date
from RHIC as far as double differential inclusive cross sections
for identified kaons in p+p interactions are concerned.
Data on central production come from STAR \cite{star1,star2,star3} using 
different identification methods and from PHENIX \cite{phenix1}, both at 
$\sqrt{s}$~=~200~GeV, as well as preliminary data from PHENIX \cite{phenix2}
at $\sqrt{s}$~=~62.4~GeV. The BRAHMS experiment has shown data at
$\sqrt{s}$~=~200~GeV with rapidities ranging from 0 to 3.3. In 
the present comparison two sets of central BRAHMS data \cite{brahms1,brahms2}
and the most forward data at rapidity 2.95 and 3.3 \cite{brahms3} 
will be addressed.

\begin{figure}[h]
  \begin{center}
  	\includegraphics[width=8.7cm]{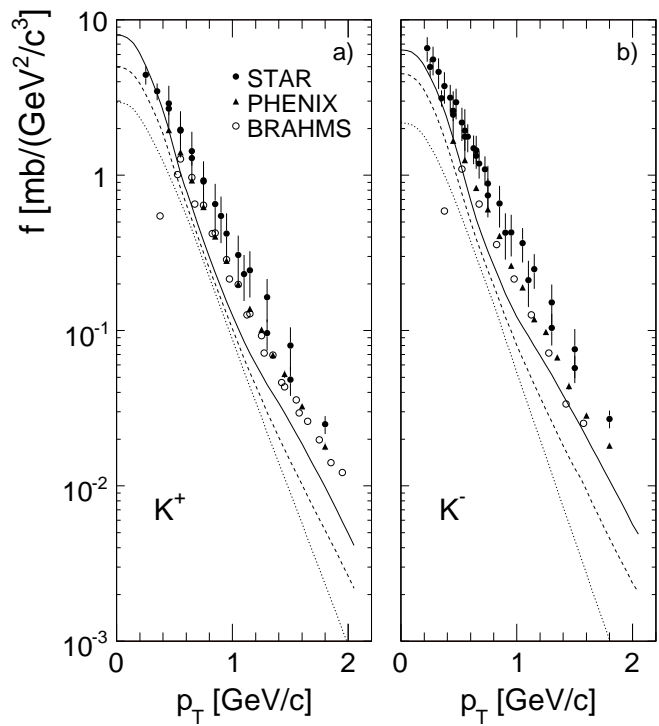}
  	\caption{Invariant kaon cross sections from RHIC at central rapidity and 
  	         $\sqrt{s}$~=~200~GeV \cite{star1,star2,star3,phenix1,brahms1,brahms2}
             as a function of $p_T$ for a) K$^+$ and b) K$^-$. The corresponding
             data from NA49 (dotted line), the interpolated data from
             the ISR at $\sqrt{s}$~=~63~GeV (dashed line) and the extrapolated 
             ISR data at $\sqrt{s}$~=~200~GeV (full line) are also shown}
  	\label{fig:rhic_ptdist}
  \end{center}
\end{figure}

The invariant cross sections at $y$~=~0 and $\sqrt{s}$~=~200~GeV, Fig.~\ref{fig:rhic_ptdist}, 
form a wide band within a margin of about a factor of 1.5--2 in the 
$p_T$ range from a lower limit at 0.25~GeV/c for STAR and about 0.4~GeV/c 
for PHENIX and BRAHMS up to the upper limit at about 2~GeV/c 
usable for the direct confrontation with the lower energy data 
in this paper.

The reason for this large internal variation of results which goes 
beyond any other data sets discussed in the preceding sections,
has to remain open for the time being. It is however clear
that, if compared to the NA49 data, to the ISR data at $\sqrt{s}$~=~63~GeV
and to the extrapolated ISR data, Sect.~\ref{sec:isr_extr} also shown in 
Fig.~\ref{fig:rhic_ptdist}, there is an evident change in shape of the $p_T$ 
distributions in particular compared to the 200~GeV extrapolation. In the 
lower $p_T$ region, below about 1~GeV/c, all RHIC data approach or cut below
the ISR extrapolation, whereas towards high $p_T$ a rather constant,
large offset of factors 2 to 3 is visible.

\begin{figure}[h]
  \begin{center}
  	\includegraphics[width=10.cm]{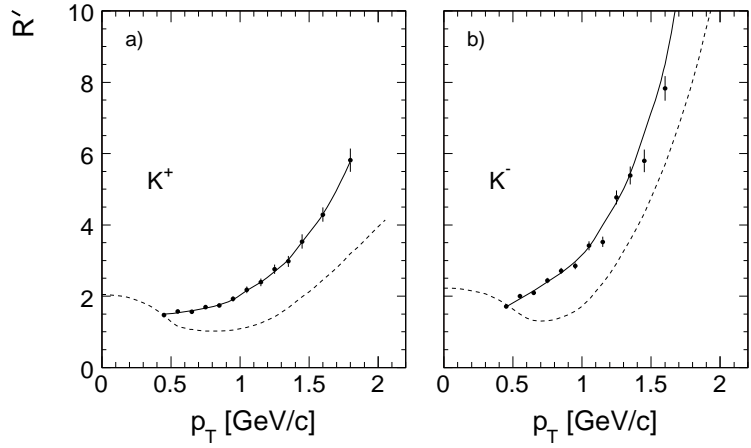}
  	\caption{Ratios $R'$, Eq.~\ref{eq:rihc_rat}, as functions of $p_T$, for the data
            from PHENIX \cite{phenix1} (full line) and the extrapolated ISR data
            at $\sqrt{s}$~=~200~GeV (dashed line). Panel a) for K$^+$, b) for K$^-$}
  	\label{fig:phe_rprime}
  \end{center}
\end{figure}

This is quantified in the ratio plots shown in Fig.~\ref{fig:phe_rprime}. Here the ratios
$R'$ of particle densities per inelastic event,

\begin{equation}
  \label{eq:rihc_rat}
   R'=\frac{f(x_F,p_T)/\sigma_{\textrm{inel}}^{\sqrt{s} = \textrm{200 GeV}}}
                {f(x_F,p_T)/\sigma_{\textrm{inel}}^{\textrm{NA49}}}              
\end{equation}
are presented in order to take out the increase of the inelastic cross
sections with energy for the PHENIX data \cite{phenix1} and the extrapolated
ISR data (Sect.~\ref{sec:isr_extr})

In Fig.~\ref{fig:phe_rprime}, two basic features of this data comparison are clearly
visible for the PHENIX data: both for K$^+$ and for K$^-$ there is
at $p_T >$~1~GeV/c a nearly constant factor of 1.5--1.8 with respect to
the extrapolated ISR data, whereas for $p_T <$~1~GeV/c the data sets
approach each other rapidly to become equal at the lower $p_T$ cut-off
of the RHIC data.

\begin{figure}[h]
  \begin{center}
  	\includegraphics[width=8.7cm]{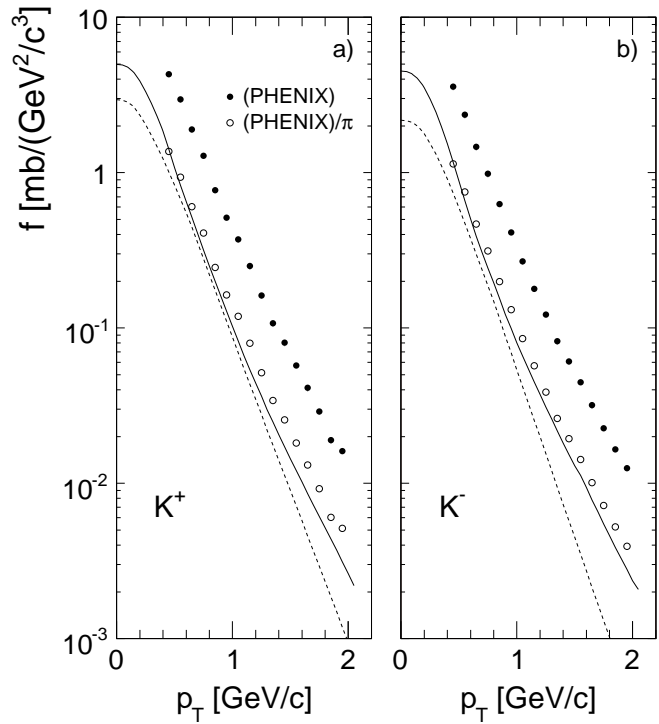}
  	\caption{Invariant kaon cross sections at $\sqrt{s}$~=~62.4~GeV
             from PHENIX \cite{phenix2}, for a) K$^+$ and b) K$^-$, as a function of $p_T$.
             The original data points (full points) and the data divided by $\pi$ (open points)
             are indicated. The data from the ISR \cite{alper,guettler} at $\sqrt{s}$~=~63~GeV 
             are also shown, together with the NA49 data, as the full and dashed lines, respectively}
  	\label{fig:phe63_ptdist}
  \end{center}
\end{figure}

At this point it might be useful to look at the central kaon data
from PHENIX at $\sqrt{s}$~=~62.4~GeV \cite{phenix2} that is, in the immediate 
neighbourhood of the ISR data \cite{alper,guettler} at $\sqrt{s}$~=~63~GeV. Here, the 
PHENIX experiment gives a $p_T$ distribution with the same lower cut-off 
as at 200~GeV, at $p_T$~=~0.45~GeV/c as shown in Fig.~\ref{fig:phe63_ptdist}.

\begin{figure}[h]
  \begin{center}
  	\includegraphics[width=7cm]{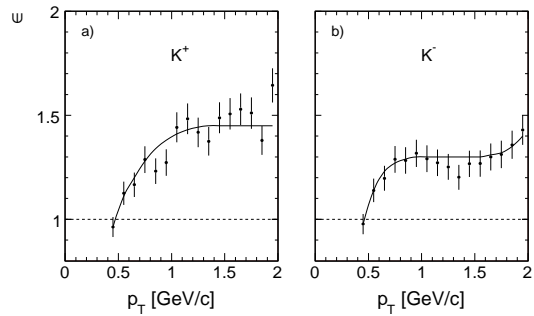}
  	\caption{Cross section ratio $\epsilon(p_T)$ as a function
             of $p_T$ for a) K$^+$ and b) K$^-$. The full lines give a representation
             of $\epsilon$ as a constant offset at $p_T >$~1~GeV/c combined
             with an efficiency loss towards lower $p_T$}
  	\label{fig:phe_eps}
  \end{center}
\end{figure}

A glance at the RHIC data points in Fig.~\ref{fig:phe63_ptdist} shows that they are far
above the ISR data by a factor of 3--5 varying with $p_T$, and even well
above the PHENIX data at $\sqrt{s}$~=~200~GeV \cite{phenix1}. As the same large 
factors apply for pions and baryons, it has been concluded that
a factor of 1/$\pi$ has probably been dropped in the cross sections 
given in \cite{phenix2}, a fact that is not uncommon in the definition of 
rapidity densities. Tentatively applying this factor, the cross 
sections move down to the lower data points shown in Fig.~\ref{fig:phe63_ptdist}. 
In direct comparison to the ISR data one may define the ratio

\begin{equation}
 \label{eq:epsilon}
 \epsilon(p_T,x_F=0) = \frac{f^{\textrm{RHIC}}(p_T,x_F=0,\sqrt{s}=63)}{f^{\textrm{ISR}}(p_T,x_F=0,\sqrt{s}=63)}
\end{equation}
shown in Fig.~\ref{fig:phe_eps}.

Both for K$^-$ and for K$^+$ two features are emerging from this plot.
Below $p_T \sim$~1~GeV/c there is a sharp drop of $\epsilon$ reaching values
below 1, corresponding to cross sections below the ISR data 
at the lower $p_T$ cut-off at 0.45~GeV/c. This looks like an apparative
loss of efficiency for kaon detection towards low $p_T$. At $p_T >$~1~GeV/c
on the other hand there is an offset which is approximately $p_T$
independent at a value of about 1.3 for K$^-$ and 1.45 for K$^+$. 
Tentatively regarding the ISR data as a reference this may be 
translated into a correction factor to be applied to the PHENIX 
data as a function of $p_T$ indicated by the full lines in Fig.~\ref{fig:phe_eps}
which combine a constant offset determined at $p_T >$~1~GeV/c, with 
an efficiency drop towards lower $p_T$. The latter effect, if of 
apparative origin, might be expected to hold for all reactions and 
all interaction energies studied by this experiment, in particular 
also for the measurements at $\sqrt{s}$~=~200~GeV both for elementary 
and nuclear collisions. The overall offset, on the other hand, 
could well depend on different experimental constraints as for 
example vertex distributions and/or trigger efficiency, and 
thereby be $s$ and reaction dependent. In particular the trigger 
conditions are largely different for elementary and nuclear reactions. 
One critical factor in comparing p+p interactions between ISR and RHIC 
experiments is given by the fraction of inelastic events picked up by 
the trigger arrangements, with trigger efficiencies approaching 100\% 
at the ISR as compared to typically 60--70\% at RHIC which favours small
impact parameters and thereby will tend to enhance strangeness yields.  

\begin{figure}[h]
  \begin{center}
  	\includegraphics[width=10.cm]{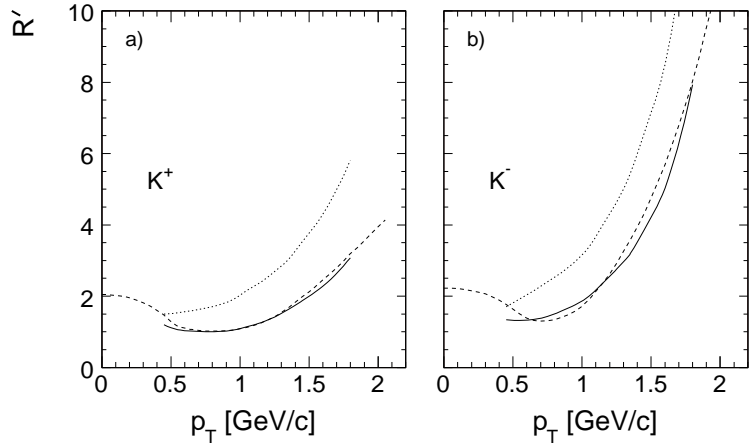}
  	\caption{Ratios $R'$ as functions of $p_T$, for the data from PHENIX \cite{phenix1} 
  	         divided by 1.3$\cdot\epsilon$ (full line) and the extrapolated ISR data
            at $\sqrt{s}$~=~200~GeV (dashed line). Panel a) for K$^+$, b) for K$^-$}
  	\label{fig:phe_rprime_eps}
  \end{center}
\end{figure}

Coming back now to the situation at $\sqrt{s}$~=~200~GeV, Fig.~\ref{fig:phe_rprime}, one
may try to apply the correction factor $\epsilon(p_T,x_F=0)$, Eq.~\ref{eq:epsilon}, as
determined from the PHENIX data at $\sqrt{s}$~=~62.4~GeV, to the higher
energy data, allowing only for an additional constant overall factor
corresponding to a variation of the offset term. As shown in Fig.~\ref{fig:phe_rprime_eps} 
an additional factor of 1.3 applied to $\epsilon(p_T,x_F=0)$ both for K$^+$ and K$^-$
brings the RHIC data sets into close agreement, within a 10\% margin, 
with the extrapolated ISR data.

This admittedly rather daring procedure might nevertheless bring some 
consistency into an experimental situation which otherwise would appear
distressingly incoherent within large factors. 

As far as the STAR results are concerned, they seem to indicate 
a similar combination of droop at low $p_T$ and a constant, very large 
overall offset. In view of the sizeable internal inconsistencies 
between the different publications from this experiment, a comparable 
study has however not been tried here.

The central BRAHMS data follow the PHENIX cross sections rather
closely for K$^+$ down to a lower cut-off in $p_T$ at 0.55~GeV/c \cite{brahms2}. 
Concerning the K$^+$ and K$^-$ data shown in \cite{brahms1} there is however a 
rather dramatic and unphysical drop at the given lower $p_T$ limit at 
0.375~GeV/c indicating an efficiency loss very similar to the one 
observed for PHENIX. In \cite{brahms1} the data for both charges fall below 
the PHENIX values by about 20\% in the overlapping $p_T$ region.

\begin{figure}[b]
  \begin{center}
  	\includegraphics[width=5cm]{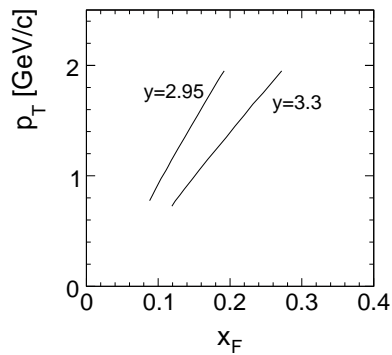}
  	\caption{Correlation between $x_F$ and $p_T$ for the two
             rapidities 2.95 and 3.3 of \cite{brahms3} at $\sqrt{s}$~=~200~GeV/c}
  	\label{fig:ptxf}
  \end{center}
\end{figure}

The final data sample addressed in this comparison concerns the
forward measurements from BRAHMS at rapidities 2.95 and 3.3 and 
$\sqrt{s}$~=~200~GeV \cite{brahms3}. As shown in the $x_F$/$p_T$ correlation plot
of Fig.~\ref{fig:ptxf} these data start from a lower limit at about 0.7~GeV/c
in $p_T$ and correspond to an $x_F$ range between about 0.1 to 0.3
for the combined two rapidity values.

\begin{figure}[h]
  \begin{center}
  	\includegraphics[width=7.5cm]{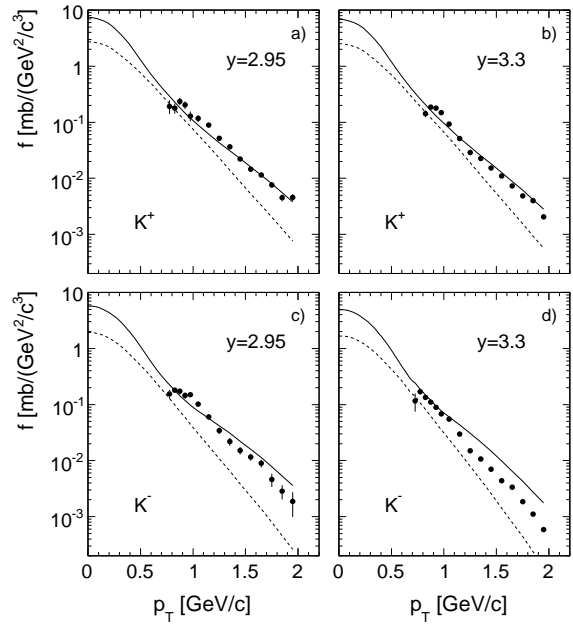}
  	\caption{Invariant cross sections as a function of $p_T$
             for rapidity 2.95 and 3.3 for K$^+$ (panels a and b) and K$^-$
             (panels c and d), respectively. Also shown are the NA49
             data (dashed lines) and the extrapolated ISR data (full
             lines) at these rapidities}
  	\label{fig:brahms_forw}
  \end{center}
\end{figure}

This coverage is comparable to the intermediate data of Capiluppi
et al. \cite{capi} at the ISR, see Sect.~\ref{sec:capi}. These results have
been shown, Figs.~\ref{fig:capi_kaplus} to \ref{fig:capi_kaminus_sdep}, to be 
compatible with the application of the $s$-dependence observed at $x_F$~=~0 to the 
NA49 data in the corresponding $x_F$ and $p_T$ ranges. It is therefore 
interesting to confront the forward BRAHMS data both with the NA49 data and 
with the extrapolation of the ISR results to $\sqrt{s}$~=~200~GeV, Sect.~\ref{sec:isr_extr},
as shown in Fig.~\ref{fig:brahms_forw}.

\begin{figure}[h]
  \begin{center}
  	\includegraphics[width=7.5cm]{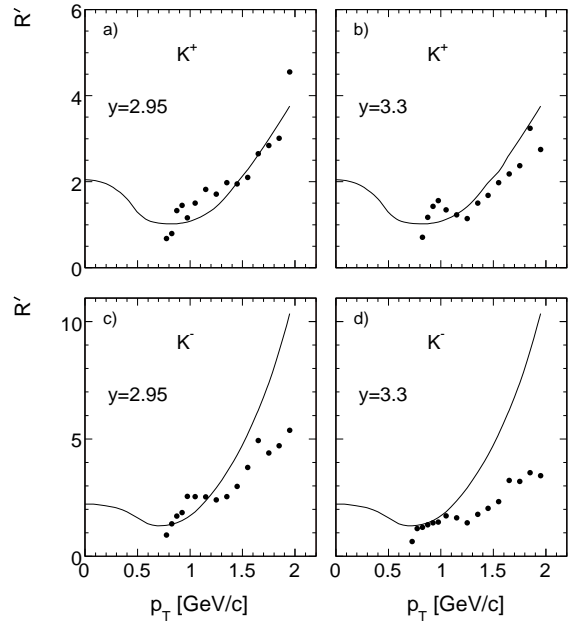}
  	\caption{Ratio $R'$ of kaon densities per inelastic event 
             as a function of $p_T$ for K$^+$ (panels a and b) and K$^-$ (panels c
             and d). The ratios $R'$ corresponding to the extrapolated ISR
             data are shown as the full lines}
  	\label{fig:brahms_rprime}
  \end{center}
\end{figure}

Evidently the K$^+$ data from BRAHMS are rather close to the extrapolated
ISR data for $p_T >$~1.2~GeV/c, whereas the K$^-$ data show offsets by
factors of about 0.6 at $y$~=~2.95 and 0.5 at $y$~=3.3 in the same $p_T$
range. Below $p_T \sim$~1.2~GeV/c the BRAHMS data increase rapidly up
to a local maximum at $p_T \sim$~0.8--0.9~GeV/c which is evidently
non-physical. The sharp drop of the cross sections below this
maximum to values even below the NA49 data indicates again the
loss of kaon detection efficiency below $p_T \sim$~1~GeV/c which seems
to be common to all RHIC data which have been discussed in this 
section. These features are quantified in Fig.~\ref{fig:brahms_rprime} where the ratio
of kaon densities per inelastic event $R'$ (Eq.~\ref{eq:rat_inel}) is plotted as
a function of transverse momentum.

In conclusion to this section it appears that the RHIC data discussed
here seem to indicate not only problems with absolute normalization
evident in the comparison of different experiments at the same
energy and to extrapolations from the ISR range, but in addition a 
common drop in kaon efficiency in the approach to their lower $p_T$ 
cut-off, see also Sect.~\ref{sec:coll_200}. This cut-off is, with about 0.4 to 0.8~GeV/c, 
uncomfortably high with respect to an eventual determination of  $\langle p_T \rangle$. 
The use of these data as a reference for nuclear interactions, in particular
concerning eventual "nuclear modification" or "jet quenching"
effects widely claimed by the RHIC community, is therefore to be
seen with some concern.    

%
%
\subsection{Data from p+$\overline{\textrm{p}}$ colliders}
\vspace{3mm}
\label{sec:coll}

Three experiments at the CERN p+$\overline{\textrm{p}}$ collider have given kaon
cross sections: UA5 \cite{alpgard,alner1,ansorge1,alner2,ansorge2}, 
UA2 \cite{ua2} and UA1 \cite{ua1} in the range of $\sqrt{s}$ 
from 200 to 900~GeV. At the Fermilab Tevatron, two groups, CDF \cite{cdf1,cdf2} 
and E735 \cite{e735} have produced kaon data from $\sqrt{s}$~=~300 to 1800~GeV.
These data are generally centered at central rapidity, within a range of
1.5 to 5 units. From refs.\cite{cdf2} and \cite{e735} 
only unnormalized yields are available. For charged kaons, 
the statistical significance is limited to a few dozen to a
few hundred identified particles, whereas for K$^0_S$ a wide range
from a few hundred up to 60k reconstructed decays is covered.
Due to the isospin configuration of the initial state and the
limited acceptance of all experiments in $x_F$, only the mean
charged yields, (K$^+$ + K$^-$)/2 are given. As also the equality:

\begin{equation}
  \label{eq:isosym}
  \textrm{K}^0_S = \frac{\textrm{K}^+ + \textrm{K}^-}{2}
\end{equation}
is at least within the quoted errors fulfilled for this energy
region, see the following Sect.~\ref{sec:kzero} for a more detailed argumentation,
both the mean charged kaon and the K$^0_S$ data are combined in this 
section in an attempt to link the results to the lower energy
regime discussed above.

As all experiments use double-arm triggers with a limited coverage
in the extreme forward direction, the trigger cross sections 
correspond in general not to the total inelastic cross section
but to a fraction of the so-called "non single-diffraction"
cross section. This fraction is quoted as 93\% (E735), 95\% (UA5), 
96\% (UA1) and 98\%(UA2). Since single diffraction makes up about
15\% of the total inelastic cross section, the experiments trigger
on about 80\% of $\sigma_{\textrm{inel}}$. If compared to the NA49 and ISR data
including the extrapolation to 200~GeV which are obtained in
relation to the full $\sigma_{\textrm{inel}}$, a correction for the trigger losses
is in principle necessary. Only the UA5 collaboration has estimated
this correction \cite{ansorge1} to about -16\% at $\sqrt{s}$~=~200~GeV and -12\%
at $\sqrt{s}$~=~900~GeV. In the following subsections all comparisons 
are carried out including the necessary correction to the full 
inelastic cross section. In addition all data are given as invariant 
densities by dividing the invariant cross sections, if given in mb, 
by the inelastic cross section. 

%
%
\subsubsection{Data at $\sqrt{s}$~=~200~GeV}
\vspace{3mm}
\label{sec:coll_200}

\begin{figure}[h]
  \begin{center}
  	\includegraphics[width=11.cm]{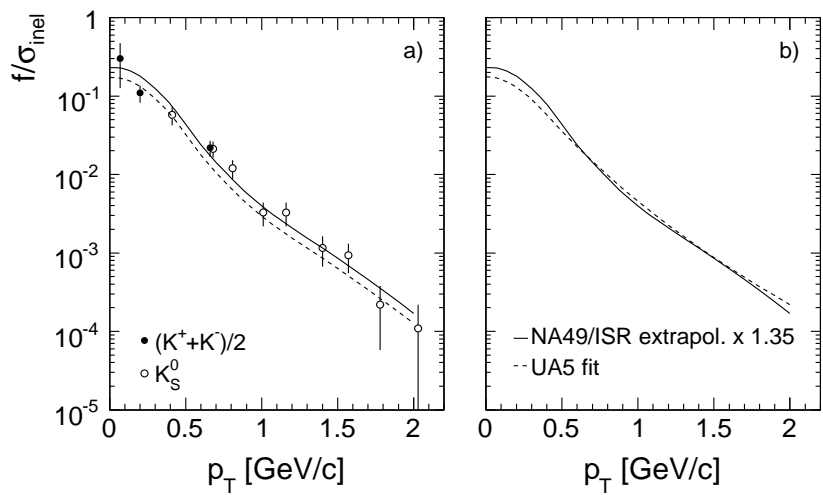}
  	\caption{a) Invariant kaon density from UA5 compared to the
             NA49/ISR data extrapolation (dashed line) and to the 
             extrapolation multiplied by 1.35 (full line) b) UA5 fit to
             their data (dashed line) compared to the data extrapolation multiplied
             with 1.35 (full line)}
  	\label{fig:ua5}
  \end{center}
\end{figure}

The UA5 experiment \cite{ansorge2} gives cross sections for K$^0_S$ and (K$^+$ + K$^-$)/2
which may be compared to the extrapolation from NA49 and ISR data to
this energy, Sect.~\ref{sec:isr_extr}. As the UA5 data are given over their
rapidity interval of $\pm$2.5 units of rapidity in the form
$1/\sigma_{\textrm{NSD}} d^2\sigma/dp_T^2$ a transformation into invariant density
$1/(2\pi p_T\sigma_{\textrm{inel}}) d^2\sigma/dydp_T$ has been performed including
the correction for trigger losses, see above. This results in the
data shown in Fig.~\ref{fig:ua5}a compared to the extrapolated NA49/ISR data.

Evidently the UA5 data are on average higher than the data extrapolation
(dashed line in Fig.~\ref{fig:ua5}a) by about 35\% as shown by the full line in
Fig.~\ref{fig:ua5}a. This systematic difference is certainly compatible with
the uncertainty inherent in the data extrapolation and with the
$\sim$20\% uncertainty given for the normalization of the UA5 data. What
is interesting here is that the shape of the extrapolated distribution
after renormalization is compatible within the statistical errors with 
the UA5 data over the full range of $p_T$ from 0.07 to 2~GeV/c. 

UA5 has performed a fit to their data of the double form:

\begin{equation}
   \label{eq:ua5fit}
   \frac{1}{\sigma_{\textrm{NSD}}} \frac{d\sigma}{dp_T^2} = 
   \begin{cases}   
      A e^{-b\, m_T},        & \textrm{for~$p_T \leq$~0.4~GeV/c}  \\
      A' \left(\frac{p_0}{p_T + p_0}\right)^n,  & \textrm{for~$p_T >$~0.4~GeV/c}
   \end{cases}
\end{equation}

The first form at low $p_T$ is necessitated by the unphysical behaviour
of the second form through $p_T$~=~0; it is motivated by the idea of
thermal behaviour at low transverse momentum. The fit parameters \cite{ansorge1}
at $\sqrt{s}$~=~200~GeV are $A$~=~10.9, $b$~=~8.2, $A'$~=~0.60, $n$~=~8.8 and $p_0$~=~1.3~GeV/c.
The inverse $m_T$ slope of 0.12~GeV/c implied by the parameter $b$ is
however rather low and corresponds to the non-thermal behaviour
observed for the ISR data, see Fig.~\ref{fig:slope_isrxf0}. As stated above, the fit values
have been reduced by 16\% for comparison at the full inelastic 
cross section.

The UA5 fit is compared to the renormalized data extrapolation 
(factor 1.35 introduced above) in Fig.~\ref{fig:ua5}b where good 
agreement is visible down to $p_T \sim$~0.5~GeV/c. The deviation towards
$p_T$~=~0~GeV/c leads to a difference of about 20~MeV/c in $\langle p_T \rangle$, 
see Sect.~\ref{sec:coll_mpt} below. As far as integrated yields are concerned, UA5 gives
a rapidity density of 0.12 per inelastic event \cite{ansorge1} which corresponds
to the integrated cross section $F = \sigma_{\textrm{inel}}/\pi \cdot dn/dy$~=~1.608~mb and
compares to 1.277~mb for the NA49/ISR extrapolation. This is a 26\%
difference which agrees, taking into account the different
shape of the $p_T$ distributions, with the renormalization shown in
Fig.~\ref{fig:ua5}a. For the total K$^0_S$ yield, UA5 extrapolates to full phase
space using model assumptions \cite{ansorge1}. This leads to a total K$^0_S$ multiplicity
of 0.68 \cite{ansorge2} or 0.72 \cite{ansorge1} per inelastic event at $\sqrt{s}$~=~200~GeV. 
For the NA49/ISR data extrapolation this number is 
$\langle n_{(\textrm{K}^+ + \textrm{K}^-)/2} \rangle$~=~0.735 
per inelastic event. This agreement to within 5\% is of course to be 
regarded as fortuitous in view of the large uncertainties involved 
in both attempts to estimate total yields.

\begin{figure}[h]
  \begin{center}
  	\includegraphics[width=16cm]{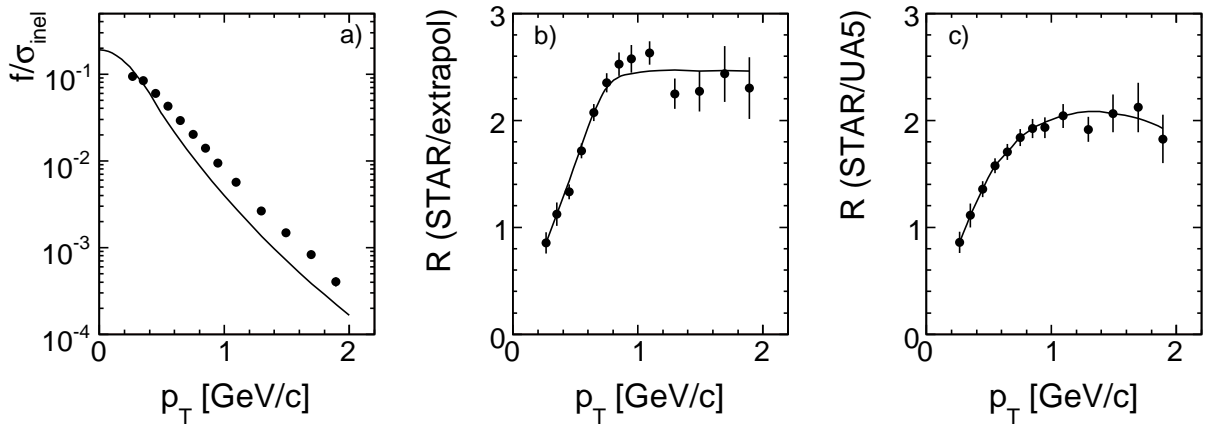}
  	\caption{Comparison of K$^0_S$ data from STAR at RHIC with
             the (K$^+$ + K$^-$)/2 extrapolation at $\sqrt{s}$~=~200~GeV. Panel a)
             invariant rapidity densities, panel b) ratio between the
             two results, panel c) ratio between the STAR results
             and the UA5 data fit}
  	\label{fig:star_ua5}
  \end{center}
\end{figure}

A further interesting comparison is offered by the K$^0_S$ data from
STAR \cite{star2} concerning invariant densities per event $1/(2\pi p_T) \cdot d^2N/dydp_T$.
These data are shown in comparison to the NA49/ISR extrapolation
of (K$^+$ + K$^-$)/2 in Fig.~\ref{fig:star_ua5}a.

At $p_T >$~0.8~GeV/c there is a large offset of more than a factor of 2
between the STAR data and the extrapolation. This offset reduces
rapidly towards lower $p_T$ until the STAR yields fall below the
extrapolation at their lowest measured $p_T$. The ratio between
STAR and extrapolation is given in Fig.~\ref{fig:star_ua5}b. The observed
behaviour with an offset factor of 2.4 and a rapid decrease 
of the ratio below $p_T \sim$~1~GeV/c reproduces the features seen for 
charged kaons, see Sect.~\ref{sec:rhic}. The comparison of the STAR data
with the UA5 data fit, Fig.~\ref{fig:star_ua5}c, shows that these data are also
in disagreement with the UA5 results obtained at the same energy.
  
%
%
\subsubsection{The $\sqrt{s}$ region of 540--630~GeV}
\vspace{3mm}
\label{sec:coll_540}

Five measurements are available in this region at $\sqrt{s}$ of
540 and 630~GeV: (K$^+$ + K$^-$)/2 and K$^0_S$ from UA5, (K$^+$ + K$^-$)/2 from
UA2 and (K$^+$ + K$^-$)/2 from E735 at $\sqrt{s}$~=~540; (K$^+$ + K$^-$)/2 and
K$^0_S$ from UA1 and K$^0_S$ from CDF in two different data sets at 
$\sqrt{s}$~=~630 GeV. As the central rapidity density $dN/dy$ changes
only by 4.8\% and the inelastic cross section by 1.6\% between
these two energies, the results may be compared by introducing the
resulting small correction. In the following all results will be 
referred to $\sqrt{s}$~=~540~GeV. As in Sect.~\ref{sec:coll_200}, invariant densities
will be obtained from mb cross sections, whenever given,  
dividing by the inelastic cross section. In addition a reduction
of 14\% is introduced to refer the data to the full inelastic
cross section.

\begin{figure}[h]
  \begin{center}
  	\includegraphics[width=6.5cm]{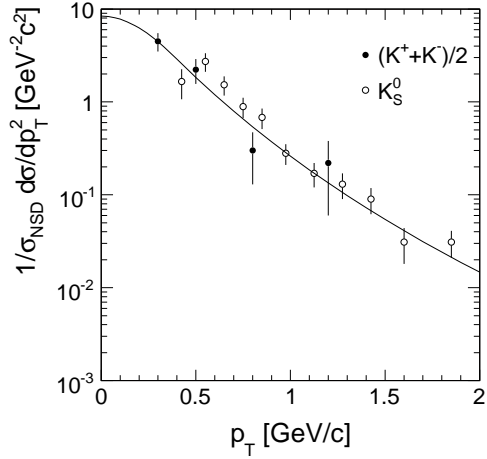}
  	\caption{UA5 data at $\sqrt{s}$~=~540~GeV. The full line
             corresponds to the fit described in the text}
  	\label{fig:ua5_540}
  \end{center}
\end{figure}

The fit (\ref{eq:ua5fit}) to the UA5 data \cite{ansorge1} has been chosen as an absolute 
reference for the subsequent data comparison. At $\sqrt{s}$~=~540~GeV
the parameters are $A$~=~7.09, $b$~=~7.5, $A'$~=~0.508, $n$~=~7.97 and 
$p_0$~=~1.3~GeV/c. The comparison of the UA5 data \cite{alner1} with this 
fit is shown in Fig.~\ref{fig:ua5_540}.

\begin{figure}[h]
  \begin{center}
  	\includegraphics[width=12cm]{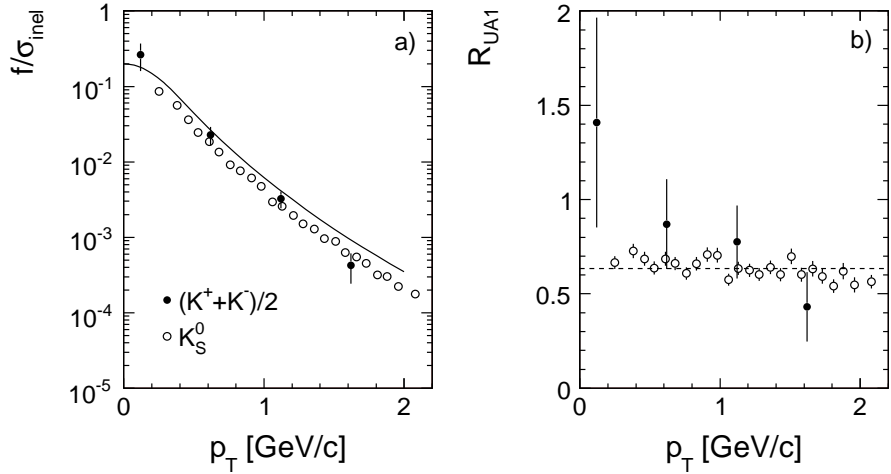}
  	\caption{a) K$^0_S$ and (K$^+$ + K$^-$)/2 data from UA1 \cite{ua1} in
             comparison to the fit of the UA5 data (full line) b)
             ratio $R_{\textrm{UA1}} = \left(\frac{f}{\sigma_{\textrm{inel}}}\right)_{\textrm{UA1}}\Big/
                       \left(\frac{f}{\sigma_{\textrm{inel}}}\right)_{\textrm{UA5 fit}}$. The mean
             offset factor of 0.634 for K$^0_S$ is indicated in panel b) with dashed line }
  	\label{fig:ua1}
  \end{center}
\end{figure}

UA1 data with about 60k K$^0_S$ and 3000 charged kaons are available \cite{ua1}. 
These data are shown in comparison to the UA5 fit in 
Fig.~\ref{fig:ua1} up to $p_T$~=~2.08~GeV/c.

For the K$^0_S$ there is a mean offset by a factor of 0.634, 
with an excellent reproduction of the shape of the
UA5 fit as a function of $p_T$. This is quantified in panel b)
of Fig.~\ref{fig:ua1} where the ratio between the UA1 data and the fit
is presented as a function of $p_T$. The fluctuation around
the mean ratio, with a standard deviation of about 7\%, is
compatible with the errors quoted by UA1. The four given data
points for charged kaons, with substantially larger errors,
are on average higher than the K$^0_S$ data and fluctuate to
within one standard deviation around the UA5 fit.

\begin{figure}[h]
  \begin{center}
  	\includegraphics[width=11.5cm]{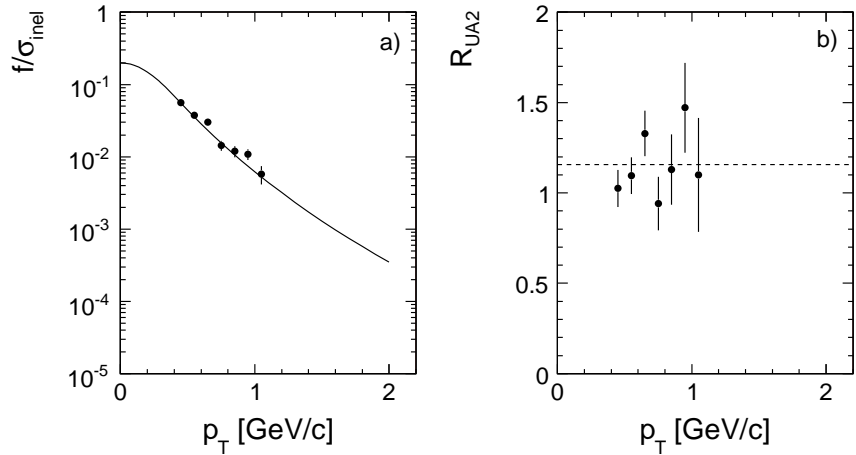}
  	\caption{a) (K$^+$ + K$^-$)/2 data from UA2 \cite{ua2} at 
             $\sqrt{s}$~=~540 GeV compared to the UA5 fit, b) ratio
             $R_{\textrm{UA2}} = \left(\frac{f}{\sigma_{\textrm{inel}}}\right)_{\textrm{UA2}}\Big/
                       \left(\frac{f}{\sigma_{\textrm{inel}}}\right)_{\textrm{UA5 fit}}$. The mean
             offset factor of 1.13 is indicated in panel b) with dashed line}
  	\label{fig:ua2}
  \end{center}
\end{figure}

UA2 \cite{ua2} has published seven data points on (K$^+$ + K$^-$)/2 yields
which are compared to the UA5 fit in Fig.~\ref{fig:ua2}.

As is visible from the ratio as a function of $p_T$ in Fig.~\ref{fig:ua2}b
there is good agreement between the two data sets, with a
mean offset of only +13\% of the UA2 data with respect to the
UA5 fit. Again the shape of the $p_T$ distribution is well
reproduced.

\begin{figure}[h]
  \begin{center}
  	\includegraphics[width=15.5cm]{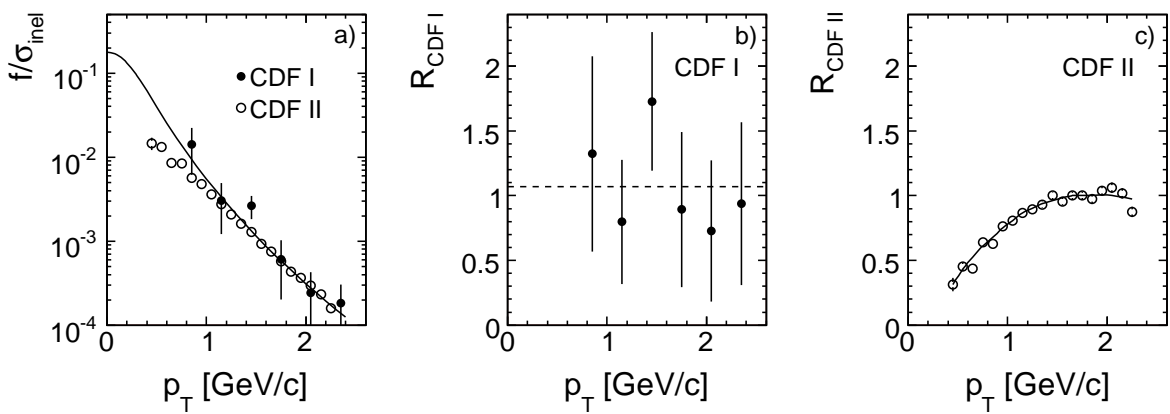}
  	\caption{Comparison of K$^0_S$ data from CDF with the
             UA5 fit, a) cross sections as a function of $p_T$.
             CDF I data (full circles), re-normalized CDF II data 
             (open circles); Ratios
             b) $R_{\textrm{CDF I}} = \left(\frac{f}{\sigma_{\textrm{inel}}}\right)_{\textrm{CDF I}}\Big/
                \left(\frac{f}{\sigma_{\textrm{inel}}}\right)_{\textrm{UA5 fit}}$,
             c) $R_{\textrm{CDF II}} = \left(\frac{f}{\sigma_{\textrm{inel}}}\right)_{\textrm{CDF II}}\Big/
                \left(\frac{f}{\sigma_{\textrm{inel}}}\right)_{\textrm{UA5 fit}}$. The mean
             offset of CDF I data from UA5 fit is indicated in panel b) with dashed line}
  	\label{fig:cdf}
  \end{center}
\end{figure}

The CDF collaboration at the Fermilab Tevatron has published
two data sets concerning K$^0_S$ production at $\sqrt{s}$~=~630~GeV.
The first set (CDF I), with only 27 K$^0_S$ measured, yields 6 
absolutely normalized data points compared in Fig.~\ref{fig:cdf} to 
the UA5 fit. The second set (CDF II) with the very large
statistics of 32k K$^0_S$ is not absolutely normalized. It
has been re-normalized to the UA5 fit at $p_T$~=~1.55~GeV/c. 

As quantified in the ratio plots, Fig.~\ref{fig:cdf}b and c, the absolute
data CDF I fluctuate around the UA5 fit, with a mean 
offset of only a couple of percent, in the range 0.85~$< p_T <$~2.35~GeV/c. 
The CDF II data on the other hand, which cover a larger
$p_T$ range starting at 0.45~GeV/c, show after re-normalization to
the UA5 fit large systematic deviations from the UA5 fit which
increase sharply below $p_T \sim$~1.5~GeV/c, as presented in Fig.~\ref{fig:cdf}c.  
 
\begin{figure}[h]
  \begin{center}
  	\includegraphics[width=12cm]{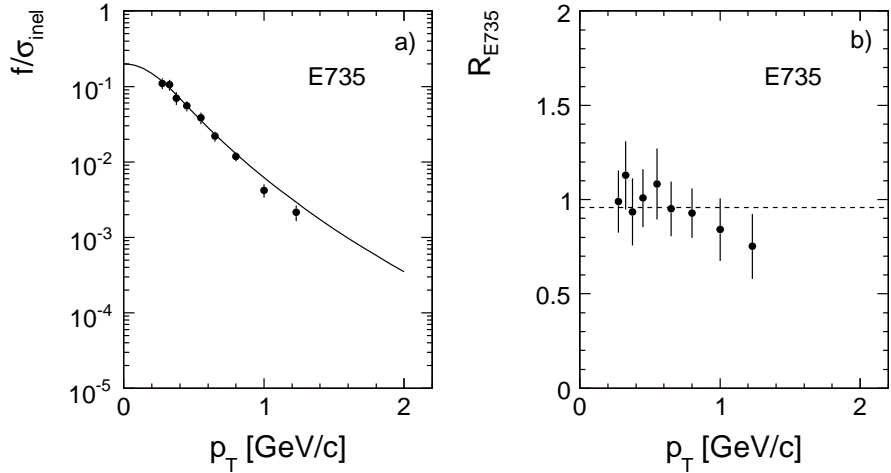}
  	\caption{Comparison of the (K$^+$ + K$^-$)/2 data from
             E735 with the UA5 fit, a) data and fit as a function
             of $p_T$, b) ratio 
             $R_{\textrm{E735}} = \left(\frac{f}{\sigma_{\textrm{inel}}}\right)_{\textrm{E735}}\Big/
                \left(\frac{f}{\sigma_{\textrm{inel}}}\right)_{\textrm{UA5 fit}}$ as a function of $p_T$.
             The mean offset factor of 0.958 is indicated in panel b) with dashed line}
  	\label{fig:e735}
  \end{center}
\end{figure}

In this situation the (K$^+$ + K$^-$)/2 data from the E735 experiment 
at the Tevatron \cite{e735}, although not absolutely normalized, 
give important information in this lower $p_T$ region as to the 
shape of the $p_T$ distribution, Fig.~\ref{fig:e735}.

The 9 data points given, after re-normalization to the UA5 fit
at $p_T$~=~0.45~GeV/c, clearly support the shape of the
UA5 fit in the region 0.2~$< p_T <$~1.2~GeV/c, as compared to the
deviating CDF II data. This is quantified in Fig.~\ref{fig:e735}b with
a mean deviation by a factor of 0.958 and fluctuations which
comply with the given error bars.

%
%
\subsubsection{Data at $\sqrt{s}$=~1800~GeV}
\vspace{3mm}
\label{sec:coll_1800}

\begin{figure}[h]
  \begin{center}
  	\includegraphics[width=15.5cm]{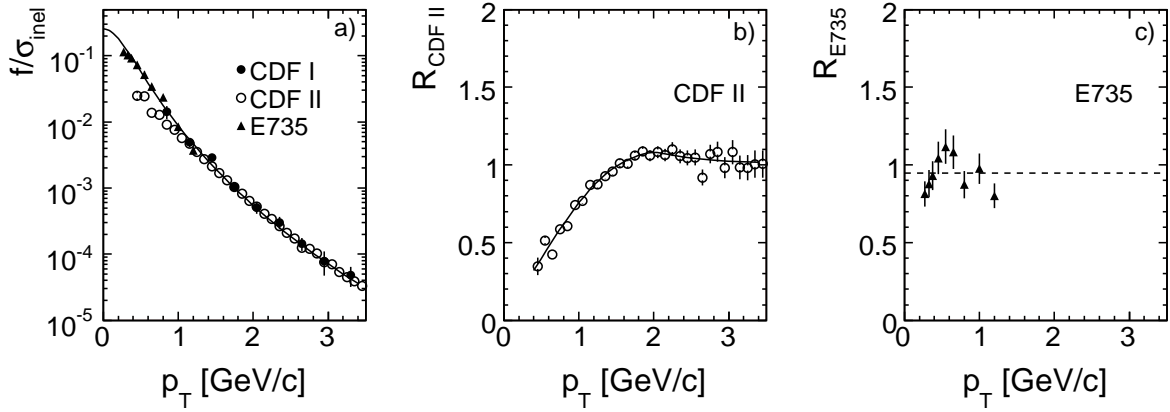}
  	\caption{Kaon data at $\sqrt{s}$~=~1800~GeV; a) Full line
             fit to the CDF I data \cite{cdf1}. Full circles: CDF I data. Open
             circles: CDF II data re-normalized to the CDF I fit at
             $p_T$~=~1.55~GeV/c. Triangles: E735 data re-normalized to the
             CDF I fit. b) Ratio between the re-normalized CDF II data
             and the fit as a function of $p_T$ c) Ratio between the
             re-normalized E735 (K$^+$ + K$^-$)/2 data and the fit. The mean
             offset of E735 data from CDF I fit is indicated in panel c) with dashed line}
  	\label{fig:fnl}
  \end{center}
\end{figure}

Only the CDF experiment, again with two data sets (CDF I \cite{cdf1}
and CDF II \cite{cdf2}) and the E735 collaboration \cite{e735} have published 
kaon data at the highest Tevatron energy of 1800~GeV. 
Here the fit to the CDF I data, transformed to kaon densities
by dividing by the inelastic cross section, and corrected by 
-14\% for the trigger losses, is used as a reference. 
The fit has the form  $f/\sigma_{\textrm{inel}}=C/(p_0+p_T)^n$ with $C$~=~5.38, $n$~=~7.7 and 
$p_0$~=~1.3~GeV/c. As shown in Fig.~\ref{fig:fnl} it has been modified at 
$p_T <$~0.4~GeV/c following the shape of the UA5 fit in this $p_T$ region, 
in order to avoid the unphysical behaviour of this form at low $p_T$.

The 9 data points given for the CDF I sample, corresponding
to about 450 K$^0_S$, are given as full dots in Fig.~\ref{fig:fnl}a. The data
points from CDF II (open circles) deviate again from
the fit for $p_T <$~1.5~GeV/c. This deviation, Fig.~\ref{fig:fnl}b, reproduces 
exactly the phenomenon observed at $\sqrt{s}$~=~630~GeV, see Fig.~\ref{fig:cdf}c, 
thus indicating a systematic problem in the CDF II data analysis.
On the other hand the re-normalized E735 data trace the CDF I
fit rather well as a function of $p_T$, Fig.~\ref{fig:fnl}c, supplementing 
the $p_T$ scale of CDF I which is limited to $p_T >$~0.8~GeV/c, towards low $p_T$.

%
%
\subsubsection{Mean transverse momenta}
\vspace{3mm}
\label{sec:coll_mpt}

\begin{figure}[h]
  \begin{center}
  	\includegraphics[width=9cm]{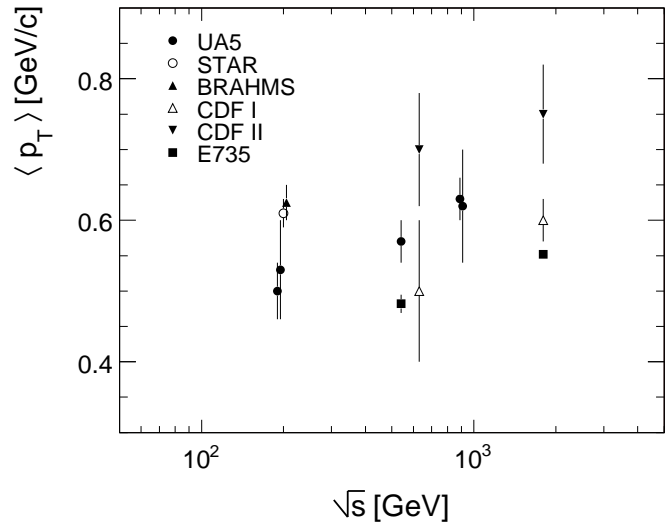}
  	\caption{$\langle p_T \rangle$ in the $\sqrt{s}$ range from 200 to 1800~GeV
             from different experiments}
  	\label{fig:mpt}
  \end{center}
\end{figure}

Given the uncertainties and partial inconsistencies of the
collider (and RHIC) data discussed in the preceding sections,
especially concerning the general lack of coverage and the
evident systematic deviations in the low $p_T$ region, it is not
surprising to perceive large variations in the first moments
of the $p_T$ distributions. Indeed, if the mean transverse 
momentum of K$^0_S$ or (K$^+$ + K$^-$)/2 is plotted as a function of 
$\sqrt{s}$ in the RHIC and p+$\overline{\textrm{p}}$ collider energy range, 
Fig.~\ref{fig:mpt}, a rather disturbing overall picture emerges.

The data which have been published in a time window from 1985 to
2008 span an extremely wide band of typically 0.2~GeV/c at
each of the 5 available energies. For each of the experiments
certain assumptions about the shape of the $p_T$ distributions
have to be made (see the preceding section for some examples)
and in all cases some extrapolation either towards low $p_T$ or
towards high $p_T$ has to be established.

In order to bring the evaluation of the mean transverse momentum
and the data comparison on a more quantitative level, the
following definitions have been used.

The kaon density from which the mean transverse momentum is
derived may be defined in bins of $x_F$ as a function of $x_F$ and 
in bins of $y$ as a function of $y$:

\begin{equation}
   \frac{dn}{dx_Fdp_T} \quad \textrm{and} \quad \frac{dn}{dydp_T}.
\end{equation}

The corresponding mean $p_T$ values are:

\begin{align}
\label{eq:mpt_xf}
\langle p_T \rangle_{x_F} &= \frac{\displaystyle \int p_T \,\frac{dn}{dx_Fdp_T} \,dp_T}
                                  {\displaystyle \vphantom{\sum_{a}}\int \frac{dn}{dx_Fdp_T}\, dp_T} 
                           = \frac{\displaystyle \int \frac{p_T^2}{E} \,f\, dp_T}
                                  {\displaystyle \int \frac{p_T}{E} \,f \,dp_T} \\ 
\label{eq:mpt_y}
\langle p_T \rangle_{y}   &= \frac{\displaystyle \vphantom{\sum^{a}}\int p_T \,\frac{dn}{dydp_T}\, dp_T}
                                  {\displaystyle \vphantom{\sum_{a}}\int \frac{dn}{dydp_T}\, dp_T} 
                           = \frac{\displaystyle \int p_T^2 \,f \,dp_T}
                                  {\displaystyle \int p_T\, f\, dp_T} , 
\end{align}
where $E$ is the kaon energy and $f$ the invariant inclusive cross section, 
Sect.~\ref{sec:corr}.

\begin{figure}[h]
  \begin{center}
  	\includegraphics[width=11cm]{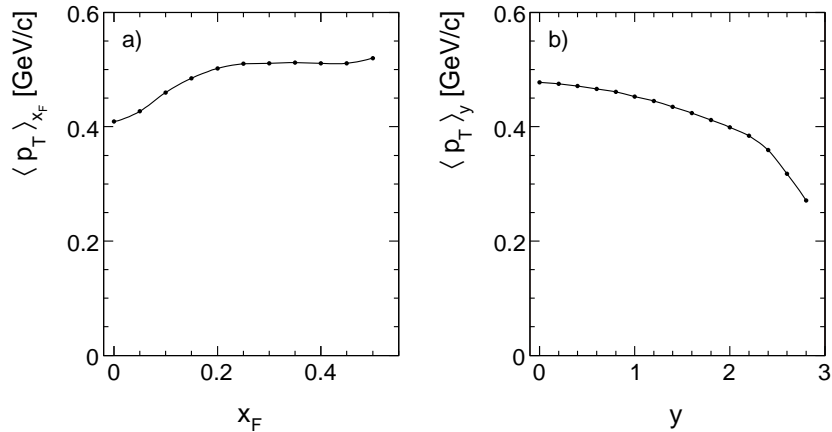}
  	\caption{a) $\langle p_T \rangle_{x_F}$ as a function of $x_F$ and b) $\langle p_T \rangle_y$ 
              as a function of $y$ at $\sqrt{s}$~=~17.2~GeV}
  	\label{fig:mpt17}
  \end{center}
\end{figure}

Evidently there is a difference between the two definitions
given by the energy factor in $\langle p_T \rangle_{x_F}$ (Eq.\ref{eq:mpt_xf}). This 
term will enhance the contribution from low $p_T$ and reduce the contribution at 
high $p_T$ to the mean value in $x_F$ as compared to the $y$ binning. 
In addition at $y$ unequal to 0 the longitudinal dependence of the 
cross sections will couple into the mean value $\langle p_T \rangle_y$ as well as the 
kinematic limit in $p_{\textrm{tot}}$ which will truncate the $p_T$ distribution 
at small angles. The resulting $x_F$ and $y$ dependences of $\langle p_T \rangle_{x_F}$ and 
$\langle p_T \rangle_y$ are shown in Fig.~\ref{fig:mpt17} for the case of the NA49 experiment 
at $\sqrt{s}$~=~17.2~GeV. 

Clearly $\langle p_T \rangle_y$ is bigger than $\langle p_T \rangle_{x_F}$ at $x_F$~=~$y$~=~0, 
at this energy by about 70~MeV/c. One may question the extension of $\langle p_T \rangle_y$ to
$y >$~0 as then a rather complex interplay of transverse and
longitudinal dependences intervenes. The $y$ distribution
of $\langle p_T \rangle_y$ therefore decreases steadily with $y$ whereas 
$\langle p_T \rangle_{x_F}$ shows a characteristic increase with $x_F$ ("seagull" effect).
  
The $\langle p_T \rangle$ values shown in Sect.~\ref{sec:ptint_dist} above are defined in 
Feynman $x_F$, whereas all $\langle p_T \rangle$ values at collider energies shown 
in Fig.~\ref{fig:mpt} are defined in rapidity bins. In addition, the $p_T$ 
integration for the lower energy data has been established in 
the range 0~$< p_T <$~2~GeV/c. Hence these results are not directly 
comparable. 

\begin{figure}[h]
  \begin{center}
  	\includegraphics[width=8.5cm]{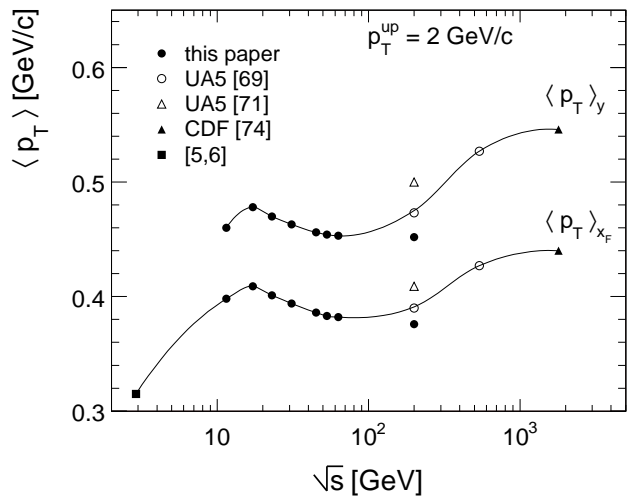}
  	\caption{$\langle p_T \rangle_{x_F}$ at $x_F$~=~0 and $\langle p_T \rangle_y$ at $y$~=~0 
  	         as a function of $\sqrt{s}$. The lines are drawn to guide the eye}
  	\label{fig:mpt_sdep}
  \end{center}
\end{figure}

The following procedure has therefore been adopted. In the collider energy 
range 200--1800~GeV the results with doubtful cross section behaviour at low $p_T$ 
\cite{star1,cdf2} are not considered for their $\langle p_T \rangle$ values. 
The fits to the UA5 \cite{ansorge1} and CDF I \cite{cdf1} K$^0_S$ data are used 
at $\sqrt{s}$~=~200, 540 and 1800~GeV for the determination of both $\langle p_T \rangle_{x_F}$ 
and $\langle p_T \rangle_y$. For the lower energy data (Sect.~\ref{sec:ptint_dist}) 
$\langle p_T \rangle_y$ is calculated in addition to $\langle p_T \rangle_{x_F}$
for (K$^+$ + K$^-$)/2 including the data extrapolation to $\sqrt{s}$~=~200~GeV. 
In a first step, the integration is carried out in the range 0~$< p_T <$~2.0~GeV/c 
in order to obtain for all data a comparable basis in this lower $p_T$ range.The 
resulting $\langle p_T \rangle_{x_F}$ and $\langle p_T \rangle_y$ values are shown 
in Fig.~\ref{fig:mpt_sdep} as a function of $\sqrt{s}$.

This Figure exhibits a smooth behaviour of $\langle p_T \rangle$ in the
energy range from $\sqrt{s}$~=~11.5 to 63~GeV, also including the
extrapolation to 200~GeV, with a variation of only 30~MeV/c
for $\langle p_T \rangle_{x_F}$ and $\langle p_T \rangle_y$. 
There is, however, a clear offset of about 10~MeV/c for $\langle p_T \rangle_{x_F}$ 
and 20~MeV/c for $\langle p_T \rangle_y$ between the extrapolation of the 
lower energy data to $\sqrt{s}$~=~200~GeV and the trend of the 
collider data which cannot be imputed to high $p_T$ tails in this 
integration window. It is rather the different behaviour at low 
$p_T$, see Fig.~\ref{fig:ua5}b, which can explain the difference. Given 
the general uncertainty of the collider data in the $p_T$ range
below 0.5~GeV/c, the observed offset may still be regarded
as compatible with the published errors which are on the level
of 30 to 40~MeV/c \cite{ansorge1}. Another interesting feature is the
rather small increase of $\langle p_T \rangle$ which is only on the order of 
50~MeV/c for $\langle p_T \rangle_{x_F}$ and 70~MeV/c for $\langle p_T \rangle_y$
between $\sqrt{s}$~=~200 and 1800~GeV, always in the $p_T$ range below 2~GeV/c.   

\begin{figure}[h]
  \begin{center}
  	\includegraphics[width=6cm]{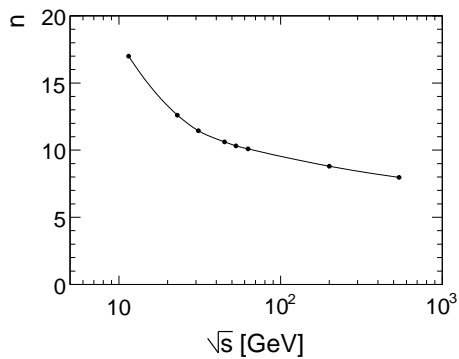}
  	\caption{Fitted exponent $n$ as a function of $\sqrt{s}$. The full
             line is shown to guide the eye}
  	\label{fig:ns}
  \end{center}
\end{figure}

In order to quantify the dependence of $\langle p_T \rangle$ on the upper
integration limit in $p_T$, this limit has been increased from
2~GeV/c to 6~GeV/c. For this study the published polynomial 
fits of the collider data have been used. For the lower energy 
data the following procedure has been chosen. The polynomial form  

\begin{equation}
  f(p_T) = A\left(\frac{p_0}{p_T+p_0}\right)^n
\end{equation}
has been fitted to the high $p_T$ region $p_T >$~1.5~GeV/c with
$p_0$ fixed at 1.3~GeV/c. This procedure is possible for $\sqrt{s}$~=~11.5~GeV 
where the data reach up to $p_T$~=~4.2~GeV/c and in the
ISR energy range where data up to $p_T$~=~4~GeV/c are available.
The corresponding exponents $n$ are plotted in Fig.~\ref{fig:ns} as
a function of $\sqrt{s}$.

A consistent and smooth drop in the exponent $n$ from about 16 at 
$\sqrt{s}$~=~11.5~GeV to 8 at $\sqrt{s}$~=~1800~GeV is evident, describing
the flattening of the $p_T$ distributions with increasing $\sqrt{s}$.
This allows for the interpolation to $n$~=~14 at $\sqrt{s}$~=~17.2~GeV 
where the NA49 data do not reach beyond $p_T$~=~1.7~GeV/c.

For the $\sqrt{s}$ range below ISR energies, the kinematic limit in $p_T$ 
at $x_T=2p_T/\sqrt{s}$~=~1 has to be taken into account. This limit
influences the measured yields progressively from $x_T$~=~0.5 upwards. This necessitates
a downward correction of the polynomial fit at $p_T >$~3~(4.5)~GeV/c for 
$\sqrt{s}$~=~11.5~(17.2)~GeV, respectively.  

\begin{figure}[h]
  \begin{center}
  	\includegraphics[width=14.cm]{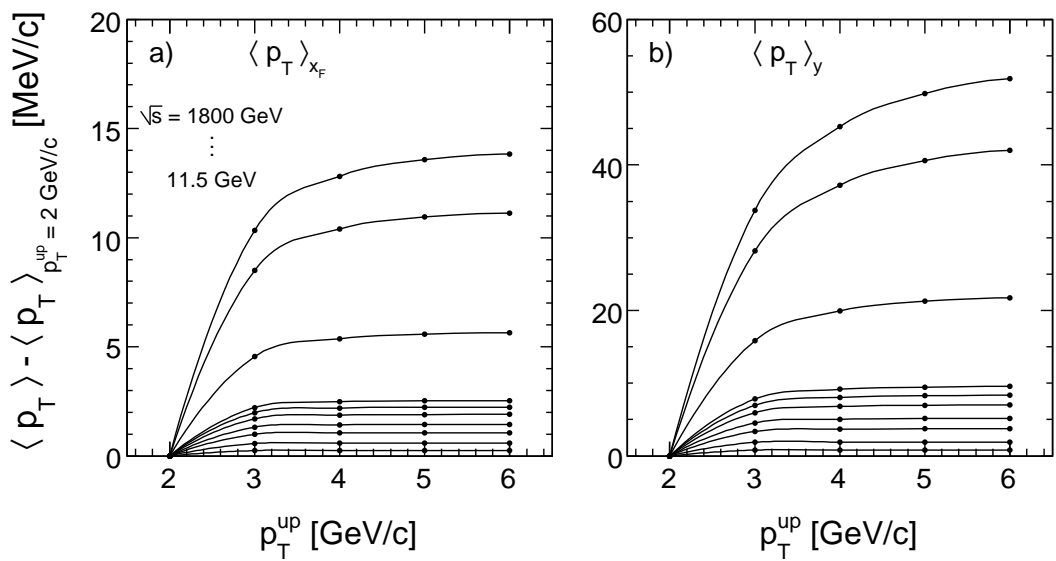}
  	\caption{$\langle p_T \rangle$-$\langle p_T \rangle_{p_T^{\textrm{up}\textrm{~=~2~GeV/c}}}$ 
             for a) $\langle p_T \rangle_{x_F}$ and
             b) $\langle p_T \rangle_y$. The $\sqrt{s}$ values range from 11.5 to 1800~GeV. 
             The curves are shown to guide the eye}
  	\label{fig:diff2gev}
  \end{center}
\end{figure}

The increase of the mean $p_T$ values as a function of the upper
integration limit from 2 to 6~GeV/c is given in Fig.~\ref{fig:diff2gev} where the difference 
$\langle p_T \rangle$-$\langle p_T \rangle_{p_T^{\textrm{up}\textrm{~=~2~GeV/c}}}$
is shown for $\sqrt{s}$ from 11.5 to 1800~GeV both defined in $x_F$ and in $y$ bins.

\begin{figure}[h]
  \begin{center}
  	\includegraphics[width=10cm]{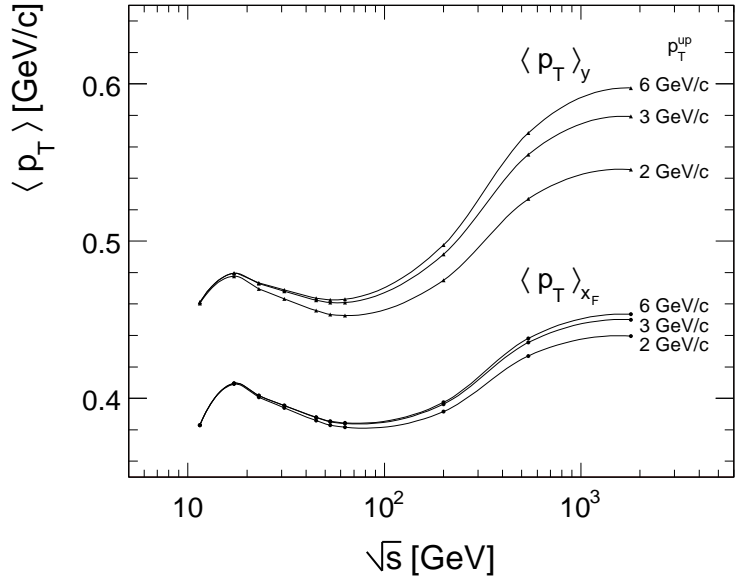}
  	\caption{$\langle p_T \rangle_{x_F}$ at $x_F$~=~0 and $\langle p_T \rangle_y$ at $y$~=~0
  	         as a function of $\sqrt{s}$ for different values of the upper integration limit
             from 2 to 6~GeV/c}
  	\label{fig:mpt_xy}
  \end{center}
\end{figure}

The $\langle p_T \rangle$ values saturate rapidly at an upper integration limit
of 2.5 to 3.5~GeV/c between Serpukhov and ISR energies, with
a total increase of less than 3~MeV/c (10~MeV/c) for $\langle p_T \rangle_{x_F}$
and $\langle p_T \rangle_y$, respectively, in this energy range. At collider
energies the saturation limit moves up to beyond 6~GeV/c,
with very substantial increases of more than 15~MeV/c and
more than 50~MeV/c in $\langle p_T \rangle_{x_F}$ and $\langle p_T \rangle_y$, respectively.
$\langle p_T \rangle_{x_F}$ and $\langle p_T \rangle_y$ are shown in 
Fig.~\ref{fig:mpt_xy} as a function of $\sqrt{s}$ for the upper integration values from
2 to 6~GeV/c.

The rather complex dependence of the mean $p_T$ values both on
$\sqrt{s}$ and on the upper integration limits, in addition to
the apparent systematic effects in passing from the ISR 
to collider energies, calls for some remarks:

\begin{itemize}
  \item A precision measurement of $\langle p_T \rangle$ with an absolute error of less
        than 20~MeV/c in the region above $\sqrt{s}$~=~100~GeV/c is still 
        missing. This fact is mostly due to uncertainties in the low $p_T$ region.
  \item The sizeable difference between $\langle p_T \rangle_{x_F}$ and $\langle p_T \rangle_y$ 
        has to be taken into consideration whenever results on $\langle p_T \rangle$ are to
        be compared for different experiments and $\sqrt{s}$ regions.
  \item The large dependence of $\langle p_T \rangle$ on the upper integration limit
        in the collider energy range, especially for $\langle p_T \rangle_y$, is a
        reason for concern. It may be asked whether the definition of an average 
        quantity which depends strongly on a high $p_T$ tail more than a factor 
        of 10 above its value, makes any sense.
  \item In fact at least part of the increase of $\langle p_T \rangle$ with $\sqrt{s}$ 
        is to be imputed to the extension of the available transverse
        phase space. The kinematic limit in $p_T$ is below 6~GeV/c 
        at Serpukhov energy and it must be recalled that this limit
        is influencing the particle yields already at $x_T$~=~2$p_T$/$\sqrt{s} >$~0.5, 
        this means above $p_T \sim$~3 GeV/c at this energy.
  \item The above remarks are especially applying for the dependence
        of $\langle p_T \rangle$ on additional constraints, as for instance on the
        total hadronic multiplicity. Also in this case it might be
        advisable to separate clearly the behaviour in the lower
        $p_T$ region from the increasing high $p_T$ tails.
\end{itemize}

%
%
\section{The $\boldsymbol{s}$-dependence of K$^0_S$ production and its relation to charged kaons}
\vspace{3mm}
\label{sec:kzero} 

A sizeable number of experiments 
\cite{eisner,jaeger1,blobel,fesef,boggild,alpgard1,bartke,bogo,ammosov,
      alston,chapman,brick,allday,jaeger2,sheng,lopinto,dao,bailly,kass,kichimi}
have addressed neutral kaon production from $\sqrt{s}$~=~3~GeV to 
$\sqrt{s}$~=~27.6~GeV. This ensures coverage from close to 
threshold up to well into the ISR energy range.
Essentially all these measurements come from Bubble 
Chambers. This has the consequence that the total number 
of reconstructed K$^0_S$ is usually rather limited to a range 
between a few hundred and a few thousand. This handicap 
is however offset by the superior quality of the Bubble 
Chamber technique in terms of reconstruction efficiency, 
control of systematic effects and corrections, and above 
all a well-defined absolute normalization. It is in
particular interesting to compare the K$^0_S$  to the
average charged kaon yields discussed above, as the Eq.~\ref{eq:isosym}
is generally assumed to hold based on isospin symmetry \cite{alner1}
although it is not fulfilled for instance for $\phi$  and Charm
decay.

\begin{figure}[h]
  \begin{center}
  	\includegraphics[width=10.2cm]{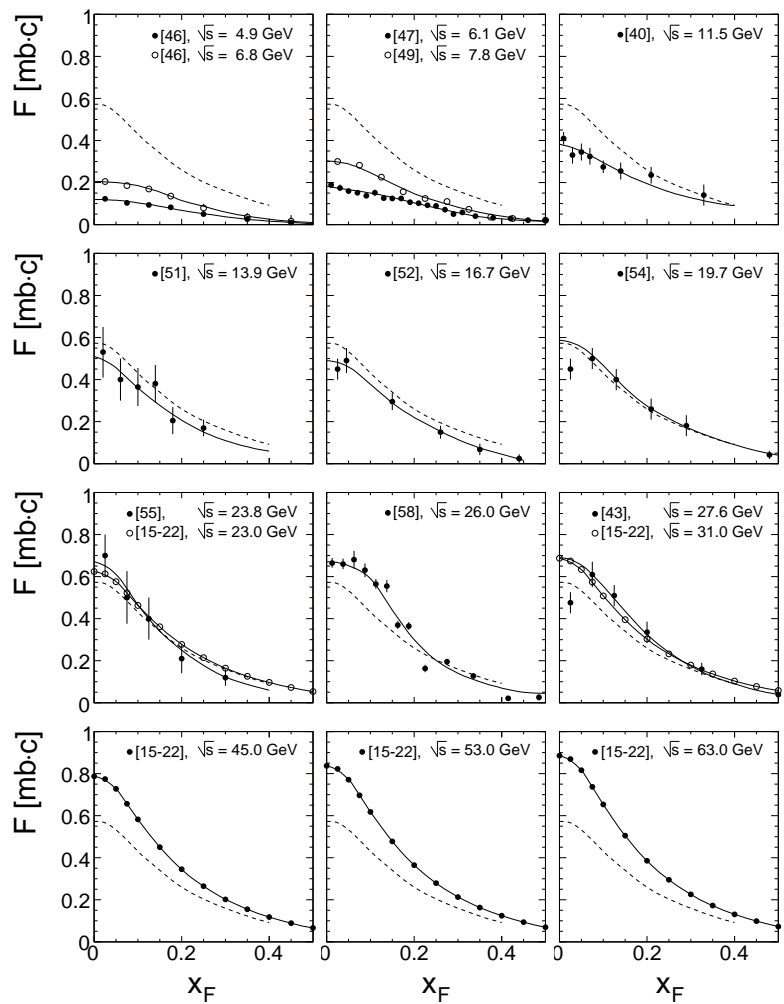}
  	\caption{$p_T$ integrated invariant K$^0_S$  cross sections $F$ as a function 
  	         of $x_F$  for various values of $\sqrt{s}$. Independent hand-interpolations 
  	         at each energy are given as full lines. The dashed lines correspond to 
             $F((\textrm{K}^+ + \textrm{K}^-)/2)$ from NA49, 
             Sect.~\ref{sec:ptint_dist}, Table~\ref{tab:integr}. The interpolated results from ISR
             \cite{albrow1,albrow2,albrow3,albrow4,albrow5,capi,alper,guettler}
             for $F((\textrm{K}^+ + \textrm{K}^-)/2)$ are also presented}
  	\label{fig:k0_xfdist}
  \end{center}
\end{figure}

Due to the low event statistics, double differential cross
sections are not available from any of the experiments with
the exception of \cite{blobel}. The $p_T$ integrated invariant cross
section $F$ (see Eq.~\ref{eq:int}) has however been given by 10 experiments
between $\sqrt{s}$~=~4.9 and $\sqrt{s}$~=~27.6~GeV. These data are plotted
in Fig.~\ref{fig:k0_xfdist} as a function of $x_F$ .

\begin{figure}[h]
  \begin{center}
  	\includegraphics[width=8.cm]{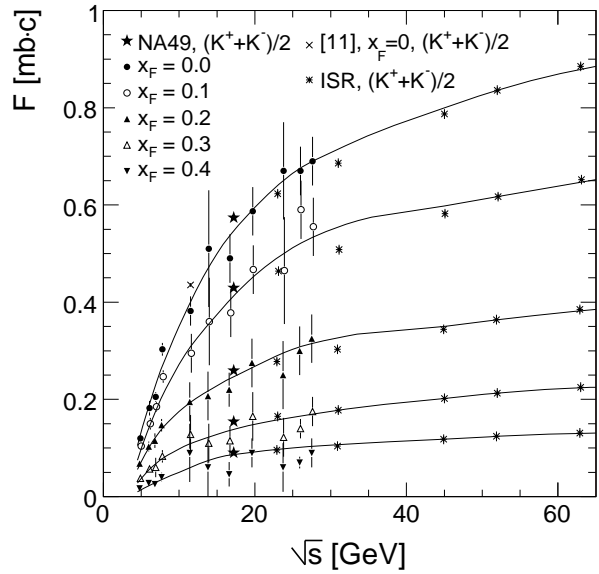}
  	\caption{$p_T$ integrated invariant K$^0_S$  cross sections  
             $F$ as a function of $\sqrt{s}$, interpolated to fixed values
             of $x_F$  from 0 to 0.4 using the hand-fits shown in Fig.~\ref{fig:k0_xfdist}.
             The error bars are an estimation of the uncertainties of
             the full lines in Fig.~\ref{fig:k0_xfdist}. The lines are drawn to guide the eye}
  	\label{fig:k0_sdist}
  \end{center}
\end{figure}

Independent hand-interpolations at each energy have been performed
in order to allow for the evaluation of the $s$-dependence at fixed
values of $x_F$  as shown in Fig.~\ref{fig:k0_sdist}.

Also shown in this Figure is $F((\textrm{K}^+ + \textrm{K}^-)/2)$ as derived above
for the $p_T$-extrapolated Serpukhov data \cite{abramov} at $\sqrt{s}$~=~11.5~GeV, 
the NA49 data and the interpolated ISR data, Sect.~\ref{sec:isr}.   
Evidently there is agreement, within the experimental
uncertainties, with the interpolated K$^0_S$  data at all $x_F$ 
values. This might lend some credibility to the assumptions
contained in the evaluation of the ISR data over the full phase
space in Sect.~\ref{sec:sdep}.

\begin{figure}[b]
  \begin{center}
  	\includegraphics[width=11cm]{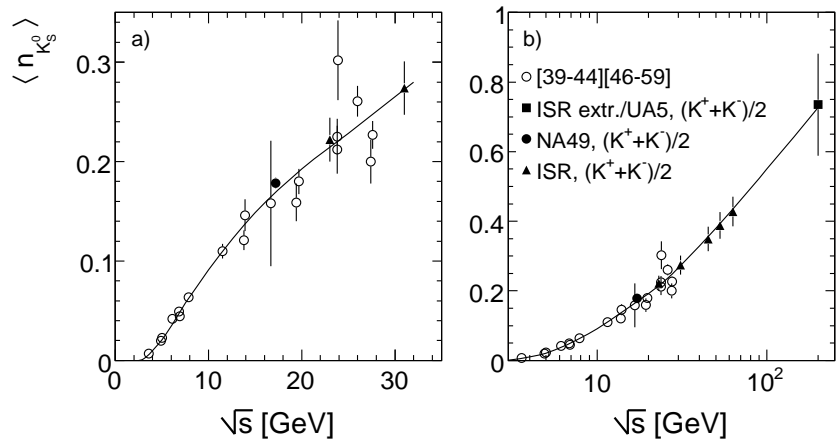}
  	\caption{Total integrated K$^0_S$  yields as a function of $\sqrt{s}$ 
  	        \cite{eisner,jaeger1,blobel,fesef,boggild,alpgard1,bartke,bogo,ammosov,
                  alston,chapman,brick,allday,jaeger2,sheng,lopinto,dao,bailly,kass,kichimi} (open circles).
             The total mean charged kaon yields for NA49 (full circle), ISR and extrapolation to 200 GeV 
             (triangles) are also shown.The scale in 
  	         panel b) is extended up to $\sqrt{s}$~=~200~GeV. The lines are drawn to
             guide the eye}
  	\label{fig:k0_tot}
  \end{center}
\end{figure}

A similar procedure may be performed for the total integrated
K$^0_S$  yields per inelastic event in comparison to the total
mean charged kaon yields. This comparison is shown in Fig.~\ref{fig:k0_tot}
using the total number of K$^0_S$  per inelastic event given by the
19 experiments refs.
\cite{eisner,jaeger1,blobel,fesef,boggild,alpgard1,bartke,bogo,ammosov,
      alston,chapman,brick,allday,jaeger2,sheng,lopinto,dao,bailly,kass,kichimi}.

Also for the total kaon yields the equality (\ref{eq:isosym}) is fulfilled
within errors. Note that the NA49 data have a 3\% error bar
corresponding to the estimated systematic uncertainty, whereas
the interpolated ISR data have been tentatively given a 10\%
error bar.

%
%
\section{Some remarks about contributions from resonance decay}
\vspace{3mm}
\label{sec:reson} 

The evolution of the observed kaon yields with transverse
momentum and interaction energy described in the preceding 
sections is characterized by rather complex patterns which are 
not easily describable by straight-forward parametrizations 
as they might follow from parton dynamics or thermal models.
It seems therefore reasonable to evoke for illustration
the contribution from the decay of some known 
resonances to the inclusive kaon cross sections. Three 
resonances, the $\phi$(1020), the $\Lambda$(1520) and the charmed
mesons D(1865) have been selected here as they give an idea
about the build-up of kaon yields at low $p_T$ for the two
former cases, and towards high $p_T$ for the latter one. In
this context it should be recalled here that most if not all 
final state hadrons are known to be created by the decay of 
mesonic and baryonic resonances \cite{ehs,grassler}. Indeed, the 
estimations quoted in \cite{ehs,grassler}, using only a limited set of
mesonic and baryonic resonances, arrive at fractions of
60-80\% from resonance decay for all studied final state
hadrons. See also \cite{site:S8} for a more recent study based on
two-body decays of 13 known resonances.

%
%
\subsection{$\boldsymbol{\phi}$(1020) and $\boldsymbol{\Lambda}$(1520) production and decay}
\vspace{3mm}
\label{sec:philam}

\begin{figure}[h]
  \begin{center}
  	\includegraphics[width=11cm]{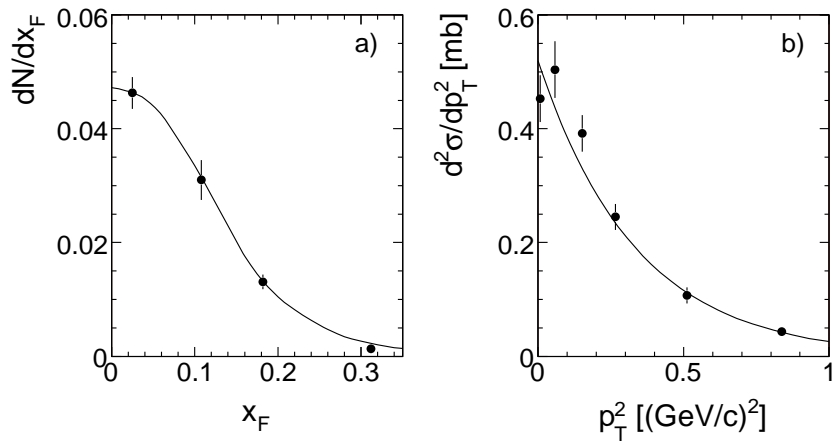}
  	\caption{a) $x_F$ and b) $p_T^2$ distributions of $\phi$}
  	\label{fig:phi_dist}
  \end{center}
\end{figure}

The $\phi$ production has been measured by a number of experiments
for p+p interactions in the SPS energy range \cite{ehs,daum}. Results
from the NA49 experiment \cite{afanasiev} are being used here to obtain
the inclusive $dn/dx_F$ and $d^2\sigma/dp_T^2$ distributions shown
in Fig.~\ref{fig:phi_dist}.

These distributions integrate to $\langle n_{\phi} \rangle$~=~0.0143 per inelastic
event or an inclusive cross section of 0.453~mb in good agreement
with other measurements.

\begin{figure}[h]
  \begin{center}
  	\includegraphics[width=11cm]{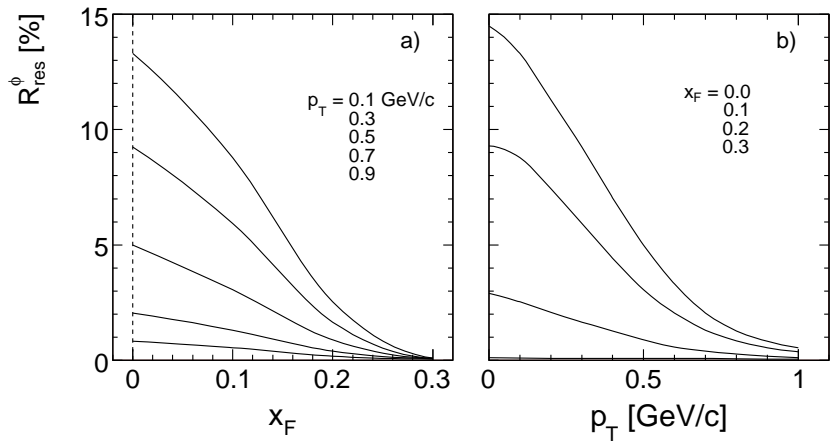}
  	\caption{Ratio 
             $R_{\textrm{res}}^{\phi} = \textrm{K}^-_{\phi}/\textrm{K}^-_{\textrm{incl}}$ 
             as a function of a) $x_F$ for different $p_T$ and b) $p_T$ for different $x_F$}
  	\label{fig:phi_r2inc}
  \end{center}
\end{figure}

Due to the very low $Q$ value (32~MeV) of the $\phi$ decay into two kaons,
the resulting $p_T$ and $x_F$ distributions are narrow compared to 
the inclusive kaon cross sections. This is reflected in the
ratios $R_{\textrm{res}}^{\phi}$ of K$^-$ mesons from $\phi$ decays to inclusive K$^-$ shown
in Fig.~\ref{fig:phi_r2inc}.

Evidently this contribution is very sharply peaked at small $p_T$
and vanishes at $x_F >$~0.3. The given percentages have to be regarded
as lower limits, as $\phi$ production is known to be accompanied by
additional kaons in most if not all cases \cite{daum}. These additional
kaons come partially from double-$\phi$ production \cite{booth} with
again small $Q$ values since the four-K mass spectrum has
a steep threshold enhancement in the mass range from 2.1 to 2.3~GeV \cite{booth}. 
This would mean that the effective contributions,
Fig.~\ref{fig:phi_r2inc}, could increase by as much as a factor of 1.5, see below.  

\begin{figure}[h]
  \begin{center}
  	\includegraphics[width=11cm]{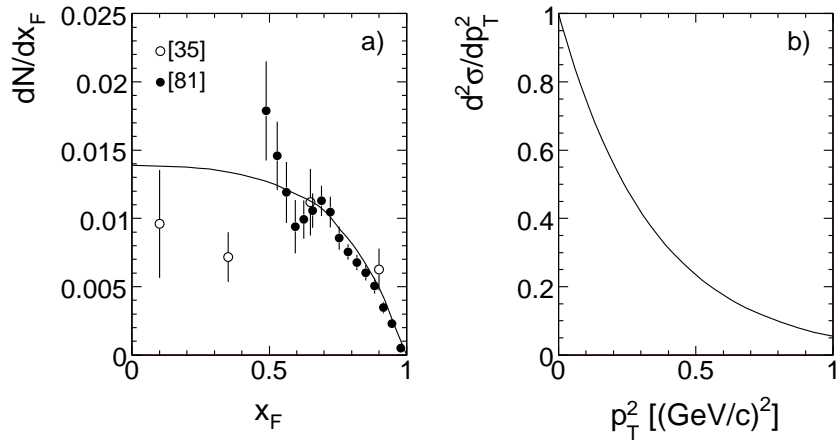}
  	\caption{a) $x_F$ and b) $p_T^2$ distributions of $\Lambda$(1520)}
  	\label{fig:lam_dist}
  \end{center}
\end{figure}

Another candidate resonance for low-$p_T$ kaon production is the
$\Lambda$(1520) in the N$\overline{\textrm{K}}$ decay channel with its small $Q$ value 
of 87~MeV. Measurements at ISR energy \cite{bobbink} and at the SPS \cite{ehs} 
have been combined in Fig.~\ref{fig:lam_dist}a to obtain an approximate $dn/dx_F$ 
distribution. 

The full line in Fig.~\ref{fig:lam_dist}a is a hand-interpolation of 
these data that has been used in the Monte Carlo simulation.
It integrates to the number $\langle n_{\Lambda \textrm{(1520)}} \rangle$~=~0.0219 per inelastic
event or a total inclusive cross section of 0.697~mb. Since no
data on the corresponding $p_T^2$ distribution are available the fit
to $f(x_F,p_T^2) = e^{-Bp_T^2}$, with $B$~=~2.9 as given in \cite{bobbink}
and shown in Fig.~\ref{fig:lam_dist}b has been used.

\begin{figure}[h]
  \begin{center}
  	\includegraphics[width=11cm]{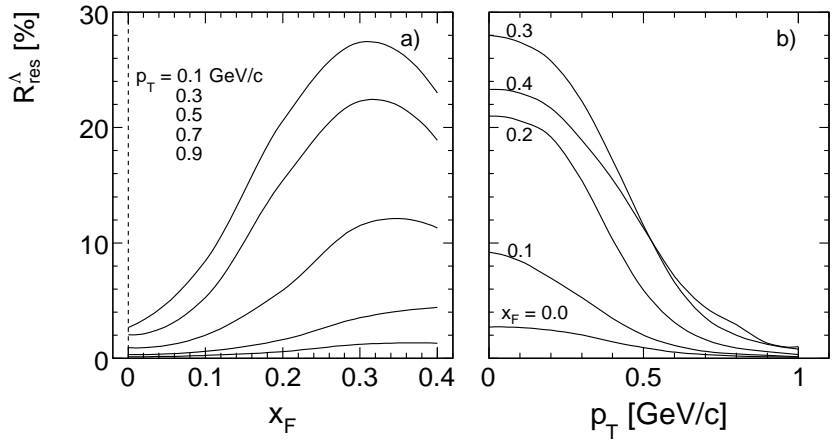}
  	\caption{Ratio 
             $R_{\textrm{res}}^{\Lambda} = \textrm{K}^-_{\Lambda}/\textrm{K}^-_{\textrm{incl}}$
             as a function of a) $x_F$ for different $p_T$ and b) $p_T$ for different $x_F$}
  	\label{fig:lam_r2inc}
  \end{center}
\end{figure}

Due to the rather flat $x_F$ distribution of the $\Lambda$(1520) which
is typical of neutral strange baryons, the ratio between decay 
and inclusive K$^-$ shows a characteristic increase from $x_F$~=~0 to a 
maximum at $x_F \sim$~0.3 as shown in Fig.~\ref{fig:lam_r2inc}.

\begin{figure}[h]
  \begin{center}
  	\includegraphics[width=11cm]{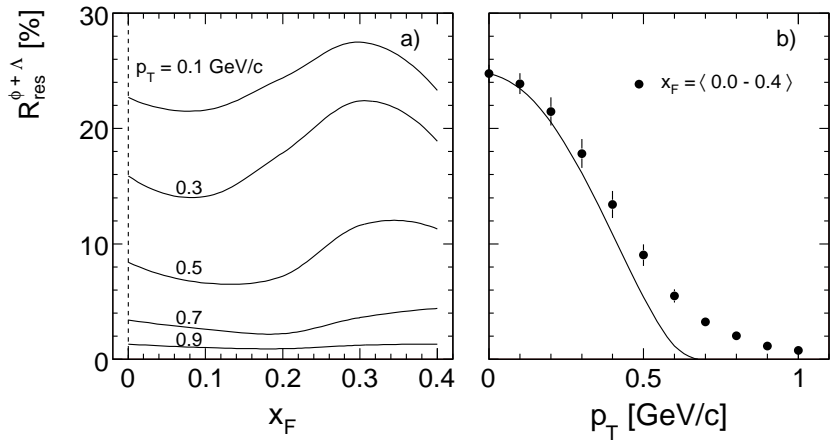}
  	\caption{Ratio 
             $R_{\textrm{res}}^{\phi+\Lambda} = \textrm{K}^-_{\phi+\Lambda}/\textrm{K}^-_{\textrm{incl}}$
             a) as a function of $x_F$ for different $p_T$ and b) as a function of $p_T$ averaged over 
             the $x_F$ range 0~$< x_F <$~0.4. In panel b) the error bars give the variation with 
             $x_F$ around the average. The full line represents the relative increase of the K$^-$ cross
             section as a function of $\sqrt{s}$, Fig.~\ref{fig:isr_factor}b, normalized at $p_T$~=~0~GeV/c}
  	\label{fig:philam_r2inc}
  \end{center}
\end{figure}

Again, as for the $\phi$ decay, the very sharp enhancement of $R_{\textrm{res}}$ 
towards low $p_T$ is evident whereas the contribution to the inclusive
kaon yield vanishes at about $p_T$~=~1~GeV/c. On the other hand the
$x_F$ distribution of $R_{\textrm{res}}^{\Lambda}$ is complementary to the one from $\phi$  
decay in its $x_F$ dependence such that the sum of the two contributions 
becomes rather $x_F$ independent. This is evident in the combined ratio 
$R_{\textrm{res}}^{\phi+\Lambda} = \textrm{K}^-_{\phi+\Lambda}/\textrm{K}^-_{\textrm{incl}}$
shown in Fig.~\ref{fig:philam_r2inc}, where the 
K$^-$ yield from $\phi$  has been multiplied by a factor 1.5 in order to 
make up for the production of additional kaons in $\phi$  production, 
see above.

As the combined contributions from $\phi$ and $\Lambda$(1520) decay reach
about 25\% of the total K$^-$ yield at low $p_T$ and $\sqrt{s}$~=~17.2~GeV
this discussion shows again the importance of resonance decay 
for the understanding of inclusive hadron production, in this 
particular case for the low $p_T$ behaviour of the kaon cross sections.
This is also evident in the $p_T$ distribution shown in Fig.~\ref{fig:philam_r2inc}b
which is very similar to the low $p_T$ enhancement with $\sqrt{s}$
shown in Fig.~\ref{fig:isr_factor}. The $s$-dependence in the region below $p_T \sim$~1~GeV/c
will be determined by the $s$-dependence of $\phi$ and $Y^*$ production
with respect to the other contributing resonances. If the latter
contributions rise as little as in the region 0.8~$< p_T <$~1~GeV/c
$\phi$ and $Y^*$ decays will become dominant in the ISR energy range. 
  
%
%
\subsection{D(1860) decay}
\vspace{3mm}
\label{sec:d1860}
 
In complement to the discussion of the low $p_T$ area of kaon
production in the preceding section, it is interesting to
look for resonance decay contributions in the high $p_T$ region
of $p_T >$~1~GeV/c. Here high mass mesonic resonances with sizeable
decay branching fractions into 2 or 3 body final states
including kaons will contribute. Although there is a large number 
of non-strange and strange resonances in the mass range above
1.5~GeV fulfilling this criterion, the charm mesons $\textrm{D}^{0\pm}$(1860)
will be regarded here as an example of heavy flavour production 
and decay. In fact the charm production threshold is crossed in
the SPS energy range and the charm yields will start to saturate
at p+$\overline{\textrm{p}}$ collider energies where beauty meson production will
give access to still higher transverse momentum ranges.

Close to 100\% of all charmed meson decays end up in final state
kaons, either in semi-leptonic or hadronic decay modes. 
Most of these are few body decays with large $Q$ values, 
like Kl$\nu$,K$^*$l$\nu$ in the semi-leptonic
and K$\pi$, K$\pi\pi$ and K$\pi\pi\pi$ in the hadronic case. Given the high
D mass, the addition of one or two pions in the final state 
will not change the phase space distribution of the kaons
appreciably. The two body decay of charmed mesons into K$^-$ will
therefore be studied in the following.

\begin{figure}[h]
  \begin{center}
  	\includegraphics[width=11cm]{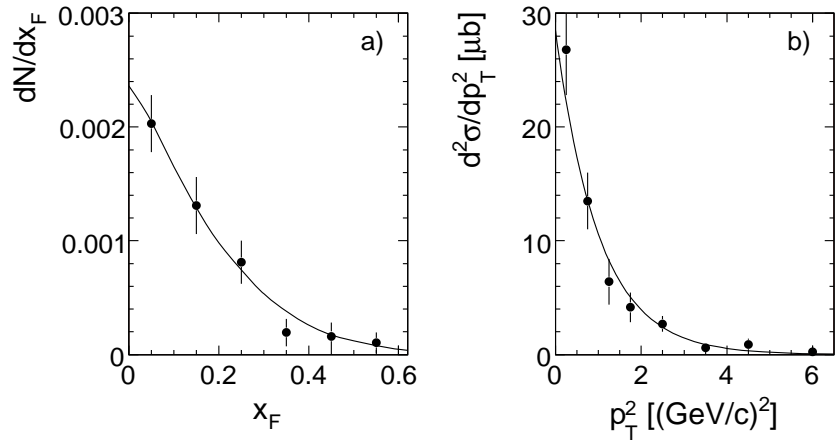}
  	\caption{a) $dn/dx_F$ as a function of $x_F$ and b) $d\sigma/dp_T^2$ as a function of $p_T^2$ of D(1860)}
  	\label{fig:d_dist}
  \end{center}
\end{figure}

One of the rare measurements of charm production in p+p 
interactions by the LEBC-EHS collaboration \cite{aguilar} at the
CERN SPS will be used to establish the input $x_F$ and $p_T$ 
distributions as shown in Fig.~\ref{fig:d_dist}.

The inter/extrapolated $dn/dx_F$ distribution and the fitted
Gaussian $d\sigma/dp_T^2 = Ae^{-0.99p_T^2}$ are shown in
panels a) and b) as full lines. The integration of this
parametrization yields $\langle n_{\textrm{D}^{0\pm}} \rangle$~=~0.000944 per inelastic
event corresponding to a cross section of 29~$\mu$b.
This cross section, at $\sqrt{s}$~=~27~GeV, contradicts an
upper limit of less than 10~$\mu$b established from the study of muon
pair production \cite{hgf} at this energy. This discrepancy
notwithstanding, the effective cross section has been reduced
to 20~$\mu$b taking into account the steep $s$-dependence
for the following comparison to inclusive K$^-$ cross sections at
the energy of the NA49 experiment, $\sqrt{s}$~=~17.2~GeV. The
effective branching fraction of charm meson pairs into K$^-$ may be
estimated from \cite{aguilar} to about 47\%. In order to take
into account the softening of the decay kaon spectra in
multibody decays, a conservative value of 30\% has been
used for the following two-body decay simulation.

\begin{figure}[h]
  \begin{center}
  	\includegraphics[width=14cm]{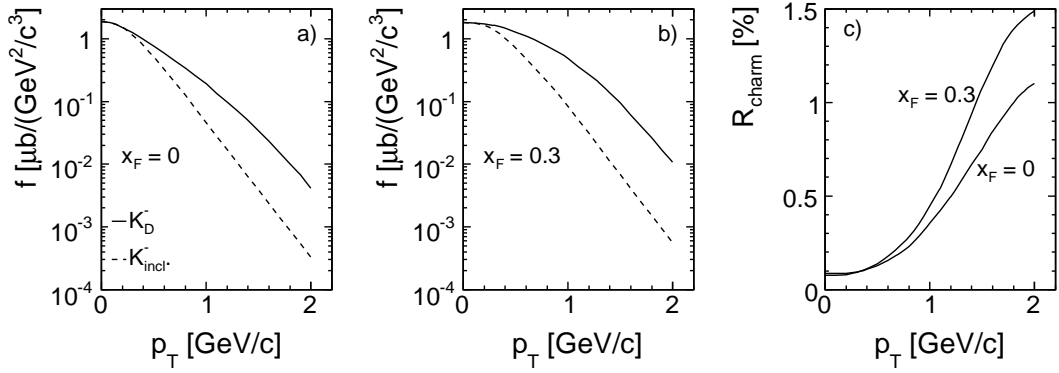}
  	\caption{Invariant K$^-$ cross sections from charm meson
             decay a) for $x_F$~=~0, b) for $x_F$~=~0.3 as functions of $p_T$ at 
             $\sqrt{s}$~=~17.2~GeV. The corresponding total inclusive K$^-$ yields (dashed lines) are
             shown for comparison normalized to the decay distributions at
             $p_T$~=~0~GeV/c. Panel c) shows the absolute ratio $R_{\textrm{charm}}$ in percent as 
             a function of $p_T$ for the two $x_F$ values}
  	\label{fig:d_r2incl}
  \end{center}
\end{figure}

Typical $p_T$ distributions of the invariant K$^-$ cross section from
charm meson decay are shown in Fig.~\ref{fig:d_r2incl} for two values of $x_F$.
These distributions are compared to the total inclusive K$^-$ yields
normalized to the decay distribution at $p_T$~=~0~GeV/c.

\begin{figure}[h]
  \begin{center}
  	\includegraphics[width=5.5cm]{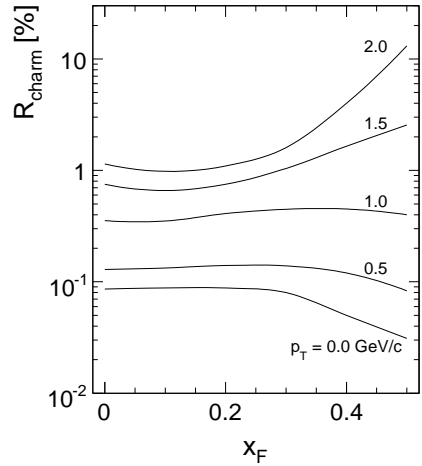}
  	\caption{$R_{\textrm{charm}}$ as a function of $x_F$ for fixed values of $p_T$, in \%}
  	\label{fig:rcharm}
  \end{center}
\end{figure}

The decay kaons evidently show a much wider $p_T$ distributions than
the inclusive K$^-$. The relative increase of the ratio

\begin{equation}
  R_{\textrm{charm}} = \frac{\textrm{K}^-_{\textrm{D}^{0\pm}}}{\textrm{K}^-_{\textrm{incl}}}
\end{equation}
is shown in Fig.~\ref{fig:d_r2incl}c as a function of $p_T$ for the two $x_F$ values
of 0 and 0.3. $R_{\textrm{charm}}$ increases steeply with $p_T$ from values of less than
0.1\% at low $p_T$ to more than 1\% at $p_T$~=~2~GeV/c, whereas the ratio of
the total inclusive K$^-$ cross sections is of order 0.15\%. This
increase will clearly continue at $p_T >$~2~GeV/c. The situation is
quantified for the complete $x_F$/$p_T$ plane in Fig.~\ref{fig:rcharm} which
shows $R_{\textrm{charm}}$ as a function of $x_F$ for different values of $p_T$.

A characteristic pattern emerges. At low $x_F$ $R_{\textrm{charm}}$ gains about 
one order of magnitude between $p_T$~=~0 and $p_T$~=~2~GeV/c. This gain
increases with $x_F$ and reaches more than two orders of magnitude
at $x_F$~=~0.5 (see also the discussion of two body decays in \cite{site:S8}).
In view of the sizeable experimental uncertainties still involved
with charm production in p+p interactions, the percentage scale
of the observed pattern should be taken as an indication rather
than a precise prediction. Scale variations of up to a factor of two 
are easily possible should more precise measurements become
available. It is the relative evolution with $x_F$ and $p_T$ which
is unavoidably involved with heavy flavour decay given the
precisely measured, large branching fractions into few body decays. 
Taking into account the rapid increase of the total charm cross 
section with $\sqrt{s}$ there is no doubt that heavy flavour decay 
will represent an important contribution to the total kaon yields at 
large $p_T$ and at large $x_F$ already in the ISR energy range. 

%
%
\subsection{Non-thermal behaviour of the decay products}
\vspace{3mm}
\label{sec:nonthermal}

\begin{figure}[h]
  \begin{center}
  	\includegraphics[width=5.5cm]{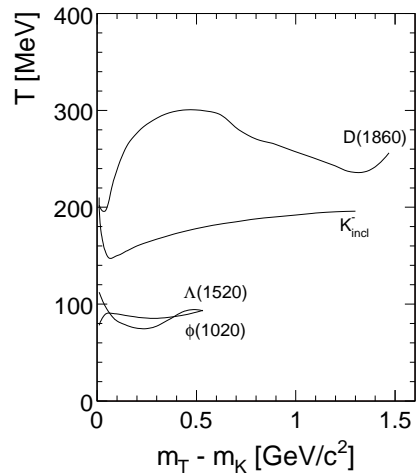}
  	\caption{Inverse slopes of K$^-$ from the decay of
             $\phi$(1020), $\Lambda$(1520) and D(1860) as a function of
             $m_T-m_K$. The result for inclusive K$^-$ production is also shown }
  	\label{fig:inverse_res}
  \end{center}
\end{figure}

Transverse mass distributions of the inclusively produced kaons
have been presented in Sect.~\ref{sec:rap} above. The inverse slopes
of both K$^+$ and K$^-$ show a strong variation with ($m_T-m_K$) from
about 150~MeV at low $m_T-m_K$ to 200~MeV at the upper limit of
$m_T-m_K$ available in this experiment. In this context it is
interesting to have a look at the inverse slope parameters of
the decay kaons from the $\phi$(1020), $\Lambda$(1520) and D(1860)
discussed above and shown in Fig.~\ref{fig:inverse_res} as a function of $m_T-m_K$.

Evidently the inverse slopes of K$^-$ from $\phi$(1020) and $\Lambda$(1520)
decay cluster around the low "temperature" values of 80--100~MeV,
whereas K$^-$ from charm decay show inverse slopes between 200 and
300~MeV. This is of course a result of the largely different $Q$ 
values of the respective decays convoluted with the sizeable
transverse momentum of the parent particles which gives them
a mean transverse velocity $\langle \beta_T \rangle \sim$~0.3--0.4.

In thermal models such anomalies are not a priori foreseen, as
all secondary hadrons are supposed to have Boltzmann-type 
distributions in $m_T$ with a unique inverse slope characteristic
of the hadronic reaction involved. In Hagedorn's thermodynamic
model for instance this "black body" radiation of hadrons
happens from "fireballs" which are not allowed to have transverse
momentum. Hagedorn \cite{hagedorn} has in fact realized that decay products 
are non-thermal in the above sense for the decays $\Delta$(1230) $\rightarrow$ N$\pi$
and $\rho$(770) $\rightarrow$ $\pi\pi$, albeit in a non-complete argumentation as
these parent resonances were in fact allowed transverse momentum 
and were taken as $\delta$-functions at their PDG mass values 
(see \cite{site:S8} concerning the importance of the proper resonance 
mass distribution). If, as argued above, the majority of final
state hadrons stem from the few-body decay of resonances which 
have important transverse degrees of freedom, the concept of a 
unique hadronic "temperature" in p+p interactions becomes an
artefact. The fact that this concept is not able to explain the 
evolution of particle yields towards high transverse momentum
and high $\sqrt{s}$, and, by the way, towards nuclear interactions
without the introduction of ad-hoc concepts like the Quark-Gluon
Plasma, see \cite{hagedorn}, has its origin in the same deficiency.

%
%
\section{Data summary}
\vspace{3mm}
\label{sec:over}
After the detailed discussion of charged and neutral kaon
yields in the preceding sections it is now mandatory to
summarize the obtained results and to compare them to
existing studies of global kaon production. The single
differential, $p_T$ integrated invariant cross sections
$F(x_F=0)$, see Eq.~\ref{eq:int}, and the total yields elaborated
in Sects.~\ref{sec:sdep} and \ref{sec:kzero} are listed in Table~\ref{tab:total} 
for K$^+$, K$^-$and K$^0_S$.

\begin{table}[h]
\scriptsize
\renewcommand{\tabcolsep}{0.3pc} 
\renewcommand{\arraystretch}{1.25} 
\begin{center}
\begin{tabular}{|c|c|c|c|c|c|c|c|}
\hline
   ref. & 
   $\sqrt{s}$ [GeV] & 
   $F_{\textrm{K}^+}(0)$ [mb] & 
   $F_{\textrm{K}^-}(0)$ [mb] & 
   $F_{\textrm{K}^0_S}(0)$ [mb] &
   $\langle n_{\textrm{K}^+} \rangle$ & 
   $\langle n_{\textrm{K}^-} \rangle$ & 
   $\langle n_{\textrm{K}^0_S} \rangle$ \\ \hline
   \cite{louttit}   
     &2.9   &       &       &         &0.00462  &         &0.00082  \\
   \cite{hogan,reed}  
     &2.9   &0.042  &       &         &0.00481  &         &         \\  
   \cite{alex}  
     &3.45  &       &       &         &0.00802  &         &0.00294  \\
   \cite{eisner}   
     &3.59  &       &       &         &         &         &0.00670  \\
   \cite{fire}   
     &4.04  &       &       &         &0.01760  &0.00080  &0.00719  \\
   \cite{blobel}   
     &4.9   &       &       &0.120    &         &         &0.0190   \\
   \cite{fesef}   
     &4.9   &       &       &         &0.0473   &0.00747  &0.0198   \\
   \cite{boggild}  
     &6.1   &       &       &0.185    &         &         &0.0420   \\
   \cite{blobel}   
     &6.8   &       &       &0.206    &         &         &0.0410   \\
   \cite{fesef}   
     &6.8   &       &       &         &0.0999   &0.0330   &0.0493   \\
   \cite{allaby1}   
     &6.84  &0.440  &0.120  &(0.280)  &0.107    &0.0262   &(0.0666) \\
   \cite{bogo}   
     &7.8   &       &       &0.300    &         &         &0.0636   \\
   \cite{abramov} 
    &11.5   &0.549  &0.322  &(0.435)  &         &         &         \\ 
   \cite{ammosov} 
    &11.5   &       &       &0.375    &         &         &0.109    \\
   \cite{alston}  
    &13.8   &       &       &         &         &         &0.121    \\
   \cite{chapman}  
    &13.9   &       &       &0.505    &         &         &0.146    \\
   \cite{brick}  
    &16.7   &       &       &0.490    &         &         &0.158    \\
  NA49  
    &17.2   &0.672  &0.477  &(0.575)  &0.227    &0.130    &(0.179)  \\
   \cite{jaeger2}  
    &19.7   &       &       &0.590    &         &         &0.181    \\
   \cite{albrow2,albrow3,albrow4,albrow5,capi,alper,guettler} 
    &23.0   &0.718  &0.547  &(0.633)  &0.273    &0.171    &(0.222)  \\
   \cite{sheng}  
    &23.8   &       &       &0.670    &         &         &0.224    \\
   \cite{lopinto}  
    &23.8   &       &       &         &         &         &0.212    \\
   \cite{bailly} 
    &25.7   &       &       &0.670    &         &         &0.262    \\
   \cite{kass}  
    &27.4   &       &       &         &         &         &0.200    \\
   \cite{kichimi}  
    &27.6   &       &       &0.680    &         &         &0.232    \\
   \cite{albrow2,albrow3,albrow4,albrow5,capi,alper,guettler}
    &31.0   &0.767  &0.614  &(0.691)  &0.327    &0.220    &(0.274)  \\
   \cite{albrow2,albrow3,albrow4,albrow5,capi,alper,guettler}
    &45.0   &0.861  &0.714  &(0.788)  &0.409    &0.290    &(0.350)  \\
   \cite{albrow2,albrow3,albrow4,albrow5,capi,alper,guettler}
    &52.0   &0.907  &0.766  &(0.837)  &0.448    &0.328    &(0.388)  \\
   \cite{albrow2,albrow3,albrow4,albrow5,capi,alper,guettler}
    &63.0   &0.959  &0.811  &(0.885)  &0.493    &0.363    &(0.428)  \\
   \cite{albrow2,albrow3,albrow4,albrow5,capi,alper,guettler}
   &200.0   &1.361  &1.192  &(1.277)  &0.819    &0.651    &(0.735)  \\
   \cite{ansorge1} 
   &200.0   &       &       &1.680    &         &         &0.700    \\
   \cite{ansorge1}
   &546.0   &       &       &2.306    &         &         &1.000    \\ \hline
\end{tabular}
\end{center}
\caption{Single differential, $p_T$ integrated cross sections
         $F(x_F=0)$ in mb and total yields for K$^+$, K$^-$ and K$^0_S$ for 31 values
         of $\sqrt{s}$. The values in brackets for K$^0_S$ are derived from the
         cross sections and yields for the charged kaons under the assumption
         $\langle n_{\textrm{K}^0_S} \rangle = 
         0.5(\langle n_{\textrm{K}^+} \rangle + \langle n_{\textrm{K}^-} \rangle)$ }
\label{tab:total}
\end{table}

A look at this Table shows that the K$^0_S$ yields present by far the
most dense and consistent coverage of the $\sqrt{s}$ scale from
threshold up to collider energies, as compared to the results for
charged kaons. This has already been evoked in Sect.~\ref{sec:kzero}, 
see Fig.~\ref{fig:k0_tot}.

%
%
\subsection{Total kaon yields}
\vspace{3mm}
\label{sec:over_tot}

\begin{figure}[h]
  \begin{center}
  	\includegraphics[width=9.5cm]{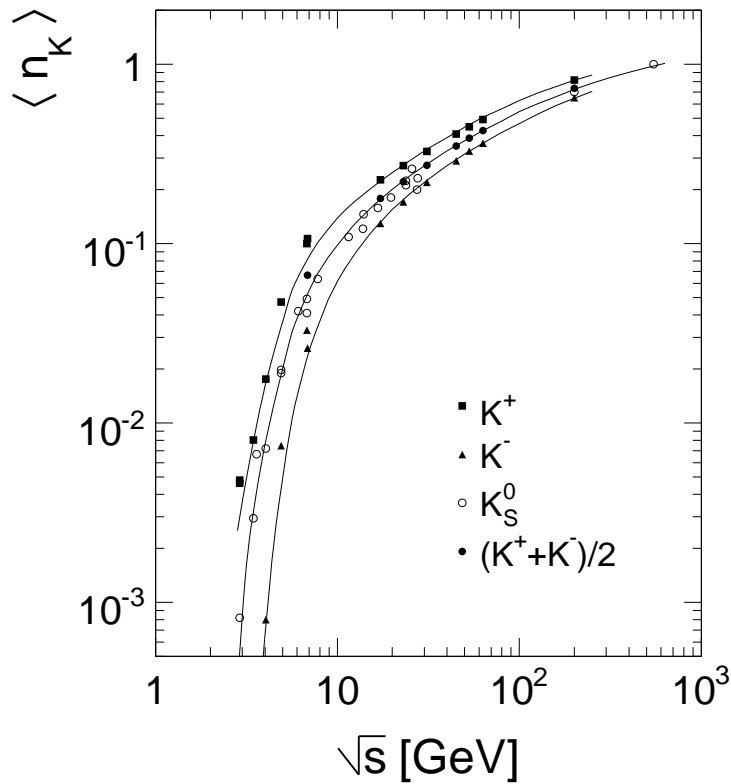}
  	\caption{Total yields $\langle n_{\textrm{K}^+} \rangle$, 
             $\langle n_{\textrm{K}^-} \rangle$ and $\langle n_{\textrm{K}^0_S} \rangle$ as a 
             function of $\sqrt{s}$. The full line through the K$^0_S$ results is an eyeball
             fit, the lines through the K$^+$ and K$^-$ data are derived from Fig.~\ref{fig:kaon_rat}
             below. The full circles in the K$^0_S$ data correspond to
             $0.5(\langle n_{\textrm{K}^+} \rangle + \langle n_{\textrm{K}^-} \rangle)$ 
             established at the corresponding $\sqrt{s}$ values}
  	\label{fig:total_kaon}
  \end{center}
\end{figure}

In a first attempt at establishing a consistent $s$-dependence
from these data, the total yields $\langle n_{\textrm{K}^+} \rangle$, 
$\langle n_{\textrm{K}^-} \rangle$ and $\langle n_{\textrm{K}^0_S} \rangle$ will
be treated. These quantities are shown in Fig.~\ref{fig:total_kaon}.

\begin{figure}[h]
  \begin{center}
  	\includegraphics[width=7cm]{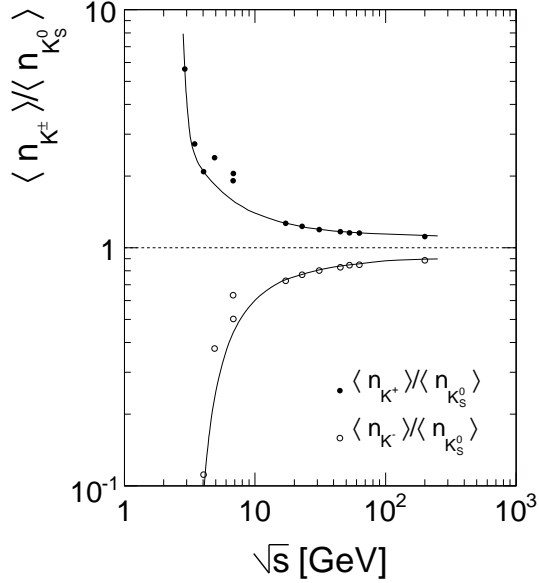}
  	\caption{Ratios 
             $\langle n_{\textrm{K}^+} \rangle$/$\langle n_{\textrm{K}^0_S} \rangle$ 
             and $\langle n_{\textrm{K}^-} \rangle$/$\langle n_{\textrm{K}^0_S} \rangle$ as a 
             function of $\sqrt{s}$. The full lines are eyeball fits through
             the data at $\sqrt{s} <$~4.9~GeV and $\sqrt{s} >$~6.8~GeV}
  	\label{fig:kaon_rat}
  \end{center}
\end{figure}

Whereas the coverage in the $\sqrt{s}$ scale is dense and continuous for 
$\langle n_{\textrm{K}^0_S} \rangle$, the corresponding data for the charged kaons show wide gaps 
in the range 7~$< \sqrt{s} <$~17~GeV and above $\sqrt{s}$~=~63~GeV. In this
upper energy range, the extrapolation from ISR to RHIC energy has been
evaluated in Sect.~\ref{sec:isr}. The situation towards lower energies is 
confounded by the fact that the available data at $\sqrt{s}$~=~4.9 and 
6.8~GeV are evidently doubtful by internal inconsistency, see 
Sect.~\ref{sec:ps-ags}. The following procedure has therefore been followed
to come to a consistent description of the $s$-dependence. In a first 
step an eyeball fit through the K$^0_S$ data is established over the
full $\sqrt{s}$ scale, see the full line in Fig.~\ref{fig:total_kaon}. This fit gives a 
consistent description of the situation within point-by-point 
variations of typically 10--20\%. In a second step the ratios 
$\langle n_{\textrm{K}^+} \rangle$/$\langle n_{\textrm{K}^0_S} \rangle$ and 
$\langle n_{\textrm{K}^-} \rangle$/$\langle n_{\textrm{K}^0_S} \rangle$ are 
obtained from the available data, see Table~\ref{tab:total}. These ratios are 
presented in Fig.~\ref{fig:kaon_rat}.

A smooth $\sqrt{s}$ dependence is imposed on these data points between
the low energy range $\sqrt{s} <$~4.8~GeV and the higher energies 
$\sqrt{s} >$~6.8~GeV/c using the fact that the cross sections in the 
PS energy range have been shown to deviate upwards (Sect.~\ref{sec:ps-ags}).
The $\langle n_{\textrm{K}^+} \rangle$/$\langle n_{\textrm{K}^0_S} \rangle$ and 
$\langle n_{\textrm{K}^-} \rangle$/$\langle n_{\textrm{K}^0_S} \rangle$ ratios 
thus obtained are then used to produce the smooth lines in Fig.~\ref{fig:total_kaon} 
through the K$^+$ and the K$^-$ data by multiplying with the K$^0_S$ interpolation.  

\begin{figure}[h]
  \begin{center}
  	\includegraphics[width=7cm]{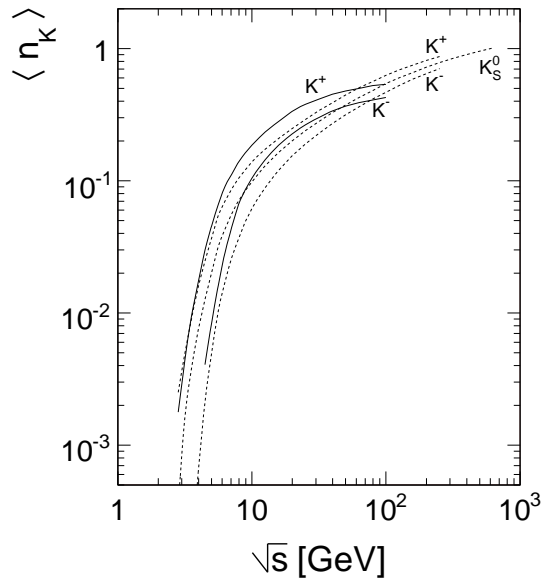}
  	\caption{$\langle n_{\textrm{K}^+} \rangle$ and $\langle n_{\textrm{K}^-} \rangle$ 
             from ref. \cite{rossi} as a function
             of $\sqrt{s}$, full lines, compared to the interpolated K$^+$, K$^-$
             and K$^0_S$ yields Fig.~\ref{fig:total_kaon} (broken lines)}
  	\label{fig:rossi_comp}
  \end{center}
\end{figure}

It is interesting to compare these interpolated results with the
global study of kaon yields by Rossi et al. \cite{rossi} which dates from  
1975 but is still widely used today \cite{kraus}. Their fits to the 
K$^+$ and K$^-$ data are shown in Fig.~\ref{fig:rossi_comp} in comparison with the present
results.

\begin{figure}[h]
  \begin{center}
  	\includegraphics[width=8cm]{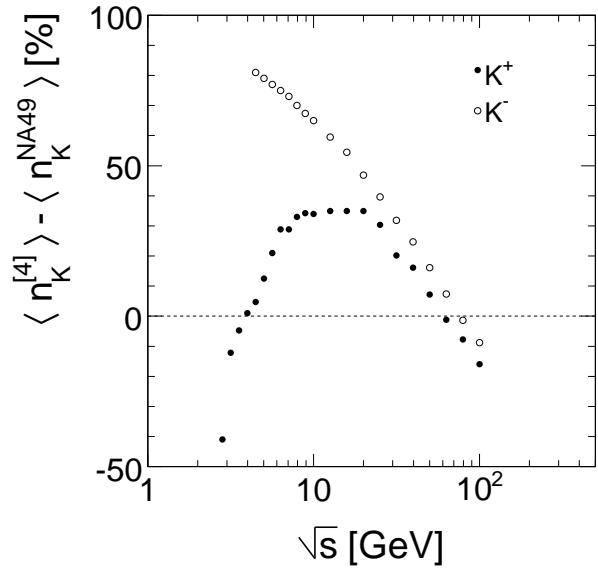}
  	\caption{Relative deviations between the present study
             of charged kaon yields and \cite{rossi} in percent as a function
             of $\sqrt{s}$}
  	\label{fig:rossi_dif}
  \end{center}
\end{figure}

There are evidently major deviations over the full $\sqrt{s}$ range.
Especially flagrant is the fact that the K$^-$ yields of \cite{rossi}
are above the K$^0_S$ multiplicities between $\sqrt{s}$~=~10 and 50~GeV.
This is clearly unphysical. The relative deviations between the two
attempts are shown in Fig.~\ref{fig:rossi_dif} on a percent scale.

These deviations are in the range from -40 to +80\% and demonstrate
which order of magnitude of systematic effects must be expected
if using existing parametrizations. The K$^-$ analysis of \cite{rossi} is
particularly questionable in the range 8~$< \sqrt{s} <$~12~GeV by using
proton-nucleus data from a Serpukhov experiment \cite{bushin} with protons 
on Aluminium in an $x_F$ range above 0.3 and for only 3 fixed lab
angles between 0 and 12~mrad. It remains a mystery, considering
the poor state of knowledge about proton-nucleus interactions and
their normalization, especially in the strange sector, how these 
data could be translated into total kaon yields in p+p collisions. 

%
%
\subsection{$\boldsymbol{p_T}$ integrated invariant cross sections at $\boldsymbol{x_F}$~=~0}
\vspace{3mm}
\label{sec:over_ptint}

\begin{figure}[h]
  \begin{center}
  	\includegraphics[width=9cm]{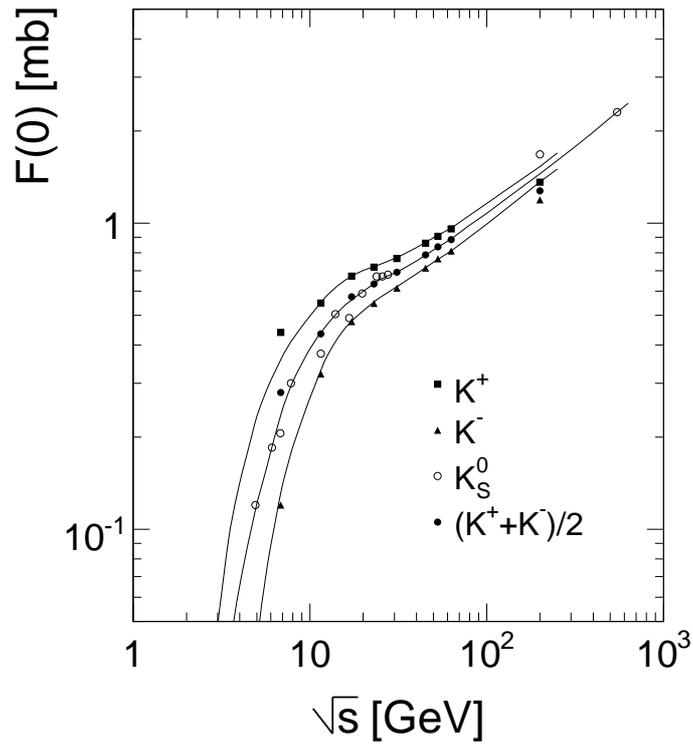}
  	\caption{$F(x_F=0,\sqrt{s})$ as a function of $\sqrt{s}$ for K$^0_S$
             (open circles: direct measurements, full circles:
             obtained as average from K$^+$ and K$^-$), K$^+$ (squares) and K$^-$
            (triangles)}
  	\label{fig:f_kaon}
  \end{center}
\end{figure}

As far as the $p_T$ integrated invariant cross sections at $x_F$~=~0,
$F(x_F=0)$ are concerned, the experimental situation is similar
to the one for the total kaon yields with the exception that 
only very scarce data below $\sqrt{s}$~=~6.8~GeV for charged kaons
and below 4.9~GeV for K$^0_S$ are available. Again the K$^0_S$ data
may be used as a reference in establishing a consistent $\sqrt{s}$
dependence as shown in Fig.~\ref{fig:f_kaon}.

\begin{figure}[h]
  \begin{center}
  	\includegraphics[width=7cm]{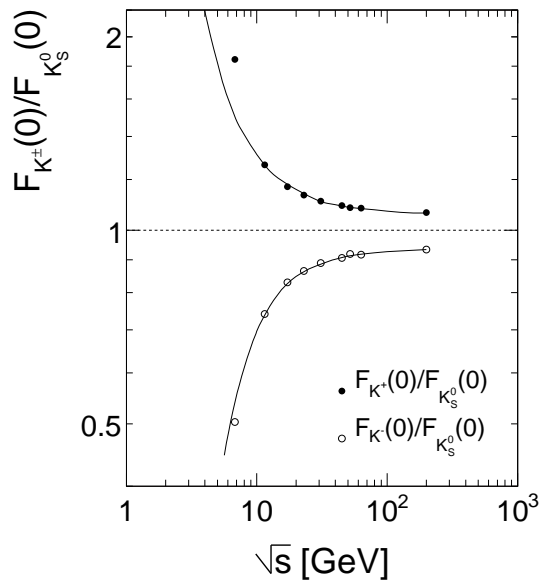}
  	\caption{Ratio between K$^+$ and interpolated K$^0_S$ data
             (full circles) and between K$^-$ and interpolated K$^0_S$ data
             (open circles) as a function of $\sqrt{s}$}
  	\label{fig:f_rat}
  \end{center}
\end{figure}

The full line through the K$^0_S$ data is an eyeball fit down to
$\sqrt{s}$~=~4.9~GeV. As the UA5 data \cite{ansorge1} and the ISR extrapolation
at $\sqrt{s}$~=~200~GeV show a 26\% difference, see Sect.~\ref{sec:coll_200},
the fit has been chosen to pass 13\% above the data extrapolation 
and 13\% below the UA5 data which allows a smooth continuation to
$\sqrt{s}$~=~540~GeV. In order to obtain a fit through the charged kaon
data, a reference to K$^0_S$ has been used by plotting the ratio
between the charged kaon and interpolated K$^0_S$ data as presented
in Fig.~\ref{fig:f_rat}.

These ratios may be smoothly connected by an eyeball fit between
$\sqrt{s}$~=~200~GeV (both K$^+$ and K$^-$ shifted upwards by 13\%) and
$\sqrt{s}$~=~11.5~GeV. As already visible for the case of total yields,
Fig.~\ref{fig:kaon_rat}, the data \cite{allaby1} at $\sqrt{s}$~=~6.8~GeV do not fall on the
smooth extrapolation below $\sqrt{s}$~=~11.5~GeV indicated in Fig.~\ref{fig:f_rat}, 
with a deviation of about 60\% for K$^+$ and 25\% for K$^-$.
This complies with the discussion of these data in Sect.~\ref{sec:ps-ags}.
In view of this the following attempt to nevertheless obtain
an approximate description of $F(x_F=0)$ at low energy has been
followed. The fit to the K$^0_S$ data together with the K$^+$/K$^0_S$
and K$^-$/K$^0_S$ ratios has been used to obtain the charged kaon
cross sections down to $\sqrt{s}$~=~4.9~GeV. The line through the
K$^+$ data has then been continued to the single measured K$^+$
cross section at $\sqrt{s}$~=~2.9~GeV. The fit for K$^-$ has been
discontinued at $\sqrt{s}$~=~4.9~GeV. This admittedly rather daring
procedure produces nevertheless a consistent overall picture
with systematic errors below $\sqrt{s} \sim$~5~GeV on the level of
about 20\%. Only new precision measurements in this energy
region may help to improve on this unsatisfactory situation.

%
%
\subsection{K$^+$/K$^-$ ratios}
\vspace{3mm}
\label{sec:over_ratio}

\begin{figure}[h]
  \begin{center}
  	\includegraphics[width=7cm]{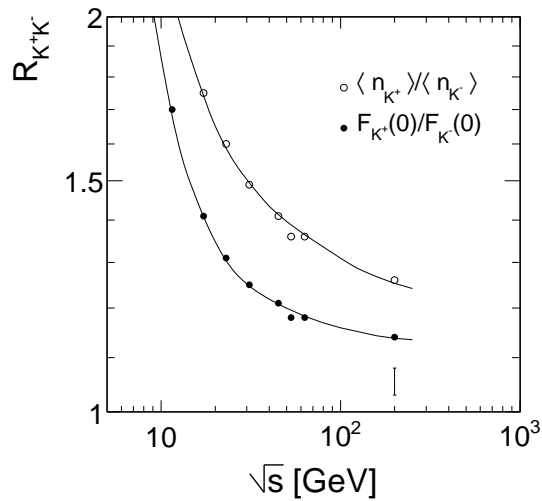}
  	\caption{$R_{\textrm{K}^+\textrm{K}^-}$ ratios for total yields (open circles)
             and $F(x_F=0)$ (closed circles) as a function of $\sqrt{s}$. The error bar 
             at $\sqrt{s}$~=~200~GeV marks the range of $F_{\textrm{K}^+}(0)/F_{\textrm{K}^-}(0)$
             of published results by the RHIC experiments}
  	\label{fig:tot_p2m}
  \end{center}
\end{figure}

The data interpolation described above allows also an overview
of the K$^+$/K$^-$ ratios both for the total yields and for the
$p_T$ integrated cross sections at $x_F$~=~0. This is shown in Fig.~\ref{fig:tot_p2m},
where the full lines refer to the eyeball fits in Figs.~\ref{fig:total_kaon}
and \ref{fig:f_kaon} and the data points to Table~\ref{tab:total}.

One feature of these dependences is the rather slow approach to
K$^+$/K$^-$~=~1 for $F_{\textrm{K}^+}(0)$/$F_{\textrm{K}^-}(0)$ with increasing 
$\sqrt{s}$. For $\sqrt{s}$~=~200~GeV the ratio
of the $p_T$ integrated cross sections at $x_F$~=~0 is 1.14 for the
extrapolated ISR data as compared to a range of values between 
1.03 and 1.08 published by the RHIC experiments 
\cite{star3,phenix1,phenix2,brahms1,brahms2,brahms3}. The
reason for this further discrepancy might be again the low
trigger cross sections at RHIC with respect to the total inelastic 
cross section which enhances central collisions and hence kaon 
ratios closer to unity.  

%
%
\section{Conclusions}
\vspace{3mm}
\label{sec:concl}

The new data on the inclusive production of charged kaons 
in p+p interactions at SPS energy presented here complete 
a detailed study on charged secondary hadrons including
pions and baryons in the framework of the NA49 experiment
at the CERN SPS. These data offer the possibility to
check the sum rules of mean charged multiplicity and charge
conservation. It is demonstrated that both constraints are
fulfilled within the tight error limits of about 2\%
as they have been quoted for the systematic uncertainties
in the independent evaluation of inclusive cross sections
for the different particle types.

The extended coverage of the data in the $x_F$/$p_T$ plane, from
$x_F$~=~0 to 0.5 and from $p_T$~=~0.05 to 1.7~GeV/c, allows a precise
study of particle ratios from the same experiment, both for
K/$\pi$ and K/baryon ratios. A detailed comparison to existing 
data in the SPS/Fermilab energy range shows in general good
agreement in the limited phase space regions available, in
particular also for the complete set of particle ratios,
with some exceptions essentially due to normalization
problems.

As the interaction energy of $\sqrt{s}$~=~17.2~GeV is located at a
strategical point between threshold-dominated and scaling
phenomena at lower and higher energies, respectively, a new
and complete study of the $s$-dependence of kaon production
including K$^0_S$ has been attempted using the new NA49 data
as a reference. This study covers the energy range 
3~$< \sqrt{s} <$~1800~GeV and aims at establishing an internally 
consistent picture of kaon production as far as this is 
possible with the often restricted and contradictory
available data. Throughout, the use of data parametrization
with simple arithmetic formulations has been avoided in
order to take the rather complex dependence of the 
measured particle yields on the kinematic variables
fully into account.

This study reveals basic weaknesses in the existing data
base both at lower and higher $\sqrt{s}$. At PS/AGS energies
the charged kaon data suffer from large systematic 
inconsistencies, and the almost complete absence of 
differential data at Serpukhov energies renders the establishment 
of integrated yields hazardous to say the least. The extension 
into the ISR energy range on the other hand, using all available 
data, gives new insights into the complex evolution of 
strangeness yields as functions of $x_F$ and, in particular,
$p_T$ resulting in explicitly non-thermal transverse distributions.
These findings are discussed in connection with some typical
examples of resonance production and decay which are relevant
to this phenomenology.

In addition to the charged kaon cross sections, it has been
found useful and necessary to also look at the evolution of
the K$^0_S$ yields. In fact the relation 
$\textrm{K}^0_S = 0.5(\textrm{K}^+ + \textrm{K}^-)$ 
which is found valid within the experimental precision at least in
the energy range $\sqrt{s} >$~5~GeV provides a strong
constraint on the overall data consistency. In this context
the early bubble chamber work up to $\sqrt{s} \sim$~28~GeV proves to
be essential due to its internal consistency and its superior
precision concerning the overall normalization. 

As far as the extension of the study to the RHIC and p+$\overline{\textrm{p}}$
collider energies is concerned, a rather disturbing overall
picture emerges. Evidently, the published results do not
represent a decisive improvement as far as precision and 
internal consistency are concerned in comparison to the
lower energy data which in most cases date back by more 
than 30 years. There are several reasons for this situation:

\begin{itemize}
  \item The study of soft hadronic production in elementary collisions
        is certainly not at the heart of the experimental programs at 
        collider energies. On the contrary it is the discovery 
        potential for "new"  physics either in Heavy Ion interactions 
        (RHIC) or within and beyond the Standard Model 
        (p+$\overline{\textrm{p}}$ colliders) 
        which defines the main priorities.
  \item Precision studies of elementary hadronic production call for
        specific constraints both concerning accelerator layout
        and operation (as for instance vertex distributions and 
        stability) and the experimental set-ups (trigger efficiency, 
        data normalization, material budget at low total momenta).
  \item The large multi-purpose detectors generally set up at the 
        colliders are not really optimized for these constraints
        and small-size, dedicated experiments as they have been
        used in practically all the preceding lower energy work, 
        are neither available nor planned.
\end{itemize}

In view of these problems in a first step the extrapolation
of the ISR results to $\sqrt{s}$~=~200 GeV has been attempted in 
order to obtain a common point of comparison between ISR extension,
RHIC and the lowest available p+$\overline{\textrm{p}}$ collider energy. At this energy
the UA5 streamer chamber data turn out to offer a reliable
reference although the overall statistical errors are sizeable.
This is reminiscent of the bubble chamber work at lower energies
which definitely benefits from the application of optical methods 
in terms of reconstruction efficiency and normalization.
In contrast all RHIC data show large systematic offsets and a
general weakness towards the lower cut-off in $p_T$ which lies
in general in the region 0.4--0.7~GeV/c.

This new study of the $s$-dependence of charged and neutral kaon
production results in smoothed interpolations of the central, $p_T$ 
integrated invariant yields $F(x_F=0,\sqrt{s})$ and of the total
kaon multiplicities as they are presented in Sect.~\ref{sec:over} of this
paper.

\section*{Acknowledgements}
\vspace{3mm}
This work was supported by
the Polish State Committee for Scientific Research (P03B00630),
the Polish Ministry of Science and Higher Education (N N202 078735),
the Bulgarian National Science Fund (Ph-09/05),
the EU FP6 HRM Marie Curie Intra-European Fellowship Program,
the Hungarian Scientific Research Fund OTKA (T68506) and
the Hungarian OTKA/NKTH A08-77719 and A08-77815 grants.

\vspace{2cm}


\begin{thebibliography}{[20]}
\bibitem{pp_pion} C.~Alt et al., Eur. Phys. J. {\bf C45} (2006) 343
\bibitem{pp_proton} T.~Anticic et al., Eur. Phys. J. {\bf C65} (2010) 9
\bibitem{nim} S.~Afanasiev et al., Nucl. Instrum. Meth. {\bf A430} (1999) 210
\bibitem{rossi} A.~M.~Rossi et al., Nucl. Phys. {\bf B84} (1975) 269
\bibitem{hogan} W.~J.~Hogan, P.~A.~Pirou\'e and A.~J.~S.~Smith, Phys. Rev. {\bf 166} (1968) 1472
\bibitem{reed} J.~T.~Reed et al., Phys. Rev. {\bf 168} (1968) 1495
\bibitem{allaby1} J.~V.~Allaby et al., CERN 70-12 (1970)
\bibitem{allaby2} U.~Amaldi et al., Nucl. Phys. {\bf B86} (1975) 403 
\bibitem{akerlof} C.~Akerlof et al., Phys. Rev. {\bf D3} (1971) 645
\bibitem{dekkers} D.~Dekkers et al., Phys. Rev. {\bf 137} (1965) B 962
\bibitem{abramov} V.~V.~Abramov et al., Nucl. Phys. {\bf B173} (1980) 348
\bibitem{cronin} D.~Antreasyan et al., Phys. Rev. {\bf D19} (1979) 764
\bibitem{brenner} A.~E.~Brenner et al., Phys. Rev. {\bf D26} (1982) 1497
\bibitem{johnson} J.~R.~Johnson et al., Phys. Rev. {\bf D17} (1978) 1292
\bibitem{albrow1} M.~G.~Albrow et al., Nucl. Phys. {\bf B56} (1973) 333
\bibitem{albrow2} M.~G.~Albrow et al., Phys. Lett. {\bf B42} (1972) 279
\bibitem{albrow3} M.~G.~Albrow et al., Nucl. Phys. {\bf B140} (1978) 189
\bibitem{albrow4} M.~G.~Albrow et al., Nucl. Phys. {\bf B51} (1973) 388
\bibitem{albrow5} M.~G.~Albrow et al., Nucl. Phys. {\bf B73} (1974) 40
\bibitem{capi} P.~Capiluppi et al., Nucl. Phys. {\bf B70} (1974) 1 \\
               P.~Capiluppi et al., Nucl. Phys. {\bf B79} (1974) 189
\bibitem{alper} B.~Alper et al., Nucl. Phys. {\bf B100} (1975) 237
\bibitem{guettler} K.~Guettler et al., Nucl. Phys {\bf B116} (1976) 77
\bibitem{star1} B.~I.~Abelev et al., Phys. Rev. {\bf C79} (2009) 034909
\bibitem{star2} B.~I.~Abelev et al., Phys. Rev. {\bf C75} (2007) 064901
\bibitem{star3} J.~Adams et al., Phys. Lett. {\bf B616} (2005) 8
\bibitem{phenix1} S.~S.~Adler et al., Phys. Rev. {\bf C74} (2006) 024904
\bibitem{phenix2} T.~Chujo, Talk at Quark Matter 2006, J. Phys. {\bf G34} (2007) S893
\bibitem{brahms1} H.~Yang, J. Phys. {\bf G32} (2006) S491 
\bibitem{brahms2} H.~Yang, J. Phys. {\bf G34} (2007) S619 
\bibitem{brahms3} I.~Arsene et al., Phys. Rev. Lett. {\bf 98} (2007) 252001
\bibitem{pai} W.~W.~M.~Allison, J.~H.~Cobb, Ann. Rev. Nucl. Part. Sci. 30:253-298 (1980) \\
              J.~Berkowitz, Atomic and molecular photoabsorption, 1: Absolute total cross section,
              Academic press (2002)
\bibitem{pc_pion} C.~Alt et al., Eur. Phys. J. {\bf C49} (2007) 897
\bibitem{site} http://cern.ch/spshadrons
\bibitem{site:S8} http://cern.ch/spshadrons, document S8 \\
                  A.~Rybicki et al., Int. J. Mod. Phys. {\bf A24} (2009) 385
\bibitem{ehs} M.~Aguilar-Benitez et al., Z. Phys. {\bf C50} (1991) 405
\bibitem{ansorge} R.~E.~Ansorge et al., Nucl. Phys. {\bf B103} (1976) 509
\bibitem{alex} G.~Alexander et al., Phys. Rev. {\bf 154} (1967) 1284
\bibitem{bierman} E.~Bierman et al., Phys. Rev. {\bf 147} (1966) 922
\bibitem{jaeger1} K.~Jaeger et al., Phys. Rev. {\bf D11} (1975) 1756
\bibitem{ammosov} V.~V.~Ammosov et al., Nucl. Phys. {\bf B115} (1976) 269
\bibitem{fesef} H.~Fesefeldt et al., Nucl. Phys. {\bf B147} (1979) 317
\bibitem{alpgard1} K.~Alpg{\aa}rd et al., Nucl. Phys. {\bf B103} (1976) 234
\bibitem{kichimi} H.~Kichimi et al., Phys. Rev. {\bf D20} (1979) 37 \\
                  T.~Okusawa et al., Europhys. Lett. {\bf 5} (1988) 509 
\bibitem{eisner} R.~L.~Eisner et al., Nucl. Phys. {\bf B123} (1977) 361
\bibitem{oh} B.~Y.~Oh et al., Nucl. Phys. {\bf B49} (1972) 13
\bibitem{blobel} V.~Blobel et al., Nucl. Phys. {\bf B69} (1974) 454
\bibitem{boggild} H.~B{\o}ggild et al., Nucl. Phys. {\bf B57} (1973) 77 
\bibitem{bartke} J.~Bartke et al., Nuovo Cim. {\bf 29} (1963) 8
\bibitem{bogo} M.~Yu.~Bogolyubsky et al., Sov. J. Nucl. Phys. {\bf 50} (1989) 424, Yad. Fiz. {\bf 50} (1989) 683
\bibitem{alston} M.~Alston-Garnjost et al., Phys. Rev. Lett. {\bf 35} (1975) 142
\bibitem{chapman} J.~W.~Chapman et al., Phys. Lett. {\bf B47} (1973) 465
\bibitem{brick} D.~Brick et al., Nucl. Phys. {\bf B164} (1980) 1
\bibitem{allday} J.~Allday et al., Z. Phys. {\bf C40} (1988) 29
\bibitem{jaeger2} K.~Jaeger et al., Phys. Rev. {\bf D11} (1975) 2405
\bibitem{sheng} A.~Sheng et al., Phys. Rev. {\bf D11} (1975) 1733
\bibitem{lopinto} F.~LoPinto et al., Phys. Rev. {\bf D22} (1980) 573
\bibitem{dao} F.~T.~Dao et al., Phys. Rev. Lett. {\bf 30} (1973) 1151
\bibitem{bailly} J.~L.~Bailly et al., Z. Phys. {\bf C31} (1986) 367
\bibitem{kass} R.~D.~Kass et al., Phys. Rev. {\bf D20} (1979) 605
\bibitem{wroblewski} A.~Wr\'oblewski, Acta Phys. Pol. {\bf B15} (1984) 785
\bibitem{whitmore} J.~Whitmore, Phys. Rep. {\bf 10} (1974) 273
\bibitem{na61} Proposal, CERN-SPSC-P-330 (2006)
\bibitem{louttit} R.~I.~Louttit et al., Phys. Rev. {\bf 123} (1961) 1465
\bibitem{fire} M.~Firebaugh et al., Phys. Rev. {\bf 172} (1968) 1354
\bibitem{landolt} A.~N.~Diddens and K.~Schl\"upmann, Landolt-B\"ornstein -- Group I, {\bf Vol. 6} (1972) 79--164
\bibitem{zabrodin} E.~E.~Zabrodin et al.,Phys. Rev. {\bf D52} (1995) 1316
\bibitem{alpgard} K.~Alpg{\aa}rd et al., Phys. Lett. {\bf B115} (1982) 65 
\bibitem{alner1} G.~J.~Alner et al., Nucl. Phys. {\bf B258} (1985) 505
\bibitem{ansorge1} R.~E.~Ansorge et al., Phys. Lett. {\bf B199} (1987) 311
\bibitem{alner2} G.~J.~Alner et al., Phys. Rept. {\bf 154} (1987) 247
\bibitem{ansorge2} R.~E.~Ansorge et al., Z. Phys. {\bf C41} (1988) 179
\bibitem{ua2} M.~Banner et al., Phys. Lett. {\bf B122} (1983) 322
\bibitem{ua1} G.~Bocquet et al., Phys. Lett. {\bf B366} (1996) 441 
\bibitem{cdf1} F.~Abe et al., Phys. Rev. {\bf D40} (1989) 3791
\bibitem{cdf2} D.~Acosta et al., Phys. Rev. {\bf D72} (2005) 052001
\bibitem{e735} T.~Alexopoulos et al., Phys. Rev. {\bf D48} (1993) 984
\bibitem{grassler} H.~Gr\"assler et al., Nucl. Phys. {\bf B132} (1978) 1 \\
                   G.~Jancso et al., Nucl. Phys. {\bf B124} (1977) 1
\bibitem{daum} C.~Daum et al., Phys. Lett. {\bf B98} (1981) 313
\bibitem{afanasiev} S.~V.~Afanasiev et al., Phys. Lett. {\bf B491} (2000) 59 \\
                    T.~Sammer, PhD thesis MPI Munich(2000)
\bibitem{booth} P.~S.~L.~Booth et al., Nucl. Phys. {\bf B273} (1986) 677 \\
                A.~Etkin et al., Phys. Lett. {\bf B201} (1988) 568
\bibitem{bobbink} G.~J.~Bobbink et al., Nucl. Phys. {\bf B217} (1983) 11
\bibitem{aguilar} M.~Aguilar-Benitez et al., Z. Phys. {\bf C40} (1988) 321
\bibitem{hgf} H.~G.~Fischer and W.~Geist, Z. Phys. {\bf C19} (1983) 159
\bibitem{hagedorn} R.~Hagedorn, Riv. Nuovo Cim. {\bf vol. 6}, {\bf n. 10} (1983) 1
\bibitem{kraus} I.~Kraus et al., arXiv:0910.3125v1 (2009)
\bibitem{bushin} Y.~B.~Bushin et al., Phys. Lett. {\bf B29} (1969) 48 \\
                 F.~Binon et al., Phys. Lett. {\bf B30} (1969) 506
\end{thebibliography}
\end{document}